PRESSURE BROADENING AND PRESSURE SHIFT

OF DIATOMIC IODINE AT 675 NM

by

ERICH N. WOLF

A DISSERTATION

Presented to the Department of Chemistry
and the Graduate School of the University of Oregon
in partial fulfillment of the requirements
for the degree of
Doctor of Philosophy

June 2009

"Pressure Broadening and Pressure Shift of Diatomic Iodine at 675 nm," a dissertation prepared by Erich N. Wolf in partial fulfillment of the requirements for the Doctor of Philosophy degree in the Department of Chemistry. This dissertation has been approved and accepted by:

_______________________________________________________

Dr. David Herrick, Chair of the Examining Committee

_______________________________________

Date

Committee in Charge:     Dr. David R. Herrick, Chair
                         Dr. John L. Hardwick
                         Dr. Michael G. Raymer
                         Dr. Jeffrey A. Cina
                         Dr. David R. Tyler

Accepted by:

_______________________________________________________

Dean of the Graduate School

An Abstract of the Dissertation of

Erich N. Wolf for the degree of Doctor of Philosophy

in the Department of Chemistry to be taken June 2009

Title: PRESSURE BROADENING AND PRESSURE SHIFT OF DIATOMIC

IODINE AT 675 NM

Approved: _______________________________________________

Dr. David Herrick, Chair of the Examining Committee

Doppler-limited, steady-state, linear absorption spectra of $^{127}I_2$ (diatomic iodine) near 675 nm were recorded with an internally-referenced wavelength modulation spectrometer, built around a free-running diode laser using phase-sensitive detection, and capable of exceeding the signal-to-noise limit imposed by the 12-bit data acquisition system. Observed $I_2$ lines were accounted for by published spectroscopic constants.

Pressure broadening and pressure shift coefficients were determined respectively from the line-widths and line-center shifts as a function of buffer gas pressure, which were determined from nonlinear regression analysis of observed line shapes against a Gaussian-Lorentzian convolution line shape model. This model included a linear superposition of the $I_2$ hyperfine structure based on changes in the nuclear electric quadrupole coupling constant. Room temperature (292 K) values of these coefficients were determined for six unblended $I_2$ lines in the region 14,817.95 to 14,819.45 cm$^{-1}$ for each of the following buffer gases: the atoms He, Ne, Ar, Kr, and Xe; and the molecules $H_2$, $D_2$, $N_2$, $CO_2$, $N_2O$, air, and $H_2O$. These coefficients were also determined at one additional temperature (388 K) for He and $CO_2$, and at two additional temperatures (348 and 388 K) for Ar. Elastic collision cross-sections were determined for all pressure broadening coefficients in this region. Room temperature values of these coefficients were also determined for several low-$J$ $I_2$ lines in the region 14,946.17 to 14,850.29 $cm^{-1}$ for Ar.

A line shape model, obtained from a first-order perturbation solution of the time-dependent Schrödinger equation for randomly occurring interactions between a two-level system and a buffer gas treated as step-function potentials, reveals a relationship between the ratio of pressure broadening to pressure shift coefficients and a change in the wave function phase-factor, interpreted as reflecting the "cause and effect" of state-changing events in the microscopic domain. Collision cross-sections determined from this model are interpreted as reflecting the inelastic nature of collision-induced state-changing events.

A steady-state kinetic model for the two-level system compatible with the Beer-Lambert law reveals thermodynamic constraints on the ensemble-average state-changing rates and collision cross-sections, and leads to the proposal of a relationship between observed asymmetric line shapes and irreversibility in the microscopic domain.

(The graduate school at the University of Oregon imposes a word limit [350 words] on dissertation abstracts so that the last paragraph of the abstract did not include a passage akin to the following: A modified version of the Einstein *A* and *B* Coefficient model appropriate for linear absorption is offered [Section 2.11], one that is based on the mathematical operation of convolutions and allows for a derivation of the Beer-Lambert Law for the state-changing processes in the two-level model [stimulated absorption, stimulated emission, and spontaneous emission] for the case of steady-state dynamics in the microscopic domain. This model indicates that the rates for stimulated absorption and stimulated emission are not necessarily equal, which is tantamount to the *B* coefficients for stimulated absorption and stimulated emission not necessarily being equal. Furthermore, this model indicates that the collision cross-sections for stimulated absorption and stimulated emission are not equal. As well, with regard to the well documented appearance of asymmetric line shapes in high-resolution linear absorption spectra, it would have been mentioned that such features may be more fully consistent with the reality that the universe we live in [i.e. that there is no such thing as a perfectly isolated object] is more properly described by non-Hermitian Hamiltonians; in the context of the [non-degenerate] two-level model, the Hamiltonian that describes photon absorption is not the Hermitian conjugate of the Hamiltonian that describes photon emission.

Apart from the foregoing addendum to the abstract, the inclusion of a footer on all pages of the dissertation indicating the author's name, dissertation title, and year of submission of the dissertation, and inclusion of publications in the field of organometallic chemistry, this version of the dissertation is the one that was accepted by the graduate school at the University of Oregon.)

CURRICULUM VITAE

NAME OF AUTHOR:  Erich N. Wolf

GRADUATE AND UNDERGRADUATE SCHOOLS ATTENDED:

>University of Oregon
>California State University, Northridge

DEGREES AWARDED:

>Doctor of Philosophy in Chemistry, 2009, University of Oregon
>Bachelor of Science in Physics, 1993, California State University, Northridge

AREAS OF SPECIAL INTEREST:

>Chemical Physics, Quantum Electrodynamics, Statistical Physics,
>    Thermodynamics, Solid State Physics, Classical Physics,
>    Astrophysics and Cosmology, and Mathematics
>Atomic and Molecular Spectroscopy
>Inorganic and Organometallic Chemistry and Catalysis
>Materials and Properties of Materials

PROFESSIONAL EXPERIENCE:

>Research Assistant, Department of Chemistry, University of Oregon, Eugene,
>    June 2001 – September 2007

>Teaching Assistant, Undergraduate Physical Chemistry Laboratory,
>    Department of Chemistry, University of Oregon, Eugene,
>    September 2001 – June 2002, September 2003 – June 2004,
>    and September 2005 – June 2006

>Research Assistant, Department of Chemistry, California State University,
>    Northridge, March 1988 – June 1992

PUBLICATIONS:

Spectroscopy:

Arteaga, S. W., C. M. Bejger, J. L. Gerecke, J. L. Hardwick, Z. T. Martin, J. Mayo, E. A. McIlhattan, J.-M. F. Moreau, M. J. Pilkenton, M. J. Polston, B. T. Robertson, and E. N. Wolf. "Line Broadening and Shift Coefficients of Acetylene at 1550 nm." *Journal of Molecular Spectroscopy* 243 (2007): 253-266.

Hardwick, J. L., Z. T. Martin, M. J. Pilkenton, and E. N. Wolf. "Diode Laser Absorption Spectra of $H^{12}C^{13}CD$ and $H^{13}C^{12}CD$ at 6500 $cm^{-1}$." *Journal of Molecular Spectroscopy* 243 (2007): 10-15.

Hardwick, J. L., Z. T. Martin, E. A. Schoene, V. Tyng, and E. N. Wolf. "Diode Laser Absorption Spectrum of Cold Bands of $C_2HD$ at 6500 $cm^{-1}$." *Journal of Molecular Spectroscopy* 239 (2006): 208-215.

Eng, J. A., J. L. Hardwick, J. A. Raasch, and E. N. Wolf. "Diode Laser Wavelength Modulated Spectroscopy of $I_2$ at 675 nm." *Spectrochimica Acta Part A* 60 (2004): 3413-3419.

Organometallic Chemistry:

Rosenberg E., S. E. Kabir, L. Milone, R. Gobetto, D. Osella, M. Ravera, T. McPhillips, M. W. Day, D. Carlot, S. Hajela, E. Wolf, K. I. Hardcastle. "Comparative Reactivity of Triruthenium and Triosmium μ 3-η 2-Imidoyls. 2. Reactions with Alkynes." Organometallics 16 (1997): 2674-2681.

Rosenberg E., L. Milone, R. Gobetto, D. Osella, K. I. Hardcastle, S. Hajela, K. Moizeau, M. Day, E. Wolf, D. Espitia. "Comparative Reactivity of Triruthenium and Triosmium μ 3-η 2-Imidoyls. 1. Dynamics and Reactions with Carbon Monoxide, Phosphine, and Isocyanide." Organometallics 16 (1997): 2665-2673.

Rosenberg E., S. E. Kabir, M. Day, K. I. Hardcastle, E. Wolf, T. McPhillips. "Chemistry of Nitrogen Donors with μ 3-Imidoyl Triosmium Clusters: Dynamics of a Monometallic Site in a Trimetallic Cluster." Organometallics 14 (1995): 721-733.

Rosenberg E., M. Day, W. Freeman, K. I. Hardcastle, M. Isomaki, S. E. Kabir, T. McPhillips, L. G. Scott, E. Wolf; "Comparative study of the reactions of diazomethane with $\mu$ 3-imidoyl and $\mu$ 3-butyne trinuclear clusters", Organometallics 11 (1992): 3376-3384.

Rosenberg E., M. W. Day, S. Hajela, S. E. Kabir, M. Irving, T. McPhillips, E. Wolf, K. I. Hardcastle, L. Milone, et al.; "Reactions of tertiary amines with trinuclear clusters. 3. Reactions of N-methylpyrrolidine with $Ru_3(CO)_{12}$ and $Os_3(CO)_{10}(CH3CN)_2$." Organometallics 10 (1991): 2743-2751.

Rosenberg E., S. E. Kabir, K. I. Hardcastle, M. Day, E. Wolf; "Reactions of secondary amines with triosmium decacarbonyl bis(acetonitrile): room-temperature carbon-hydrogen bond activation and transalkylation." Organometallics 9 (1990): 2214-2217.

## ACKNOWLEDGMENTS

- Thomas Dyke for research opportunity, academic guidance, and financial assistance.

- John Hardwick for research opportunity and academic guidance.

- David Herrick, Jeffrey Cina and Michael Raymer for academic guidance.

- Finally, and most importantly, Vivian Ding, without whom this dissertation would not be possible.

x

I bow to all the high and holy lamas.

# TABLE OF CONTENTS

LIST OF FIGURES

LIST OF TABLES

CHAPTER I

INTRODUCTION 1 – CONTEXT AND OVERVIEW

## 1.1 Overview of Chapter I

This chapter begins with a brief historical review of linear absorption spectroscopy, which seeks to provide a context for the measurement of pressure broadening and pressure shift coefficients. A brief overview of the experimental methods of linear absorption spectroscopy and wavelength-modulated linear absorption spectroscopy are presented. The role of such methods in training the next generation of scientists is briefly explored. The last section on error propagation is essential in the attempt to make this manuscript more complete and self-contained.

## 1.2 Historical Overview of Absorption Spectroscopy

In the mid-17$^{th}$ century, Newton constructed a low resolution spectrometer using a glass prism, which he then used to investigate the properties of the light emitted by the sun (Figure 1.1). Passing sunlight through a single prism, Newton observed the (spatial) dispersion of white light into an ordered band of colors, the same progression observed in rainbows for time immemorial. Newton also reported on the reversibility (at the macroscopic level) of this dispersion process by using a second prism to recombine the light dispersed by the first prism to form a beam of white light.

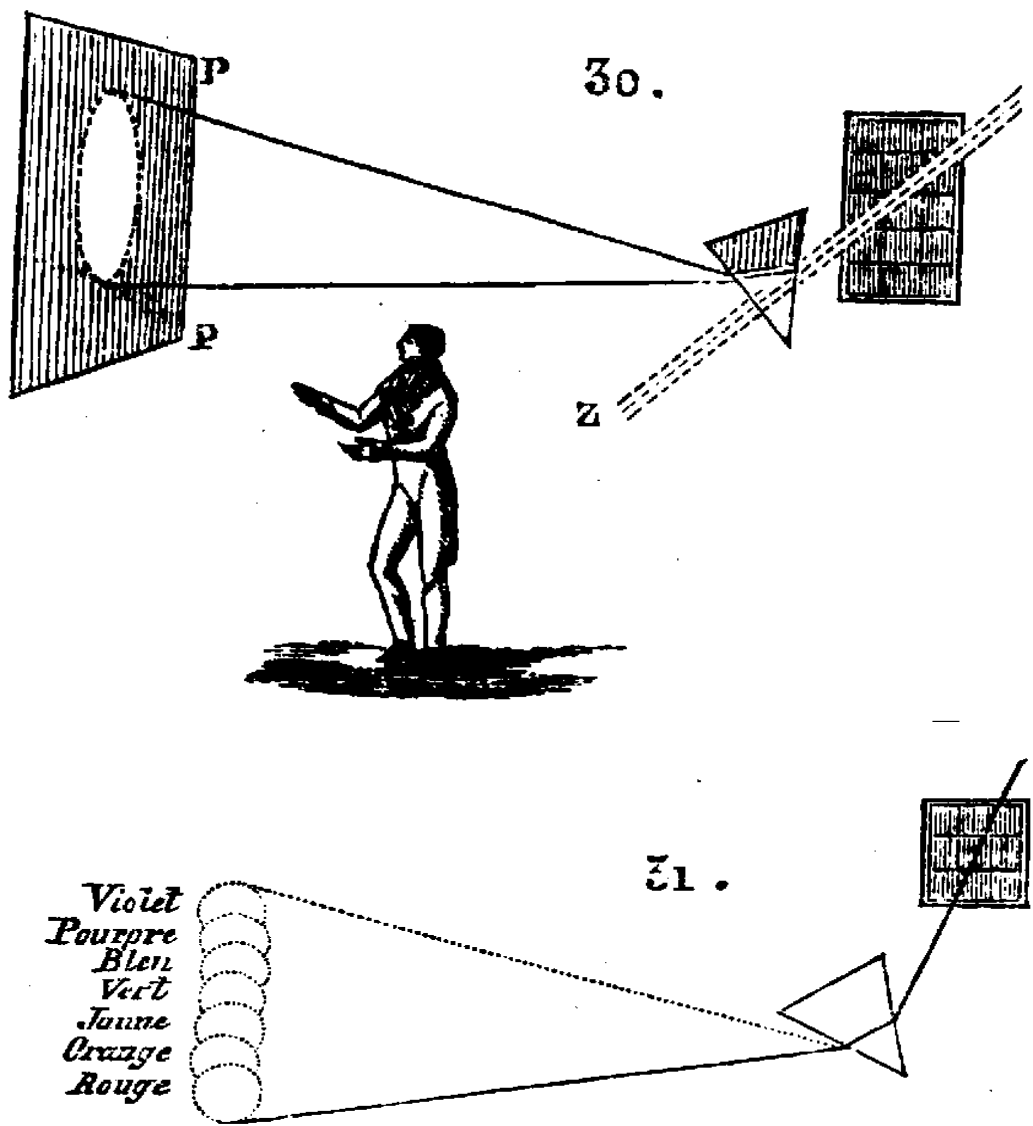

Figure 1.1 Depiction of Newton's experiments of dispersing (decomposing) sunlight into its constituent colors using a glass prism. (Digitized images from Voltaire's *Elements de la Philosophie de Newton,* published in 1738.)

Roughly 150 years later the design and construction of spectrometers had improved in (wavelength) resolution and (detection) sensitivity so that discrete absorption features (a.k.a. lines) in the spectrum of the sun were revealed (Figure 1.2). The introduction of the diffraction grating for light dispersion was the source of these improvements. As well, spectrometers were constructed for observing the emission spectra of burning samples (i.e. a relatively fast reduction-oxidation reaction), which also revealed the existence of discrete spectral features, or lines. It was eventually recognized that the features observed in absorption and emission spectra are related to each other, that the dark lines of an absorption spectrum occur at the same wavelength as the bright lines of an emission spectrum.

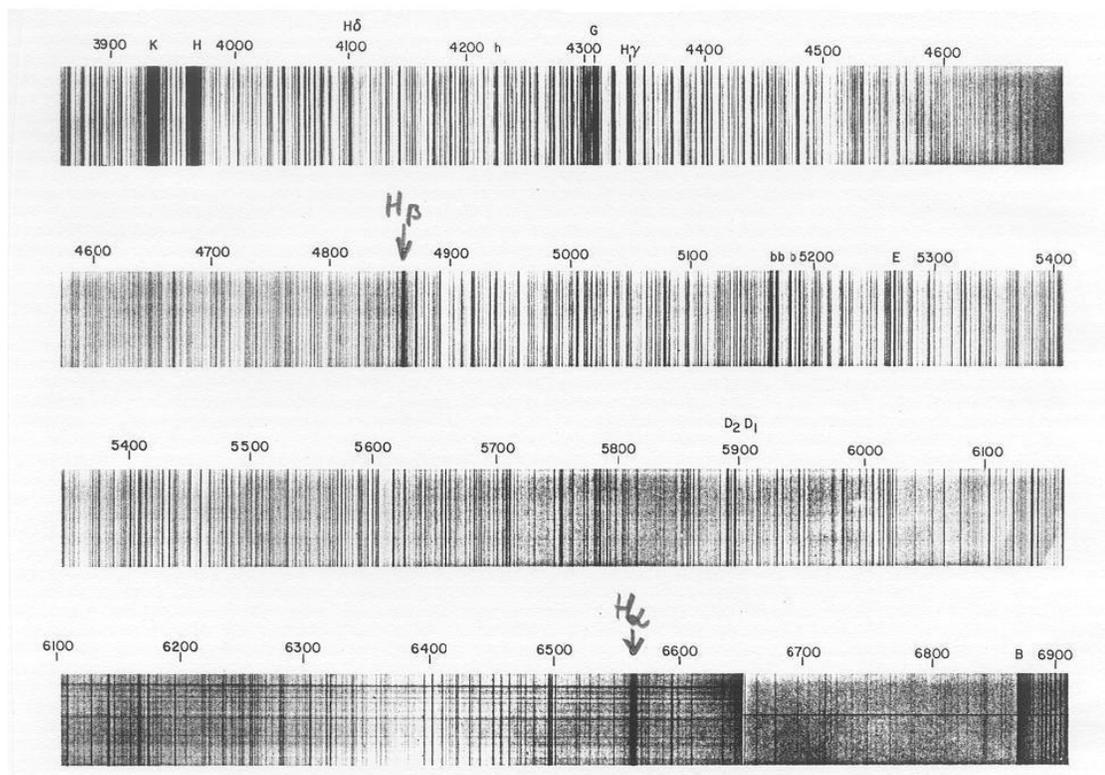

Figure 1.2 Photographic image of the solar spectrum at optical wavelengths obtained with a large grating spectrograph. Wavelengths are given in units of Angstrom at the top of each image-strip. The labels B, E, F, G, H, and K correspond to Fraunhofer's original designations, while C has been changed to $D_1$ and $D_2$, which refer to the sodium D-lines, and D has been changed to $H\alpha$ of the Balmer sequence. Also labeled are the hydrogen Balmer lines $H\beta$, $H\gamma$, and $H\delta$ (Mt. Wilson Observatory, Carnegie Institution of Washington.)

In the late 19[th] century, using a spectrometer of sufficient resolution and sensitivity, Michelson and Morley appear to have "observed" (i.e. make note of) the hyperfine structure of metallic vapors [Michelson]. As well, the experimental and theoretical knowledge-base in (classical) electricity and magnetism continued to grow and mature through the 19[th] century, so that by the end of this century physicists began to offer meaningful theoretical models of spectral lines [Allen 1].

In the early 20[th] century, the resolution and sensitivity of spectrometers operating at near-infrared and visible wavelengths revealed the existence of systematic changes in the intensity of the resonance fluorescence (emission) spectrum of, for example, diatomic iodine ($I_2$) as a function of total gas pressure [Franck]. Using photographic film as the medium for detecting the light transmitted through the absorption medium, new lines appeared at longer wavelengths (relative to the excitation wavelength) in these spectra as the buffer gas pressure was increased. In the late 1950's, electronic photo-detectors (photo-multiplier tubes) began to replace film as a light detection medium for absorption and emission studies, which has proven to be particularly convenient for quantitative characterization of spectral features [Steinfeld].

The resolution of an absorption spectrum of a (relatively low pressure) gas-phase sample in the visible region of the electromagnetic spectrum is inherently limited by the Doppler Effect. In the last several decades, narrow bandwidth lasers (narrower than the width of the atomic and molecular transitions being studied) and high-speed computers have allowed for the development of techniques that "see" beyond the Doppler-limit. The experimental methods used in this project belong to this path of development [Allard].

The changes in line-shape (i.e. width) and line-center position (of "individual" lines) as a function of buffer gas pressure (in absorption and emission spectra) have long been recognized as being important [Margenau]. Empirical observation of the linear changes in line shape and in line-center position (with respect to buffer gas pressure), and attempts to theoretically model these effects can be traced at least as far back as the 1930s. The linear rate of change of the width and line-center position (of an "individual" line) as a function of buffer gas pressure are known respectively as the pressure broadening and pressure shift coefficients. The use of an internally-referenced high-resolution spectrometer, along with the ability to "see" beyond the Doppler-limit has made such measurements at visible wavelengths more accessible and reliable.

## 1.3 Linear Absorption Spectroscopy

Of particular note is the empirical observation and statement of linear absorption (a.k.a. Beer-Lambert law, Beer-Lambert-Bouguer law, Beer's law, Lambert's law, or Bouguer's law) in the 18th and 19th centuries, that the change in light intensity with respect to the change in distance traveled through a particular absorption medium is proportional to a constant value, often referred to as the linear absorption coefficient.

An absorption medium is generally conceptualized as being composed of a very large number of microscopic constituents (i.e. atoms or molecules). Current theoretical descriptions at the microscopic domain indicate that the process of absorption and emission of photons from the radiation field is accompanied by state-changes (a.k.a. transitions) in the microscopic constituents of the absorption medium (a.k.a. chromophore). These state-changes involving absorption and emission of photons are not the only type of interactions that occurs between a chromophore and radiation field (e.g. scatter and phase-altering), but they are essential in the observation of a linear absorption spectrum. The microscopic models that describe absorption and emission of photons in an absorption medium make use of the equality of the energy of a single photon and the difference in energies between two (non-degenerate) quantum states (a.k.a. levels) in a chromophore (a.k.a. transition energy). In mathematical terms, $\Delta E = E_{photon} = \hbar\omega$, where $\Delta E$ is the transition energy between quantum states, $E_{photon}$ is the (particle-picture) energy of a single photon, $\hbar$ is the Planck constant, and $\omega$ is the (wave-picture) frequency of the radiation field (i.e. light). The detection of the changes in the (cycle-averaged) intensity of the radiation transmitted through an absorption medium (as a function of frequency $\omega$) is used to probe these state-changing processes.

The relationship between wave number ($k$) and wavelength ($\lambda$) can be defined as $k \equiv 1/\lambda$; the relationship between wavelength and frequency ($\omega = 2\pi\nu$) is deduced from the classical theory of electricity and magnetism as $c = \lambda\nu$, where $c$ is the speed of light (i.e. the distance traveled per unit time by electro-magnetic waves) in a vacuum. The speed of light is approximately $3 \times 10^8$ m s$^{-1}$ (a.k.a. meters per second). The units of wave number often encountered in spectroscopy are cm$^{-1}$ (a.k.a. reciprocal centimeters).

In a typical linear absorption experiment, a well-collimated beam of light is directed to pass through a gas cell that contains an absorption medium (e.g. diatomic iodine). The intensity of the light beam transmitted through the absorption medium is recorded point-by-point with a computer-based data acquisition system. Each point of the intensity profile corresponds to a different ("single") wavelength of the radiation source. Thus, an absorption spectrometer used to obtain quantitative results requires calibration of the wavelength of the detected light beam and the ability to tune (or scan) this radiation source through a continuous segment of the electromagnetic spectrum.

The calibration of each point of the intensity profile of a point-by-point continuous spectrum to a frequency scale is achieved through comparisons to simultaneously recorded (on a separate channel of the data acquisition system) primary and/or secondary absolute (atomic or molecular) frequency standards (e.g. diatomic iodine atlases). The calibration procedure can also be made more precise by the simultaneous recording (on a separate channel of the data acquisition system) of the spectral features (a.k.a. fringes) produced by an optical element, such as an etalon. The spacing between fringes produced by an etalon reflects to a high degree of accuracy the relative changes in frequency between data points in the recorded spectrum [Yariv].

Each data point in a (continuous) spectrum is acquired over a time-interval that is much longer than the ensemble-average time-interval between state-changing events of a chromophore.  And the radiation detection system is configured to integrate the detected radiation for a time-interval that is much longer than the ensemble-average time-interval between state-changing events in a (single) chromophore.  Thus, the recorded spectra are thought of as being a reflection of steady-state dynamics for the interaction between light and an absorption medium.  The mathematical relationship for a steady-state condition of this state-changing interaction is that the (total) rate of photon absorption is equal to the (total) rate of photon emission.  The steady-state model can be interpreted as saying that the total light beam at a particular "single" (i.e. monochromatic) wavelength is (to a first approximation) the sum of the photons that reach the detector plus the photons that are held in the absorption medium (due to the absorption and emission process) plus those photons that are re-emitted by the absorption medium in random directions (and thus never reach the detector).

## 1.4 Wavelength Modulated Linear Absorption Spectroscopy

Modulating the wavelength of a nearly monochromatic light source is a technique used to improve the signal-to-noise (S/N) ratio of an absorption experiment. This technique is useful for conditions where it is not practical or convenient to obtain the necessary path length and/or chromophore number density that would utilize the full dynamic range of the detection electronics, which is the situation encountered for diatomic iodine at room temperature and approximately 0.2 torr.  The term "wavelength modulation" (as opposed to "frequency modulation") refers to the situation that the wavelength of the radiation source is modulated at a rate that is considerably slower than the state-changing rate of the chromophore [Silver].

The process of modulating the wavelength of a relatively narrow line-width light source as it is slowly tuned across a chromophore line-shape is equivalent to taking a first derivative of the (direct) absorption spectrum described in the previous section [Demtröder]. The wavelength-modulation method eliminates the constant offset signal (a.k.a. off-resonance base-line signal) introduced by the radiation incident on the gas cell, thus allowing the full dynamic range of the detection (and thus data acquisition) system to be utilized in an experimental configuration that is relatively easy to realize. In conjunction with signal-averaging techniques (e.g. detection bandwidth narrowing achieved with the use of phase-sensitive detection, often referred to as lock-in amplifier detection) signal-to-noise ratios (S/N) can surpass the limits imposed by a 12-bit data acquisition system (i.e. better than 1 part in $2^{12} = 4096$).

However, these levels of resolution and sensitivity do not necessarily indicate that the chromophore line-shape was accurately recorded. Experience suggests that systematic artifacts routinely appear in the recorded spectral line-shape; e.g. etalon fringes unintentionally introduced into the sample or reference spectrum by optical elements such as gas cell windows. Identifying the sources of these artifacts and mitigating their effects is not trivial. The result is that the values of the pressure broadening and pressure shift coefficients obtained for a particular buffer gas might be accurate to a single significant figure, while the relative values between different buffer gasses might be accurate to about two significant figures.

## 1.5 Research and Education

Universities are well situated with regard to physical resources and human expertise in carrying out the mission of training the next generation of scientists. In addition to a curriculum filled with many of the concepts and models used in the sciences, there is also the very important task of coming to understand, appreciate, and make proper use of the Scientific Method. For undergraduate students, this task is generally regarded as being best achieved through the many required hands-on laboratory courses. Graduate students play an important role in facilitating the training that takes place in the undergraduate laboratory courses, which in turn reinforces and expands the knowledge base of the graduate student. I had the good fortune during my years at the University of Oregon to have frequent involvement in the undergraduate physical chemistry laboratory course and more or less continual involvement in the training of several undergraduate students.

Such undergraduate laboratory courses are time-intensive for both the students and instructors, and generally require a considerable investment of physical resources. The relatively simplicity and low monetary cost of linear absorption spectroscopy is well suited to such an environment. As well, under the supervision of John Hardwick, a master of such spectroscopic methods, it has been possible to provide undergraduate students with the opportunity of being involved in fundamental research. John Hardwick, the lead instructor for the undergraduate physical chemistry laboratory course, has been quite inventive and flexible in allowing for a wide range of undergraduate and graduate student participation in these projects [Hardwick 1, 2, 3, and 4].

In this sense, the equipment used in much (if not most) of this project was applied toward fundamental research and the training of undergraduate students. Furthermore, the monetary costs were mitigated by purchasing much of the necessary equipment through on-line second-hand auction-based websites (e.g. ebay) at about 10 to 20% the price of comparable new items. (A partial list of equipment acquired in this manner includes external cavity diode lasers, lock-in amplifiers, oscilloscopes, photo-detectors, and various optical components.) And sharing equipment, within a given department in the university, and between different areas of the university, is another method of reducing the monetary costs; for example, many personal computers used in this project and in the undergraduate physical chemistry laboratory were acquired second-hand from different areas of the university.

However, in spite of the joy of learning about science and what it has to say about how the universe operates, we are still left to ponder the difference between training and educating the next generation. As the physicist Victor Weisskopf noted: "Human existence is based upon two pillars: Compassion and knowledge. Compassion without knowledge is ineffective; knowledge without compassion is inhuman." Educating the next generation would include considerable attention to such ideas, whereas merely training the next generation appears to have the tendency of avoiding a more balanced consideration of such wisdom.

## 1.6 Error Propagation of Uncorrelated Parameters

This section could perhaps have been left as an appendix, but such an approach would seem to underemphasize the central role played by error analysis in the Scientific Method.

Measurement any physical quantity is *always* uncertain by some finite amount. By virtue of these measurement uncertainties, the quantitative values obtained from such measurements are also uncertain by a quantifiable amount [Young]. The theory of the statistical treatment of experimental data provides a model for the propagation of error due to uncorrelated parameters ($x$, $y$, …) such that a derived value $f(x, y, …)$ has a variance given by $\sigma_f^2 = (\partial f/\partial x)^2 \sigma_x^2 + (\partial f/\partial y)^2 \sigma_y^2$…. The square root of the variance gives the standard error $\sigma$ (a.k.a. standard deviation).

For the Gaussian (or Normal) distribution model of measurement error, the standard deviation about an average value characterizes the (statistically) anticipated result of about 78% of such measurements. In this sense, the standard deviation is a quantitative representation of measurement uncertainty (a.k.a. experimental uncertainty).

## 1.7 Endnotes for Chapter I

[Allard]　　　　　N. Allard and J. Keilkopf; "The effect of neutral nonresonant collisions on atomic spectral lines", Reviews of Modern Physics, 54, 1103-1182 (1982).

[Allen 1]　　　　　L. Allen and J. H. Eberly, *Optical Resonance and Two-Level Atoms*, pages 1-15; Dover Publications, Mineola, New York (1987); ISBN 0-486-65533-4.

[Demtröder]　　　W. Demtröder; *Laser Spectroscopy: Basic Concepts and Instrumentation*, Third Edition, pages 374-378; Springer-Verlag, Berlin (2003); ISBN 3-540-65225-6.

[Franck]　　　　　J. Franck and R. W. Wood; "Influence upon the Fluorescence of Iodine and Mercury Vapor of Gases with Different Affinities for Electrons", Philosophical Magazine, 21, 314-318 (1911).

[Hardwick 1]　　　J. A. Eng, J. L. Hardwick, J. A. Raasch and E. N. Wolf; "Diode laser wavelength modulated spectroscopy of $I_2$ at 675 nm", Spectrochimica Acta, Part A, 60, 3413-3419 (2004).

[Hardwick 2]　　　J. L. Hardwick, Z. T. Martin, E. A. Schoene, V. Tyng and E. N. Wolf; "Diode laser absorption spectrum of cold bands of $C_2HD$ at 6500 cm$^{-1}$", Journal of Molecular Spectroscopy, 239, 208-215 (2006).

[Hardwick 3]　　　J. L. Hardwick, Z. T. Martin, M. J. Pilkenton and E. N. Wolf; "Diode laser absorption spectra of $H^{12}C^{13}CD$ and $H^{13}C^{12}CD$ at 6500 cm$^{-1}$", Journal of Molecular Spectroscopy, 243, 10-15 (2007).

[Hardwick 4]　　　S. W. Arteaga, C. M. Bejger, J. L. Gerecke, J. L. Hardwick, Z. T. Martin, J. Mayo, E. A. McIlhattan, J.-M. F. Moreau, M. J. Pilkenton, M. J. Polston, B. T. Robertson and E. N. Wolf; "Line broadening and shift coefficients of acetylene at 1550 nm", Journal of Molecular Spectroscopy, 243, 253-266 (2007).

[Margenau]　　　H. Margenau and W. W. Watson; "Pressure Effects on Spectral Lines", Reviews of Modern Physics, 8, 22-53 (1936).

[Michelson]     Nobel Lecture in Physics given in 1907 by A. A. Michelson;
                "Recent Advances in Spectroscopy"; *Nobel Lectures, Physics1901-
                1921*; Elsevier, Amsterdam (1967); archived at the web site
                http://nobelprize.org/nobel_prizes/physics/laureates/ (2009).

[Silver]        J. A. Silver; "Frequency-modulation spectroscopy for trace species
                detection: theory and comparison among experimental methods",
                Applied Optics, 31, 707-717 (1992).

[Steinfeld]     J. I. Steinfeld and W. Klemperer; "Energy-Transfer Processes in
                Monochromatically Excited Iodine Molecules. I. Experimental
                Results", Journal of Chemical Physics, 42, 3475-3497 (1965).

[Yariv]         A. Yariv; *Optical Electronics*, Third Edition, Chapter 4; CBS
                College Publishing, New York (1985); ISBN 0-03-070289-5.

[Young]         H. D. Young; *Statistical Treatment of Experimental Data*; McGraw-
                Hill Book Co., New York (1962); LCCN 62-16764.

# CHAPTER II

# INTRODUCTION 2 – BACKGROUND KNOWLEDGE

## 2.1 Overview of Chapter II

This chapter is mostly concerned with outlining and/or summarizing some background knowledge pertinent to this project. The first half of this chapter (Sections 2.2 through 2.6) is a review of the stationary-state quantum mechanical description (i.e. solutions of the time-independent Schrödinger equation) of the chromophore (diatomic iodine). The second half of this chapter (Sections 2.7 through 2.14) covers a variety of material relevant to spectroscopic methods in both the time and frequency domains, and the information content of the spectra obtained from them. The material in Section 2.11 presents what appears to be a new formulation of the steady-state state-changing kinetics for a two-level system model.

## 2.2 Electronic States, Hund's Coupling, and Selection Rules

For the conditions encountered in this research project diatomic iodine is a homonuclear diatomic molecule, which is generally written symbolically as "$I_2$". At typical room temperature conditions diatomic iodine sublimes to a pressure of about 0.2 torr [Tellinghuisen]. It is relatively easy to handle in a typical physical chemistry laboratory setting, which helps to minimize the risk of exposure to amounts that could be toxic [Merck]. A naturally occurring sample of diatomic iodine is nearly 100 percent isotope $^{127}I_2$ [CRC].

Among the more prominent absorption features (spectral lines) of diatomic iodine are the transitions between the B and X electronic states from about 500 nm to 700 nm in the visible region of the electromagnetic spectrum.  There is also a corresponding bright emission of fluorescence extending from the visible far out into the infrared wavelengths (and beyond) that has received a considerable amount of attention in research efforts.  The earliest high resolution absorption spectra of diatomic iodine date back to about 1911 [Franck].  Such spectra of atomic and molecular systems were used to discover and confirm many of the descriptions offered in the emerging modern theory of quantum mechanics.

The B and X electronic states of diatomic iodine are often designated with term symbols as $B^3\Pi_{u,0}{}^+$ and $X^1\Sigma_g{}^+$, respectively [Hougen].  The first symbol, a capital letter, indicates the electronic state.  The letter X is reserved for the ground electronic state.  The next electronic state with an energy minimum above that of the ground electronic state is usually designated by the letter "A", and so on in alphabetical order for increasing potential energy minimum of the electronic states (e.g. see Figure 2.1).  However, in the case of diatomic iodine, the discovery of the B electronic state came before the modern quantum theory and so there are many A states distinguished from one another by labeling them as A′, A″ and so on [Field].  Analogous with s and p orbitals in atoms, $\Sigma$ and $\Pi$ indicate that the magnitude of the total electronic orbital angular momentum ($\Lambda$) are respectively zero and one (in units of $h \div (2\pi) \equiv \hbar$, where $h$ is the Planck constant).  The superscripts 1 and 3 refer to the electron spin multiplicity, 2S + 1, where S is the magnitude of the vector sum of the spins of the individual electrons.  The electronic wave-function for a homonuclear diatomic molecule has either even (g) or odd (u) symmetry with respect to inversion through the center of mass of the molecule.  Allowed electric dipole transitions (i.e. selection rules) for a discrete single photon event require that one state have g symmetry and the other have u symmetry.  A more detailed analysis of the spin wave-functions leads to the superscript symbols "+" and "–" that account for symmetric and anti-symmetric reflection at any plane through the internuclear axis [Herzberg 1].

The subscript 0 for the B state term symbol ($B^3\Pi_{u,0}^+$) indicates that the total electronic angular momentum about the internuclear axis $\Omega$ is zero [Zare 1]. Alternatively, the B electronic state can be written as $B^3 0_u^+$. This notation emphasizes that the electronic state has selection rules similar to that of the $\Sigma$ electronic state, which becomes relevant when calculating Hönl-London factors and nuclear spin statistics for transitions between these two electronic states (see Sections 2.4.4 and 2.4.5).

It is also necessary to take into account the manner in which rotation and electronic motions influence each other [Zare 2]. For diatomic iodine the relative strength of the couplings that result from electronic orbital angular momentum ($V_{el}$), electronic spin angular momentum ($V_{so}$), and nuclear rotation angular momentum ($V_{rot}$) in the body fixed axis system for both the B and X electronic states are $V_{so} > V_{el} > V_{rot}$. It is also common to list these with slightly more precision; for diatomic iodine the qualifiers are that $V_{so}$ is strong, $V_{el}$ is intermediate, and $V_{rot}$ is weak. This relative ordering of these couplings is classified as Hund's Case C. In this coupling case $\Omega$ is the only "good" quantum number for this diatomic molecule. Determination of $\Omega$ in this coupling case is done by projecting the vector sums of **$\Lambda$** (total electronic orbital angular momentum) and **S** (total electronic spin angular momentum) onto the internuclear axis. For this coupling case the selection rule for dipole allowed electronic transitions is $\Delta\Omega = 0, \pm 1$ [Herzberg 2]. Since $\Omega = 0$ for both the B and X electronic state the selection rule for these transitions is $\Delta\Omega = 0$ and $\Delta J = J' - J'' = \pm 1$. The result is that the spectrum of the B–X electronic transitions will contain P and R rotation branches, but no Q rotation branch. The rotation quantum number symbols $J'$ and $J''$ designate the upper (B electronic state) and lower (X electronic state) states, respectively. Also, $\Delta J = -1$ for a P branch, $+1$ for an R branch, and 0 for a Q branch.

The intensity distribution within a band is of some importance in the results presented in this dissertation with regard to verifying the assignment of quantum numbers to the observed transitions. The maximum probability for the room temperature (292 K) thermal equilibrium rotational distribution occurs for $J \simeq 52$ [Herzberg 3]. It is possible to observe values of the rotation quantum number $J$ as large as 140 this temperature. There are approximately 115 and 90 vibration levels in the X and B electronic states, respectively. Typical spectroscopy experiments conducted near room temperature with diatomic iodine in the gas phase are sensitive enough to observe the first 5 to 10 vibration levels of the ground-state. A rough estimate of the number of transitions per $cm^{-1}$ can thus be made for a B electronic state well depth of 4274 $cm^{-1}$: $(90 \times 10 \times 140) \div 4274 \simeq 29$ transitions per $cm^{-1}$.

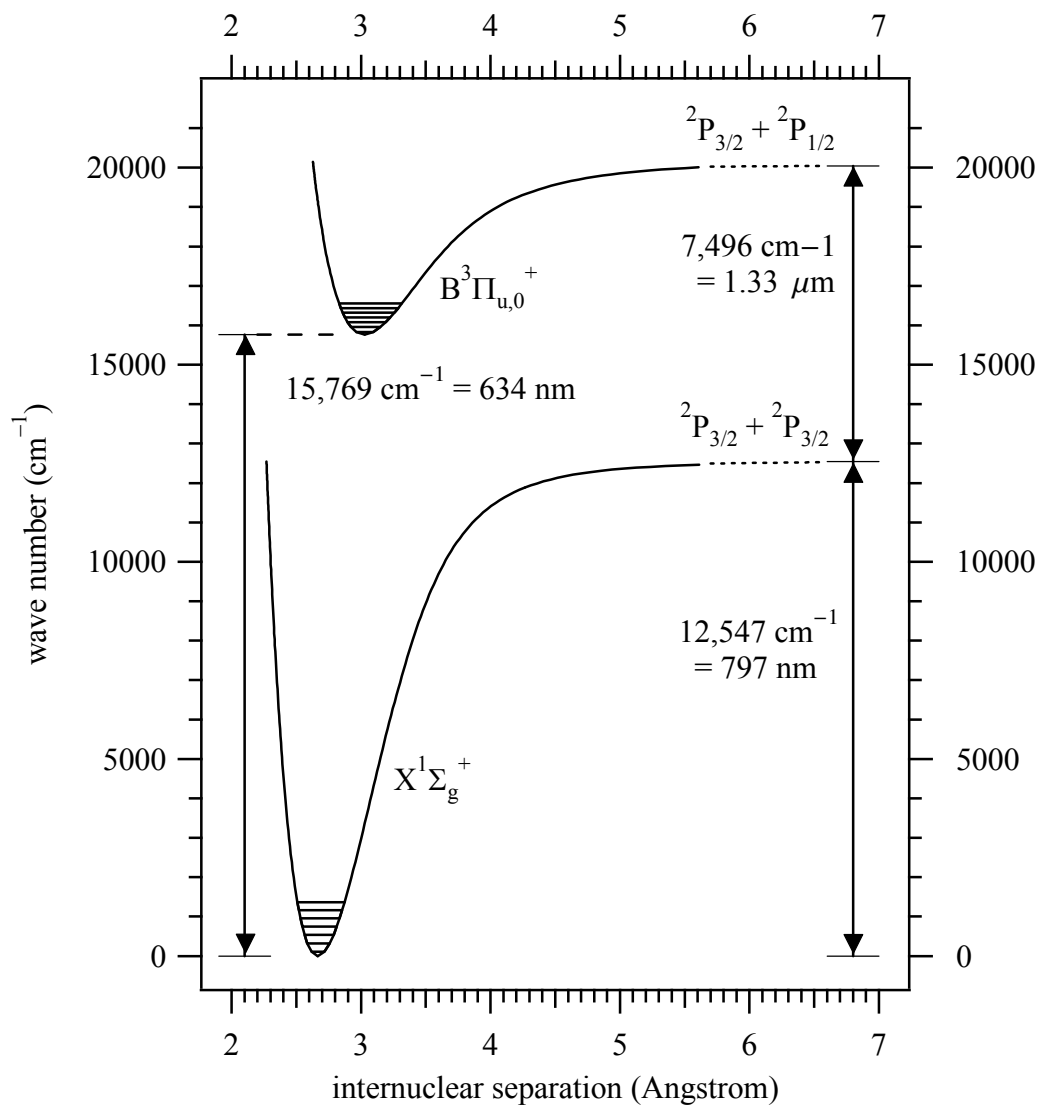

Figure 2.1 Potential energy surfaces (PES) for the X and B electronic states of diatomic iodine (I₂) calculated with the use of LeRoy's "RKR1" program [LeRoy 1] and the spectroscopic constants from Gerstenkorn [Hutson] and Bacis [Martin]. The ground vibration level internuclear separation $R$ for the X and B electronic states are 2.667 Å and 3.027 Å, respectively. The calculated B electronic state PES span a range of $R$ from 2.63 to 17.2 Å. The calculated X electronic state PES spans a range of $R$ from 2.27 to 9.08 Å. The state-designations $^2P_{3/2}$ and $^2P_{1/2}$ correspond to the separated iodine atoms at infinite internuclear separation $R$.

## 2.3 Spectroscopic Constants

An important goal in nearly all spectroscopic investigations is the assignment of the quantum numbers for the states involved in the transitions of an observed spectrum. These quantum numbers are generally represented in a compact form known as spectroscopic constants. These spectroscopic constants are obtained from high-resolution frequency domain spectra by various linear and non-linear regression analysis strategies in which the quantum numbers (for electronic, vibration, and/or rotation) are the dependent variable and the observed transition energies are the dependent variable. The model used in the regression analysis is a stationary-state Hamiltonian that is a function of the quantum numbers.

The spectroscopic constants determined through a regression analysis are the terms of an expansion representation (typically) in terms of rotation and vibration quantum numbers of the relatively smoothly varying potential energy surface of a given electronic state. These spectroscopic constants can then be used to re-calculate the transition energies observed in the original spectrum (or spectra); each electronic state will have its own unique set of spectroscopic constants determined to the limit of the precision of the original energy measurements. One such tabulation of spectroscopic constants that accounts for both vibration and rotation in a given electronic state through a double Taylor series with the expansion coefficients referred to as the Dunham coefficients [Dunham]. A variation on this expansion representation is to simply tabulate the vibration band head energies and give a set of spectroscopic constants for the rotation energies for each vibration band, which are often referred to as centrifugal distortion constants [Herzberg 4]. A common technique that is employed to determine separately the spectroscopic constants for both the upper and lower is known as combination differences [Herzberg 5].

Several high resolution studies of the B−X (read "B to X" or "X to B") electronic transitions of diatomic iodine across the entire range of roughly 500 to 700 nm have been carried out over the last roughly one hundred years. The earliest studies were performed at modest resolution and did not resolve the rotational structure. These early investigations of diatomic iodine (making use of the modern quantum theory) employed a model Hamiltonian in a linear regression analysis that was a function of only the vibration quantum number [Loomis]. Over the years, as the spectroscopic instrumentation improved, allowing for resolution up to and even beyond the Doppler-limit, and increased sensitivity in detecting relatively weak transitions, the model Hamiltonians used have become functions of vibration and rotation quantum numbers, with ever more transitions being included in the linear regression analysis [Hutson; Martin]. Today there is a consensus that the vibration and rotation assignments of the B−X system are well understood, down to the level of nuclear quadrupole coupling.

## 2.4 Line Intensities in Vibronic Bands

Reconstructing (i.e. simulating) the observed spectrum from a "trusted" set of spectroscopic constants is relevant to this project for three primary reasons. The first is to be confident (beyond the measurements of a wave meter) in the wave number calibration of an observed spectrum. Second, the rotation quantum numbers $J$ are necessary as input data for use in the hyperfine structure model of the line shape model used in the nonlinear regression analysis; a simplified high-$J$ model only requires knowledge of whether $J''$ is odd or even, while a more complete hyperfine model requires knowledge of the actual $J$ values for both levels. And third, a general goal of chemical physics projects is to discern patterns in the observed spectra as a function of the stationary-state quantum numbers for the two levels in a transition.

An important tool used in deducing the quantum number assignments of a spectrum is the relative (and sometimes absolute) intensities of the observed transitions. A set of such transitions are often referred to as a vibronic band. An element of this set is designated by its value of $J''$ and is referred to as a ro-vibronic transition. The word "ro-vibronic" is derived from **ro**tation + **vib**ration + elect**ronic**. The factors that determine the observed spectral intensity of molecular lines are of fundamental importance in spectroscopy. In the remainder of this section the use of relative intensities (which are used as an aide in deducing quantum numbers assignments for vibration and rotation levels of the observed molecular transitions in an absorption spectrum) will be presented. In the case of the relatively congested and fairly dense B–X electronic transitions of diatomic iodine it is necessary to determine if the observed lines are a blend of more than one line or if they are reasonably well isolated from neighboring lines to be suitable for a line-shape analysis; in some instances a blend of two well isolated lines is still suitable for a line shape analysis, which is somewhat common for low-$J$ B–X lines of diatomic iodine. The quantum number assignments are made by comparing a simulated spectrum to the measured line positions and relative intensities in an observed spectrum. The simulated spectrum consists of calculated transition energies and calculated relative intensities. The band-origin of the vibration levels for what are predicted to be the most intense bands in the spectral regions investigated in this project are shown in Figure 2.2; see also Section 4.5.

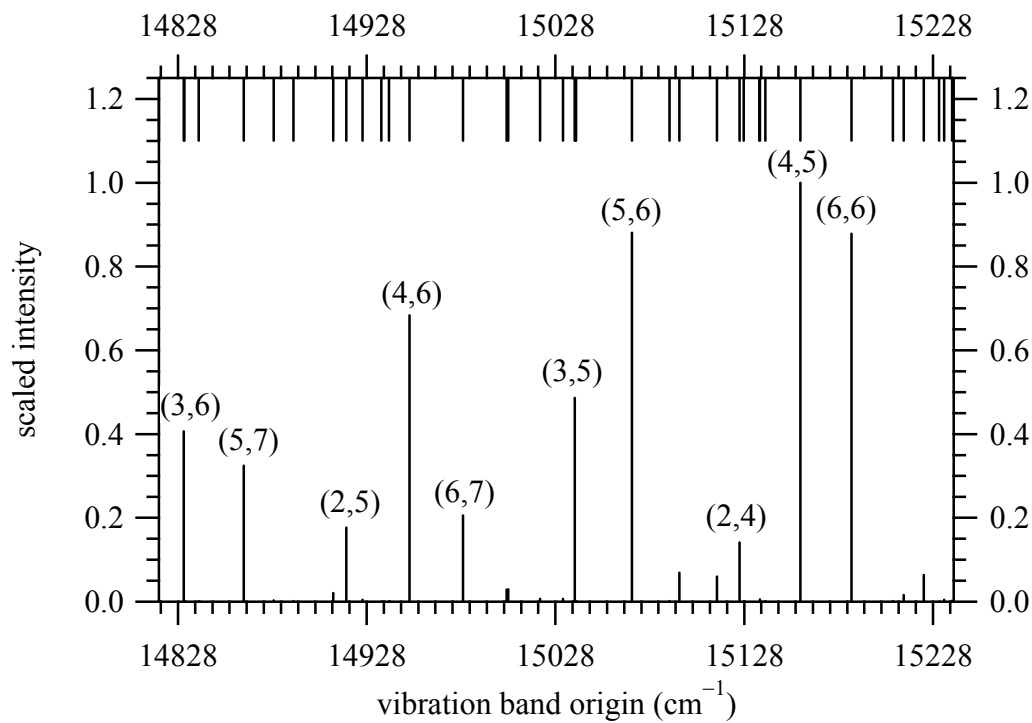

Figure 2.2 Simulation of the relative intensities of the vibration levels present in the regions explored during the course of this project for B–X electronic transitions in diatomic iodine. The thermal distribution ($N_v(v'')$ and $N_J(J'')$) and Franck-Condon factors (*FCF*) are included in the weighting of the relative intensities in the lower portion of the plot for the more intense spectral features in the region 14,818 to 15,240 cm$^{-1}$ (674.85 to 656.17 nm). The labeling of the stronger vibration band-origins follows the usual convention of ($v'$, $v''$) where the upper and lower energy levels are given respectively by $v'$ and $v''$. All 39 vibration band-origins in this spectral region are shown in the upper plot as equal length sticks (i.e. without consideration to thermal distribution or Franck-Condon factors). The spectroscopic constants used in computing this simulation are from Gerstenkorn [Hutson] and Bacis [Martin]. See also Section 4.5.

A typical simulated "stick-spectrum" (without line-width) for the rotational fine structure of a typical band in the B–X system of diatomic iodine is depicted for the (4, 6) band in Figure 2.3. (The vibration quantum number symbols $v'$ and $v''$ designate the upper (B electronic state) and lower (X electronic state) states, respectively.) Due to the relatively small rotation constant ($B_e$) values for the X and B electronic states (0.03735 cm$^{-1}$ and 0.02920 cm$^{-1}$, respectively), and their relatively small difference, the R branch turns around after $J'' = 3$ and follows the P branch into the region of decreasing transition energies (i.e. lower wave numbers) [Herzberg 6].

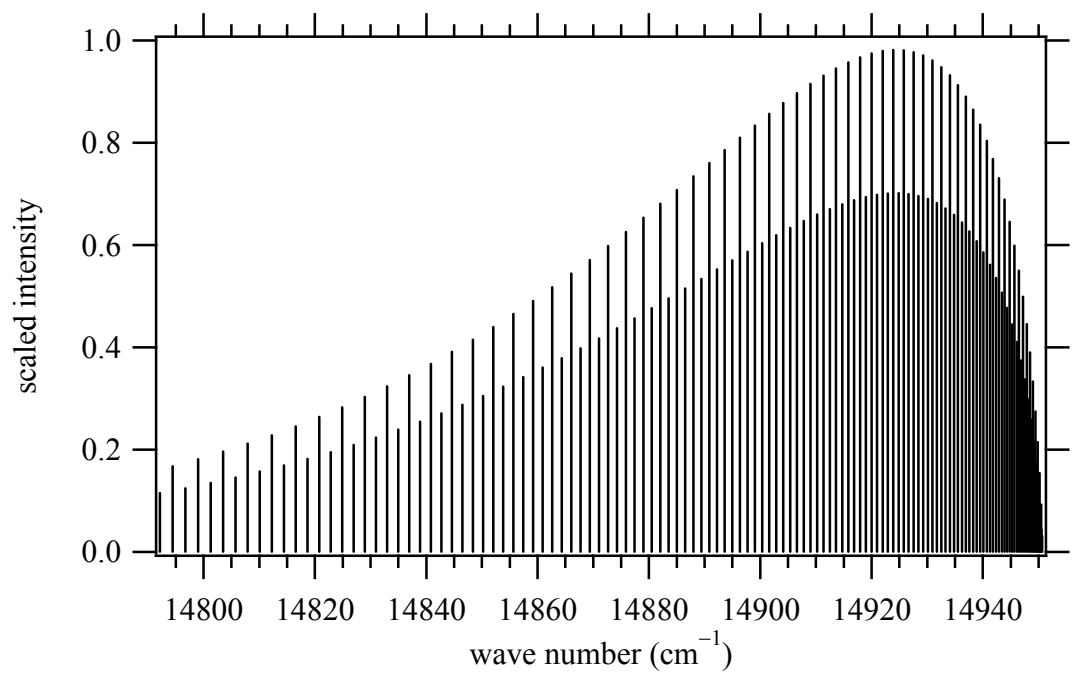

Figure 2.3 Simulation of the relative intensities of the P branch rotational progression for the $(v', v'') = (4, 6)$ vibration transition for B−X electronic transitions in diatomic iodine. The simulation is weighted for thermal distribution, Hönl-London factor, and nuclear spin statistics for odd and even $J''$. The spacing between the $J'' = 0$ and $J'' = 1$ transitions is 0.089 cm$^{-1}$ (near the large wave number limit of the spectrum) and monotonically spreads out to 2.3 cm$^{-1}$ (in going toward the low wave number end of the spectrum) where the last two values of $J''$ are (from right to left) 133 and 134. The spectroscopic constants used in computing this simulation are from Gerstenkorn [Hutson] and Bacis [Martin]. See also Section 4.5.

After computing the transition energies of previously assigned lines from a "trusted" set of spectroscopic constants it is then necessary to compute the theoretically predicted relative intensities $I_{rel}$ of these lines in order to make a meaningful comparison of the observed and predicted spectra for the purpose of assigning quantum numbers to the observed spectra. Six multiplicative factors are generally considered to be sufficient in accounting for most of the relative intensities of lines in a simulated spectrum of transitions between the B and X electronic states of diatomic iodine. These factors are the Boltzmann weighting of the ground-state levels for both vibration and rotation, the square of the electronic transition moment, the Franck-Condon factor, the Hönl-London factor, and nuclear spin statistics [Herzberg 7].

The relative intensity $I_{rel}$ of a given electronic transition due to these six multiplicative factors (taken in the same order as mentioned in the previous sentence above) can be expressed as:

$$I_{rel}(v', J', v'', J'') = N(v'') \, N(J'') \, S_{el}(v', v'')$$

$$\times FCF(v', v'') \frac{S_{rot}(J', J'')}{2 J'' + 1} \, I_{ns} \tag{2.1}$$

where $(v'', J'')$ refer to the lower (or initial) state and $(v', J')$ to the upper (or excited) state. Each of the factors in equation 2.1 will be defined and described further in the sub-sections below.

The electronic transition moment ($S_{el}$), Franck-Condon factor ($FCF$), and Hönl-London factor ($S_{rot}$) usually result from a quantum mechanical derivation in the context of the Born-Oppenheimer approximation in which the solutions to the time-independent Schrödinger equation for a molecular system are separated into electronic, vibration, and rotation wavefunctions, respectively. The nuclear spin statistics ($I_{ns}$) are a "purely" quantum mechanical effect, which does not depend on spatial coordinates, and does not have a classical mechanics counter-part.

For the purpose of assigning quantum numbers to the observed spectra, a possible electronic degeneracy factor and some other physical constants can be neglected when simulating the relative intensities of transitions between the B and X electronic states diatomic iodine. Similarly, for the present purpose, changes in the electronic transition moment (as a function of rotation or vibration quantum numbers) can often be neglected, as was done in this project [Tellinghuisen].

### 2.4.1 Boltzmann Weighting Factors

In equation 2.1, the first two terms are the Boltzmann weighting factors $N_v(v'')$ and $N_J(J'')$ for vibration and rotation, respectively. These terms account for the fraction of molecules in a given vibration and rotation quantum state in the X electronic state (lower energy level) at thermal equilibrium. (The probability of finding diatomic iodine in the B electronic state is being neglected, which is appropriate for the experimental conditions used in this project.)

The thermal distribution of quantum states $N_v(v'')$ dependent on the vibration quantum number $v''$ is then given by [Herzberg 3]:

$$N_v(v'') \;\equiv\; \frac{N(v'')}{N} \;=\; Q(v)^{-1} \exp\!\left(-\frac{E(v'')\,h\,c}{k_B\,T}\right)$$

$$(2.2)$$

In equation 2.2 (without loss of generality as compared to the derivations by Herzberg [Herzberg 3]), $N(v'')$ is the number density (i.e. number per unit volume) for the vibration level $v''$ and $N$ is the total number density; the vibration energy of the $v''$ level is given by $E(v'')$ (generally in wave number units of $cm^{-1}$); $h$ is the Planck constant, $c$ is the speed of light (generally in cm/sec), $k_B$ is the Boltzmann constant, and $T$ is the absolute temperature.

The partition function (state sum) $Q(v)$ is given by:

$$Q(v) \; = \; \sum_{v''=0}^{\max} \exp\left(- \frac{E(v'') \, h \, c}{k_B \, T}\right)$$

(2.3)

In equation 2.3, "max" refers to the highest vibration level quantum number consistent with a bound state of the diatomic molecule (i.e. a finite internuclear separation in Figure 2.1), and $E(0)$ is often referred to as the zero point energy.

Determining the thermal distribution of rotational states $N_J(J'')$ is similar to the above case for the vibration states except that an additional factor of $(2J + 1)$ is included to account for degeneracy that exists in the absence of external electric or magnetic fields. It is common practice to consider the case of a rigid rotor for which the energy as a function of rotation quantum number is $E(J) = BJ(J + 1)$ with $B =$ rotation constant (in units of $cm^{-1}$). It can then be readily shown that the thermal distribution of rotational states $N_J(J'')$ are is well approximated by:

$$N_J(J'') \; \equiv \; \frac{N(J'')}{N} \; = \; \frac{B \, h \, c \, (2 \, J'' + 1)}{k_B \, T} \exp\left(- \frac{B \, J''(J'' + 1) \, h \, c}{k_B \, T}\right)$$

(2.4)

## 2.4.2 Electronic Transition Moment

The electronic transition moment is the interaction term between the molecular charge distribution and an electromagnetic field. Inasmuch as the electronic wavefunctions are parameterized by the internuclear separation, $R$, the electronic transition moment is also a function of this variable. In "bra-ket" notation the matrix elements for the square of the electronic transition moment is often written as $S_{el}(v',v'') = \langle v'|$ $\mu_e(R) |v''\rangle^2 = |\mu_e(R)|^2$. For diatomic iodine the variation of this quantity as a function of $R$ and excitation wavelength has been partially mapped out by various researchers for transitions between the B and X electronic states of diatomic iodine [Tellinghuisen]. However, a meaningful functional description of the electronic transition moment of diatomic iodine based on vibration and rotation quantum numbers does not yet exist.

Since the electronic transition moment is typically a relatively slowly varying quantity, and is also rather difficult to measure with any great precision or accuracy, it is common to use an average value. While this approach is not rigorously accurate, it is expected to suffice for the purpose of comparing a simulated spectrum to an observed spectrum, at least when the main purpose of such a comparison is to deduce the quantum number assignments of the states involved in the observed lines. The agreement in comparing the relative intensities between a simulated spectrum and an observed spectrum is expected to be reasonably accurate for transitions that span only a few adjacent vibration levels in each of the electronic states. However, this agreement is likely to falter a bit when using a single approximate value of the transition moment for all rotational levels in these vibration manifolds. The relatively congested and overlapping nature of the diatomic iodine spectrum means that a single $2 \text{ cm}^{-1}$ slice of the B–X spectrum of diatomic iodine can have transitions that differ in their rotational quantum number $J''$ by more than 100.

## 2.4.3 Franck-Condon Factor

The Franck-Condon factor is often referred to as the square of the overlap integral of the vibration wave-functions between the two levels involved in a given electronic transition. In modern notation this can be written as $FCF(v',v'') = \langle v'|v''\rangle^2$. Since the vibration wave functions (a.k.a. eigenfunctions) are on two different electronic surfaces they are not necessarily orthogonal to each other and so in general this integral is not zero. In consideration of the numerous excellent expositions on the Franck-Condon principle [Condon] it hardly seems necessary to elaborate in any great detail on its foundations and physical interpretations. It should be noted, however, that insofar as a given vibration wave-function will change slightly in its spatial extent as a function of rotation level due to centrifugal distortion the Franck-Condon factor is also a function of the rotation quantum number $J$. To a first approximation, though, it is not uncommon to ignore the $J$ dependence of the Franck-Condon factor, instead making the approximation that the values of this factor for the non-rotating molecule ($J = 0$) do not differ significantly across all values of $J$.

It is also worth noting that readily available algorithms that make use of spectroscopic constants allow for computation of various spectroscopic parameters, including the Franck-Condon factors. One such set of programs freely provided by Professor Robert J. LeRoy accomplishes this task of computing Franck-Condon factors by first numerically computing the potential energy surface using the semi-classical Rydberg-Kline-Rees inversion procedure in a program named "RKR1" [LeRoy 1]. This representation of the potential surface of a given diatomic molecule is then readily used in a second program (referred to as "Level 7.4") [LeRoy 2] to numerically solve based on the Cooley-Cashion-Zare computer algorithm the radial portion of the time-independent Schrödinger equation from which the eigenvalues $E(v, J)$, and the eigenfunctions $|v, J\rangle$ are obtained. These eigenfunctions are then used to numerically compute the Franck-Condon factors.

Using Professor LeRoy's computer programs (and the commercially available personal computer program SimgaPlot) it was possible to produce a slightly improved graphical representation of the Frank-Condon factors of diatomic iodine. A previous version [Martin] of the plot in Figure 2.4 did not include the contours indicating the energy difference between the vibration band origins for states involved in a particular transition. The Franck-Condon factor for a particular vibronic transition is proportional to the area of the circle. The lack of visible circles in the lower right quadrant of this plot would appear to be due to the Franck-Condon overlap integrals being much smaller in magnitude than those that have a visible diameter at this scale.

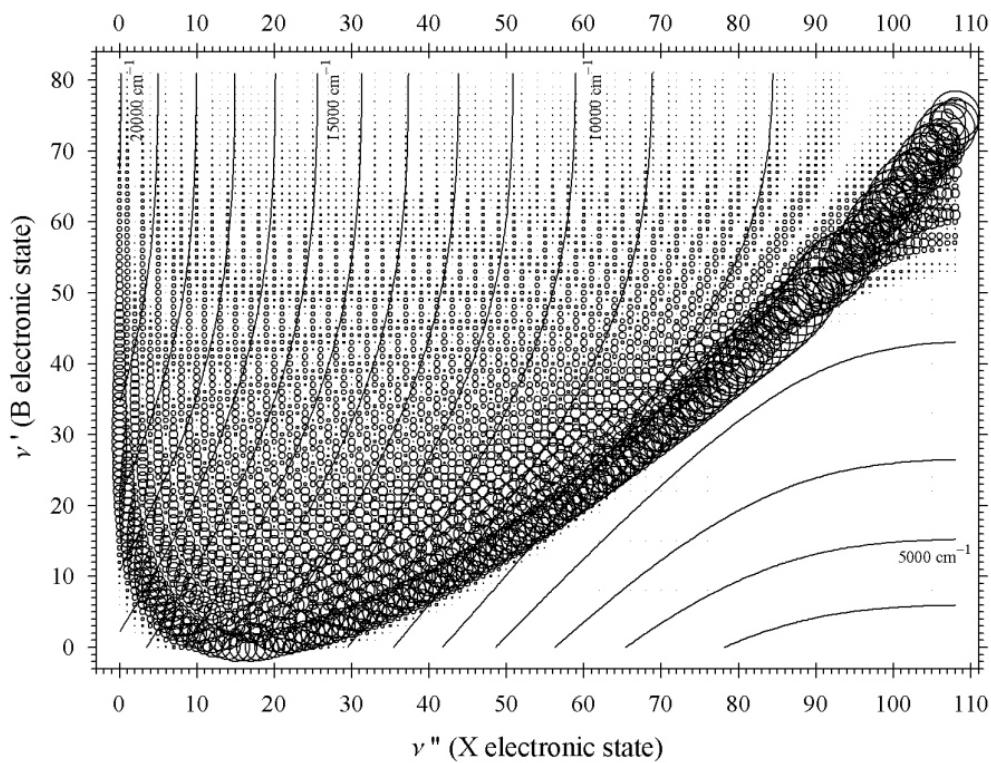

Figure 2.4 Franck-Condon factors for B−X transitions in diatomic iodine ($I_2$) as a function of vibration quantum number calculated with the use of LeRoy's "Level 7.4" program [LeRoy 2]. The Franck-Condon factor is proportional to the area of the corresponding circle at the coordinates ($v''$,$v'$). The rotational quantum number $J$ was set equal to zero in the calculations. The X electronic state spectroscopic constants are from Bacis [Martin]. The B electronic state spectroscopic constants that were included with LeRoy's Level 7.4 program are from Gerstenkorn, but it is uncertain (per Professor LeRoy in a private correspondence) if this set has been published elsewhere.

## 2.4.4 Hönl-London Factor

The Hönl-London factor, $S(J', J'')$, accounts for the spatial orientation of the transition moment of a rotating molecule with respect to a specific polarization direction of an electromagnetic field involved in inducing transitions between the lower and upper electronic states. This must be done for each of the $(2J'' + 1)$ magnetic sub-levels $m_{J''}$ of a given rotation state. The Hönl-London factor is determined by adding together the $(2J'' + 1)$ contributions from the individual magnetic sub-levels for a given rotation state. In order to not over count the contribution of this summation on the magnetic sub-levels it is necessary to divide by the degeneracy factor $(2J'' + 1)$, which is why the term containing the Hönl-London factor in equation 2.1 is written as $S(J', J'') \div (2J'' + 1)$. Dividing by the degeneracy factor of $(2J'' + 1)$ will simply cancel the degeneracy factor appearing in the rotation Boltzmann weighting factor above in equation 2.4.

When computing relative intensities $I_{\text{rel}}$ of diatomic iodine for the purpose of assigning quantum numbers to a small slice of the spectrum (approximately 2 cm$^{-1}$) it will prove to be sufficient to use the formulas that were tabulated by Herzberg [Herzberg 8] for the case that $\Delta\Lambda = 0$; for the R branch $S^{\text{R}}(J', J'') = (J'' + 1)$ and for the P branch $S^{\text{P}}(J', J'') = J''$. More detailed considerations for deriving Hönl-London factors can be pursued in the monograph written by Jon T. Hougen [Hougen].

## 2.4.5 Nuclear Spin Statistics

In order to satisfy the Pauli principle (interchange of indistinguishable particles in the theory of quantum mechanics), homonuclear diatomic molecules in which the two nuclei have the same isotope number will exhibit an intensity alternation between odd and even rotation levels [Herzberg 8; Landau; Atkins]. Furthermore, for $\Sigma$ to $\Sigma$ electronic transitions (in which $\Lambda = 0$) the intensity alternation between "strong" and "weak" transitions will be in the ratio of $(I + 1)/I$ where $I$ is the quantum number of the atomic nuclear spin. Both the B and X electronic states of diatomic iodine have $\Lambda = 0$, and for the 127 isotope of iodine the nuclear spin quantum number of the nucleus is $I = 5/2$, so that the intensity alternation for odd and even $J''$ is in the ratio of 7/5. In calculating relative intensities of diatomic iodine the "strong" transitions occur for the odd values of $J''$ (i.e., the lower level has $J = 1, 3, 5…$) and it is reasonable then to set $I_{nss} = 7$. The "weak" transitions occur for even values of $J''$ (i.e., the lower level has $J = 0, 2, 4…$) for which we then use $I_{nss} = 5$ [Strait].

Accounting for this intensity alternation is achieved through consideration of the nuclear spin statistics and selection rules in conjunction with an analysis of the nuclear hyperfine interactions (between the nuclei and electrons) in diatomic iodine. In addition to splitting a single ro-vibronic transition into a multiplet of closely spaced transitions that have slightly different line-centers, the nuclear spin is fundamental in determining the relative intensity of odd and even lines. An analysis of the quantum statistics and selection rules for B−X transitions in diatomic iodine, where both iodine nuclei have spin $I = 5/2$, gives the number of possible total nuclear spin spatial orientations (i.e. sum over degeneracy) as 21 for odd $J''$ and 15 for even $J''$ [Kroll].

## 2.5 Nuclear Hyperfine Structure

By the late 19[th] century the resolution of (electromagnetic) spectrometers had increased to the point of revealing even finer details of the line-shape. Some of the earliest observations of what came to be known as hyperfine structure of metallic vapors appears to have been made by Michelson and Morley in the 1880s in their studies of the line shape for light emitted by metallic vapors in an electric discharge [Michelson]. A (modern) quantum theory of atomic nuclear hyperfine interactions due to the nuclear electric quadrupole moment was first proposed by Casimir in 1936 [Casimir]. In 1940, building on Casimir's model, Kellogg [Rabi] used molecular beam resonance spectroscopy on the diatomic molecules HD and $D_2$ to determine the nuclear electric quadrupole moment of the deuterium atom. In 1945, Feld and Lamb [Feld] provided a theoretical model for the nuclear electric quadrupole interaction of a herteronuclear diatomic molecule in a magnetic field as a function of rotation quantum number J. It was pointed out by Feld and Lamb that in the high-$J$ limit (i.e. large values of $J$) the interaction matrix has a diagonal asymptotic limit. A couple of years later, Foley [Foley 1] built on the results of Feld and Lamb to provide matrix elements for the nuclear electric quadrupole interactions in homonuclear diatomic molecules.

The nuclear hyperfine interaction in an atom or molecule is a small perturbation caused by higher-order electromagnetic interactions among the nuclei and electrons. The nuclear hyperfine structure is the manifestation of these interactions on the stationary-state energy spectrum. The total electromagnetic interaction among all electrons ($m$) and nuclei ($n$) in a system can be expressed [Cook], respectively, as integrated charge-charge ($\rho$) and current-current ($j$) terms in the Hamiltonian:

$$H_E = \int_n \int_m \frac{\rho_n(r_n)\,\rho_m(r_m)}{|r_m - r_n|}\,d\mathrm{v}_m\,d\mathrm{v}_n$$

(2.5)

$$H_M = -\frac{1}{c^2} \int_n \int_m \frac{j_n(r_n)\,j_m(r_m)}{|r_m - r_n|}\,d\mathrm{v}_m\,d\mathrm{v}_n$$

(2.6)

In equations 2.5 and 2.6, the charge-charge and current-current terms are functions of position ($r = (x, y, z)$) and the integrals are over all space ($d\mathrm{v} = dx \times dy \times dz$). The speed of light $c$ appears in the pre-factor of equation 2.6. The signs of $H_E$ and $H_M$ are opposite, because like charges repel each other and parallel currents attract each other. The denominator in equations 2.5 and 2.6 can be expressed as a multi-pole expansion using spherical harmonic functions $Y_i^{(j)}$:

$$|r_m - r_n| = \sum_{k=0}^{\infty} \left( \frac{4\,\pi}{2\,k+1} \right) \frac{r_n^k}{r_m^{k+1}} \sum_{k=0}^{\infty} (-1)^q \, Y_q^{(k)}(\theta_m, \, \phi_m) \, Y_{-q}^{(k)}(\theta_n, \, \phi_n)$$

$$(2.7)$$

In equation 2.7 terms other than the Coulomb interaction ($k = 0$) are considered to be part of the nuclear hyperfine interactions (i.e. $k > 0$).

   A quantum mechanical derivation [Ramsey] proceeds by applying the expansion of equation 2.7 to the Hamiltonians in equation 2.5 and 2.6. Each $k$-term in $H_E$ and $H_M$ is then re-expressed as the scalar product of two irreducible tensors of rank $k$ with one tensor containing only electron coordinates and the other only the nuclear coordinates. The Wigner-Eckart theorem is then used to derive the quantum mechanical matrix elements of the nuclear hyperfine interactions for each value of $k$. These matrix elements are by necessity given in a coupled representation so that it is necessary to diagonalize the appropriate matrix (for a given $k$-term) on each pass through a nonlinear regression analysis of the chromophore line-shape. The non-vanishing off-diagonal matrix elements in the coupled representation exist because of the selection rules that "prevent transitions between dissimilar nuclear-spin states" [Schawlow].

   From parity considerations it is possible to deduce that all odd-$k$ terms for the electric interaction and all even-$k$ terms for the magnetic interaction vanish. The leading term of the expansion (equation 2.7) for magnetic interactions is a dipole at $k = 1$. This term typically arises from the magnetic dipole moment of one nucleus interacting with the magnetic dipole created by the rotating molecule (spin-rotation interaction). The leading term of the expansion (equation 2.7) for electric interactions is a quadrupole at $k = 2$.  This electric nuclear terms account for the "static" (i.e. the illusion of a time-independent "stationary-state" consistent with a time-averaged configuration) structure of the diatomic molecule.

## 2.6 Nuclear Hyperfine Structure of Diatomic Iodine

The nuclear hyperfine structure of diatomic iodine can be modeled in several different ways, the choice of which depends on the value of $J''$ and the nature of the experiment in terms of resolution and the signal-to-noise ratio (often abbreviated as "S/N" and generally recognized as a useful indication of the detection sensitivity).

Increasingly more accurate models of the nuclear hyperfine interaction are obtained from considering higher-order effects.  The necessary higher-order corrections (for all values of $J''$) come from accounting for the next $k$-term in equations 2.7 commensurate with the resolution and sensitivity the spectrometer being used to acquire such data.  In 1971, Hanes, et al. [Bunker] transferred (and transformed) Foley's 1947 results [Foley 1] into a more modern notation.  Subsequent studies of the nuclear hyperfine structure have made use of the matrix elements given by Hanes and even found a simpler (but equivalent) expression for Hanes, et al. equation A4  [Borde 1; Borde 2]; that is, considerable amount of effort has gone into verifying the quantum mechanical derivation of nuclear hyperfine interactions to this level of description ($k = 1$ and 2).

An additional correction for small-$J$ ($J'' \lesssim 20$) values must also be taken into account for the perturbations due to the nearby rotation levels separated by two quanta in each of the ro-vibronic levels [Bunker].  The contribution to the observed intensity for transitions that result from perturbations due to these nearby rotation levels can be expected to grow as $J$ decreases, but exactly what the relative intensity, as compared to the perturbations due to the same value of $J$ as the ro-vibronic level involved in the transition, is a more difficult problem to solve.  As a result, we did not attempt to account for the effects of these nearby rotation levels in the models used in the line-shape analyses in this project.

In many experiments the primary goal is to "directly" measure the energy of the nuclear hyperfine components without regard for recording an accurate line-shape; the best relative precision frequency measurements for B–X transitions of diatomic iodine is approaching the sub-kHz level of resolution.  The increasing resolving power and sensitivity of spectrometers has provided the opportunity to investigate the nuclear hyperfine structure of diatomic iodine to higher-order corrections.  In 1955, Schwartz [Johnson] began the exploration of higher-order ($k > 2$) atomic nuclear hyperfine interactions, beyond the electric quadrupole and magnetic dipole terms.  In 1978, Broyer, et al. [Lehmann] gave a comprehensive derivation of nuclear hyperfine interactions in homonuclear diatomic molecules, including higher-order terms ($k = 1$, 2, 3, and 4), and applied these results to diatomic iodine.  In 2000, Kato, et al. [Kato] published a nuclear hyperfine-resolved atlas of diatomic iodine for the wave-number region 15,000 – 19,000 cm$^{-1}$.  In 2002, Bodermann, et al. provided interpolation formulae (accurate to about 50 kHz) for calculating the nuclear hyperfine splitting in the region 12,195 – 19,455 cm$^{-1}$ [Knöckel].  The usefulness of wavelength standards in physics can not be easily over emphasized; Kato is simply taking the next leap in resolution with regard to compiling a "diatomic iodine atlas" (i.e. an atlas for diatomic iodine) [Luc 1; Luc 2; Salami].

In many instances a high-$J$ ($J'' \gtrsim 20$) diatomic iodine line is the subject of investigation, which is often modeled using a "nuclear interaction" matrix in the diagonal asymptotic limit [Kroll; Schawlow].  This situation is particularly favorable for a line-shape analysis [Hardwick 1] since it collapses the 21 (odd $J''$) or 15 (even $J''$) nuclear hyperfine transitions into a six groups of transitions related to each other (in energy) by a single undetermined parameter, the difference in the nuclear electric quadrupole coupling constant $\Delta eQq$ between the X and B electronic states of diatomic iodine. The majority of the results presented in this dissertation (and certainly the more reliable results with regard to precision and accuracy) made use of this model based on the high-*J* asymptotic limit in the line-shape analyses. (See also Section 5.4.)

## 2.7 Frequency and Time in Spectroscopy

Modern spectroscopic methods are often broadly divided between two categories. In one category are the time-resolved methods such as fluorescent decay and optical photon-echo. These experiments explore dynamical properties of an ensemble of chromophores in a non-equilibrium state prepared by (somewhat powerful) pulses of light (a.k.a. radiation) that are comparable to or considerably shorter in duration ($\Delta t$) than the state-changing rate of the events being observed (i.e. measured). Decay of the (macroscopic) non-equilibrium state (usually back to thermal equilibrium) is monitored by changes in the fluorescence emission as a function of time.

The other category of experimental methods, which includes linear absorption and fluorescence excitation, uses nearly monochromatic ($\Delta \nu$) light sources (i.e. narrow bandwidth) that are tuned (i.e. scanned) in a relatively slow manner through a (continuous) range of frequencies to record a steady-state frequency domain spectrum (i.e. line shape). The condition of steady-state is achieved by configuring the observation time-interval (i.e. integration time-interval of detection electronics) of these frequency domain experiments to be much longer than the ensemble-average time-interval between state-changing events. Information on the dynamical properties of an ensemble of chromophores can be obtained by an analysis of the recorded line shape from a steady-state frequency domain spectrum.

A couple of important qualitative relationships can be discerned based on the properties of Fourier transforms and Dirac delta-functions [Kauppinen; Butkov]. At one (purely conceptual) limit is the perfectly monochromatic ($\Delta \nu = 0$) radiation source that appears (at a fixed location in space) to oscillate at a single frequency $v$ for an infinite period of time ($\Delta t \rightarrow \infty$). At the other limit is the infinitely short pulse ($\Delta t = 0$) of light that contains all the frequencies of the electromagnetic spectrum ($\Delta v \rightarrow \infty$, or perhaps somewhat more precisely, $v = 0$ to $\infty$). Of course, in practice, neither limit is realizable; bandwidth and duration are finite quantities in any real world situation.

It is perhaps also worth noting that the technological development of radiation sources (and thus spectrometers) has been (steadily) advancing toward these two limits, and such advances are generally dependent on understanding the various processes (e.g. the state-changing events) that take place in the microscopic domain.

## 2.8 Line-Shape Model for Steady-State Frequency Domain Spectra

A steady-state high resolution frequency domain absorption spectrum of a large ensemble of chromophores in the gas phase (i.e. an absorption medium) contains distribution-like features that have a finite width and are commonly referred to as (spectral) lines. To a first-order approximation, a chromophore in the gas phase spends most of its time relatively well-isolated from external influences (e.g. forces due to other atomic systems) and so some portion of an observed spectral line represents discrete transitions between two ro-vibronic quantum states; these two quantum states – a lower energy level and an upper energy level – correspond to stationary-state solutions of the time-independent Schrödinger equation. Furthermore, observation of such spectral features in a (typical) linear absorption experiment is predicated on the existence of these ro-vibronic transitions. It is important to remember, though, that the "object" being observed in steady-state high resolution frequency domain absorption spectrum is the radiation field (intensity) after traversing the absorption medium.

These spectral features can be modeled as being composed of a homogeneous (Lorentzian) component and an inhomogeneous (Guassian) component. A common method for modeling an individual line (for its line-shape at fixed number density, temperature, and radiation source intensity) is through the convolution (or de-convolution) of these homogeneous and inhomogeneous components to form a Voigt profile (i.e. distribution function) [Bernath]. Such modeling is essentially an effort to mimic the manner in which Nature "convolves" independent microscopic domain processes to provide a composite line shape for an observer in the macroscopic domain. The Voigt distribution was used for modeling the observed (i.e. recorded) line-shapes during the course of this project; see also Sections 5.2 and 5.3 for a more mathematically oriented description of this convolution method and its use in the context of linear absorption (i.e. the Beer-Lambert law).

When seeking to understand the measurement process, the concept of convolution appears to provide a vital link between the microscopic and macroscopic domains. But it is an area for which there may not yet be a firm consensus on the appropriate (i.e. fundamental) method to employ, so that there are variations on this theme of constructing a model line-shape that accounts for the "mixing" of homogeneous and inhomogeneous contributions, such as the Rautian and Galatry line-shape functions, as well as the generalized complex-valued form of the Voigt distribution function [Lepère].

The homogeneous contribution to an observed high-resolution absorption line shape is that portion due to incoherent (i.e. stochastic or random) processes that have the same effect on all of the microscopic constituents (e.g. molecules in a given ro-vibronic state) in the macroscopic absorption medium. The homogeneous portion of the (frequency domain) line-shape is generally considered to have the functional form of a Lorentz (a.k.a. Cauchy) distribution. The mathematically-oriented presentation in Section 6.3 on the random nature of the state-changing process and material presented in the next three sections of this chapter suggests the following relatively brief summary of the contributions to an observed line shape.

First, there is a state-changing process that can be described as stimulated photon absorption and emission (i.e. energy exchange between the chromophore and the radiation field). Absorption and emission of photons is an energy exchange process between the chromophore and the radiation field, exposing the particle-nature of the radiation field in the amplitude of the observed line shape. However, it is also widely recognized that this state-changing process contributes to the width of the observed line shape (which implies that it is also a phase-changing event). This stimulated state-changing process can be further divided into those state-changes that involve a collision with a buffer gas ($c_{abs}$ and $c_{em}$) and those that do not ($r_{abs}$ and $r_{em}$), where the parenthetical comments refer to the associated ensemble-average rates of the processes depicted in Figure 2.5. Of course, it is possible to propose a mechanism of pressure-independent photon absorption and emission of photons that is due to internal collisions of the constituents of an atomic system (e.g. its electrons). Also, it is perhaps still not entirely clear if the direction of propagation of the radiation field (if any) changes during those state-changing processes that involve stimulated emission of a photon, but it is common to presume that the propagation directions are (at a minimum) perfectly parallel to the stimulating radiation field.

Second, there is a state-changing process that can be described as a phase-change of the radiation field, which contributes only to the width of the observed line shape and not to its amplitude (or perhaps more precisely, its area); see also Section 2.10. While a collision-induced phase-change due to a collision between a chromophore and buffer gas are relatively easy to envisage, the notion of collisions between the internal constituents of the chromophore (e.g. its electrons) could be used to recognize a pressure-independent contribution to this state-changing process. The phase-change state-changing process can be thought of as exposing the wave-nature of the radiation field, except that these events occur randomly in time and space (as opposed to coherently) and so do not give rise to (constructive and destructive) interference patterns. (c.f. Bohr Complementarity principle.)

Third, especially in a molecule, there are state-changes that involve emission of photons at energies that are different from (and typically smaller than) those of the incident radiation field. And fourth, there is a spontaneous emission process, which radiates in random directions away from the (typically well collimated) incident light beam. In what manner (if at all) do these last two cases appear in an observed line shape? Is the third case much like the first in which the stimulated emission is parallel to the stimulating radiation field? In the fourth case, by what mechanism would the total loss of a photon from the well collimated radiation field affect the observed line shape? The tacit assumption adopted throughout this project is that all such processes will, at a minimum, contribute to the shape (i.e. width and line-center shift) of the radiation field transmitted through an absorption medium [Barut 1].

The inhomogeneous contribution to the observed (frequency domain) line-shape is the result of processes that are not the same for all of the constituents in the absorption medium. For molecules in the gas phase at a relatively low pressure, this contribution appears to be described quite well by Doppler broadening, for which a quantitative model is obtained by consideration of the Maxwell-Boltzmann distribution of velocities, which can be related to the distribution of resonant (line-center) frequencies $\omega_0 = 2\nu_0$ in the gas sample for thermal equilibrium at a temperature $T$. In the laboratory frame the radiation frequency is un-shifted, and the distribution of molecular velocities parallel to the propagation direction of the radiation field can be modeled by a shift in the resonant frequency such that the resonant frequency is given by $\omega_0' = (1 \pm v/c)\,\omega_0$ in the reference frame of the molecules, where v is the speed of a molecule along the propagation direction of the radiation field, $c$ is the speed of light, and $\omega_0'$ is the frequency of the radiation field in the laboratory reference frame. The inhomogeneous width of the line-shape is then found to have the functional form of a Gaussian (Normal) distribution [Bernath]. It is also common (in time domain models) for this width to be referred to as a decay time-interval $T_2^*$, which can then be expressed as the decay rate $1/T_2^*$ [Allen 2]. Conceptually, the inhomogeneous decay rate $1/T_2^*$ can be thought of as the rate at which the line-center frequencies in the ensemble of transition moments (i.e. large collection of chromophores) de-phase (i.e. precess) relative to each other during the period of "interaction-free" translational motion between all types of collision events, including those collision events that do not result in an appreciable, or perhaps perceptible, state-change of the radiation field.

In the case of transitions in the visible (a.k.a. optical) portion of the electromagnetic spectrum the inhomogeneous contribution (Doppler/Gaussian) to the line-shape is generally much larger than the homogeneous (Lorentzian) portion for relatively low total gas pressure (up to a few tens of torr in the case of diatomic iodine). The homogeneous width is typically observed to increase linearly as the pressure of buffer gas is increased. At about 100 torr of buffer gas, the homogeneous width of diatomic iodine is comparable in magnitude to the constant Doppler width. In the case of diatomic iodine, the theoretical Doppler width (full width at half maximum intensity) is 341 MHz at room temperature (292 K) and transition energy of 14,818 $cm^{-1}$ (corresponding to a radiation wavelength of approximately 675 nm); see also Section 5.2 for more details. (However, due to the rather large nuclear hyperfine splitting, a high resolution spectrum of a well isolated ro-vibronic transition at a pressure of about 0.2 torr has an (observed) Doppler-broadened width of roughly 600 MHz.)

## 2.9 The Two-Level System Model

The basic two-level system model offers a description of energy exchange processes between a chromophore and radiation field. There are three distinct processes shown in Figure 2.5. The two processes on the left-hand side of Figure 2.5 – stimulated absorption and emission of radiation and spontaneous emission (of photons) – do not depend on collisions with a buffer gas and so do not depend on the pressure (a.k.a. pressure-independent). The right-hand side of Figure 2.5 depicts the collision-induced stimulated photon absorption and emission process and it does depend on the pressure (a.k.a. pressure-dependent). As usual, the stimulated processes are presumed to be "coherent" with the incident radiation field: it is presumed (with or without a collision event) that the photon emitted during stimulated emission has the same phase as the radiation field involved in the stimulation process, and that it propagates in the same direction as this stimulating field; it is presumed that a stimulated absorption event has a strong orientation effect on the transition moment of the chromophore. And it is presumed that the spontaneous emission events occur in completely random spatial directions and have no phase relationship with the (typically well collimated) radiation field.

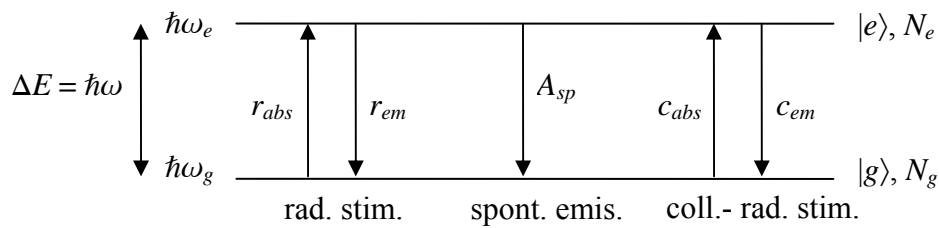

Figure 2.5 The two-level system model depicting three state-changing processes (from left to right): pressure-independent stimulated absorption and emission of photons ("rad. stim."), spontaneous emission of photons ("spont. emis."), and pressure-dependent (i.e. collision-induced) stimulated absorption and emission of photons ("coll.-rad. stim."). On the left-hand side of this figure are the energies of the stationary-states ($\hbar\omega_g$ and $\hbar\omega_e$) and the transition (and photon) energy $\Delta E$. On the far right-hand side of this figure is the "ket" designation of the steady-states ($|g\rangle$, and $|e\rangle$), and the number densities of these two states ($N_g$ and $N_e$) in the ensemble (integrated over all frequencies $\omega$). The ensemble-average photon absorption and emission rates of the three processes are given (from left to right) by $r_{abs}$, $r_{em}$, $A_{sp}$, $c_{abs}$, and $c_{em}$, with the subscripts "*abs*" and "*em*" respectively indicating absorption and emission, and "sp" for spontaneous emission.

Upon turning-off the radiation field, the spontaneous emission process will continue to operate in its usual manner. The photons emitted by spontaneous emission might provide the photons for collision-induced stimulated emission, or (perhaps) an entirely different (but similar) collision-induced emission process could be operating.

With respect to the spontaneously emitted photons, the pressure-independent stimulated absorption and emission and the collision-induced stimulated absorption processes could still be active, but their effects are expected to be rather negligible. Whatever the events that lead to decay, the propagation of photons will appear to occur in random spatial directions. The ensemble of chromophores will decay towards a distribution of states consistent with thermal equilibrium; at optical frequencies this equilibrium state has nearly all of the chromophores in the lower energy level (designated as $|g\rangle$ in Figure 2.5).

In time-resolved experiments using relatively short duration pulses of light (at or near visible frequencies), the non-equilibrium state is presumed to be (coherently) prepared by the pressure-independent stimulated absorption process (given by the rate $r_{abs}$ on the left-hand side of Figure 2.5). The ensemble-average decay rate of the prepared collection of chromophores in the upper energy level is often characterized symbolically as $1/T_1$, which can be sub-divided into a contribution from the (pressure-independent) spontaneous emission process and another from the (pressure-dependent) collision-induced emission process: $1/T_1 = A_{sp} + c_{em}$. An example of an experiment that can measure the random events that contribute to $1/T_1$ is fluorescent decay [Paisner].

The two-level system model is fairly ubiquitous and is generally regarded as being rather useful. As a complement to this point of view, an outline of a kinetic model based on ensemble-average transition rates for steady-state linear absorption will be presented in Section 2.11, one which might prove useful in the measurement of transition moments. However, the two-level system model described in this section only accounts for energy exchange between the chromophore in one particular total stationary-state (i.e. ro-vibronic and translational) and the radiation field. Even though a collision-induced photon absorption and emission mechanism (i.e. a cause) has been included, this model is not able to account for energy exchange due to collisions with a buffer gas. Furthermore, the two-level system model does not consider in a more general manner the nature of collision-induced phase-changes of the radiation field, which will be addressed in the next section by the addition of two more "levels" to this model.

## 2.10 The Four-Level System Model

In order to account for a process that is perhaps more fundamentally related to the wave-nature of the radiation field (i.e. phase-change, as opposed to energy exchange in the particle-nature point of view), two levels have been added to the two-level system model to form the basis for a four-level model. The four levels are depicted in Figure 2.6. To simplify the model, the $|g'\rangle$ and $|e'\rangle$ states are here taken as being respectively in the same "ro-vibronic" (or electronic) states as the $|g\rangle$ and $|e\rangle$ states, but representing different translational states. The transition from $|g\rangle$ to $|g'\rangle$ or $|e\rangle$ to $|e'\rangle$ (or vice versa) occurs through a collision with a buffer gas, which can be envisaged as resulting in a change in the velocity (in a vector sense) of the chromophore.

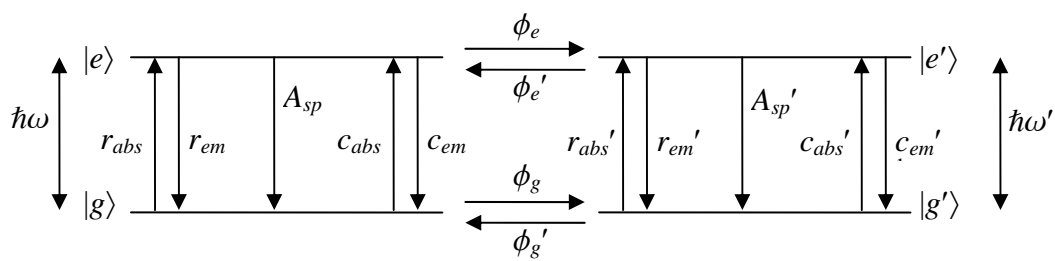

Figure 2.6 The four-level model depicting phase-changing state-changes between $|g\rangle$ and $|g'\rangle$ and transitions between $|e\rangle$ to $|e'\rangle$ due to collisions with a buffer gas. Each pair of states ( $|g\rangle$, $|e\rangle$ ) and ( $|g'\rangle$, $|e'\rangle$ ) satisfies the two-level system model of Figure 2.5. The ensemble-average phase-changing rates (of the radiation field) are given by $\phi_g$, $\phi_g'$, $\phi_e$, and $\phi_e'$. In general, it is expected that $\omega \neq \omega'$.

It can be envisaged that at some point in time and space during the course of a collision (perhaps in a deterministic sense) the phase of the radiation is "suddenly" disrupted (i.e. changed to a meaningful extent in an almost discontinuous manner). This phase-change of the radiation field is generally modeled as involving a significant change in the orientation of the transition moment, such that (some form of) an interaction between the chromophore and radiation field is disrupted. The change in velocity of the chromophore (due to a collision event) also means that its transition energy is slightly altered (relative to the laboratory frame due to the Doppler Effect); i.e. $\omega \neq \omega'$. As Figure 2.7 indicates, by including (more explicitly) the possibility for velocity-changes during a collision event, the four-level model, even in this limited form, makes more transparent the possible range of pressure-dependent processes.

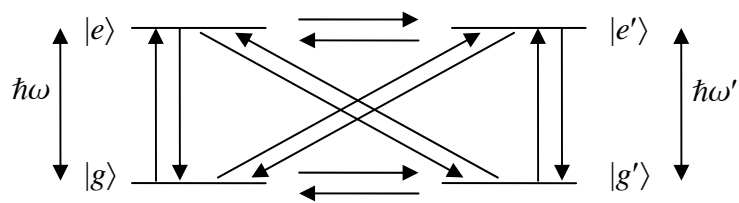

Figure 2.7 The four-level model depicting (some of the) possible paths for the state-changing process.

In time-resolved (non-equilibrium) experiments (using relatively short duration pulses of light), the total decay rate due to homogeneous processes can be expressed (i.e. modeled) as the sum of two ensemble-average rates, the photon emission rate for fluorescent decay $1/T_1$ (mentioned at the end of the last section) and the rate of collision-induced (i.e. pressure-dependent) phase-changes of the transition moment $\phi_e$ for $|e\rangle \rightarrow |e'\rangle$ transitions, which can be written symbolically as $1/T_2' = 1/T_1 + \phi_e$. An example of an experiment that measures (i.e. observes) this decay time is the optical photon echo [Dantus].

## 2.11 Steady-State Kinetic Model

An important objective in building two-level and four-level system models is to decipher the information content encoded in an observed steady-state linear absorption line shape (i.e. spectrum). In this section we will seek to justify a kinetic model for the two-level system and then explore its implications, especially with regard to the underlying dynamics of the observed steady-state linear absorption line shape. This model differs slightly from the usual "$A$ and $B$ Coefficient" model [Einstein], but perhaps we should keep in mind that the model developed by Einstein was intended for a perfectly isolated (but non-existent) black-body radiator based on the ratio of (single temperature $T$ thermal equilibrium) Boltzmann distributions for the two energy levels (in the two-level system model). Perhaps it is fair (or reasonable) to claim that the $A$ and $B$ Coefficient model describes a realm of steady-state state-changing behavior from a point of view that is not quite consistent with the steady-state state-changing condition created by directing a well-collimated external source of relatively low-intensity, narrow bandwidth radiation to pass through an absorption medium.

A meaningful goal of any two-level system model (e.g. Figures 2.5) is to derive (or re-derive) the Beer-Lambert law from the microscopic point of view. The Beer-Lambert law is an empirical observation about steady-state linear absorption of radiation by a large ensemble of chromophores, often expressed in the differential form as $dI(z)/dz = -\kappa I(z)$, which leads to the integrated form:

$$I((\omega_0 - \omega_0''), z) = I((\omega_0 - \omega_0''), 0) \exp(-\kappa z)$$

$$\equiv I(z) = I(0) \exp(-\kappa z) \qquad (2.8)$$

In equation 2.8, $I(z) \equiv I((\omega_0 - \omega_0''), z)$ is the line-integrated radiation field intensity (i.e. the total intensity of the radiation field across all frequencies $\omega$) after passing though an absorption medium of length $z$; the radiation field initially incident on the absorption medium (at $z = 0$) has a line-integrated intensity $I(0) \equiv I((\omega_0 - \omega_0''), 0)$; and $\kappa$ is the absorption coefficient (a characteristic constant of the absorption medium at a given pressure and temperature). The radiation field line-center frequency $\omega_0''$ is often the same as the observation frequency (as was the case in this project). The intensity of the radiation field is thus properly described by the difference in the chromophore line-center frequency $\omega_0$ and the observation frequency $\omega_0''$.

Intensity has units of energy per unit area per unit time. One of the hallmarks of linear absorption (worthy of considerable emphasis) is that the ratio $I((\omega_0 - \omega_0''), z) \div I((\omega_0 - \omega_0''), 0)$ is constant for a broad range of initial intensities $I((\omega_0 - \omega_0''), 0)$. Also, without loss of generality, the following descriptions will use the simplifying assumption that the incident radiation field intensity $I((\omega_0 - \omega_0''), 0)$ is equal to a constant value across a broad range of radiation field line-center frequencies $\omega_0''$ (i.e. the ideal case of the radiation field being independent of the absorption medium); see also Section 5.3.

Whatever the experimental conditions may be, and whatever state-changing processes might be envisaged as operating, a steady-state condition is achieved in the two-level system model when the differential change in the population densities of the ground and excited states with respect to time has vanished:

$$\frac{dN_e}{dt} = -\frac{dN_g}{dt} = 0$$

(2.9)

In equation 2.9, the total number densities for a particular radiation field line-center frequency $\omega_0''$ (a.k.a. line-integrated number density) for the two-level system model are given by $N_g$ and $N_e$.

From the point of view of the chromophore, application of equation 2.9 (for the two-level system model of described in Section 2.9 and Figure 2.5), where the sum of ensemble-average stimulated photon absorption and emission rates are given respectively by $k_{abs} = r_{abs} + c_{abs}$ and $k_{em} = r_{em} + c_{em}$, gives:

$$-A_{sp}\,N_e \;-\; k_{em}\,N_e \;+\; k_{abs}\,N_g \;=\; 0 \tag{2.10}$$

From the point of view of the externally applied (and typically well collimated) radiation field, the observed change in the total (line-integrated) intensity as a function of path length can be expressed as:

$$\frac{d\,I(z)}{d\,z} \;=\; (-A_{sp}\,N_e + k_{em}\,N_e - k_{abs}\,N_g)\,\hbar\,\omega_0 \tag{2.11}$$

The observed line-center frequency $\omega_0$ is being used (on the right-hand side of equation 2.11) on the pretense that the energy of a photon ($E = \omega\hbar$) does not change to an appreciable extent across a relatively narrow frequency-interval of the Doppler and collision broadened radiation field line shape at optical frequencies.

Substitution of equation 2.10 into equation 2.11 gives:

$$\frac{d\,I(z)}{d\,z} \;=\; -2\,\hbar\,\omega_0\,A_{sp}\,N_e \;=\; 2\,\hbar\,\omega_0\,(k_{em}\,N_e - k_{abs}\,N_g) \tag{2.12}$$

The next step is to reconstruct equation 2.12 from the point of view of the microscopic processes that give rise to $I(z) \equiv I((\omega_0 - \omega_0''),\, z)$ and $N_e \equiv N_e(z) \equiv N_e((\omega_0 - \omega_0''),\, z)$. This reconstruction will be achieved through the convolution of independent processes (i.e. homogeneous and inhomogeneous line broadening mechanism) [Demtröder, Butkov]. Two successive convolutions will be performed, so it is useful to note that the order in which convolutions are performed can be arbitrarily chosen. In the language of mathematics, the convolution procedure obeys the commutative, associative, and distributive properties. The first two properties are relevant to what follows in this section, namely that $f * g = g * f$ and $f * (g * h) = (f * g) * h$ [Kauppenin].

The intensity of the radiation field per unit bandwidth interval $I_\omega(((\omega_0 - \omega_0'') - \omega), z)$ can be described by the following distribution function:

$$I_\omega(((\omega_0 - \omega_0'') - \omega), z)$$

$$= I((\omega_0 - \omega_0''), z) \, \text{F}_\text{R} \, g_\text{R}((\omega_0 - \omega_0'') - \omega)$$

$$\equiv I(z) \, \text{F}_\text{R} \, g_\text{R}((\omega_0 - \omega_0'') - \omega) \qquad (2.13)$$

Equation 2.13 describes the radiation field intensity per unit bandwidth interval relative to the difference of the observed line-center frequency $\omega_0$ and the radiation field line-center frequency $\omega_0''$ as a fraction of the total radiation field intensity $I((z) \equiv I((\omega_0 - \omega_0''), z)$ (across all frequencies $\omega$) at a particular position $z$ in the absorption medium. The distribution function $g_\text{R}((\omega_0 - \omega_0'') - \omega)$ has a normalization factor given by $\text{F}_\text{R}$:

$$\text{F}_\text{R} \int_0^\infty g_\text{R}((\omega_0 - \omega_0'') - \omega) \, d\omega = 1$$

$$(2.14)$$

Standard (or classical) treatments of electromagnetic wave propagation indicate that the cycle-averaged radiation field intensity $I((z) \equiv I((\omega_0 - \omega_0''), z)$ is related to the electric field strength $E(z) \equiv E((\omega_0 - \omega_0''), z)$ by:

$$I(z) = \frac{1}{2} c \eta \epsilon_0 |E(z)|^2$$

(2.15)

In equation 2.15, $c$ is the speed of light in a vacuum, $\eta$ is the index of refraction, and $\epsilon_0$ is the vacuum permittivity.

The anticipated steady-state probability $|c_e(\omega_0 - \omega)|^2$ of finding a given chromophore in the excited-state contains a contribution from the square of the radiation field strength $|E((\omega_0 - \omega_0''), z)|^2 \equiv |E(z)|^2$ (e.g. see equation 7.36). Equation 2.15 then says that this contribution is directly proportional to the radiation field intensity $I(z) \equiv I((\omega_0 - \omega_0''), z)$. However, a more realistic description of a radiation field acknowledges the nonexistence of such fields being perfectly monochromatic, which is achieved with a distribution function of finite width and having units of intensity per unit bandwidth interval (e.g. equation 2.13).

Likewise, the distribution for the excited-state probability $|c_e(\omega_0 - \omega)|^2$ is converted to a probability density characterizing the excited-state probability per unit bandwidth interval. The portion pertaining to the radiation field has been factored out of $|c_e(\omega_0 - \omega)|^2$, so that the remaining core portion of this distribution ( $|c_e'(\omega_0 - \omega)|^2$ ) can be expressed as:

$$|c_e'(\omega_0 - \omega)|^2 = \frac{2 \mu^2}{c \eta \epsilon_0 \hbar^2} F_e \, g_e(\omega_0 - \omega)$$

(2.16)

In equation 2.16, the distribution function $g_e(\omega_0 - \omega)$ is normalized by $F_e$:

$$F_e \int_0^\infty g_e(\omega_0 - \omega)\, d\omega = 1$$

(2.17)

The transformation of the descriptions of the radiation field and excited-state probability into probability density functions (of finite widths that describe the associated quantities per unit bandwidth interval) then leads to the steady-state probability $|c_e''((\omega_0 - \omega_0'') - \omega)|^2$ of finding a given chromophore in the excited-state per unit bandwidth interval:

$$|c_e''((\omega_0 - \omega_0'') - \omega)|^2$$

$$= \int_0^\infty |c_e'(\omega_0' - \omega)|^2\, I(((\omega_0 - \omega_0'') - \omega_0'), z)\, d\omega_0'$$

$$= \frac{2\,\mu^2\, I(z)}{c\,\eta\,\epsilon_0\,\hbar^2} \int_0^\infty F_e\, g_e(\omega_0' - \omega)$$

$$\times\; F_R\, g_R((\omega_0 - \omega_0'') - \omega_0')\, d\omega_0'$$

$$\equiv \frac{2\,\mu^2\, I(z)}{c\,\eta\,\epsilon_0\,\hbar^2}\, F_H\, g_H((\omega_0 - \omega_0'') - \omega)$$

(2.18)

In equation 2.18, $\mu$ is the electronic dipole transition moment. The subscript "H" on $g_H((\omega_0 - \omega_0'') - \omega)$ is in anticipation of this distribution function being commensurate with the homogeneous contribution to the observed line shape.

The normalization factors $F_R$ and $F_e$ in equations 2.14 and 2.17 require considerable care when proceeding through this analysis. The normalization factor obtained by the convolution of distribution functions is in general a function of the normalization factors of the convolved components. In equation 2.18, the explicit expression of this point can be given as $F_H = F_H(F_R, F_e)$. For the case that the $g_R$ and $g_e$ are both described by Lorentzian distributions, it can be readily shown that the normalization factor is given by $F_H = F_R + F_e$, in which $F_R$ and $F_e$ are simply related to the full width at half-maximum height of $g_R(\omega_0 - \omega_0'') - \omega)$ and $g_e(\omega_0 - \omega)$, respectively [Loudon 1; Bernath; see also equations 5.3 and 5.4].

The second convolution is between the result of the convolution in equation 2.18 and the inhomogeneous contribution to the observed line shape, in this case taken to be well described by the Doppler profile. In a typical gas cell the total number density (of chromophore) per unit bandwidth interval $N(\omega_0 - \omega)$ is distributed in the frequency domain according to the normalized Doppler profile $F_D \, g_D(\omega_0 - \omega)$ [Bernath]:

$$N(\omega_0 - \omega) = N \, F_D \, g_D(\omega_0 - \omega)$$

$$= \frac{2\pi N}{\omega_0} \left( \frac{m c^2}{2\pi k_B T} \right)^{1/2} \exp\left( -\frac{m c^2}{2 k_B T} \left( \frac{\omega_0 - \omega}{\omega_0} \right)^2 \right)$$

$$(2.19)$$

In equation 2.19, $N$ is the total number density of the chromophore (and $N = N_g + N_e$), $m$ is the mass of the chromophore, $c$ is the speed of light in a vacuum, $T$ is the temperature, and $k_B$ is the Boltzmann constant; $g_D$ is given by the exponential factor [Bernath].

The excited-state number density per unit bandwidth interval $N_e((\omega_0 - \omega_0'') - \omega)$ at a particular radiation field line-center frequency $\omega_0''$ is then given by the convolution of equations 2.18 and 2.19:

$$N_e((\omega_0 - \omega_0'') - \omega)$$

$$= \int_0^\infty |c_e''((\omega_0 - \omega_0'') - \omega_0')|^2 \, N(\omega_0' - \omega) \, d\omega_0'$$

$$= \frac{2 \, N \, \mu^2 \, I(z)}{c \, \eta \, \epsilon_0 \, \hbar^2} \int_0^\infty F_H \, g_H((\omega_0 - \omega_0'') - \omega_0')$$

$$\times F_D \, g_D(\omega_0' - \omega) \, d\omega_0'$$

$$\equiv \frac{2 \, N \, \mu^2 \, I(z)}{c \, \eta \, \epsilon_0 \, \hbar^2} F_V \, g_V((\omega_0 - \omega_0'') - \omega)$$

$$(2.20)$$

In equation 2.20, the subscript "V" on $g_V((\omega_0 - \omega_0'') - \omega)$ is in anticipation of this distribution function being well modeled as a Voigt profile. And it is worth noting that the normalization factor for the Voigt profile is dependent on the underlying distribution functions: $F_V = F_V(F_H, F_D)$. In this project, for which the radiation field line shape and temperature were held constant, changes in buffer gas pressure affect the width of $g_e((\omega_0 - \omega))$ and thus $F_e$., which in turn affects the normalization factor $F_V$ of the Voigt profile; see also Section 5.3.

The more general form of equation 2.12 for per-bandwidth-interval distributions (e.g. probability density functions) is given by:

$$\frac{d\!I_\omega(((\omega_0 - \omega_0'') - \omega), z)}{dz} = -2\,\hbar\,\omega_0\,A_{sp}\,N_e((\omega_0 - \omega_0'') - \omega)$$

$$(2.21)$$

Using equations 2.13 and 2.14, integration of the left-hand side of equation 2.21 across all frequencies $\omega$ gives:

$$\int_0^\infty \left( \frac{d\!I_\omega(((\omega_0 - \omega_0'') - \omega), z)}{dz} \right) d\omega$$

$$= \frac{d\!I((\omega_0 - \omega_0''), z)}{dz} \int_0^\infty F_R\, g_R((\omega_0 - \omega_0'') - \omega)\, d\omega$$

$$= \frac{d\!I((\omega_0 - \omega_0''), z)}{dz} \equiv \frac{d\!I(z)}{dz}$$

$$(2.22)$$

Using equation, 2.20, integration of the right-hand side of equation 2.21 across all frequencies $\omega$ gives:

$$-2\,\hbar\,\omega_0\,A_{sp}\int_0^\infty N_e((\omega_0\,-\,\omega_0'')\,-\,\omega)\,d\!\!\!/\,\omega$$

$$=\,-2\,\hbar\,\omega_0\,A_{sp}\,N_e$$

$$=\,-\frac{4\,\omega_0\,A_{sp}\,N\,\mu^2\,\mathrm{F_V}\,I(z)}{c\,\eta\,\epsilon_0\,\hbar}\int_0^\infty g_\mathrm{V}((\omega_0\,-\,\omega_0'')\,-\,\omega)\,d\!\!\!/\,\omega$$

$$\equiv\,-\mathrm{K}\,I(z)\int_0^\infty g_\mathrm{V}((\omega_0\,-\,\omega_0'')\,-\,\omega)\,d\!\!\!/\,\omega$$

$$(2.23)$$

Substitution of equations 2.22 and 2.23 into equation 2.21, followed by rearrangement and integration yields the Beer-Lambert law:

$$I(z)\,=\,I(0)\exp\!\left(-\mathrm{K}\left(\int_0^\infty g_\mathrm{V}((\omega_0\,-\,\omega_0'')\,-\,\omega)\,d\!\!\!/\,\omega\right)z\right)$$

$$(2.24)$$

Equation 2.24 indicates that for each value of the radiation field line-center frequency $\omega_0''$ the distribution of transmitted light follows a Voigt-like profile $g_\mathrm{V}\,((\omega_0-\omega_0'')-\omega)$ defined in equation 2.20. As the radiation field line-center frequency $\omega_0''$ is scanned across the absorption profile, the integration of this underlying Voigt-like distribution across all frequencies $\omega$ is traced out in the recorded spectrum; recall that $I((z)\equiv I((\omega_0-\omega_0''),z)$ and $I((0)\equiv I((\omega_0-\omega_0''),0)$.

Not shown explicitly in the preceding developments (in equations 2.18 through 2.24) is a convolution of the detector response with the transmitted radiation field. To the extent that a detector has a flat response across a relatively broad spectral region that includes and is much broader than the spectral region of interest (i.e. the transition line shape encoded in the transmitted radiation field), the detector response $g_d(\omega - \omega_d)$ can be factored out of the corresponding convolution integral in the (approximate) form $\gamma \times (\Delta\omega_d)^{-1}$, where $\Delta\omega_d$ is the bandwidth of the detector, $\gamma$ is a dimensionless instrument function, and $\omega_d$ is the center frequency of the detector. Integration of the left-hand side of equation 2.21 then gives:

$$\int_0^\infty \left( \frac{d\,I_\omega((\omega - \omega_0'') - \omega),\ z)}{d\,z} \right) d\,\omega$$

$$= \int_0^\infty \left( \frac{d}{d\,z} \int_0^\infty g_d(\omega - \omega_d)\, I(z)\, F_R\, g_R((\omega_0 - \omega_0'') - \omega)\, d\,\omega \right) d\,\omega$$

$$\cong \frac{\gamma}{\Delta\omega_d}\, \frac{d\,I(z)}{d\,z}\, \left( \int_0^{\Delta\omega_d} d\,\omega \right) \int_0^\infty F_R\, g_R((\omega_0 - \omega_0'') - \omega)\, d\,\omega$$

$$= \frac{\gamma}{\Delta\omega_d}\, \frac{d\,I(z)}{d\,z}\, (\Delta\omega_d)\, 1\ =\ \gamma\, \frac{d\,I(z)}{d\,z}$$

$$(2.25)$$

Integration of the right-hand side of equation 2.21 treats the Voigt profile $g_V((\omega_0 - \omega_0'') - \omega)$ as a Dirac-delta function, which filters out its value at $\omega = \omega_0 - \omega_0''$ (i.e. $g_V(\omega_0 - \omega_0'')$):

$$-2\,\hbar\,\omega_0\,A_{sp}\,N_e((\omega_0 - \omega_0'') - \omega)$$

$$= -2\,\hbar\,\omega_0\,A_{sp}\int_0^\infty\left(\int_0^\infty g_d(\omega - \omega_d)\,N_e((\omega_0 - \omega_0'') - \omega)\,d\omega\right)d\omega$$

$$\cong -K\,\frac{\gamma\,I(z)}{\Delta\omega_d}\left(\int_0^{\Delta\omega_d}d\omega\right)\int_0^\infty g_V((\omega_0 - \omega_0'') - \omega)\,d\omega$$

$$\cong -K\,\frac{\gamma\,I(z)}{\Delta\omega_d}\,(\Delta\omega_d)\,g_V(\omega_0 - \omega_0'')$$

$$= -K\,\gamma\,I(z)\,g_V(\omega_0 - \omega_0'')$$

$$(2.26)$$

Combining equations 2.25 and 2.26 (and recalling that $I(z) \equiv I((\omega_0 - \omega_0''), z)$) leads to the more familiar form of the Beer-Lambert law given by equation 2.12 (as opposed to equation 2.21):

$$\gamma\,\frac{d\,I((\omega_0 - \omega_0''),\ z)}{d\,z} = -2\,\hbar\,\omega_0\,A_{sp}\,\gamma\,N_e(\omega_0 - \omega_0'')$$

$$= -K\,\gamma\,I((\omega_0 - \omega_0''), z)\,g_V(\omega_0 - \omega_0'')$$

$$(2.27)$$

In equation 2.27, the observation frequency is given by the radiation field line-center frequency $\omega_0''$, which corresponds to the experimental configuration used in this project. Rearrangement and integration of equation 2.27 then leads to the often used form of the Beer-Lambert law in high-resolution linear absorption experiments (c.f. equation 2.8), especially for the task of performing a line shape analysis:

$$I((\omega_0 - \omega_0''), z) = I(0) \exp(-K\, g_{\mathrm{V}}(\omega_0 - \omega_0'')\, z) \tag{2.28}$$

See also Sections 5.2 and 5.3, especially equation 5.3, and recall that $\omega = 2\pi\nu$.

As equation 2.24 (or equation 2.28) indicates, one reason for considering the details of deriving the Beer-Lambert law from the microscopic point of view is to relate measurements of line shape and line-integrated radiation field intensity to the transition moment $\mu$ of the chromophore. However, such measurements are not pertinent to the results presented in this dissertation. Another reason for deriving a model of the Beer-Lambert law based on microscopic processes of equations 2.10 and 2.11 is to gain some confidence in their validity.

The main reason for pursuing the above developments in this section has to do with a central question in spectroscopy: what is the information content encoded in an observed line shape? Equation 2.12 (or equation 2.21) indicates that the observed line shape will be influenced by both the absorption and emission processes [Barut 1]. Furthermore, the steady-state kinetic model developed in this section is consistent with classical equilibrium thermodynamics; such considerations lead to results (described below) that will be important in the analysis of state-changing processes later in this dissertation, specifically Section 6.3 and Chapter VII.

If we tacitly recognize that the kinetic model for the two-level system presented so far accounts for an essential portion of the physics of linear absorption, then comparison of equation 2.10 to the Boltzmann distribution gives:

$$\frac{N_g}{N_e} = \frac{A_{sp} + k_{em}}{k_{abs}} = \exp\left(\frac{\hbar\,\omega_0}{k_B\,T}\right) > 1$$

$$\implies A_{sp} + k_{em} > k_{abs}$$

$$(2.29)$$

The temperature $T$ appearing in equation 2.29 is a characteristic temperature of the steady-state state-changing process, which is not the same as the usual sense of this parameter (as used throughout this dissertation) for describing the center-of-mass motion (a.k.a. translation) of an atomic system.

The time domain view of the inequality in equation 2.29 is depicted in Figure 2.8. This inequality indicates that the ensemble-average time-interval $\Delta t_{abs}$ between $|g\rangle \to |e\rangle$ (photon absorption) state-changing events is longer than the time-interval $\Delta t_{em}$ between $|e\rangle \to |g\rangle$ (photon emission) state-changing events; in Chapters VI and VII, the notation for this inequality on the time-intervals between state-changing events will appear as $\tau_{em} \neq \tau_{abs}$.

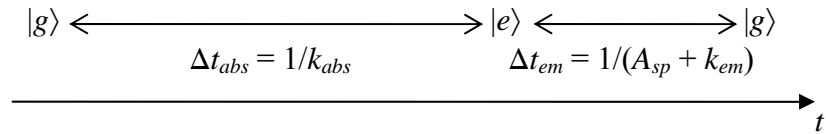

Figure 2.8 Ensemble-average state-changes in the time domain depicting the inequality of equation 2.17 in which $\Delta t_{abs} > \Delta t_{em}$. The ensemble-average time-interval between $|g\rangle \to |e\rangle$ and $|e\rangle \to |g\rangle$ state changes are given respectively by $\Delta t_{abs}$ and $\Delta t_{em}$.

Collision cross-sections are described in more detail in Sections 6.3 and 7.5. The concept of a collision-cross section will be utilized in the last few paragraphs of this section. So, for pedagogical purposes, perhaps it is useful to mention that a collision cross-section is essentially "an (effective) area representative of a collision reaction between atomic or nuclear particles or systems, such that the number of reactions that occur equals the product of the number of target particles or systems and the number of incident particles or systems which would pass through this area if their velocities were perpendicular to it [McGraw-Hill]." This concept is often expressed as $1/\tau_0 = n\sigma v$, where $\tau_0$ is the ensemble-average time-interval between collision events, $n$ is the number density of atomic systems, $\sigma$ is the collision cross-section, and v is the ensemble-average relative speed between these atomic systems [Reif].

For a situation in which the radiation stimulated photon absorption and emission (and spontaneous emission) are the dominant state-changing processes, collision cross-sections for photon absorption ($\sigma_{abs}$) and emission ($\sigma_{em}$) can be constructed from the rate of change of $r_{em}$ and $r_{abs}$ with that of the radiation field intensity. In the abstract sense of considering a collision-cross-section function of the form $r_{abs} = \sigma_{abs}f(I(z))$ and $r_{em} = \sigma_{em}f(I(z))$ (i.e. linearization of $I(z)$ with respect to the pressure-independent photon absorption and emission rates), then for $\partial A_{sp}/\partial(f(I(z)) = 0$ (a typical assumption that may require further consideration [Allen 3]) equation 2.29 leads to:

$$\frac{\left(\dfrac{\partial r_{em}}{\partial f(I(z))}\right)}{\left(\dfrac{\partial r_{abs}}{\partial f(I(z))}\right)} = \left(\frac{\sigma_{em}}{\sigma_{abs}}\right)_{\text{rad. stim.}} > 1$$

$$(2.30)$$

In equation 2.30, the symbol "$\partial$" indicates a partial derivative; e.g. the denominator means taking the partial derivative of $r_{em}$ with respect to $f(I(z))$.

Similarly, and of relevance to the pressure-dependent studies presented in this dissertation (see also Sections 6.3 and 7.5), if the radiation field intensity is held constant and the pressure $P$ is changed, then $\partial r_{abs}/\partial P = \partial r_{em}/\partial P = \partial A_{sp}/\partial P = 0$. The use of an ideal gas-kinetic model for the collision-induced collision cross-section (applied to the two-level system model of Section 2.9) gives $c_{abs}$ and $c_{em}$ as being linearly related to the pressure $P$, so that equation 2.29 leads to:

$$\frac{\left(\dfrac{\partial c_{em}}{\partial P}\right)}{\left(\dfrac{\partial c_{abs}}{\partial P}\right)} = \left(\frac{\sigma_{em}}{\sigma_{abs}}\right)_{\text{coll.−rad.stim.}} > 1$$

$$(2.31)$$

Only the change in width of the line shape is relevant when measuring pressure-dependent (elastic) collision cross-sections ($\sigma_{abs}$ and $\sigma_{em}$) and so the increase in the ratio of $I(z)$ to $I(0)$ can be ignored when measuring the self-quenching collision cross-section. In this case, the pressure $P$ is due entirely to the number density $n$ of chromophores (through the ideal gas law), and so the number density $n$ of chromophores must be changed in order to affect a change in pressure $P$. And equation 2.31 is expected to be valid for the case of holding the chromophore number density constant and relatively small compared to the number density $n$ of a buffer gas, as was done for the measurement of collision cross-sections in this project.

Equation 2.31 is a statement from classical (equilibrium) thermodynamics, which says explicitly that $\sigma_{em} > \sigma_{abs}$. The importance of this inequality is substantial. And it is perhaps still an open question as to whether or not a dynamical theory like quantum mechanics can or should be able to account for the "conclusions" obtained from classical thermodynamics.

It should be understood that the use of the same notation for the collision cross-sections ($\sigma_{abs}$ and $\sigma_{em}$) in equations 2.30 and 2.31 is not meant to imply that these quantities are the same for these two different processes, which is why the subscripts "rad. stim." and "coll.–rad. stim." have been appended to these ratios in the above respective equations; see also Figures 2.5 and 2.6.

An obvious question arises: what might be the connection between the model developed in this section for linear absorption and the well respected $A$ and $B$ Coefficient model? The energy density of the radiation field $W(x, y, z)$ is related to the intensity of the radiation field $I(x, y, z)$ by $I(x, y, z) = cW(x, y, z)$, where $c$ is the speed of light in a vacuum. The Einstein model for a steady-state black-body radiator can be expressed as $A_{sp}N_e + B_{eg}I(z)N_e - B_{ge}I(z)N_g = 0$, and it is generally assumed that $B_{ge} = B_{eg}$. Furthermore, for the black-body radiator model used by Einstein (i.e. the Plank law for the radiation field energy density in a perfectly isolated black-body radiator, or Planck radiation law), the single temperature $T$ of the black-body radiator would appear to imply that the radiation field intensity $I(x, y, z)$ is the same at all locations $(x, y, z)$, which can be expressed as $I(x, y, z) = $ constant. This condition of constant radiation field intensity $I(x, y, z)$ does not appear to be consistent with linear absorption as described by equation 2.8 (i.e. $dI(z)/dz = -\kappa I(z) \neq 0$). It is beyond the scope of this dissertation to consider modifying the underlying Planck radiation law used in the Einstein $A$ and $B$ Coefficient model to account for the change in radiation field intensity $I(x, y, z)$ as a function of position $(x, y, z)$, therefore, as hinted at the beginning of this section, it has been tacitly assumed that it is possible to approach the Einstein $A$ and $B$ Coefficient model in a way that is consistent with the Beer-Lambert law for steady-state linear absorption. It is worth noting, though, that the inequality $k_{em} \neq k_{abs}$ (suggested by equation 2.29) would appear to be related to the suggestion that $B_{ge} \neq B_{eg}$ [Mompart].

Further comparison of the Einstein $A$ and $B$ Coefficient model and the Beer-Lambert law does not appear to be appropriate or necessary at this time. However, the question (of mechanism) has been raised as to how a black-body radiator progresses towards a state of thermal equilibrium with a material system (e.g. a two-level system) [Irons]. An answer to this question (in terms of the concepts of irreversibility and dissipation of heat energy) is offered at the end of Chapter VII.

## 2.12 Collision Processes

Atoms and molecule are complex objects, each one being composed of a finite number of electrons and nuclei.  In the ideal but non-existent limit of a perfectly isolated atomic (or molecular) system, time-independent quantum mechanics can be characterized by stationary-states, which is a useful description of the properly time-averaged internal dynamical properties of such a system.  However, contained within the idea of a perfectly isolated stationary-state is the reality that such an object would reside in a universe where there is "a time without time", or perhaps more accurately, "a time without change".  A more generally approach, then, would seek to model the interactions between atomic systems through considerations of time-dependent dynamics.

As described in Sections 2.9 and 2.10, some portion of the interactions between a chromophore and buffer gas particle can be characterized by collision-induced state-changing rates.  The notion of a purely elastic collision (often modeled as instantaneous collisions between infinitely-hard (i.e. non-deformable) spheres) is useful in a conceptual sense, but it is only an idealization that does not appear to exist in Nature; it appears that all collisions are, to some degree or another, inelastic.  It is also common to associate the phrase "inelastic (state-changing) collisions" with changes in the internal time-independent stationary-states of an atomic system; e.g. the state-changing processes that involve photon absorption and emission [Steinfeld].  These stationary-states are consistent with the solutions obtained from solving the time-independent Schrödinger equation for the case of a perfectly isolated chromophore.  However, with or without such internal state-changes, the inelastic nature of collisions can be expected to contribute to changes in the translational motion (i.e. trajectories) of the collision partners (e.g. chromophore and buffer gas), and thus changes in the spatial (and perhaps temporal) orientation of the chromophore transition moment with respect to, say, a linearly polarized radiation field (e.g. like the ones used in this project).

The "traditional" solution of the time-dependent Schrödinger equation for the two-level system model does not include a cause that leads to the effect of photon absorption and emission [Bernath]. In Chapter VII of this dissertation the "traditional" two-level system model will be expanded upon by taking into account (in a relatively simple manner) the time-dependent (deterministic) interactions that occur between the chromophore and buffer gas before and after a collision-induced photon absorption or emission during a state-changing event. The time-dependent interactions that occur during these collisions give rise to short-lived non-equilibrium perturbations that are notoriously difficult to (precisely or accurately) characterize in either (theoretical) models or experiments. In Chapter VII the modeling of these time-dependent interactions will introduce (what appears to be) a new parameter, the ensemble-average change in the wave function phase-factor ($\Delta\alpha$) for a state-changing event. This model is essentially a continuation of a line of thought begun several decades ago by Foley [Foley 2 and 3]. The results presented in Chapter VII for this model may be of fundamental importance to quantum theory, especially with regard to the discussion of "cause and effect" (a.k.a. causality) in the state-changing processes in the quantum domain [Zajonk].

## 2.13 Pressure Broadening and Pressure Shift Coefficients

The primary focus of this project was to extract pressure broadening ($B_p$) and pressure shift ($S_p$) coefficients by observing spatially and temporally isolated two-body (not including the photon) collision-induced state-changing events of a chromophore as a function of buffer gas pressure in high-resolution Doppler-limited frequency domain linear absorption spectra, with all other relevant parameters (such as temperature, incident radiation field intensity on the gas cell, and chromophore number density) held constant. It is quite common to observe a linear increase of both the Lorentz width (pressure broadening) and shift of the line-center (pressure shift) as the buffer gas pressure is increased, a combination of effects that have been observed since at least the 1930s [Margenau]. It is also generally thought that the pressure broadening and pressure shift coefficients will, in some meaningful manner, characterize the time-dependent forces (i.e. interactions) occurring before and after a collision-induced state-changing event [Foley 2]. The development of conceptual foundations and theoretical models attempting to make use of the pressure broadening and pressure shift coefficients has a long history and is an area of tremendous effort among physical scientists [Allard]. A meaningful line shape model must, of course, account for both of these observations – pressure broadening and pressure shift – and it should also be able to account for (i.e. predict) the observation that line-shapes at modest buffer gas pressures contain systematic deviations from being perfectly symmetric, and so are not in perfect accord with the perfectly symmetric Lorentzian distribution line-shape obtained from the "traditional" two-level system model; e.g. the line shape model of Section 6.3.

Although the spectral data obtained during the course of this project was not investigated for line shape asymmetry, such details will be briefly considered in Sections 7.6 and 7.7. In this project, the analysis of the observed line shapes used an approximate model for the hyperfine transitions, which can also be a source of systematic deviations of the residuals between the model and observed line shapes. The model line shape used in the nonlinear regression analysis – an analytic approximation to the Voigt distribution function [Humlíček] – is expected to be highly symmetric with regard to the convolution of the Lorentzian and Gaussian distributions, to better than one part in ten thousand.

The dependence on buffer gas pressure (or number density) contained within the pressure broadening coefficient (often expressed in units of MHz/torr) is then removed through calculation of a collision cross-section $\sigma$ (often expressed in units of Angstsrom squared $\text{Å}^2$); the model for this conversion and the elastic collision cross-sections obtained in this project are presented in Section 6.3. The presentation in Chapter VII offers a slightly refined view of the collision process by accounting for the interactions (i.e. forces acting over time) before and after a collision-induced state-changing event. Such a characterization of the collision process (as presented in Chapter VII) leads to a new parameter, the ensemble-average change in the wave function phase-factor ($\Delta\alpha$) during a collision-induced state-change, which is found to be related to knowledge (i.e. measurement) of both the pressure broadening and the pressure shift coefficients. The parameter $\Delta\alpha$ in turn leads to a refinement of the calculated values of $\sigma$ (i.e. $\sigma_{\text{inelastic}} \neq \sigma_{\text{elastic}}$).

It is perhaps worthwhile to offer a conceptual point of view on the origin of the observed broadening of spectral lines (a.k.a. pressure broadening) and the changes in their line-center frequency (a.k.a. pressure shift) as a function of buffer gas pressure. According to gas kinetic theory, at a fixed temperature an increase in pressure (and thus number density) of buffer gas implies that the time-interval between collision-induced state-changing events is decreasing. The Fourier transform of a shorter lifetime in the time domain is a larger line width in the frequency domain, and hence the frequency domain line width increases as the pressure of the buffer gas is increased.

Collisions with a buffer gas will lead to differential perturbations of the lower state and upper state energy levels, which appear as a change (i.e. shift) in the transition energies between the ground and excited states of the two-level system. As the pressure is increased there will be more interactions (e.g. more collision-induced state-changing events) per unit time, which will lead to larger observed shifts of the line-center. This increase in line-center shift with increasing pressure can perhaps be thought of as resulting from the chromophore, in a relative sense, spending more time interacting with the buffer gas during the measurement (i.e. observation) process.

For the results presented in this dissertation, the buffer gas pressure is set to be sufficiently large so that the ensemble-average collision-induced (i.e. pressure-dependent) state-changing rate due to this buffer gas is predicted to provide the dominant contribution to the homogeneous (a.k.a. Lorentzian) component of the observed line widths. The spontaneous (a.k.a. natural) photon emission process is assumed to be independent of the buffer gas pressure in the gas cell; while this may not be an accurate assumption [Allen 3], consideration of their relatively small values, as compared to the sum of the ensemble-average rates of the collision-induced state-changing processes, suggests that it is likely to be a reasonable approximation for our purposes here. (See also Sections 2.9 and 2.10.)

The spontaneous emission rates for the B electronic state of diatomic iodine are reported to be in the range of 0.1 to 10 $\mu$sec [Paisner], which gives a spontaneous emission rate in the range of 0.1 to 10 MHz. Estimates (i.e. back-of-the-envelope calculations) based on gas-kinetic arguments for an ideal gas at room temperature (292 K) with a collision cross-section of 50 Å$^2$, and a range of buffer gas pressure of 5 to 100 torr will result in a total collision-induced state-changing rate that is roughly one to three orders of magnitude greater than the spontaneous emission rate [Reif]; see Section 6.3 for more details on the calculation of elastic collision-cross sections. Similarly, since the radiation field intensity initially incident on the gas cell is constant and relatively weak for all pressures of buffer gas, the rates of the pressure-independent stimulated photon absorption and emission processes are assumed to be nearly constant and insignificant and have thus been neglected; see also Section 5.3 for more detail on the relative "strength" of the radiation field (from a classical physics point of view).

## 2.14 The Hermitian Hamiltonian

The description of dynamical processes in the macroscopic domain using the theory of classical mechanics makes use of an object known as a Hamiltonian [Marion]. This object is also central to the description of dynamical processes in the microscopic domain when using the theory of quantum mechanics. The Hamiltonians used in quantum mechanics are composed of operators and the properties of these operators are of some importance. Of particular concern is the Hermitian property in which a given operator is equal to its adjoint. The role of a Hermitian Hamiltonian (as any text book on quantum mechanics will describe) is to provide real-valued (as opposed to complex-valued) eignevalues for the energy in the solutions to the time-independent Schrödinger equation (i.e. the energies of the perfectly isolated stationary-states).

With an eye toward later developments on the modeling of state-changing processes (in Chapter VII), the use of a time-dependent Hermitian Hamiltonian is briefly considered in this section. In Chapter VII the Generalized Rotating Wave Approximation will be used [Barut 2 and Loudon 2], which implicitly contains (but keeps partially hidden) the raising and lowering operators obtained from the quantized description of the radiation field in terms of discrete photons. The raising ($a^{\dagger}$) and lowering operators ($a$) model the photons added to or removed from the radiation field, respectively. (The raising operator corresponds to emission of a photon by the chromophore and the lowering operator corresponds to the absorption of a photon by the chromophore.)

When acting on a general stationary-state given in the "ket" notation as $|n\rangle$, the raising and lowering operators follow the relations:

$$a^{\dagger} |n\rangle = \sqrt{n + 1} \ |n + 1\rangle \qquad \text{(raising)}$$

$$a |n\rangle = \sqrt{n} \ |n - 1\rangle \qquad \qquad \text{(lowering)} \qquad (2.32)$$

The label "$n$" appearing in $|n\rangle$ is referred to as a "quantum index", which in this case corresponds to the mode number for the quantized radiation field. The mathematical operation of finding the adjoint of an operator is indicated by the "$\dagger$" symbol, so that the raising and lowering operators were specifically constructed so as to be adjoint to each other. (Similarly, the adjoint of the adjointed operator is the original operator.)

As well, the "bra" and "ket" objects (given respectively by $|n\rangle$ and $\langle n|$) are ajdoint to each other (i.e. $|n\rangle^{\dagger} = \langle n|$ and $\langle n|^{\dagger} = |n\rangle$), so that the Hermitian conjugate of equations 2.32 (i.e. the adjoint) gives for the "bra" object:

$$\langle n| a = \sqrt{n+1}\ \langle n+1| \qquad \text{(raising)}$$

$$\langle n| a^{\dagger} = \sqrt{n}\ \langle n-1| \qquad \text{(lowering)} \tag{2.33}$$

The corresponding lowering and raising operators for the chromophore, given respectively by $|g\rangle\langle e|$ and $|e\rangle\langle g|$, can be combined with the raising and lowering operators of the radiation field (equations 2.32) to form a time-dependent state-changing perturbation term $V(t)$ in the total Hamiltonian that is Hermitian. It will be shown in Chapter VII that the perturbation term $V(t)$ can also account for the ensemble-average change in the wave function phase-factor $\Delta\alpha$ for a collision-induced state-changing event so that (in the Generalized Rotating Wave Approximation in the Schrödinger Picture) it has the form:

$$V(t) = a\,|e\rangle\,\langle g|\exp(i\Delta\alpha - i\omega t) + a^{\dagger}\,|g\rangle\,\langle e|\exp(-i\Delta\alpha + i\omega t) \tag{2.34}$$

While $V(t)$ is indeed equal to its adjoint (and so by definition is Hermitian), it should not go unnoticed that the time-dependent state-changing nature of equation 2.32 is not commensurate with the usual proof about obtaining real-valued eignevalues from the action of Hermitian operators on stationary-state solutions of the time-independent Schrödinger equation [Shankar]. That is to say, conclusions about the nature of a state-changing perturbation term $V(t)$ are perhaps best considered by solving the time-dependent Schrödinger equation.

Also, time $t$ appears in quantum mechanics in the form $\exp(-i\omega t)$, so that the adjoint of this object is equivalent to taking its complex conjugate (i.e. $\exp(i\omega t)$), which (it seems) is generally interpreted as saying that time $t$ is propagating in opposite directions for these two objects. However, given the results of Section 2.11 with regard to the process of absorption and emission not following identical paths, we are left to wonder if the appearance of such time-reversal in quantum mechanical calculations refers to time-reversal on the path of a single state-changing process (i.e. absorption or emission), or whether it is an indication that the dynamics on both paths (i.e. absorption and emission) are to be considered.

These mathematical results will be encountered again in Chapter VII. In the first five sections of that chapter the "traditional" interpretation will be used, namely that Hamiltonians must be Hermitian (e.g. equation 2.34). In the last two sections of that chapter an alternative interpretation will be offered with regard to the use of a non-Hermitian Hamiltonian in quantum mechanics for the situation that the Hamiltonian is explicitly time-dependent in its description of state-changing processes.

## 2.15 Endnotes for Chapter II

[Allard]        N. Allard and J. Keilkopf; "The effect of neutral nonresonant collisions on atomic spectral lines", Reviews of Modern Physics, 54, 1103-1182 (1982).

[Allen 2]       L. Allen and J. H. Eberly, *Optical Resonance and Two-Level Atoms*, Chapter 3; Dover Publications, Mineola, New York (1987); ISBN 0-486-65533-4.

[Allen 3]       L. Allen; "Are Einstein A coefficients constant?", Physics World, 3, 19 (1990), IOP Publishing, ISSN: 0953-8585.

[Atkins]        P.W. Atkins; *Quanta: A Handbook of Concepts*, pg 155; Clarendon Press, Oxford (1985); ISBN 0-19-855494-X.

[Barut 1]       P. W. Milonni; "Classical and Quantum Theories of Radiation"; *Foundations of Radiation Theory and Quantum Electrodynamics* (edited by A. O. Barut), page 2-4; Plenum Press, New York (1980); ISBN 0-306-40277-7.

[Barut 2]       L. Davidovich and H. M. Nussenzveig; "Theory of Natural Line Shape"; *Foundations of Radiation Theory and Quantum Electrodynamics* (edited by A. O. Barut), page 91; Plenum Press, New York (1980); ISBN 0-306-40277-7.

[Bernath]       P. F. Bernath; *Spectra of Atoms and Molecules*, Chapter 1; Oxford University Press, New York (1995); ISBN 0-19-507598-6.

[Bunker]        G. R. Hanes, J. Lapierre, P. R. Bunker, and K. C. Shotton; "Nuclear hyperfine structure in the electronic spectrum of $^{127}I_2$ by saturated absorption spectroscopy, and comparison with theory", Journal of Molecular Spectroscopy, 39, 506-515 (1971).

[Butkov]        E. Butkov; *Mathematical Physics*, Chapters 6 and 7; Addison-Wesley, Reading, Massachusetts (1968); ISBN 0-201-00727-4.

[Casimir]       H. B. G. Casimir; *On The Interaction Between Atomic Nuclei and Electrons*, Prize essay published by Tweede Genootschap, reprinted in *Series of Books on Physics*, Second Edition, W. H. Freeman & Co., San Francisco (1963); LCCN 62019662 /L/r842.

[Condon]        E. U. Condon; "Nuclear Motions Associated with Electron Transitions in Diatomic Molecules", Physical Review, 32, 858-872 (1928).

[Cook]          R. L. Cook and F. C. De Lucia; "Application of the Theory of Irreducible Tensor Operators to Molecular Hyperfine Structure", American Journal of Physics, 39, 1433-1454 (1971).

[CRC]           *CRC Handbook of Chemistry and Physics*; CRC Press, Cleveland, Ohio (2001); LCCN 77641777 //r85

[Dantus]        M. Comstock, V. V. Lozovoy, M. Dantus; "Femtosecond Photon Echo Measurements of Electronic Coherence Relaxation between the X($^1\Sigma_{g+}$) and B($^3\Pi_{0u+}$) States of $I_2$ in the Presence of He, Ar, $N_2$, $O_2$, $C_3H_8$"; Journal of Chemical Physics, 119, 6546-6553 (2003).

[Dunham]        J. L. Dunham; "The Energy Levels of a Rotating Vibrator", Physical Review, 41, 721-731 (1932).

[Einstein]      A. Einstein; "On the Quantum Theory of Radiation", Physikalische Zeitschrift, 18, 121 (1917); translated and reprinted by D. ter Haar; *The Old Quantum Theory*, pages 167-183, Pergamon Press, Oxford (1967); LCCN 66029628.

[Feld]          B. T. Feld and W. E. Lamb Jr.; "Effect of Nuclear Electric Quadrupole Moment on the Energy Levels of a Diatomic Molecule in a Magnetic Field. Part I. Heteronuclear Molecules", Physical Review, 67, 15-33 (1945).

[Field]         S. Churassy, F. Martin, R. Bacis, J. Verges, R. W. Field; "Rotation-Vibration Analysis of the B$0_{u+}$-a$1_g$ and B$0_{u+}$-a′$0_{g+}$ Electronic Systems of $I_2$ by Laser-Induced-Fluorescence Fourier-Transfrom Spectroscopy"; Journal of Chemical Physics, 75, 4863-4868 (1981).

[Foley 1]       H. M. Foley; "Note on the Nuclear Electric Quadrupole Spectrum of a Homonuclear Diatomic Molecule in a Magnetic Field", Physical Review, 71, 747-752 (1947).

[Foley 2]       H. M. Foley; "The Pressure Broadening of Spectral Lines", Physical Review, 69, 616-628 (1946).

[Foley 3]       F. W. Byron Jr. and H. M. Foley; "Theory of Collision Broadening in the Sudden Approximation", Physical Review, 134, A625-A637 (1964).

[Franck]        J. Franck and R. W. Wood; "Influence upon the Fluorescence of Iodine and Mercury Vapor of Gases with Different Affinities for Electrons", Philosophical Magazine, 21, 314-318 (1911).

[Hardwick 1]    J. A. Eng, J. L. Hardwick, J. A. Raasch and E. N. Wolf; "Diode laser wavelength modulated spectroscopy of $I_2$ at 675 nm", Spectrochimica Acta, Part A, 60, 3413-3419 (2004).

[Herzberg 1]    G. Herzberg; *Molecular Spectra and Molecular Structure: I. Spectra of Diatomic Molecules*, Second Edition, Chapter III.2.f; Krieger Publishing Company, Malabar (1989); ISBN 0-89464-268-5.

[Herzberg 2]    G. Herzberg; *Molecular Spectra and Molecular Structure: I. Spectra of Diatomic Molecules*, Second Edition, Chapter V; Krieger Publishing Company, Malabar (1989); ISBN 0-89464-268-5.

[Herzberg 3]    G. Herzberg; *Molecular Spectra and Molecular Structure: I. Spectra of Diatomic Molecules*, Second Edition, Chapter III.2.e; Krieger Publishing Company, Malabar (1989); ISBN 0-89464-268-5.

[Herzberg 4]    G. Herzberg; *Molecular Spectra and Molecular Structure: I. Spectra of Diatomic Molecules*, Second Edition, Chapter III.2.d; Krieger Publishing Company, Malabar (1989); ISBN 0-89464-268-5.

[Herzberg 5]    G. Herzberg; *Molecular Spectra and Molecular Structure: I. Spectra of Diatomic Molecules*, Second Edition, Chapter IV.3; Krieger Publishing Company, Malabar (1989); ISBN 0-89464-268-5.

[Herzberg 6]    G. Herzberg; *Molecular Spectra and Molecular Structure: I. Spectra of Diatomic Molecules*, Second Edition, Chapters II.1, IV.3, and Table 39; Krieger Publishing Company, Malabar (1989); ISBN 0-89464-268-5.

[Herzberg 7]    G. Herzberg; *Molecular Spectra and Molecular Structure: I. Spectra of Diatomic Molecules*, Second Edition, Chapter IV; Krieger Publishing Company, Malabar (1989); ISBN 0-89464-268-5.

[Herzberg 8]    G. Herzberg; *Molecular Spectra and Molecular Structure: I. Spectra of Diatomic Molecules*, Second Edition, Chapter IV.4.b; Krieger Publishing Company, Malabar (1989); ISBN 0-89464-268-5.

[Hougen]        J.T. Hougen; *The Calculation of Rotational Energy Levels and Rotational Line Intensities in Diatomic Molecules*, NBS Monograph 115 (June 1970); archived at the web site http://physics.nist.gov/ (March 2007).

[Humlíček]      J. Humlíček; "Optimized computation of the Voigt and complex probability functions", Journal of Quantitative Spectroscopy and Radiative Transfer, 27, 437-444 (1982).

[Hutson]    J.M. Hutson, S. Gerstenkorn, P. Luc and J. Sinzelle; "Use of calculated centrifugal distortion constants ($D_v$, $H_v$, $L_v$ and $M_v$) in the analysis of the B ← X system of $I_2$", Journal of Molecular Spectroscopy, 96, 266–278 (1982).

[Irons]     F. E. Irons; "Why the cavity-mode method for deriving Planck's law is flawed", Canadian Journal of Physics, 83, 617-628 (2005).

[Johnson]   M. H. Johnson Jr.; "Spectra of Two Electron Systems", Physical Review, 38, 1628-1641 (1931).

[Kato]      H. Kato et al.; *Doppler-Free High Resolution Spectral Atlas of Iodine Molecule 15,000 to 19,000 $cm^{-1}$*; published by Japan Society for the Promotion of Science (2000); ISBN 4-89114-000-3.

[Kauppinen] J. Kauppinen and J. Partanen; *Fourier Transforms in Spectroscopy*, Chapter 1; Wiley-VCH Verlag Berlin GmbH, Berlin (2001); ISBN 3-527-40289-6.

[Knöckel]   B. Bodermann, H. Knöckel and E. Tiemann; "Widely usable interpolation formulae for hyperfine splittings in the $^{127}I_2$ spectrum", European Physical Journal, Part D, 19, 31-44 (2002).

[Kroll]     M. Kroll and K. K. Innes; "Molecular electronic spectroscopy by Fabry-Perot interferometry. Effect of nuclear quadrupole interactions on the line widths of the $B^3\Pi_{0+}$ - $X^1\Sigma_g^+$ transition of the $I_2$ molecule", Journal of Molecular Spectroscopy, 36, 295-309 (1970).

[Landau]    L. D. Landau and E. M. Lifshitz; *Quantum Mechanics: Non-Relativistic Theory*; page 294; Pergamom Press, Massachusetts (1958); ISBN 57-14444.

[Lehmann]   M. Broyer, J. Vigué and J. C. Lehmann; "Effective hyperfine Hamiltonian in homonuclear diatomic molecules. Application to the B state of molecular iodine", Journal de Physique, 39, 591-609 (1978).

[Lepère]    M. Lepere; "Line profile study with tunable diode laser spectrometers", Spectrochimica Acta Part A, 60, 3249-3258 (2004), and references cited therein.

[LeRoy 1]        R. J. LeRoy; *RKR1: A Computer Program Implementing the First-Order RKR Method for Determining Diatom Potential Energy Curves from Spectroscopic Constants*, Chemical Physics Research Report, University of Waterloo (March 25, 1992).

[LeRoy 2]        R. J. LeRoy; *Level 7.4: A Computer Program for Solving the Schrodinger Equation for Bound and Quasibound Levels*, Chemical Physics Research Report, University of Waterloo (September, 2001).

[Loomis]        F. W. Loomis; "Correlation of the Fluorescent and Absorption Spectra of Iodine", Physical Review, 29, 112-134 (1927).

[Loudon 1]        R. Loudon; *The Quantum Theory of Light,* Second Edition, Chapter 2; Oxford University Press, New York, (1983); ISBN 0-19-851155-8.

[Loudon 2]        R. Loudon; *The Quantum Theory of Light,* Second Edition, Chapter 5; Oxford University Press, New York, (1983); ISBN 0-19-851155-8.

[Luc 1]        S. Gerstenkorn and P. Luc; Atlas du Spectre d'Absorption de la Molécule d'Iode entre 14 800–20 000 cm$^{-1}$, Laboratoire Aimé Cotton CNRS II, Orsay, (1978).

[Luc 2]        S. Gerstenkorn and P. Luc; "Absolute iodine ($I_2$) standards measured by means of Fourier transform spectroscopy", Revue de Physique Appliquee, 14, 791-794 (1979).

[Margenau]        H. Margenau and W. W. Watson; "Pressure Effects on Spectral Lines", Reviews of Modern Physics, 8, 22-53 (1936).

[Marion]        J. B. Marion; *Classical Dynamics of Particles and Systems*, Second Edition; Academic Press, Orlando, Florida (1970); LCCN 78-107545.

[Martin]        F. Martin, R. Bacis, S. Churassy and J. Vergès; "Laser-induced-fluorescence Fourier transform spectrometry of the $XO_g^+$ state of $I_2$: Extensive analysis of the $BO_u^+ \rightarrow XO_g^+$ fluorescence spectrum of $^{127}I_2$", Journal of Molecular Spectroscopy, 116, 71-100 (1986).

[McGraw-Hill]    *McGraw-Hill Dictionary of Physics* (edited by S.B. Parker); McGraw-Hill, New York (1985); ISBN 0-07-045418-3.

[Merck]    *The Merck Index: An Encyclopedia of Chemicals, Drugs, and Biologicals*; Whitehouse Station, New Jersey (1996); ISBN 0-91-191012-3.

[Michelson]    Nobel Lecture in Physics given in 1907 by A. A. Michelson; "Recent Advances in Spectroscopy"; *Nobel Lectures, Physics1901-1921*; Elsevier, Amsterdam (1967); archived at the web site http://nobelprize.org/nobel_prizes/physics/laureates/ (2009).

[Mompart]    J. Mompart and R. Corbalán; "Generalized Einstein B coefficients for coherently driven three-level systems", Physical Review A, 63, 063810-1 (2001).

[Paisner]    J. A. Paisner and R. Wallenstein; "Rotational lifetimes and selfquenching cross sections in the $B^3\Pi_{0u}^+$ state of molecular iodine-127", Journal of Chemical Physics, 61, 4317-4320 (1974).

[Rabi]    J. M. B. Kellogg, I. I. Rabi, N. F. Ramsey Jr. and J. R. Zacharias; "An Electrical Quadrupole Moment of the Deuteron The Radiofrequency Spectra of HD and $D_2$ Molecules in a Magnetic Field", Physical Review, 57, 677-695 (1940).

[Ramsey]    N. Ramsey; *Molecular Beams*, Clarendon Press, Oxford (1956); LCCN 56003940 /L/r883.

[Reif]    F. Reif; *Fundamentals of Statistical and Thermal Physics*, Chapter 12; McGraw-Hill, New York (1965); ISBN 07-051800-9.

[Salami]    H. Salami and A. J. Ross; "A molecular iodine atlas in ascii format", Journal of Molecular Spectroscopy, 233, 157-159 (2005).

[Schawlow]    M. D. Levenson and A. L. Schawlow; "Hyperfine Interactions in Molecular Iodine", Physical Review A, 6, 10-20 (1972).

[Shankar]    R. Shankar; *Principles of Quantum Mechanics*, Second Edition, Chapter 1; Plenum Press, New York (1994); ISBN 0-306-44790-8.

[Steinfeld]    J. I. Steinfeld and W. Klemperer; "Energy-Transfer Processes in Monochromatically Excited Iodine Molecules. I. Experimental Results", Journal of Chemical Physics, 42, 3475-3497 (1965).

[Strait]            L. A. Strait and F. A. Jenkins, "Nuclear Spin of Iodine from the
                    Spectrum of $I_2$", Physical Review, 49, 635-635, (1936).

[Tellinghuisen]     J. Tellinghuisen; "Transition strengths in the visible–infrared
                    absorption spectrum of $I_2$", Journal of Chemical Physics, 76, 4736-
                    4744 (1982).

[Zajonc]            G. Greenstein and A. G. Zajonc; The quantum challenge : modern
                    research on the foundations of quantum mechanics, Second Edition,
                    page 123; Jones and Bartlett Publishers, Sudbury, Massachusetts
                    (2006); ISBN 0-7637-2470-X.

[Zare 1]            R. N. Zare; *Angular Momentum: Understanding Spatial Aspects in
                    Chemistry and Physics*; Wiley and Sons, New York (1988); ISBN 0-
                    471-85892-7.

[Zare 2]            E. E. Nikitin and R. N. Zare; "Correlation diagrams for Hund's
                    coupling cases in diatomic molecules with high rotational angular
                    momentum", Molecular Physics, 82, 85-100 (1994).

CHAPTER III

DATA COLLECTION

## 3.1 Overview of Chapter III

A significant portion of time was invested in setting up and carrying out experiments that could yield an abundance of reliable and accurate data, which is quite typical of the activities in experimental chemical physics. This chapter describes the experiments that provided high-resolution linear absorption spectra (data) of diatomic iodine (as the chromophore) near 675 nm in the presence of measured (known) pressures of added buffer gases. (Diatomic iodine refers to the homonuclear molecule $I_2$, which is often colloquially referred to as iodine.) The spectra obtained from these experiments were deemed to be of sufficiently good quality (with regard to signal-to-noise ratio (S/N), reproducibility, and "true", extraneous-artifact-free representation of line shape) that considerable effort was then made to analyze them for pressure broadening and pressure shift coefficients; this analysis and its results will be a primary topic in the chapters that follow this one. (Figures of the spectra obtained from the experiments described in this chapter can be found in Chapters IV and V.)

Between the summer of 2001 and 2006 there were four major configurations of the internally referenced absorption spectrometer used to collect data. These four configurations can be correlated with four historical periods of data collection. The first three periods used a free-running Fabry-Perot laser diode system. The first period of data collection was carried out at room temperature (292 K) during the summer of 2001. The second period of data collection focused on the noble-gas atoms as the buffer gas (at room temperature) during the summers of 2003 and 2004. Both of these periods of data collection used the first spectrometer configuration described below. The results from the first and second periods of data collection were published in 2004 [Hardwick 1]. The third period of data collection was from about December 2004 to mid-summer 2005. This was the same as the first two periods (same laser system and same location in the undergraduate physical chemistry laboratory in the basement of Klamath Hall at the University of Oregon) except that the collimating lens for the laser had been removed and an off-axis parabolic reflector was instead used to collimate the laser beam. For this third period of data collection, some of the spectra using noble-gas atoms as the buffer gas (at room temperature) were re-recorded, all of the spectra using molecules (except for air and water vapor [Hardwick 1]) as the buffer gas (at room temperature) were recorded, and all of the higher temperature (348 K and 388 K) spectra were recorded. The fourth period of data collection was carried out in the summer of 2006 using an external cavity laser diode system in the research laboratory of Professor Tom Dyke on the first floor above the ground floor in Klamath Hall, KLA179. The wide tuning range of the external cavity laser diode used in this configuration of the internally referenced absorption spectrometer allowed us to begin exploring a wider range of rotational states of diatomic iodine, which was done with argon as the buffer gas. Fringe spectra (from a plane-parallel Fabry-Perot interferometer, commonly referred to as an etalon) were simultaneously recorded with the diatomic iodine spectra toward the end of the third period of data collection and for all data collected during the fourth period of data collection.

## 3.2 Internally Referenced Absorption Spectrometer

An overview of the experimental setup for the internally referenced absorption spectrometer used to investigate the broadening and line center shift of diatomic iodine as a function of pressure of various buffer gases is shown in Figure 3.1. This spectrometer design allowed for the separate and (nearly) simultaneous measurement of spectra from two different gas cells, referred to as the reference and sample gas cells (labeled RC and SC, respectively, in Figure 3.1).

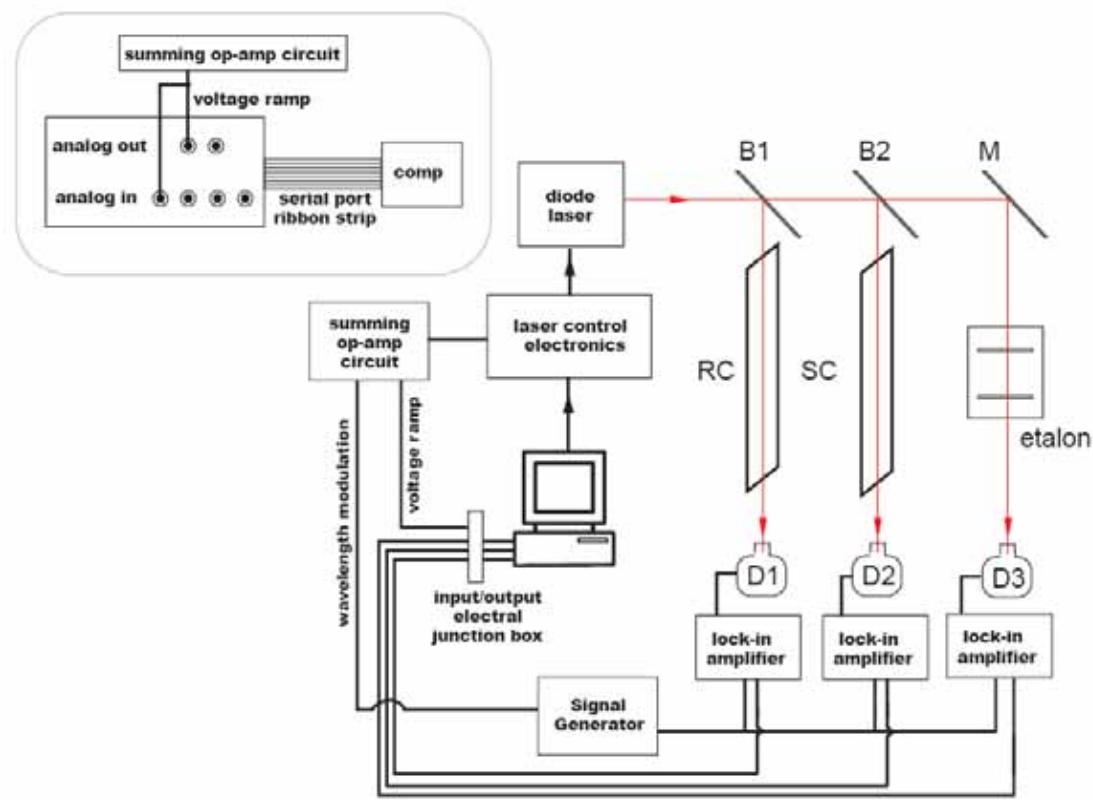

Figure 3.1 Schematic of the internally referenced absorption spectrometer used to record high resolution linear absorption spectra of diatomic iodine in the presence of known pressures of buffer gases. In this figure the computer is depicted as a pictogram that is connected to the "input/output electrical junction box" and "laser control electronics"; "RC" and "SC" are respectively the reference gas cell and sample gas cell; B1 and B2 are beam splitters; M is a mirror; D1, D2, and D3 are silicon-based photodiode detectors. An expanded view of the "input/output electrical junction box" is shown in the upper left corner of the figure; "comp" is short for computer. The "summing op-amp circuit" is described in Appendix A.

The advantage of such a configuration was the added ability to routinely determine with considerable precision the line center shift as a function of buffer gas pressure. Both the operation of the spectrometer and the recording of spectra were accomplished using a personal computer (600 MHz Pentium III using Windows 2000 operating system) with an on-board multifunction card (National Instruments model number PCI-1200 multifunction input-output device for PCI bus computers), commercially available software (Igor Pro 4, Igor Pro 5, and NIDAQ Tools from Wavemetrics), and computer device-drivers (software) to allow communication between Igor Pro and the PCI-1200 multifunction card. All electrical connections to the PCI-1200 multifunction card were made with BNC terminated co-axial cables through a junction box (National Instruments model number SC2071) connected to the personal computer by a 32-pin ribbon strip. The linear and nonlinear fitting algorithms in Igor Pro were also extensively used for wave number calibration, line shape analysis, and data analysis in general.

## 3.2.1 Philips CQL806/30 Laser Diode System Spectrometer

Most of the results presented in this dissertation were obtained from analyzing diatomic iodine spectra collected using an internally referenced absorption spectrometer (Figure 3.1) with a free-running Philips CQL806/30 laser diode providing modestly monochromatic (ca. 30 MHz) linearly polarized electromagnetic radiation (i.e. light) in the vicinity of 674.8 nm. (The Philips CQL806/30 laser diode system will generally be referred to as the Philips laser diode system. The laser alone will be referred to as the Philips laser diode; and similarly for the New Focus 6202 external cavity laser diode system.) A Newport 505 laser diode driver provided the injection current for this Fabry-Perot style laser diode. The mode settings for the Newport 505 laser diode driver were high-bandwidth and current range capability of 0 to 200 mA. The laser diode was mounted in a Newport 700 laser mount which contained a single 12.5 Watt thermoelectric module used to stabilize the temperature of the laser diode; this module was controlled by a Newport 325 temperature controller.  These components together comprise what is referred to as the Philips CQL806/30 laser diode system.

The term "free-running" laser diode indicates that no feedback, optical or electronic, was employed to stabilize the output wavelength of this laser diode system. The spectrometer used lock-in detection at the first harmonic of a wavelength-modulated absorption signal, which has the appearance of the first derivative of the corresponding direct absorption spectrum [Silver; Demtröder].

The wave number of the Philips laser diode radiation was current-tuned with a 12-bit ($2^{12}$ = 4096 data points) linear ramp that was passed in digital form point-by-point in real time from Igor Pro to the PCI-1200 multifunction card, which then output nearly instantaneously a corresponding linear tuning ramp voltage point-by-point from one of its analog output channels.  Tuning (scanning) the laser through the spectral region with the use of this point-wise linear tuning ramp voltage resulted in scans that were nearly linear in wave number.  (The linearity of the laser scans for data point number vs. wave number will be considered further in Chapter IV.)  This tuning ramp voltage was added to a small sine wave modulation voltage that modulated the laser wavelength.  The wavelength modulation voltage was provided by a signal generator (Global Specialties model number 105-2001 Signal Generator) using a homebuilt variable-attenuation summing amplifier constructed with "741 op-amps" (or compatible alternatives).  (Three different versions of this summing amplifier were utilized.  Two of these were used with the Philips laser diode system and one with the New Focus external cavity laser diode system.  Circuit diagrams and additional information about these devices are located in Appendix A.)  Using the variable gain setting on the front panel of the signal generator, it was possible to adjust the depth of the modulation voltage independent of the spectral scan range determined by the tuning ramp voltage.  The output from the summing amplifier – the summed and appropriately attenuated tuning ramp voltage and wavelength modulation voltage – was connected to the front end of the Newport 505 laser diode driver, which converted this input signal into a tuning and modulation injection current for use by the Philips laser diode (i.e. current-tuned). The tuning ramp voltage signal was split just before the input to the summing amplifier with the other piece being sent back to one of the analog data collection channels of the PCI-1200 multifunction card to be recorded (nearly) simultaneously with the signals from the beam paths. Data collection on the PCI-1200 was performed using the "multi-channel scanned data acquisition" mode at a gain setting of one, which allowed for a maximum data sampling rate of 100,000 per second with 12-bit accuracy on all four analog input channels. Also, all four input channels on the PCI-1200 junction box were configured for differential detection, a strategy meant to improve the rejection of common-mode signals on a given pair of input wires.

The output from the summing amplifier was split with one portion connected to the oscilloscope and the other portion connected to the front end of the Newport 505 laser diode driver. Measurements on this channel of the oscilloscope indicated a change in output voltage from the summing amplifier (when tuning the Philips laser diode across the $1.4 \text{ cm}^{-1}$ spectral region) of about $0.5 \pm 0.02$ V. The sine-wave modulation voltage output from the signal generator was one of the inputs to the summing amplifier, and was split just before the connection to the input, with the other portion being observed on a separate channel of the oscilloscope. Measurements on this channel of the oscilloscope indicated a peak-to-peak wavelength modulation voltage amplitude (often referred to as a modulation depth) of roughly $2.5 \pm 0.5$ mV. An estimate of the modulation depth can then be made for a single polarity tuning ramp that goes from 0 to +5 V: $(1.4 \text{ cm}^{-1}) \times (0.0025 \text{ V} \div 5 \text{ V}) \cong 0.0007 \text{ cm}^{-1} \cong 21$ MHz modulation depth.

The manufacturer specifications for the Philips laser diode indicate that a change in current of about 10 mA will be required to scan across approximately 1.4 $\text{cm}^{-1}$ (near 675 nm). The (ca. $1.4 \text{ cm}^{-1}$) single-mode scanning region chosen required an injection current from 55 mA to 65 mA. The attenuation factor of the summing amplifier was adjusted to allow the tuning ramp to generate a change in voltage at the front end of the Newport 505 laser diode driver that matched the 10 mA change in injection current required to tune across the entire 1.4 cm$^{-1}$ spectral region. The root mean square noise ripple for the Newport 505 laser diode driver for the settings used was specified to be less than 4 µA, which might result in roughly (1.4 cm$^{-1}$ ÷ 10 mA) × (0.004 mA) ≅ 0.00056 cm$^{-1}$ ≅ 17 MHz of uncertainty in the laser frequency. The short term stability (approximately 30 minutes) of the Newport 505 laser diode driver was specified to be less than 10 ppm. For a 50 mA injection current this corresponds to about 0.5 µA, which is nearly a factor of ten smaller than the ripple noise injected into the laser diode by this device. This suggests that the wavelength modulation depth of the Philips laser diode, estimated at 21 MHz for a single polarity tuning ramp, was approaching the limits of the Newport 505 laser diode driver capabilities when scanning this entire (ca. 1.4 cm$^{-1}$) spectral region, for which the laser operation was single-mode and mode-hop-free.

The 12-bit resolution of the PCI-1200 multifunction card could provide at most 4096 sequential analog output data points for the tuning ramp voltage over the region of a single acquired spectrum. The optimized step size between data points can be estimated as (1.4 cm$^{-1}$ ÷ 4096 data points) ≅ 0.00034 cm$^{-1}$ ≅ 10 MHz. It was, however, discovered during the third historical period of data collection that the optimal settings on the analog output of the PCI-1200 multifunction card was specified to be 12-bit for a bipolar scale from −5 to +5 V or a single polarity scale of 0 to +10 V. Most of the data collected with the Philips laser diode system used the above mentioned single polarity tuning ramp. Since the tuning ramps were configured to scan from 0 to +5 V (and then back down from +5 to 0 V), the analog output was operating at a non-optimized 11-bit scale, which means that the step size between data points was 20 MHz (instead of the optimized value of 10 MHz).

The only spectra obtained (and analyzed) while using a bipolar tuning ramp with 10 MHz steps between data points came from the experiments at elevated temperatures, which also contained room temperature scans. However, spectra collected with a step size of 10 MHz also (inadvertently) had the Newport 505 laser diode driver set to low frequency band pass, so that high frequency modulations are strongly attenuated. The modulation frequency was at about 88 kHz so that a much larger than usual input peak-to-peak modulation voltage (sine wave) of about $14 \pm 0.2$ mV was required to obtain the previously observed S/N ratio. Even though the modulation depth is not known for these experiments, the recorded spectra appear to be of "good" quality, with S/N ratios similar to those obtained with the 11-bit tuning ramp.

The effect of the mismatch between the 0 to +5 V single polarity tuning ramp and the analog output properties of the PCI-1200 multifunction card can be better understood by comparing recorded ramps using both the single polarity and bipolar tuning ramps. Plots of the two different recorded tuning ramp voltages as a function of data point number showed an important similarity between them: the minimum voltage increment for both recorded tuning ramps was 2.44 mV. In other words the PCI-1200 will not accept the programmed single polarity tuning ramp step size of 5 V ÷ 4096 data points $\cong$ 1.22 mV. Also, comparing tuning rate as a function of data point number (determined by computing the forward numerical derivative of these recorded tuning ramp voltages) reveals differences: the tuning rate of the single polarity tuning ramp as a function of point number was fluctuating much more than that for the bipolar tuning ramp. The 0 to +5 V single polarity tuning ramp achieved an average step size of $1.22 \pm 2.00$ mV ÷ data points based on deviations of $\pm$ two steps away from the programmed voltage for roughly 30% of the steps and $\pm$ one step away for all the other data points in a scan. The −5 to +5 V bipolar tuning ramp achieved a step size of $2.44 \pm 1.14$ mV per data point based on deviations of only $\pm$ one step away from the programmed voltage for roughly 23% of the steps in a scan and no deviation from the programmed voltage for the remaining 67% of the steps in a scan. The tuning rate (as determined from the numerically computed first-derivative of the recorded tuning ramp with respect to data point number) of the single polarity tuning ramp appeared to randomly change sign roughly 15% of the time in a scan, while the tuning rate of the bipolar tuning ramp did not change sign through the entire scan.

The electrical leads from a mounting module holding the Philips laser diode were soldered to the appropriate lead in the Newport 700 laser mount and protected (insulated) with shrink wrap. The Philips laser diode was clamped in solid physical contact inside the Newport 700 with a mounting plate that had a circular opening, allowing the laser diode to fit through, but not the mounting flange that encircles the laser diode; this assembly was held in place by two screws. Additional thin aluminum spacers (flat washers that were fractions of a millimeter in thickness) were milled to take up additional space that might remain between this mounting flange on the laser diode and mounting plate to make sure the laser diode was held firmly in place, both to maximize heat conduction and stability of the direction of propagation of the laser beam. The flange of the laser diode was held in contact with a block of aluminum (ca. 4 inch × 4 inch × 1 inch) that had its temperature controlled (stabilized) by a single 12.5 Watt thermoelectric module held in place with heat transfer grease. (The aluminum block had a small circular hole about 6 mm in diameter along the short 1 inch axis near the center of the face formed by the 4 inch sides to allow for electrical connections to the laser diode.) A precision 10 k$\Omega$ thermistor was bonded with a thermal epoxy to the inside of a hole milled in this block of aluminum and provided the feedback signal for stabilizing the temperature of the block, and thus the laser diode. The Newport 325 was electrically connected to the 10 k$\Omega$ thermistor and the thermoelectric module, which performs the feedback operation and provides power for operation of the thermoelectric module. The temperature of the aluminum block (and thus the laser diode housing) was set by manually setting the thermistor resistance (ca. 12.4 k$\Omega$) on the Newport 325 temperature controller. The temperature calibration data provided by the manufacturer (Newport) for this particular laser mount (Newport 700) indicates that the temperature was stabilized at roughly one to two degrees Kelvin above the ambient room temperature.

It has been reported [Xu] that the Philips CQL806/30 laser diode temperature tuning rate was about 0.125 nm per K. When adjusting (typically the sensitivity and phase settings) of the lock-in amplifiers it was possible to set this laser diode system to the appropriate wavelength to sit on any of the spectral features of diatomic iodine and watch over the course of several minutes. During this time interval, one phase of the wavelength-modulated absorption signal would at first be present near its maximum signal level for a few seconds and then gradually disappear over the course of several more seconds. A few seconds later the opposite phase of the wavelength-modulated absorption signal would begin to appear and several seconds later reach a maximum signal level, which will remain for several more seconds and then gradually disappear. If it is assumed that this represents a frequency stability of about 1000 MHz (for an observation time interval of several minutes) then the corresponding temperature stability of the Philips laser diode at a fixed wavelength of 675 nm was on the order of $(0.00152 \text{ nm}) \div (0.125 \text{ nm} / \text{K}) \cong 12 \text{ mK}$ (over a time interval of about one minute), which agrees well with the manufacturer specifications of the Newport 325 temperature controller.

While it was possible to collect feedback data from the 10 k$\Omega$ thermistor in the Newport 700 laser mount through the Newport 325 temperature controller during operation of the laser, no such efforts were undertaken.

The laser beam was originally collimated using an uncoated lens (or at least not optimized for antireflection at 675 nm) encased in a threaded mount that screwed into the front plate of the Newport 700 laser mount. A few additional aluminum spacer plates were milled that had a thickness of about two to three millimeters and fit between the main body of the Newport 700 laser mount and the portion that holds the collimating lens, all of which was held in place by four screws. The use of these spacer plates allowed for more flexibility in combinations of collimating lenses and laser diodes that could be used and in general allowed for more precise control of the distance between the collimating lens and the laser diode. The collimated laser beam was then split into three roughly equal portions using two partially reflecting dielectric coated mirrors as beam splitters (Figure 3.1). The power of the collimated laser beam on each of the three paths was measured to be relatively low, in the range of one to five mW, and the beam was estimated to be 5 mm in diameter. One portion of the laser beam was sent through an evacuated reference gas cell (labeled RC in Figure 3.1) containing a few mg of solid diatomic iodine that sublimed to a partial pressure in the gas phase appropriate for high-resolution spectroscopic investigations (ca. 0.18 torr at 298 K) [Tellinghuisen]. Another portion of the laser beam was sent through an evacuated sample gas cell (labeled SC in Figure 3.1) that also contained a few mg of solid diatomic iodine and was subsequently filled with a buffer gas in the pressure range of 0 to 100 torr. All reference gas cell spectra were recorded at ambient room temperature (292 ± 1 K). Most sample gas cell spectra were also recorded at this same ambient temperature. However, several experiments recorded spectra of the sample gas cell at two elevated temperatures of 75°C (348 ± 5 K) and 115°C (388 ± 5 K).

The final portion of the laser beam was sent through a plane-parallel Fabry-Perot interferometer (labeled "etalon" in Figure 3.1). The etalon was composed of two (relatively) high quality, 2 inch, flat mirrors attached to precision mounts. The precision mounts were secured with screws to the top surface of a separate 1 foot by 1 foot aluminum optical bread board that had four small rubber feet on the bottom surface in adjacent corners to provide cushioned (i.e. vibration damping) support. This optical bread board was in free-standing contact with the optical table. The etalon was at room temperature and open to the room. The inner faces of the etalon mirrors were separated from each other by about 10 inches. It was later determined by comparison with known absolute wave number values of several diatomic iodine features [Luc 1 and 2] that the plane-parallel Fabry-Perot interferometer [Yariv] had a free spectral range of about 642 ± 2 MHz when used with the Philips laser diode system. (Calibration of the free spectral range of the etalon is described in more detail in Chapter IV and Sub-Section 4.6.2.)

Detection of the laser beam intensity at the end of any of the three laser beam paths was done using a variety of silicon-based electronic photodiode detectors. Some of the sample gas cell spectra recorded with this configuration of the spectrometer used only one channel of a New Focus Nirvana auto-balancing photo receiver. (The Nirvana detector was not being used in auto-balancing mode at this time.) When using the New Focus Nirvana detector on the sample gas cell beam path, a ThorLabs 110 Volt (wall plug) powered silicon detector was used on the reference gas cell beam path and the etalon channel was blocked, as it reflected too much light back into the laser diode cavity, causing at the very least larger intensity fluctuations of the laser beam on a time scale much shorter than the time interval required to collect a single point of the spectrum. Some of the later experiments using an off-axis parabolic reflector to collimate the laser beam were recorded using two battery powered ThorLabs DET110 silicon detectors on both the sample and reference gas cell beam paths and still without the etalon being used. The ThorLabs DET110 detectors were terminated with 8 k$\Omega$ external load resistors placed in parallel on each of the output signals as specified by the manufacturer for relatively slow operation and maximum sensitivity. Most of these later experiments that used the off-axis parabolic reflector to collimate the laser beam also made use of the etalon on the third beam path. The detectors being used were the two ThorLabs DET110 and a ThorLabs 110 Volt (wall plug) powered silicon detectors. The ThorLabs DET110 detectors were mostly used on the sample and reference gas cell beam paths, but this was not the case in all recorded (and analyzed) spectra. The results obtained for the spectrometer configurations described above did not appear to depend critically on the detector being used.

The output from each photo-detector was connected to the input of a lock-in amplifier (see Figure 3.1). As mentioned above, lock-in detection at the first harmonic of a wavelength-modulated absorption signal has the appearance of the first derivative of the corresponding direct absorption spectrum. Stanford Research SR850, SR530, and SR510 lock-in amplifiers were used with the Philips laser diode system interchangeably on all the beam paths with no noticeable difference in the quality of spectra obtained. The reference frequency for lock-in detection came from the same signal generator that was used to generate the modulation voltage; an alternative output from the signal generator was used, one that provided a considerably larger voltage than that going to the summing amplifier. Typical sensitivity settings on the lock-in amplifiers were in the range of 0.5 to 10 mV. The phases of the lock-in amplifiers were optimized by tuning the laser diode to sit near the peak of a first derivative lobe of a diatomic iodine feature or etalon fringe, adjusting the phase until the output signal was set to zero, and finally rotating (or shifting) the phase by 90 degrees.

A single scan using the Philips laser diode system required anywhere from about 25 to 200 seconds. The time interval spent collecting a single data point was at least three times longer than the time constant of the lock-in amplifier. For a 25 second scan of the spectral region using the 12-bit resolution PCI-1200 multifunction card there were 4096 data points, which means that about 6 msec were spent at each data point so that the time constant was set to 2 msec or shorter. As long as an appropriate time constant was chosen the recorded line shapes did not appear to depend critically on the modulation frequency. (However, a recorded line shape is expected to depend critically on the modulation depth, which will be explored further in Sub-Section 3.2.4 and Section 5.6.)

In addition to the bandwidth narrowing technique of using a lock-in amplifier as a means of decreasing the noise levels in recorded spectra, it was also possible to reduce the $1/f$ noise and thereby increase the signal-to-noise ratio (S/N) by increasing the modulation frequency [Horowitz 1]. However, the lock-in amplifiers impose an upper limit of about 100 kHz on the modulation frequency that can be used with the Philips laser diode system. A 25 second scan with a 2 microsecond time constant would require a modulation frequency of at least 5 kHz to achieve the desired goal of at least ten modulations per time constant. The experiments using the Fabry-Perot laser diode system that were directly under my control (beginning in about the fall term of 2004) were conducted with 25 second scans at a modulation frequency of roughly 90 kHz, just under the modulation frequency upper limit of 100 kHz of the lock-in amplifiers used in these experiments. The front panel connection of the Newport 505 laser diode can accept modulation frequencies up to a rate of about 500 kHz. (Larger modulation frequencies can be achieved with the use of a bias-T, but such an approach would also require lock-in amplifiers designed to work at these larger frequencies; other issues would also have to be addressed, such as the possible generation of sidebands in the radiation field and the state-changing rate of the chromophore relative to the modulation frequency of the radiation field.)

Several wavelength-modulated linear absorption spectra were sequentially collected at each pressure of a given buffer gas by continuously tuning the laser diode back and forth from high to low wave number, low to high wave number, and so on. This process was aided and made highly automated by data collection algorithms supplied with the Wavemetrics software. The recorded spectra were later separated into single spectral scans and sorted according to the direction in wave number of the laser scan across the spectral region with custom (in-house) developed algorithms. (See Figures 4.1 and 4.2 for images of a typical raw (i.e. un-calibrated) spectrum recorded by the Philips CQL806/30 laser diode system.)

Before using the Philips laser diode system it was first necessary to establish conditions of temperature and injection current that produce single-mode laser radiation from this light source. This was done initially using a 0.75 m Czerny-Turner (grating) monochromator (Spex 1200) configured as a spectrograph by attaching a charge-coupled device (CCD) detector containing 700 pixels in a linear array with a pixel size of 0.01 mm at the focal plane. (The absolute wave number of the 0.75 m spectrometer was calibrated with the lines from a Fe-Ne hollow cathode lamp.) The dispersion of the Spex 1200 spectrometer (due to the reflection grating) was 1 nm of input radiation wavelength per mm in the output focal plane, which corresponds to a spectral recording range of $(0.01 \text{ mm / pixel}) \times (1 \text{ nm / mm}) \times (700 \text{ pixel}) = 7 \text{ nm}$. The grating was of sufficiently high quality that the CCD detector was the limiting factor in resolving power of this spectrograph. The resolution of the spectrograph was taken as $7 \text{ nm} \div 700 \text{ pixel} = 0.01 \text{ nm}$ of input radiation, which corresponds to a resolution of roughly $152 \text{ cm}^{-1} \div 700 \text{ pixel} \cong 0.2 \text{ cm}^{-1}$ in the vicinity of 675 nm. Since it is common for Fabry-Perot laser diode modes to be separated by about $2 \text{ cm}^{-1}$ (or more), this spectrograph possessed adequate resolution and covered a large enough spectral region in a single shot to distinguish between single-mode and multi-mode operation of the Philips laser diode. After setting the temperature of the Philips laser diode to just slightly above room temperature several spectrograph spectra were recorded at different values of the injection current, which allowed for identification of the appropriate current range that offered single-mode laser output.

The above steps also helped establish the beginning and ending wave number values of the single-mode scan range to a precision of about $0.2 \text{ cm}^{-1}$. The absolute wave number range was later rechecked with a wavelength meter, (TQ8325 wavelength meter from Advantest) to an accuracy of about $0.05 \text{ cm}^{-1}$. These measurements and visual inspection of fringe spectra indicate that the radiation produced by the Philips laser diode was single-mode perhaps to better than 99% across the spectral region of $14,817.95 - 14,819.45 \text{ cm}^{-1}$. Furthermore, since the tuning ramp voltage changed linearly with data point number, investigation of the fringe spacing as a function of data point number of the fringe peaks revealed that this laser diode scanned nearly linearly with changing injection current. (The linearity of laser scans is explored further in the Chapter IV.) The Philips laser diode system spectrometer was used for about five years and proved to be quite reliable; the spectra obtained during this time were highly reproducible with relatively small and infrequent changes to the front panel settings being necessary.

## 3.2.2 Collimation with an Off-Axis Parabolic Reflector

A later configuration using the Philips CQL806/30 laser diode removed the collimating lens from the Newport 700 laser diode mount and instead used an externally placed off-axis parabolic reflector to collimate the laser beam (Figure 3.2). The off-axis parabolic reflector was a specially milled piece of aluminum that was mounted so that x-y-z control of spatial orientation (relative to the laser beam) using good quality screw-driven translation stages was readily achieved and designed specifically for this purpose in a laser diode spectrometer that operates in the far-IR. The front plate of the Newport 700 was left in place in an effort to maintain good temperature control of the laser diode. Collimation with the off-axis parabolic reflector helped to further reduce the reflection of laser light back into the laser cavity, so as to mitigate the effects of an unwanted optical feedback, which resulted in a significantly quieter laser output and a larger signal-to-noise ratio (S/N) in the acquired spectra as compared to the earlier method that used a collimating lens encased in a threaded mount that screws into the front plate of the Newport 700 laser mount. A quieter laser output refers to an increased S/N as measured by a photodiode detector, which does not distinguish between amplitude and wavelength stability. In the spectrometer that did not use a parabolic mirror to collimate the laser beam, this noise increased substantially when scanning over a diatomic iodine transition line shape (as compared to an off-resonance wavelength).

This configuration of the spectrometer using an off-axis parabolic reflector with the Philips laser diode was able to surpass the 12-bit analog-to-digital conversion capability of the PCI-1200 multifunction card, which can be expressed as a maximum digital resolution of 1 part in 4096. The analog signal provided by a photodiode would surpass the 12 bit- resolution when processed through a lock-in amplifier with the time-constant set to approximately 1 second. Such an optimization would require a single spectrum scan time-interval of (3 sec / data point) × (4096 data points / scan) $\cong$ 205 minutes. Instead of investing a considerable amount of time in acquiring far fewer optimized spectra, a balance between signal-to-noise ratio (S/N) and speed of data collection was sought. The balance chosen set the lock-in amplifier time constant to a range that gave a S/N of about 400, or about 10% of the possible maximum value.

The use of an off-axis parabolic reflector for collimating a light source is a common method in the far-IR and mid-IR, where refractive optics are generally not as convenient or effective as reflective optics for manipulating and directing electromagnetic radiation.

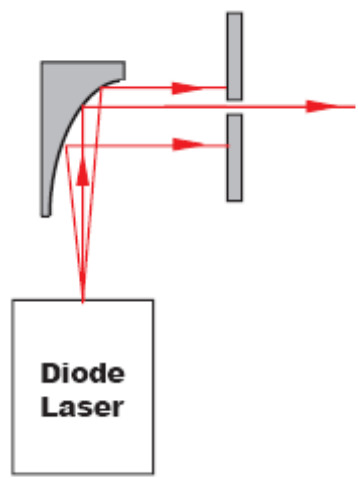

Figure 3.2 Schematic of the off-axis parabolic reflector used to collimate a divergent laser diode beam. An iris for selecting a portion of the collimated beam is depicted in the upper right portion of this figure; its diameter was set to ca. 5 mm.

The power of the laser beam on each beam path using the off-axis parabolic reflector for collimation of the laser beam (as seen by the 3.6 mm x 3.6 mm square active area of the ThorLabs DET110 detectors) was conservatively estimated to be about 250 times less than when using the collimating lens encased in a threaded mount. The reason for this reduction in laser beam intensity was that the focusing properties of the off-axis parabolic reflector require it to be placed about 4 cm away from the laser diode output, while the collimating lens encased in a threaded mount was expected to be about 2.5 mm away from the laser diode, so that in the former case the laser beam has diverged to a considerably larger diameter, about 16 times that of the latter. (This is based on the assumption of uniform beam intensity across the entire collimated beam, while a more accurate analysis would use a Gaussian profile. And since the approximately 5 mm opening of the pick-off iris was not carefully aligned with the center of the collimated beam the reduction factor might reasonably be expected to be greater than 250.) The total output power of the Philips laser diode when operated at 25 °C was specified by the manufacturer to be about 15 and 20 mW for injection currents of 50 and 60 mA, respectively. If the total laser beam power available for this configuration of the spectrometer was reduced by a factor of 250, then somewhere between 60 and 80 μW was divided into roughly equal portions for the three beam paths of this spectrometer. (Since the energy per photon at 675 nm is roughly $2.94 \times 10^{-19}$ J, a beam path power of 20 μW amounts to about 68 trillion photons per second.)

### 3.2.3 New Focus 6202 External Cavity Laser Diode System

The last few experiments conducted explored a wider range of rotational states of diatomic iodine using a New Focus 6202 external cavity laser diode, which has a linearly polarized single-mode laser output tuning range from 665.1 nm (15,035 cm$^{-1}$) to 676.1 nm (14,791 cm$^{-1}$). The total (un-split) output power emitted by this light source during these experiments was set to about 1 mW. The internally referenced absorption spectrometer built around this light source was similar to what is depicted in Figure 3.1 with some differences that will now be described. A second personal computer was used (Apple personal computer; late G3 processor using Mac OS X operating system) to set various parameters of this laser system, most particularly the start wavelength (for the first scan only of a given scan region during continuous operation of the laser), with GPIB commands sent and received using a National Instruments GPIB board that was installed internally in this personal computer, National Instruments software to operate this board, Igor 5.0 as a software platform for communicating with the laser diode system, and NIDAQ Tools as an interface between the Igor 5.0 command code and the GPIB board.  The connection between the GPIB board in the Apple personal computer (configured as a serial port) and the back panel of the New Focus laser control electronics box was achieved with a set of electrical wires bundled into a cable and terminated in the RS-232 standard.  (It was not possible to set the start wavelength with sufficient accuracy using the front panel controls for the New Focus laser diode system, and so this was accomplished by patiently tweaking (i.e. adjusting) the laser controls until a suitable starting wavelength was achieved.  The start wavelength only had to be set once for a given period of data collection; the laser diode would reliably return to the same start wavelength on each cycle of the laser scan that was software controlled through the sequentially repeated application of the voltage ramp.)  The etalon was on the more weakly reflected beam path created by the first split in the laser beam using an optically flat glass round (1 inch in diameter).  A second weakly reflected split of the laser beam (several centimeters downstream of the first split) using another flat glass round created another beam path for a fiber-optic connection to the wavelength meter (TQ8325 wavelength meter from Advantest).  The next split created a beam path for the sample gas cell using a dielectrically coated mirror as a beam splitter.  (Since the Nirvana detector was on this beam path a second split of roughly equal intensities was created with a carefully adjusted mirror creating the two beams required to operate this detector in auto-balanced mode.)  The reference gas cell was on the last beam path formed by the remaining transmitted beam.  This beam path used a battery-operated DET110 silicon detector with an 8 k$\Omega$ termination resistor placed parallel to the output signal.

This laser was exceedingly sensitive to scattered laser light reflecting back into the external cavity, so much so that a good quality optical isolator (Optics for Research model number IO-5-VIR adjustable faraday rotator, optimized for 633 nm) was necessary to isolate the laser diode from the rest of the spectrometer.  (There was not sufficient time to attempt using the off-axis parabolic reflector with the New Focus laser system.)  This laser diode system allowed for high resolution fine-tuning in segments of up to about 2.5 cm$^{-1}$ anywhere in the above stated single-mode scan region by the use of a piezoelectric element that micro-controls the position of the cavity end-mirror (a.k.a. tuning mirror).  A change in position of this highly reflective end-mirror relative to a critical pivot-point changes the length of the external cavity in such a manner that the number of waves in the cavity remains constant, which in turn results in single-mode tuning of the wavelength of this laser diode system.  Based on the manufacturer recommendation, tuning was done in only one direction, from shorter to longer wavelength.  The wavelength meter was used to measure the beginning and ending wavelengths of a given spectral region before any spectra were recorded.  After establishing the correct start wavelength using GPIB command from the second computer and readings from the wavelength meter, the Igor software was again used to continuously tune the piezoelectric element, so that the New Focus laser diode wavelength was repeatedly swept from high to low wave number at 12-bit resolution across a spectral region of interest through repeated application of a computer-controlled tuning ramp.

Due to the previously mentioned sensitivity of this laser to back-scattered laser light, the fiber optic coupling optics to the wavelength meter were covered with a black felt-like cloth while spectra were being recorded.  Three spectral regions were studied in the presence of one buffer gas (argon) with this laser system.  The region covered by the Philips laser diode was a subset of the first spectral region recorded with the New Focus (14,817.97 to 14,820.08 cm$^{-1}$), which made it possible to compare the two laser diode systems.  Also, spectra of a wider range of rotational states of diatomic iodine were recorded in the region from 14,946.17 to 14,950.29 cm$^{-1}$ with two overlapping spectral scan regions.  The two overlapping scan regions were (in the direction of the scans from high to low wave number) roughly 14,948.43 to 14,946.17 cm$^{-1}$ and 14,950.29 to 14,948.08 cm$^{-1}$.

The New Focus external cavity laser diode was repeatedly tuned (through software control) across a given spectral region with its wavelength-modulated in much the same way as the Philips laser diode.  (See Appendix A for details on changes made to the inverting summing amplifier circuit that was used with the New Focus laser diode system.)  The 12- bit PCI-multifunction board analog output of 4096 steps resulted in a step size between data points of approximately ($2.2$ cm$^{-1}$ ÷ 4096 steps) × (30,000 MHz / cm$^{-1}$) ≅ 16 MHz.  Measurements of the voltage change seen by the laser during a scan of the spectral region and the peak-to-peak voltage for the sine wave modulation voltage indicated that the modulation depth was about ($0.01$ V ÷ $5.24$ V) × $2.35$ cm$^{-1}$ ≅ $0.0045$ cm$^{-1}$ ≅ 134 MHz.

   The New Focus external cavity laser diode system, however, has an additional restriction (as compared to the Philips laser diode system) that the piezoelectric tuning element was specified to reduce its amplitude of motion by 3 dB at modulation frequency of about 2 kHz.  In order to achieve a compromise between this limitation and the reduction of $1/f$ (flicker) noise, the modulation frequency was set to about 2 kHz.  Taking as a first approximation that the change in cavity length was linear in the change in voltage applied to the piezoelectric tuning element, using a wavelength modulation frequency at the 3 dB point of the piezoelectric tuning element requires a revision of the modulation depth to roughly $134 ÷ 10^{0.3} = 95$ MHz.

   A single scan across the spectral region was set to about 256 seconds (ca. $0.0625$ seconds per data point for a 12-bit tuning ramp).  The lock-in time constant was set to $0.01$ sec.  The lock-in amplifier settings gave roughly six time constants per data point and 20 modulations per time constant.  Three lock-in amplifiers (EG&G model number 5104) were used with this laser system.

   It is also worth noting that the same Fabry-Perot etalon was used with this spectrometer except that the mirrors were positioned slightly further apart, resulting in a measured free spectral range of roughly $578 ± 2$ MHz; see also Chapter IV.  Compared to the layout used with the Philips laser system, the length of the beam path from the laser diode to the etalon was nearly doubled (to about five meters) to help further reduce backscatter of laser radiation towards the external cavity of the New

Focus laser diode. Also, the intensity on this beam path was reduced by about three orders of magnitude (using an appropriate combination of square shaped colored glass filters and a single short wavelength cut-off filter with its edge near 1000 nm) so that it is compatible with the range of the silicon-based New Focus Femtowatt detector that was used on this beam path. And finally, the etalon was enclosed in an enclosure that sat freely on the optical table supported by the sides. This cabinet had two small holes (ca. 1 inch in diameter) on opposite sides for entrance and exit of the laser beam. The etalon was covered with a cabinet in this manner to minimize the effects of air currents on the etalon signal.

The specifications provided with the New Focus external cavity laser diode state that the laser diode is temperature stabilized to 1 mK and the surrounding environment in the laser head (an enclosed mounting cabinet similar in function to the Newport 700 laser diode mount in the Philips laser system) is stabilized to 10 mK. This is done with two (Peltier) thermoelectric units; one was dedicated to the laser diode and the other to the overall temperature in the laser head. The stabilization of temperature inside the New Focus laser head to 10 mK agrees well with the (above) estimate of temperature stabilization inside the Newport 700 laser mount using a single thermoelectric unit in the Philips laser diode system.

See Figures 4.5 and 4.6 for images of a typical raw (i.e. un-calibrated) spectrum recorded by the New Focus 6202 external cavity laser diode system.

## 3.2.4 Signal-to-Noise Ratios, Modulation Depths, and Laser Line Widths

Reliable specifications of the line width or line shape of the Philips laser diode were not readily located, and a rigorous attempt to measure this quantity was not undertaken. However, the New Focus external cavity laser diode system was specified by the manufacturer to have a laser line width of about 5 MHz. Signal-to-noise ratio (S/N) and etalon finesse are used below in two different approaches to determining a semi-quantitative value of the line width for the Philips laser diode system.  The importance of knowing the laser line width lies in the fact that subsequent analysis of the molecular line shape does not account for the laser line shape, which is expected to be a reasonable approximation when the laser line width is considerably narrower than the molecular line widths.

A measure of the signal-to-noise (S/N) ratio of diatomic iodine transitions can be made by comparing the maximum total signal deflection (peak-to-peak) for a single diatomic iodine feature to the standard deviation of the measured signal for a continuous segment with two hundred off-resonance (baseline) data points (between and unaffected by molecular lines), which will be taken as the noise level.  The measure of noise obtained in this manner is expected to be predominantly related to the amplitude jitter (random fluctuations) of the laser intensity.  (It can perhaps be anticipated that the on-resonance noise will always be larger than the off-resonance noise to due a contribution from the uncertainty in wavelength; such considerations were not explored further in this project.)

Signal-to-noise (S/N) ratios for three configurations of the internally referenced absorption spectrometer are listed in Table 3.1.  These results are based on a simple analysis of the signal from only one line (Feature E; see Table 4.5) and the standard deviation (noise) of slightly less than 200 nearby baseline data points (between features E and F; see Table 4.5) of a single typical spectrum recorded with each configuration of the internally referenced absorption spectrometer.

It is tempting to speculate on the nature of the increase in modulation depth (from about 21 MHz to 95 MHz) required when switching from the wider line width of the Philips laser system to the narrow line width of the New Focus laser system (manufacturer specified to be ca. 5 MHz) in order to obtain comparable S/N ratios between the two laser systems.  Developing a simple model based on linear proportions is done by considering the case that the two laser systems produce spectra of comparable S/N ratios, which is the situation for the second and third configurations listed in Table 3.1.  In the limit that the full width at half maximum (FWHM) of the incident laser line shape goes to zero, the base population in the absorption medium capable of resonant interaction with this radiation source tends to zero, and thus the amplitude of the observed signal also tends to zero.  If we assume there exists a linear relationship between the base populations available for resonant interaction at a particular frequency for each laser system and the observed signal, then a quick estimate of the effective line width (FWHM) of the Philips laser diode necessary to give roughly the same S/N ratio (and thus the same signal levels) can be made:  (95 MHz ÷ 21 MHz) × 5 MHz ≅ 23 MHz.  (Implicit in this approximation is the equivalence of noise signatures for the two spectrometers.)

Table 3.1 Signal-to-noise (S/N) ratios for three configurations of the internally referenced absorption spectrometer.  The three configurations are classified by laser system and the method used to collimate the laser beam.  The largest S/N ratio was obtained using the Nirvana auto-balanced detector.

| Laser System | Collimation | S/N ratio | |
| --- | --- | --- | --- |
| | | Reference | Sample |
| Philips | Lens | 82 | 75 |
| Philips | parabolic reflector | 370 | 420 |
| New Focus | lens and isolator | 210 | 570 |

A second approach to providing qualitative evidence that the line width of the Philips laser diode was considerably narrower than the ro-vibronic transitions of diatomic iodine being investigated (ca. FWHM of 600 MHz) involves comparison of the apparent finesse (free spectral range ÷ FWHM of the fringe width) of the etalon for both laser diode systems (for the second and third configuration listed in Table 3.1).  Typical etalon fringe shapes for these two laser systems are shown in Figures 4.2 and 4.6.

The shape of the fringes for three etalon spectra were analyzed using a Gaussian line shape model to determine the location and width of each fringe peak (in data point number units). Two fringe spectra with about 70 fringes in each spectrum were analyzed for the Philips laser diode system; one spectrum for each scan direction (low-to-high and high-to-low wave number). A single fringe spectrum that contained 111 fringes was analyzed across the entire scan region of the New Focus external cavity laser diode system that overlapped with the region scanned by the Philips laser diode system. The Gaussian line shape model gave a residual (the observed minus calculated etalon signal) that was somewhat larger than desired for a rigorous line shape analysis; the ratio of the standard deviation of the fit residual to the fringe amplitude (used as a measure of the relative discrepancy between the model and observed line shape) was about 2%. (A more rigorous analysis of the line shape of the etalon fringes would involve a model that convolves a laser line shape with the etalon line shape.) Nonetheless, the analysis of the fringe spectra indicate that the apparent finesse of the etalon fringes when using the Philips laser diode was about $4.0 \pm 0.6$ with a free spectral range of $642 \pm 1.8$ MHz; the New Focus external cavity laser diode system finesse was about $11.8 \pm 0.3$ with a free spectral range of $579 \pm 1.5$ MHz. (A bit more detail on the etalon fringe widths can be found in Section 4.4 and calculation of the free spectral range of the etalon during wave number calibration of a spectrum is described in more detail in Section 4.6.)

The width (FWHM) of the fringes obtained when using the New Focus external cavity laser diode system are given by the free spectral range ÷ finesse = 579 ÷ 11.8 = $49 \pm 1.3$ MHz which is about an order of magnitude larger the specified width of the laser line (ca. 5 MHz). For the Philips laser diode system, this width was $642 \div 4.0 = 1.6 \times 10^2 \pm 24$ MHz. (Uncertainty in the fringe width has been included in these last two calculations using the usual uncorrelated parameter propagation of error method; see also Section 1.6.) While it is not a trivial task to predict the relative contribution of the laser line width and the etalon to the observed fringe width, these results at least qualitatively indicate that the external cavity New Focus laser diode line width was considerably narrower than the Philips free-running laser diode, perhaps by a factor of three or four, as suggested by the changes in etalon line shape and the ratio of the modulation depths. (See also Section 4.4)

## 3.3 Reference and Sample Gas Cells

The focus of this section is on the gas cells used to hold a solid sample of diatomic iodine, its sublimed vapor phase component, and buffer gases in the range of about 0 to 150 torr isolated (at a constant volume) from the rest of the universe with regard to exchange of matter. These gas cells are also equipped with windows at either end that easily allow a collimated (5 mm diameter) beam of relatively monochromatic optical wavelength light to pass directly through them. After passage through an empty and evacuated gas cell small and reproducible changes in the intensity of the detected laser beam were readily discerned. (This of course points out the usefulness in using an auto-balanced detection scheme to help remove some or all of the consistently reproducible wiggles in the spectrum that arise from imperfections in the optics.)

One of the most pressing concerns that arises in these high-resolution spectroscopy experiments that analyzes the shape of molecular lines are the induced changes in the chromophore line shape due to small, relatively smooth undulations in the baseline. This type of reproducible "noise" is often due to etalon effects e.g. in the gas cell windows. However, there are other thin parallel-face optical components in these experiments besides the gas cell windows that will internally reflect some portion of light and are thus capable of setting up etalons. Of particular concern and unchecked are what appear to be thin glass or plastic covers in front of the active element of the silicon photo-detectors, especially since they are perpendicular to the laser beam, such that a relatively slowly varying and low contrast etalon signal might be readily created when tuning the laser. If the (digitally recorded) baseline artifacts are sufficiently apparent, smooth, and slowly varying it is perhaps possible to reduce their effects in the digitally recorded spectra with the use of computer software (such as Igor 5.0; see also Figure 4.10). We did not attempt to do this on any of the spectra analyzed in this project and it proved to be non-trivial when attempting to do so in the acetylene project with spectra obtained using a multi-pass White cell [Hardwick 3]. (However, the line shape fitting algorithm did account for the typically observed constant baseline slope in linear absorption spectra that are not wavelength modulated.)

## 3.3.1 Design and Construction of Gas Cells

The reference and sample gas cells were constructed of BK7 glass, mostly cylindrical in shape, approximately 1 m long by one inch in diameter, and had a single good quality high vacuum 1-arm polytetrafluoroethylene (PTFE) valve with glass plug (body) and Viton o-ring (e.g. Chemglass model number CG-982). Three gas cells were fitted with windows at Brewster's angle [Born; Resnick]; of course, such considerations only make sense when using linearly polarized light, as is the case in this project). A fourth gas cell used wedge windows with a 3° angle mounted nearly perpendicular to the laser beam path. All windows were attached using a silicone adhesive (DuPont aquarium sealant). Each gas cell had different quality windows listed in order from most expensive to least expensive: a) water purged quartz mounted at Brewster's angle (most expensive); b) Edmond Scientific BK7 spectroscopic quality 1 inch round mounted on a thinner diameter glass cell at the Brewster angle to the laser beam; c) ThorLabs BK7 spectroscopic quality (3° angle) wedged 1 inch round; d) what might be thought of as having relatively low-grade optical properties, quality microscope cover slides (least expensive). There were no obviously noticeable or remarkable differences in the relative quality of the recorded spectra for the different windows.  The gas cell with the 3° angle wedged windows
was only used as the sample cell for the experiments at two elevated temperatures of
348 and 388 K.  (Room temperature spectra were also recorded in the sample cell
during these elevated temperature experiments.)

## 3.3.2 Preparation and Handling of Gas Cells

The reference and sample cells were allowed to fill with room air by
unscrewing the PTFE high vacuum valve from the valve body.  Also, diatomic iodine
was exposed to the same atmosphere upon opening the bottle.  A small amount (ca. 5
mg) of solid diatomic iodine was placed in the gas cells using the tip of a spatula as a
transfer tool.  After charging a given cell with diatomic iodine, the high vacuum valve
was screwed back into the valve body, the gas cell was connected to a vacuum-line
using ultra-torr connectors (and much of the time a relatively un-reactive plastic tube
of a few inches in length as a flexible bridge in the connection between the glass gas
cell and the glass vacuum-line), the gas cell was evacuated of buffer gas, the high
vacuum valve was closed, and the diatomic iodine was allowed to sublime back to its
partial pressure of about 0.18 torr.  After waiting a few minutes for out-gassing of
contaminant gases from the glass surfaces on the interior of the gas cell and the solid
piece of diatomic iodine, the gas cell was again evacuated to zero pressure and closed.
Another few minutes were allowed to pass before the mostly evacuated gas cell was
pumped down to zero pressure for a third time.  The two-stage, belt driven pump
provided a relatively fast pumping rate for the small combined volume of a gas cell
and glass vacuum-line.  According to pressure measurements made with a
thermocouple gauge, the expression "zero pressure" amounts to about 50 mtorr.  No
special procedures (e.g. flame drying) were employed to remove residual contaminant
gases that might be on the gas cell walls or to add the diatomic iodine using standard
vacuum line procedures that would have allowed for more complete purging of contaminant gases and/or complete handling under vacuum and/or inert and controlled atmospheres.

The sample gas cell was then charged with a buffer gas through a closed system comprised of a buffer gas cylinder, a sample gas cell, a vacuum line, and two pieces of tubing to connect the gas cylinder and the gas cell to the vacuum line. The sample gas cell was filled with the buffer gas to about 350 torr and evacuated to zero pressure at least two times before an appropriate high pressure quantity (typically no more than about 200 torr) was allowed into the gas cell, which was then sealed-off by the high vacuum valve. After a series of scans were made at a given pressure in the sample gas cell, the pressure was reduced by allowing the gas in the sample cell to expand into the larger volume of the sample gas cell plus the evacuated vacuum-line (including a short flexible plastic hose connecting them). This was done by reconnecting the sample gas cell to the vacuum-line, pumping down both the vacuum line and the hose connecting the vacuum line to the high vacuum valve of the sample gas cell down to zero pressure, and then closing off the vacuum line to the pump (to create a closed system composed of the vacuum line, sample gas cell, and the hose connecting them); the high vacuum valve on the sample gas cell was then opened until the pressure reading was constant, and then closed. These pressure readings also reveal that for a given set of pressure reductions the pressure reduction ratio (initial pressure ÷ reduced pressure) was roughly $1.75 \pm 0.01$ for initial pressures (in the sample cell before performing a pressure reduction) above about 10 torr. The final two pressure reductions that might be made for initial pressures below 10 torr would see a sudden drop in the pressure reduction ratio to about 1.68 for the first reduction past 10 torr and about 1.65 for the pressure reduction after that.

The gas cells were typically held securely in place on the optical table with metal clamps wrapped in a softer material like cloth or rubber tubing to isolate the metal clamp jaws from the glass gas cells. The metal clamps were attached to ca. 0.5 inch diameter metal posts that were securely anchored to the optical table. (Most of the data recorded by an undergraduate student during the second historical period of data collection used free-standing gas cells on two laboratory-jacks, with the high vacuum valve stems resting against the flat surface of the laboratory-jack offering stability against rotation of the gas cells. The laboratory-jack stands were also elevated to the appropriate height to allow the laser beam to pass though the gas cells from cell window to cell window without touching any other portion of the cell. There may have been some slight noise due to spurious motion of the gas cell in this configuration, but no attempt was made to investigate its magnitude.)

After collecting several absorption spectra of diatomic iodine at a given pressure of buffer gas the sample gas cell was carried from the table to the other side of the laboratory where the vacuum line was located to reduce the pressure of buffer gas in this gas cell. After reducing the pressure of the buffer gas an additional few minutes were allowed to pass before collecting spectra at this reduced pressure of buffer gas to allow the diatomic iodine to reach an equilibrium vapor pressure. This process was typically repeated until reaching a low pressure of between three and five torr, at which point the cell was evacuated of all buffer gases and a zero pressure spectrum was recorded. (This last step of recording zero pressure spectra of the sample gas cell was not done for two of the three spectral regions investigated with the New Focus external cavity laser diode system; $14,817.95 - 14,819.45$ cm$^{-1}$ and $14,946.17 - 14,948.43$ cm$^{-1}$.)

At each pressure of the buffer gas the lock-in amplifier on the sample gas cell beam path was adjusted to optimize the strongest signal in the spectrum to use up as much of the 12-bit dynamic range of the detection scale (i.e. $\pm 5$ V) as possible, so that it did not become the limiting factor in achieving the maximum possible signal-to-noise ratio (S/N) attainable in these experiments. The optimization involved adjusting both the phase and sensitivity settings on the lock-in amplifier.

The pressure of the vacuum line was monitored at the vacuum-line with an absolute pressure measuring gauge (SenSym model number 19CO15PA4K from Honeywell) that linearly scaled from zero to 760 torr on a zero to 100 mV scale. The manufacturer stated accuracy of this pressure gauge was $\pm 0.2\%$ of full scale, which amounts to an uncertainty in pressure of about 1.5 torr. The digital multi-meter measurements of the pressure sensor were recorded at a resolution of 10 µV.

Since the zero pressure reading obtained when completely evacuating the vacuum line was quite reproducible for several weeks at a time it is reasonable to assume the above error is dominated by systematic (as opposed to random) effects. An estimate of the possible impact on the pressure broadening and pressure shift coefficients can be readily made for the case that the pressure scale is systematically expanded or contracted by 1.5 torr. The smallest and largest pressures of buffer gas were typically around 5 and 80 torr, respectively. A rough estimate of the maximum systematic error in the pressure broadening and pressure shift coefficients for this range of pressures can then be estimated: $100 \times 1.5 \div (80 - 5) \cong 2\%$ .

### 3.3.3 Heating the Sample Gas Cell

Spectra were recorded with the sample gas cell at maintained at a constant elevated temperature (348 and 388 K, ± 5 K in both cases). The reference gas cell was maintained at room temperature in all experiments.

The glass portion of the sample gas cell was wrapped first with a layer of aluminum foil to within less than a centimeter of the wedge windows. An approximately 1 inch by 8 feet strip of flexible cloth-covered heat tape was then tightly wrapped around the long axis of the sample gas cell (on top of the aluminum foil) in a tight helical pattern down the length of the gas cell, and terminating about 0.5 inch from each end (so as not to be in contact with the sealant connection between the gas cell and the gas cell windows). Approximately 14 strips of three inch wide, six inches in length, and relatively thin (ca. 1/2 inch thick when uncompressed) fiberglass insulation, each one wrapped in aluminum foil, were wrapped on top of the heat tape layer perpendicular to the long axis of the sample gas cell with slight overlap between successive wrappings, and secured to the sample gas cell by tying them off with two pieces of string per strip of insulation. A final layer of hot-water pipe soft (pliable) foam insulation of about 0.5 inch thickness and roughly 1.75 inch diameter was set in place over this.

Both ends of the sample gas cell were surrounded by a hollow steel cylinder of about 0.125 inch thickness, 3.5 inch inner diameter, and five inches in length that was wrapped in 1 inch by 6 feet strip of heat tape. Layers of the three inch wide insulation were also wrapped around these cylinders in a similar manner as was done on the sample gas cell. These cylinders were spatially overlapped with the end of the gas cell by about 3 inches and extended beyond the windows of the sample cell by about 3 inches (Figure 3.3).

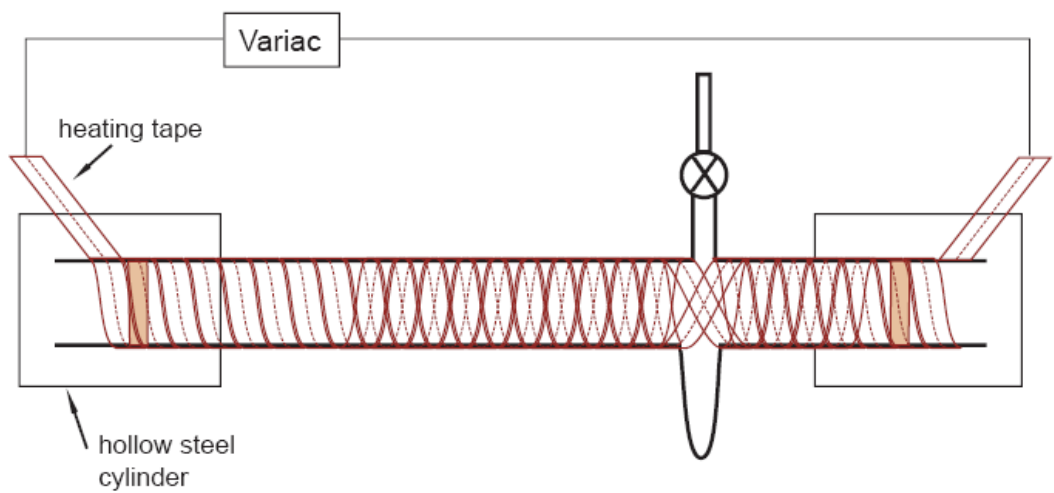

Figure 3.3 Schematic of experimental set-up used to heat the sample cell. The sample gas cell and the hollow steel cylinders surrounding both ends of the gas cell are wrapped in heat tape. The three strips of heating tape were electrically connected to separate variacs (variable transformer), each with an alternating input voltage at 120 V (wall outlet) and a maximum current handling capability of about 15 A.

The temperature of the sample gas cell was carefully monitored (two Fluke model number 51 digital thermometers) using thermocouples encased in a small diameter (ca. 3/16 inch) metal tube that were placed at both ends of the sample gas cell; these two thermocouples were placed as close as possible to the center of the heated steel cylinders without obstructing the laser beam. Flexible thermocouples were placed in direct contact with the sample gas cell wall at three locations; one thermocouple was placed near the center and two more were placed about 15 cm from each end of this gas cell.

After being added to the sample gas cell in the manner described above in Sub-Section 3.3.2, the diatomic iodine was condensed down into the cold finger (approximately 0.5 inch diameter and 2 inch depth) of the sample gas cell used in these experiments. The cold finger was about 2.5 inches from one end of the cell and sat directly below the high vacuum valve (as depicted in Figure 3.3). The cold finger was submerged about 1.5 inches into an approximately 4 inch tall, 250 mL beaker filled to near the top with room temperature water. An excess of solid diatomic iodine condensed at the bottom of the cold finger. The beaker of water remained at room temperature while the main body of the sample gas cell was maintained at one of the two elevated temperatures.

Room temperature spectra of the gas sample cell (and reference gas cell) were collected before and after elevated spectra were collected. It generally required about 30 to 45 minutes for the temperatures being measured on the sample gas cell to reach equilibrium on both heating and cooling. A large fan was generally used to help reduce the time interval required to cool the sample gas cell back to room temperature.

Since the sample gas cell was sealed off at the high vacuum valve while it was at room temperature (RT), simple application of the ideal gas proportional relationship $P_{RT} \div T_{RT} = P_{HT} \div T_{HT}$ was used to compute the pressure in the sample gas cell while it was maintained at a constant high temperature (HT). In addition to the roughly 2% uncertainty in the pressure reading obtained from the pressure sensor on the vacuum-line, an uncertainty in high and room temperatures of five and one Kelvin, respectively, and a high temperature of 388 K will result in a relative uncertainty of the pressure ($\sigma_P \div P_{HT}$) in the high temperature sample gas cell of roughly 1.7%. This was computed using the error propagation model based on the assumption of uncorrelated parameters described in Section 1.6; the uncertainty in the pressure at high temperature is given by $\sigma_P$.

## 3.4 Endnotes for Chapter III

[Born]            M. Born and E. Wolf (with contributions by A. B. Bhatia, et al.);
                  *Principles of optics; electromagnetic theory of propagation,
                  interference, and diffraction of light*; Pergamon Press, New York,
                  New York (1959); LCCN 58012496 /L.

[Demtröder]       W. Demtröder; *Laser Spectroscopy: Basic Concepts and
                  Instrumentation*, Third Edition, pages 374-378; Springer-Verlag,
                  Berlin (2003); ISBN 3-540-65225-6.

[Hardwick 1]      J. A. Eng, J. L. Hardwick, J. A. Raasch and E. N. Wolf; "Diode laser
                  wavelength modulated spectroscopy of $I_2$ at 675 nm",
                  Spectrochimica Acta, Part A, 60, 3413-3419 (2004).

[Hardwick 3]      J. L. Hardwick, Z. T. Martin, M. J. Pilkenton and E. N. Wolf;
                  "Diode laser absorption spectra of $H^{12}C^{13}CD$ and $H^{13}C^{12}CD$ at 6500
                  $cm^{-1}$", Journal of Molecular Spectroscopy, 243, 10-15 (2007).

[Horowitz 1]      P. Horowitz and W. Hill; *The Art of Electronics*, Sections 7.10 and
                  14.15; Cambridge University Press, Cambridge (1985); ISBN 0-521-
                  23151-5.

[Luc 1]           S. Gerstenkorn and P. Luc; Atlas du Spectre d'Absorption de la
                  Molécule d'Iode entre 14 800–20 000 $cm^{-1}$, Laboratoire Aimé
                  Cotton CNRS II, Orsay, (1978).

[Luc 2]           S. Gerstenkorn and P. Luc; "Absolute iodine ($I_2$) standards measured
                  by means of Fourier transform spectroscopy", Revue de Physique
                  Appliquee, 14, 791-794 (1979).

[Resnick]         R. Resnick and D. Halliday; *Physics*, Part 2, Section 48.3; John
                  Wiley and Sons, New York (1978); ISBN 0-471-34529-6.

[Silver]          J. A. Silver; "Frequency-modulation spectroscopy for trace species detection: theory and comparison among experimental methods", Applied Optics, 31, 707-717 (1992).

[Tellinghuisen]   J. Tellinghuisen; "Transition strengths in the visible–infrared absorption spectrum of $I_2$", Journal of Chemical Physics, 76, 4736-4744 (1982).

[Xu]              G. Xu; *Manipulation and Quantum Control of Ultracold Atoms and Molecules for Precision Measurements*, University of Texas at Austin, PhD dissertation (2001).

[Yariv]           A. Yariv; *Optical Electronics*, Third Edition, Chapter 4; CBS College Publishing, New York (1985); ISBN 0-03-070289-5.

CHAPTER IV

ANALYSIS 1 – ASSIGNMENT AND CALIBRATION

## 4.1 Overview of Chapter IV

At the outset of this project, only one measurement of the line-center shift of diatomic iodine as a function of foreign buffer gas pressure was found in the literature [Fletcher].  Such measurements for diatomic iodine have been rare, which is relevant to metrology [Borde 2].  In view of the lack of independent comparisons that can be made with the pressure broadening and pressure shift parameters presented in this dissertation, it became essential to characterize the limits of the experimental techniques and of the equipment used [Hardwick 1].  A significant portion of this chapter and the next one (Chapters IV and V) explores this characterization.

At the outset of this project, the beginning and end of a scan were determined approximately using a monochromator.  A more accurate wave number scale was established by assigning the diatomic iodine absorption lines in the region, then fitting the wave number of each line as a function of (tuning) ramp voltage (i.e. data point number), from which the wave number scan was found to be approximately linear in ramp voltage (or injection current).  (Diatomic iodine refers to the homonuclear molecule $I_2$, which is often colloquially referred to as iodine.)  An even more accurate wave number scale was established by recording interferometer fringes along with the spectra.  However, linear calibration of wave number (based on the wave number position of two diatomic iodine features) is considered to be sufficiently accurate in determining line width and line-center shifts (for different pressures of buffer gas in the range of roughly zero to 100 torr).  A consistent scanning range was maintained, especially in the case of the free-running laser diode system (Philips CQL806/30), by monitoring the temperature and injection current.

Visual inspection of the fringe spectrum as a function of data point number also served as evidence that the output from both laser diode systems were single-mode across the spectral regions of interest; there were no apparent discontinuities or irregularities attributable to multi-mode laser operation in the fringe spectrum. Furthermore, analysis of fringe spectra as a function of data point number obtained from both laser diode systems indicate that the fringe widths (FWHM) are relatively constant and statistically well determined across the entire spectral region, which implies that the laser line widths were relatively constant across the entire scan region.

In part, this chapter will describe the details and procedures involved in assigning quantum numbers to the states involved in the observed diatomic iodine transitions and in calibrating each spectrum such that a wave number is associated with each data point.

## 4.2 Linearity of Philips Laser Diode System Scans

Figure 4.1 shows a typical low pressure (ca. 0.18 torr at 298 K) wavelength-modulated absorption spectrum of the diatomic iodine reference gas cell and a simultaneously collected wavelength-modulated transmission fringe spectrum of the etalon obtained from scanning the Philips laser diode system across the region 14817.95 to 14819.45 cm$^{-1}$.

Figure 4.2 shows an expanded view of Figure 4.1 with the addition of a numerically integrated etalon trace. It is difficult to discern from a visual inspection the relative phase of the diatomic iodine and etalon channels; a more detailed analysis indicates that the upper integrated trace is a transmission spectrum with the high transmission at the high voltage extreme of the signal deflection. Integrating the first derivative diatomic iodine spectrum would also give such a transmission spectrum, which in both cases is due to the choice of phase settings on the lock-in amplifiers.

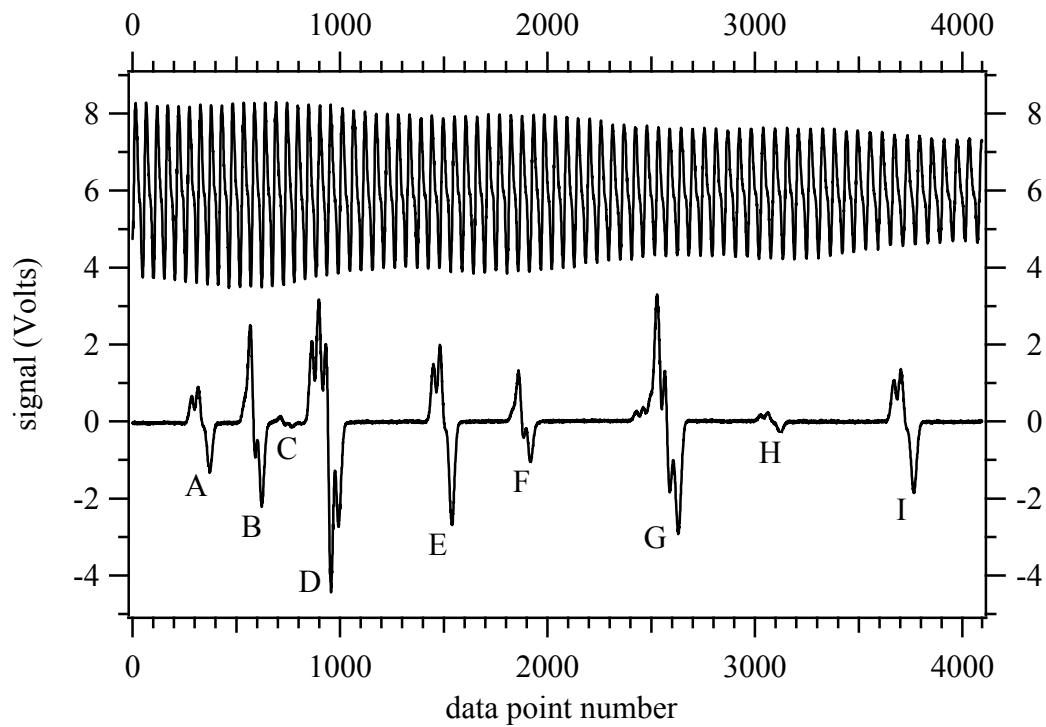

Figure 4.1 Wavelength-modulated spectrum of etalon fringes (upper trace) and diatomic iodine at room temperature (lower trace) as a function of data point number in the region 14,817.95 to 14,819.45 cm$^{-1}$ obtained with the Philips laser diode system. The laser was tuned from low to high wave number (up scan). The etalon fringe trace has been offset along the signal axis by six volts for clarity. Diatomic iodine feature labels are given by the letters A through I. The gas cell contained pure diatomic iodine at about 0.18 torr.

The extent of the linear relationship between wave number and point number for a typical spectrum obtained with the Philips laser diode spectrometer can be determined through linear regression analysis. The fringe spectrum for an "up scan" (low-to-high wave number) and a "down scan" (high-to-low wave number) were analyzed for fringe peak data point number as a function of fringe number using a polynomial model. The results for linear and quadratic fits are listed in Table 4.1.

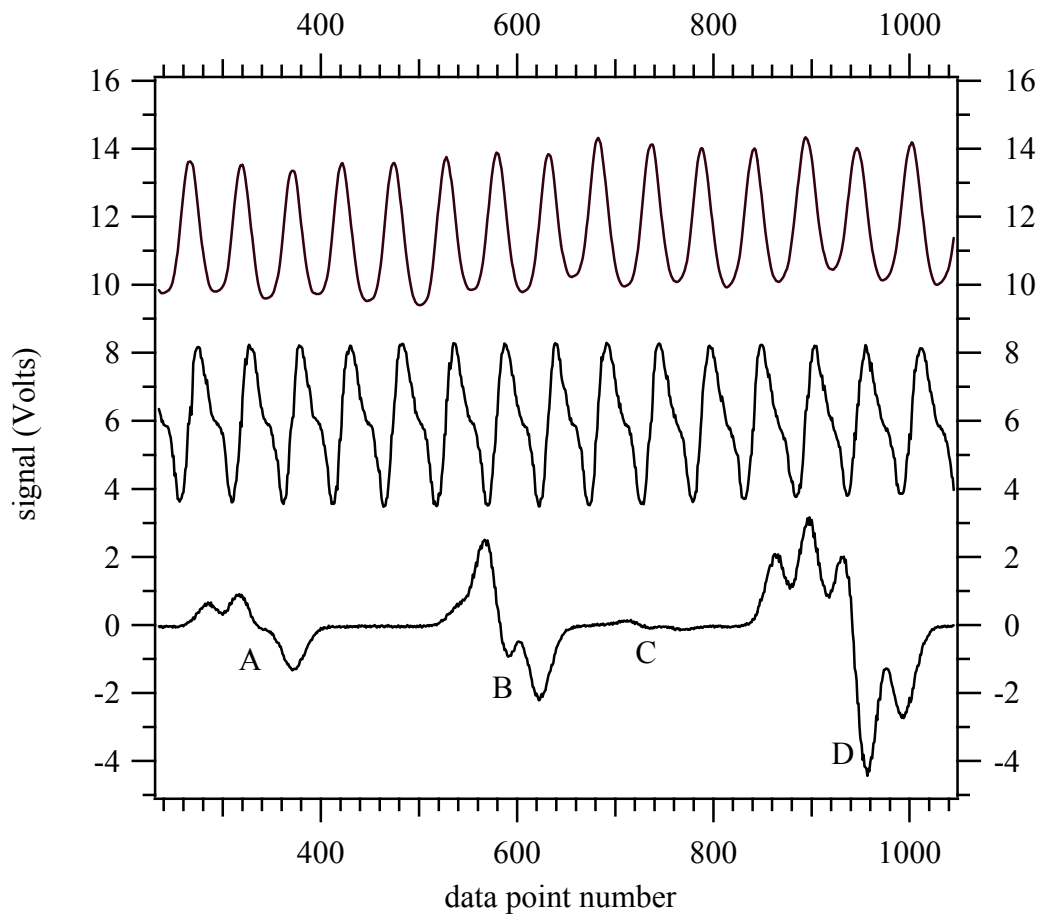

Figure 4.2 Expansion of a portion of Figure 4.1 from data point numbers 235 to 1045 with the integrated etalon signal appended as the top trace, whose amplitude has been altered and signal offset for clarity. The free spectral range of the etalon is $642 \pm 2$ MHz ($0.0214 \pm 0.0001$ cm$^{-1}$).

Figure 4.3 shows the residual ($FPPN_{observed} - FPPN_{calculated}$) plots for linear fits of the fringe peak point number ($FPPN$) as a function of the fringe number ($FN$) for one "up scan" (low-to-high wave number) and one "down scan" (high-to-low wave number); it is apparent in this figure that a quadratic correction term is warranted. No further improvement was observed in the residual plot or in the standard deviation of the residual value ($\sigma_{FPPN}$) for fits to polynomials of higher order than quadratic.

Table 4.1 Linear and quadratic fit results for observed fringe peak data point number (*FPPN*) as a function of fringe number (*FN*). The uncertainty $\sigma_{FPPN}$ is the standard deviation of the residual values ($FPPN_{observed} - FPPN_{calculated}$) for the fringe peak location in units of data point number. The "up scan" fringe spectrum contains 70 fringes and the "down scan" has 76 fringes across the approximately 1.4 cm$^{-1}$ scan region. The square of the correlation coefficient is r-squared.

$$FPPN = c \times FN^2 + b \times FN + a$$

| Scan/Fit Model | $c$ | $b$ | $a$ | $\sigma_{FPPN}$ | r-squared |
|---|---|---|---|---|---|
| Up/Linear | -- | 53.274(57) | 45.1(25) | 8.9 | 0.999932 |
| Up/Quadratic | 0.0228(10) | 51.511(74) | 411.8(11) | 3.1 | -- |
| Down/Linear | -- | 53.78(12) | 285.1(51) | 23.5 | 0.99961 |
| Down/Quadratic | 0.0533(12) | 49.598(89) | 32.9(14) | 4.3 | -- |

Let's take a moment to define and briefly describe the r-squared value, which is more accurately known as the square of the Pearson product moment correlation coefficient, and can be interpreted as the proportion of the variance in the dependent variable (*FPPN*) attributable to the variance in the independent variable (*FN*). Computation of this correlation coefficient is given by the formula:

$$r^2 = \left( \frac{\sum_i \left( x_i - \overline{x} \right)\left( y_i - \overline{y} \right)}{\sqrt{\sum_i \left( x_i - \overline{x} \right)^2 \sum_i \left( y_i - \overline{y} \right)^2}} \right)^2$$

(4.1)

Equation 4.1 contains the following pieces: $x$ is the independent variable, $y$ is the dependent variable, a bar over these variables indicates arithmetic mean (average) and the summations are performed over all data points (i). The correlation coefficient values range from zero to one, with zero indicating perfect anti-correlation (or randomness) of the variances of the independent and dependent variables, and one indicating perfect correlation of these variances; anti-correlation of the variances suggests that a linear relationship does not exist between the dependent and independent variables, while correlation of the variances suggests the opposite. The variance is the square of the standard deviation.

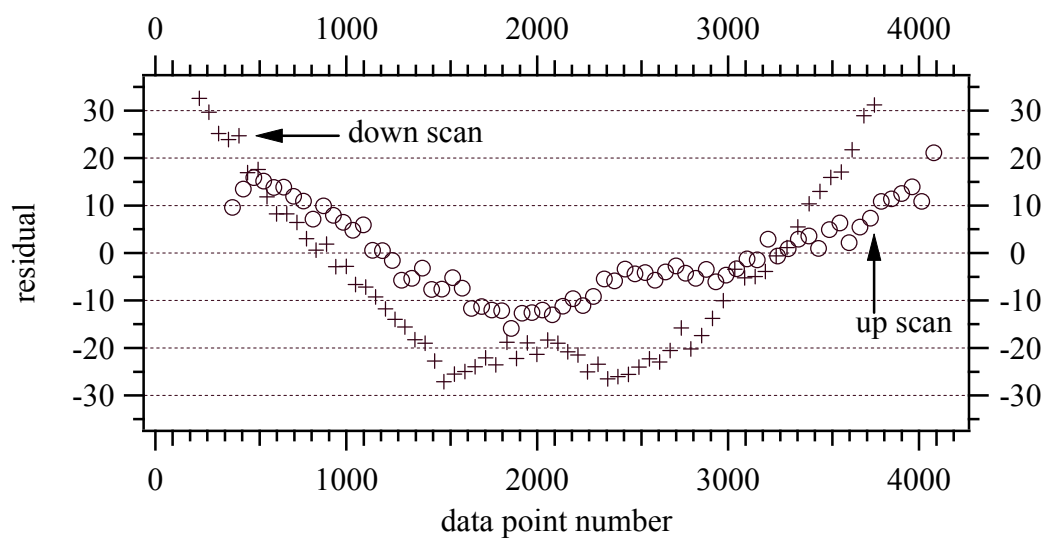

Figure 4.3 Residual plots of the linear fit of the observed fringe peak point number (*FPPN*) as a function of fringe number (*FN*) for both an "up scan" (○) and a "down scan" (+). The residual is the difference between observed and calculated fringe peak point number (*FPPN*$_{\text{obs}}$ − *FPPN*$_{\text{calc}}$) in units of data point number.

The accuracy of a relative wave number calibration based on a quadratic calibration curve for fringe peak data point number as a function of the associated fringe number can be investigated with linear fits to the numerically computed first derivative of fringe peak data point number with respect to fringe number $d(FPPN)/d(FN)$. Figure 4.4 shows the results of this computation of the derivative and the fit line. The uncertainty in the free spectral range of the etalon ($\sigma_{d(FPPN)/d(FN)}$) predicted by such a calibration procedure will be roughly 1.7 data point per fringe (Table 4.2). For the spectrum in figure 4.1 the Philips laser diode system was tuned at about 10 MHz per data point so that the uncertainty in fringe position would be roughly 17 MHz. The free spectral range of the etalon was later found to be $642 \pm 2$ MHz ($0.0214 \pm 0.0001$ cm$^{-1}$) (see Sub-Section 4.6.2 of this chapter) so that the relative accuracy of the calibration would be about $17 \div 642 \simeq 2.6$ % throughout the entire 1.4 cm$^{-1}$ scan region.

Table 4.2 Linear fit results for linear fit of first derivative of fringe peak point number (*FPPN*) with respect to fringe number (*FN*) as a function of *FN*. The uncertainty $\sigma_{d(FPPN)/d(FN)}$ is the standard deviation of the residuals, the set of which are computed using residual$_i$ = $(d(FPPN)/d(FN)_{observed} - d(FPPN)/d(FN)_{calculated})_i$.

$$d(FPPN)/d(FN) = 2c \times FN + b$$

| Scan/Fit Model | $2c$ | $b$ | $\sigma_{d(FPPN)/d(FN)}$ | r-squared |
|---|---|---|---|---|
| Up/Linear | 0.0414(11) | 51.92(44) | 1.8 | 0.18 |
| Down/Linear | 0.1177(85) | 49.47(37) | 1.6 | 0.72 |

Figure 4.4 also shows that the distance between fringes (in point number space) of the etalon slowly and monotonically increased across a scan of the laser regardless of the scan direction. Assuming that the precision of the free spectral range of the etalon far surpasses the detection capabilities of these experiments (see Sub-Section 4.6.2 of this chapter), this also implies that the mostly linear change in the laser tuning rate (slope of best fit lines in figure 4.4)) are opposite in sign in wave number space for the two possible scan directions; that is, a reproducible hysteresis affects the scan of the laser diode.

The worst case discrepancy between the observed line positions in data point number and the calculated values of a simulated spectrum in wave number units for a linear calibration based on the diatomic iodine Features D and G is about 30 data points. For a frequency step size of about 20 MHz per data point the linear wave number calibration method gives an uncertainty in line position of about 600 MHz (0.022 cm$^{-1}$). A relative error of 30 data points between two diatomic iodine features separated by about 0.7 cm$^{-1}$ would amount to a discrepancy in relative line center position between these two line of about $100 \times (0.022 \text{ cm}^{-1} \div 0.7 \text{ cm}^{-1}) \simeq 3$ %.

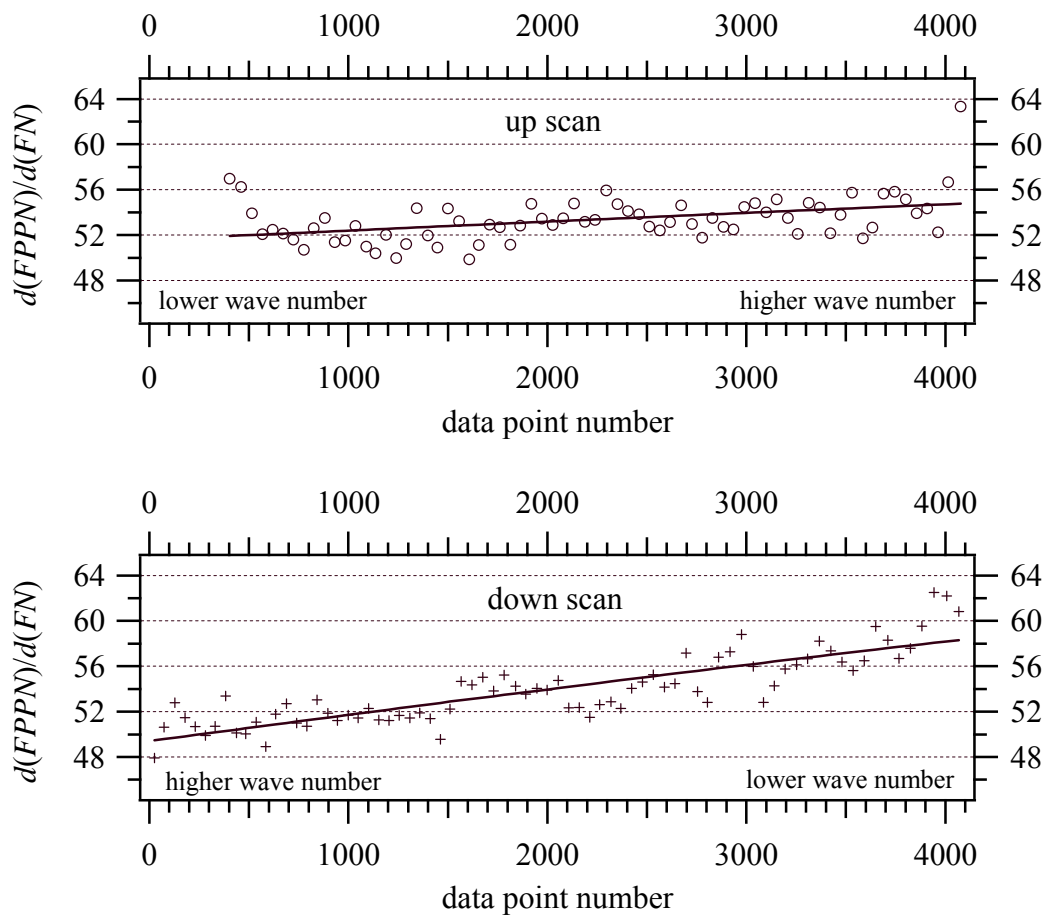

Figure 4.4 Numerical first derivative of fringe peak data point number (*FPPN*) with respect to fringe number (*FN*). The (solid) fit line is obtained from the linear regression of the first derivative of *FPPN* vs. *FN*. The coefficients and statistics of this fit are listed in Table 4.2.

Developing a calibration method that takes into account the higher order quadratic fit results can be expected to improve slightly the calibration results for the Philips laser diode system. However, there will be a remaining uncertainty in fringe position of roughly two data points. The handful of spectra obtained with the Philips laser diode spectrometer using a bipolar 12-bit tuning ramp (and thus a step size of about 10 MHz) was statistically about the same as the 20 MHz step size (unipolar 12-bit tuning ramp). That is to say, the resolution of this spectrometer is limited by either the noise of the Newport 505 laser driver or the fine tuning capability of the Philips laser diode.

In summary, the scan of the Philips laser diode system is linear enough to allow assignment of diatomic iodine transitions, but the accuracy of the wave number scale can be improved by including a quadratic term in the wave number calibration procedure. Adding terms beyond quadratic will not improve the accuracy of the wave number calibration. However, the goal of this project is to extract transition line widths and line-center shifts as a function of buffer gas pressure, so that while a simple linear contraction or expansion of the wave number scale would be of no consequence, an unaccounted for quadratic term in the calibration could have undesirable consequences on the accuracy of the pressure broadening and pressure shift coefficients calculated from spectra using a linearly calibrated wave number scale.

Along with a concern about the consistency of the laser line width across the spectral region being investigated with the Philips laser diode system, this concern about the impact of wave number calibration on the pressure broadening and pressure shift coefficients is one of the primary reasons that an effort was made to investigate the etalon spectrum. The effect of neglecting the quadratic term can be expected to depend on the spectral region in which it is located, increasing outward from the central region of a scan. While it would be hoped that a straightforward evaluation of these effects could be offered, we will see later in Chapter V that it is not such a simple (or trivial) problem. There are other factors that perhaps obscure and mitigate the effects of neglecting the quadratic calibration term. Among these factors is the overlap of neighboring lines in the somewhat congested spectrum of diatomic iodine (obscure) and the rather large nuclear quadrupole interactions (mitigate).

## 4.3 Linearity of New Focus External Cavity Laser Diode System Scans

A typical low pressure (ca. 0.18 torr at 298 K) wavelength-modulated absorption spectrum of the diatomic iodine reference gas cell and a simultaneously collected wavelength-modulated transmission fringe spectrum of the etalon obtained when using the New Focus external cavity laser diode system (to scan a region that overlaps the same one mentioned in Section 4.1) is shown in Figure 4.5. All spectra recorded with this laser system tuned the laser from high-to-low wave number.

Figure 4.6 shows an expanded view of Figure 4.5 with the addition of a numerically integrated etalon trace. It is fairly obvious for this particular spectrum that the diatomic iodine and etalon channels were maintained at opposite relative phases so that the etalon trace appears as a transmission spectrum with high transmission at low detector voltage [Horowitz 1].

A similar analysis of the nonlinear scan properties as a function of the linear tuning ramp was also carried out for the New Focus external cavity laser diode system. A fit of the fringe peak data point number vs. the fringe number is optimized for a fifth order polynomial in the independent variable (see Figure 4.7) for a fringe spectrum that contains 75 fringes covering the same spectral region as diatomic iodine Features A through I.

The New Focus external cavity laser diode system also exhibits changes in tuning rate that can be modeled with considerable accuracy using high order polynomials. The free spectral range of the etalon decreases from about 43 to 34 data point per fringe across roughly 2750 data points when scanning from high-to-low wave number.

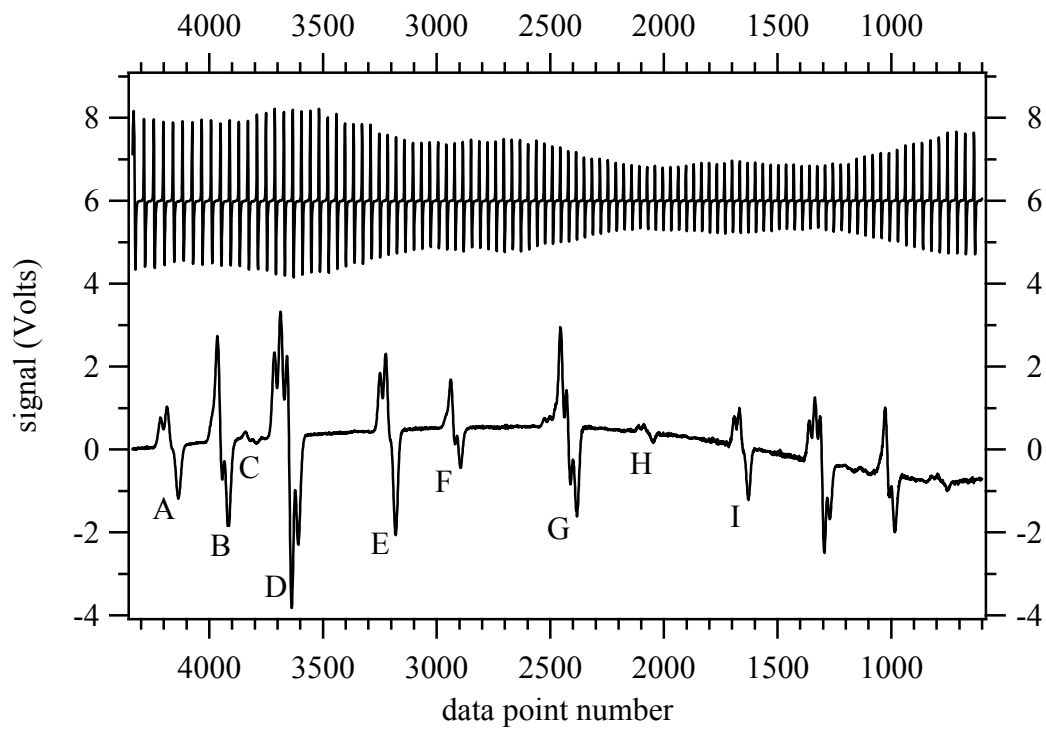

Figure 4.5 Wavelength-modulated spectrum of etalon fringes and diatomic iodine at room temperature covering the region 14,817.95 to 14,819.45 cm$^{-1}$ obtained with the New Focus external cavity laser diode system. The etalon fringe trace has been offset along the signal axis by six volts. Diatomic iodine features without labels were not used in the subsequent calibration or line-shape analysis.

The worst case discrepancy of comparing the observed line positions in data point number to the calculated values of a simulated spectrum in wave number units is about 75 data points. This laser scans at about 17 MHz per data point so that an uncertainty in line position of about 1275 MHz (0.0425 cm$^{-1}$) can be expected for some of the comparisons made between a raw (un-calibrated) spectrum in data point number units and a simulated (calculated) spectrum of relative signal amplitude and line center position in wave number units when making the quantum number assignments. A relative error of 75 data points between two diatomic iodine lines separated by 0.7 cm$^{-1}$ would amount to a discrepancy in relative line center position between these two line of about $100 \times (0.0425 \text{ cm}^{-1} \div 0.7 \text{ cm}^{-1}) = 6.1 \%$.

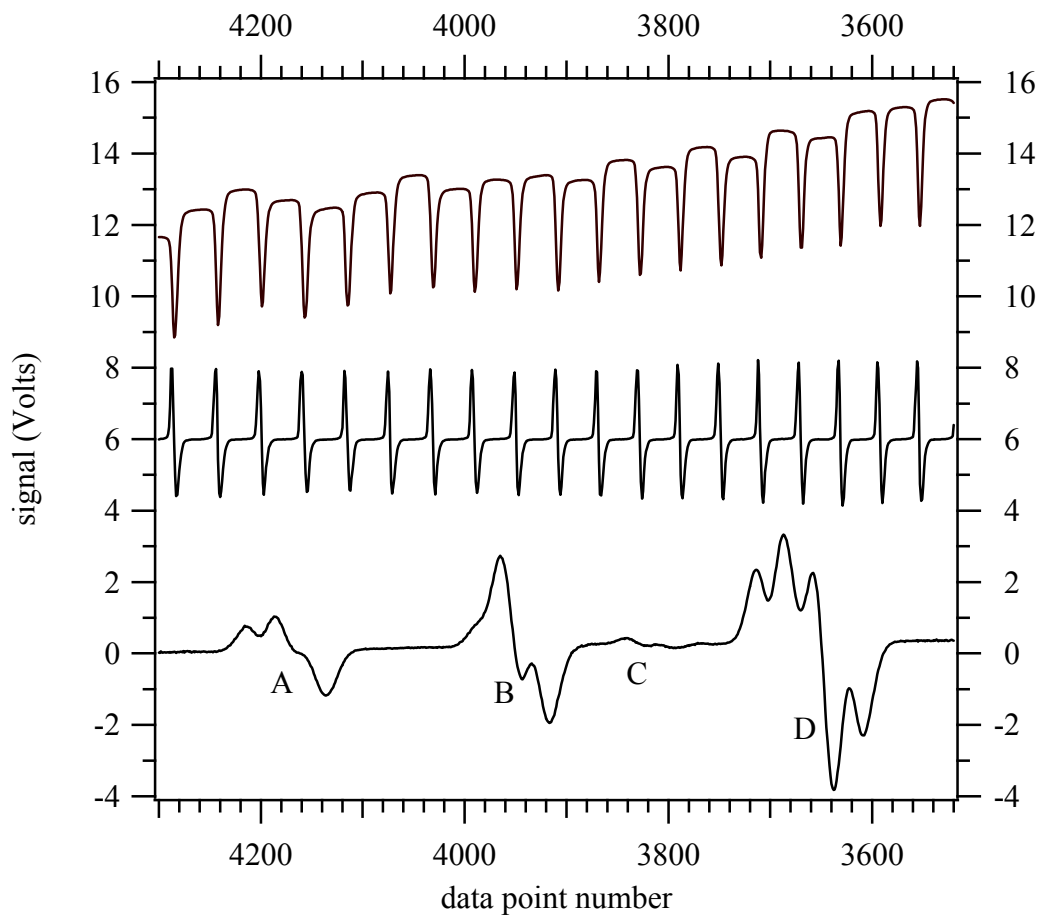

Figure 4.6 Expansion of a portion of Figure 4.5 from data point number 3500 to 4300. The New Focus external cavity laser diode system scans from high to low data point number (high to low wave number) in this spectrum. The top trace is the integrated etalon signal; the amplitude of this trace has been altered to fit the display. The observed and integrated etalon traces have been offset along the signal axis. The free spectral range of this etalon is $578 \pm 2$ MHz ($0.0193 \pm 0.0001$ cm$^{-1}$).

In summary, the scan of the New Focus external cavity laser diode is well represented by a fifth-order polynomial, and the precision of the scan is noticeably superior to that of the Philips laser diode system. We will, however, be content to implement a method based on linear interpolation of wave number between etalon fringes; the assumption being that the laser scan is well represented as being piecewise linear across the 578 MHz free spectral range of the etalon fringes.

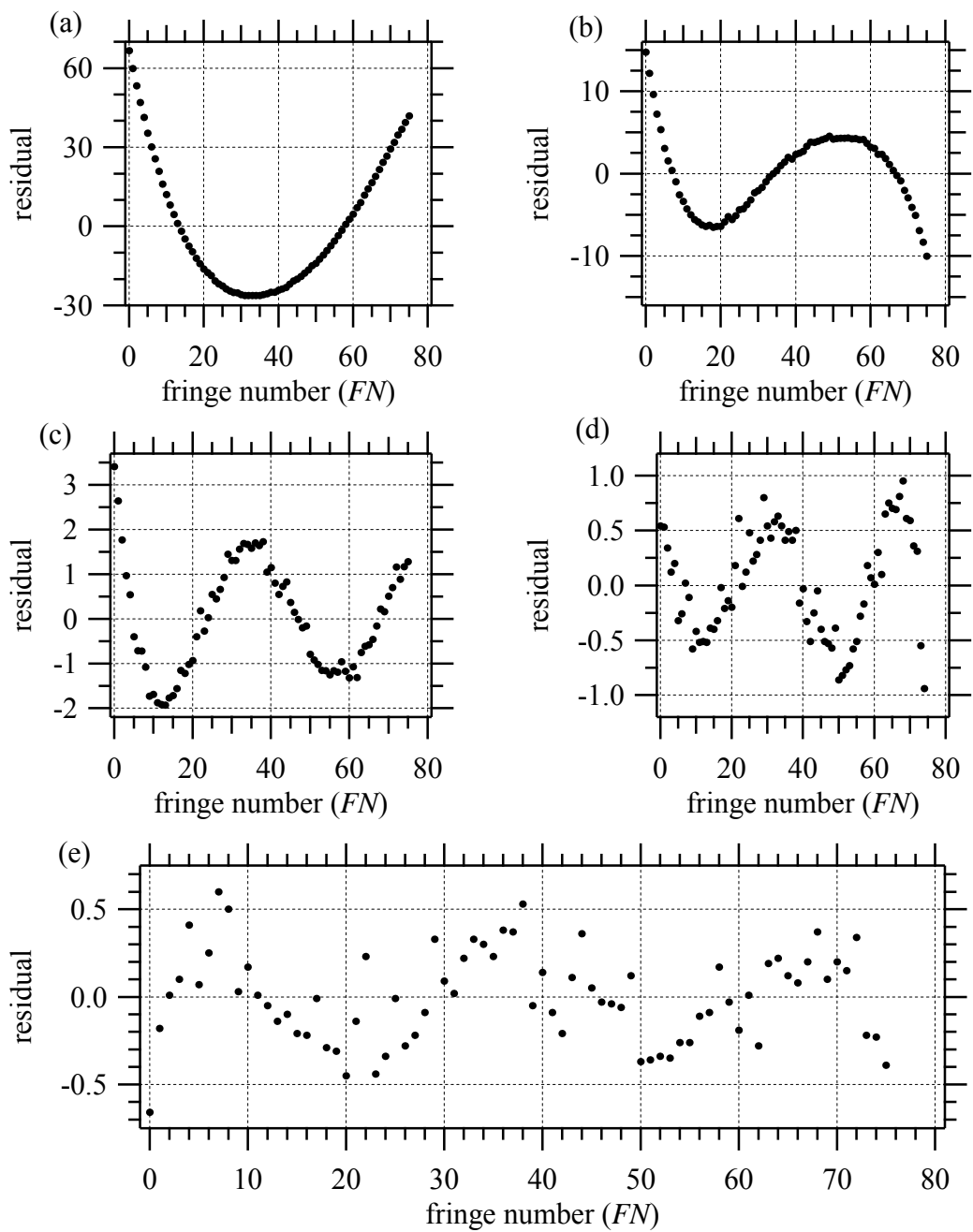

Figure 4.7 Residual plots of linear (*a*), quadratic (*b*), cubic (*c*), quartic (*d*), and quintic (*e*) fits of fringe peak data point number (*FPPN*) vs. fringe number (*FN*) to polynomials of the form $FPPN = \ldots c \times FN^2 + b \times FN + a$. The difference between calculated and observed fringe peak data point number ($FPPN_{obs} - FPPN_{calc}$) is the residual in units of data point number.

## 4.4 Etalon Fringe Widths

The previous two sections make considerable use of the number of data points between fringe peaks. Another quantity of interest, since it contains information about the radiation source (often referred to as the laser line width), is the width of the etalon fringes across the region being scanned by the laser diode systems. (As usual, the width of the laser line or the etalon fringes is considered in terms of the full width at half maximum intensity, abbreviated as "FWHM".) The characterization of fringe widths was not (from a modeling point of view) tremendously rigorous. Such efforts are meant mostly to help gain a semi-quantitative understanding of the line shape of the radiation source across the scan region and the ultimate resolution of the two spectrometers used in this project.

The (recorded) wavelength-modulated etalon spectrum was numerically integrated (with respect to point number) and a Gaussian profile was used as the line shape model for individual etalon fringes. While this is (clearly) not the correct line shape model for this problem, the ratio of the uncertainty in the fit (taken as the standard deviation of the residual) to the maximum signal deflection for the integrated fringes is about 0.025. The widths (a difference measurement) of the fringes are largely unaffected by errors in calibration since the laser scans can be well approximated as linear in wave number over short distances of about 20 data points. An overview of the results of an analysis of all the fringe widths for a single etalon spectrum from each of the three unique experimental situations is given in Figures 4.8 and 4.9. See also Sub-Section 3.2.4.

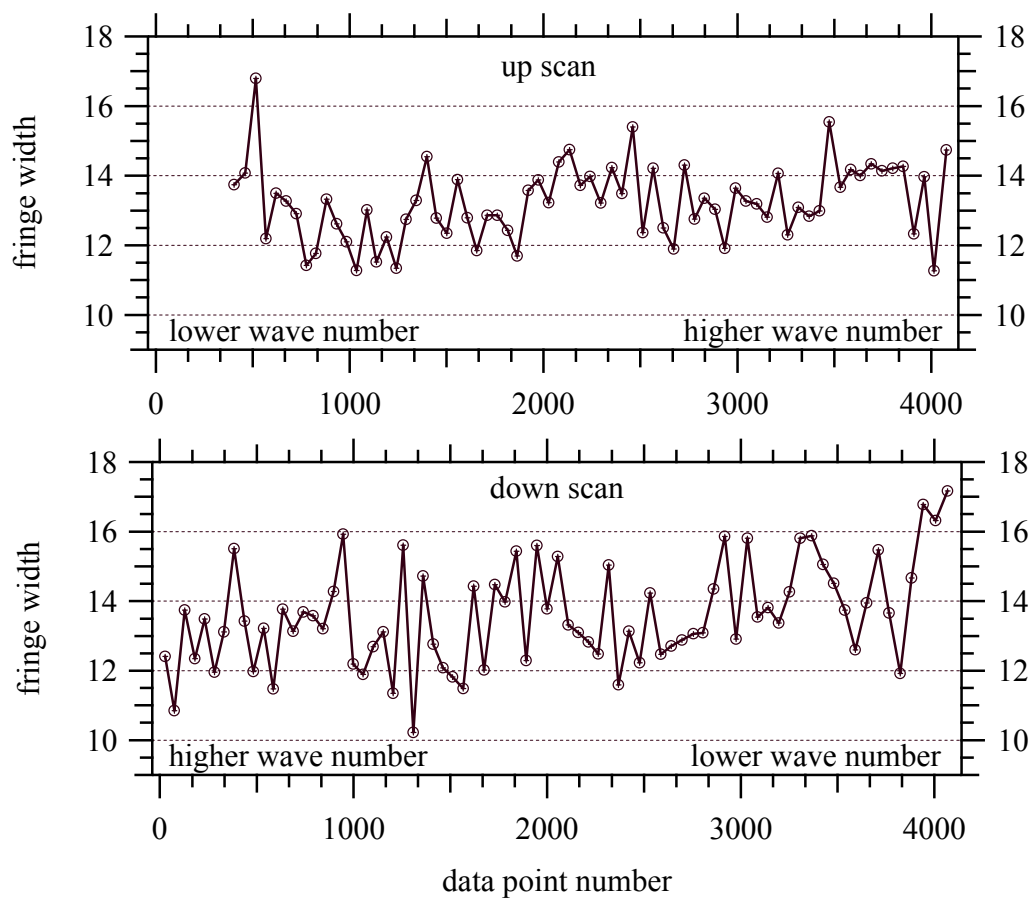

Figure 4.8 Etalon fringe width (FWHM) in data point number units for a single scan using the Philips laser diode system. The statistics of the fringe width for both scan directions are nearly the same with an average value of $13.4 \pm 1.3$ data points (FWHM).

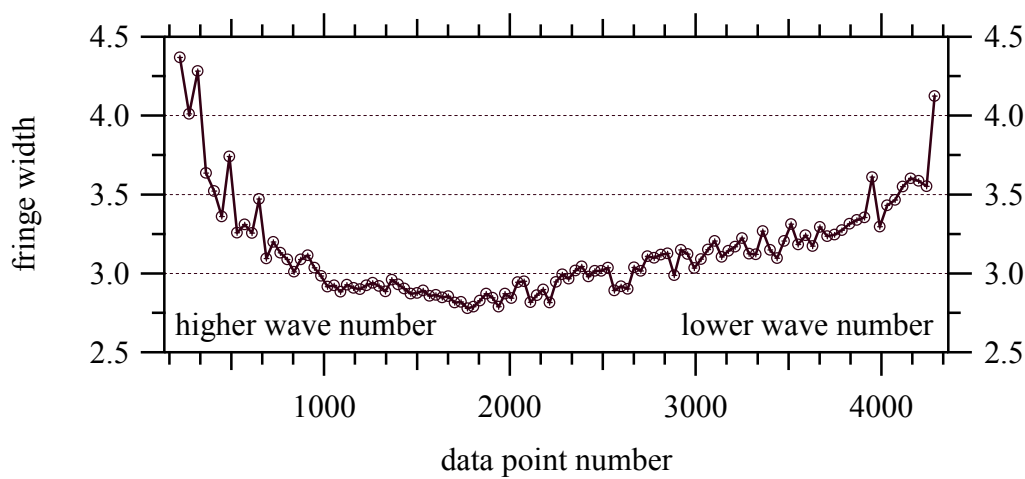

Figure 4.9 Etalon fringe width (FWHM) in data point number units for a single scan using the New Focus laser diode system. The statistics of the fringe width for this scan is 3.1 ± 0.3 data points (FWHM).

## 4.5 Assignment of Diatomic Iodine Spectral Features

The approximately linear relationship between wave number and data point number in spectra recorded with both laser systems makes it is possible to use the rough calibration results acquired with a wavelength meter or monochromator along with simulated spectra calculated from published spectroscopic constants and comparison to previously published Iodine Atlases [Luc 1; Luc 2; Hutson; Martin] to provide quantum number assignments of the observed diatomic iodine features. The process of establishing the quantum number assignments and comparison of the recorded spectra to the spectra in the Iodine Atlases helps to establish greater confidence in the absolute wave number calibration of the spectra being analyzed in this project. Quantum state assignments can also be used for a more in depth analysis of the pressure broadening and pressure shift results; even though such an analysis will not be presented in this dissertation, it is customary for the sake of completeness to include them.

The pattern of relative line positions and intensities are found to be statistically unique for any spectral region of diatomic iodine greater than about 1 cm$^{-1}$. Comparisons of the pattern of relative intensities and line-center wave number of a simulated spectrum to the point number position and relative intensities of a recorded spectrum were done to make the quantum number assignments. This process was facilitated by wavelength meter readings of the wave number at the endpoints of the scan region that agree with the simulated spectrum to better than 0.02 cm$^{-1}$. The simulated spectrum takes into account the Boltzmann weighting of the ground electronic state vibration and rotation populations, Franck-Condon and Hönl-London factors, and nuclear spin statistics (Chapter II) in order to calculate the relative intensities. The relative wave number scale of the simulated spectrum is accurate to slightly better than 150 MHz (0.005 cm$^{-1}$), while the absolute wave number scale may contain a systematic shift of about 300 MHz (0.01 cm$^{-1}$). These simulations explored (and tabulated) all of the transitions in the region 14,791 cm$^{-1}$ (676 nm) to 15,035 cm$^{-1}$ (665 nm) that satisfy the vibration quantum numbers $v'' = 0$ to 9, $v' = 0$ to 72, and the rotation quantum numbers $J''$ 0 to 175 using the spectroscopic constants of Gerstenkorn [Hutson] and Bacis [Martin].

The recorded spectrum was in point number units, which is identical in the relative separation of spectral features to a wave number calibration based on the wave number positions of two diatomic iodine features (see Section 4.4). Further comparisons are made with the Iodine Atlases, which show relative intensities and wave number positions of line centers to an absolute accuracy of about 0.01 cm$^{-1}$. Integrating the wavelength-modulated spectra obtained in this project makes for easy visual and/or semi-quantitative comparison to the Fourier-transform absorption spectra published in the Iodine Atlases. It is worth noting that the number of visible features in the Iodine Atlases is somewhat less than the laser diode spectra analyzed during the course of this project; that is to say, the laser diode spectrometer appears to be more sensitive (larger S/N ratio) than the spectrometers used to record those particular Iodine Atlases.

## 4.5.1 The Region 14,817.95 to 14,819.45 cm$^{-1}$

Since nearly all of the data for the region 14,817.95 to 14,819.45 cm$^{-1}$ were analyzed using a two-feature calibration method it is important to characterize as well as possible the limitations this imposes on the precision and accuracy of this analysis.

The observed relative intensities of diatomic iodine in the region 14,817.95 to 14,819.45 cm$^{-1}$ were determined by two successive numerical integrations of a wavelength-modulated spectrum. The first integration produces a spectrum that is linearly proportional in signal to a direct absorption signal (middle trace of Figure 4.10), except that knowledge of the off-resonance laser beam intensity can not be retrieved; this latter piece of information is not present in the original wavelength-modulated spectrum and is thought to be sufficiently low to be unimportant in this project. The diatomic iodine features in this integrated spectrum are masked out while a 10$^{th}$ order polynomial fit to the baseline is used to smooth and cancel as much as possible the residual non-zero (signal) offset. Integration of this smoothed (integrated) spectrum gives an integrated absorption spectrum that is a good measure of the relative absorption intensity (top trace of Figure 4.10). The change in vertical height corresponding to a wavelength region that spans an individual absorption feature is the measure of integrated absorption of individual ro-vibronic transitions used in Table 4.3.

Quantum state assignments of the observed diatomic iodine lines in the region 14,817.95 to 14,819.45 cm$^{-1}$ are listed in Table 4.3 and can be correlated with the feature labels in Figure 4.10. The calculated line positions, $\nu_{calc}$, are for all known features in this part of the spectrum [Hutson; Martin]. Only the most prominent diatomic iodine lines were observed; these are indicated by the feature labels and observed intensity in the first and last column of Table 4.3, respectively.

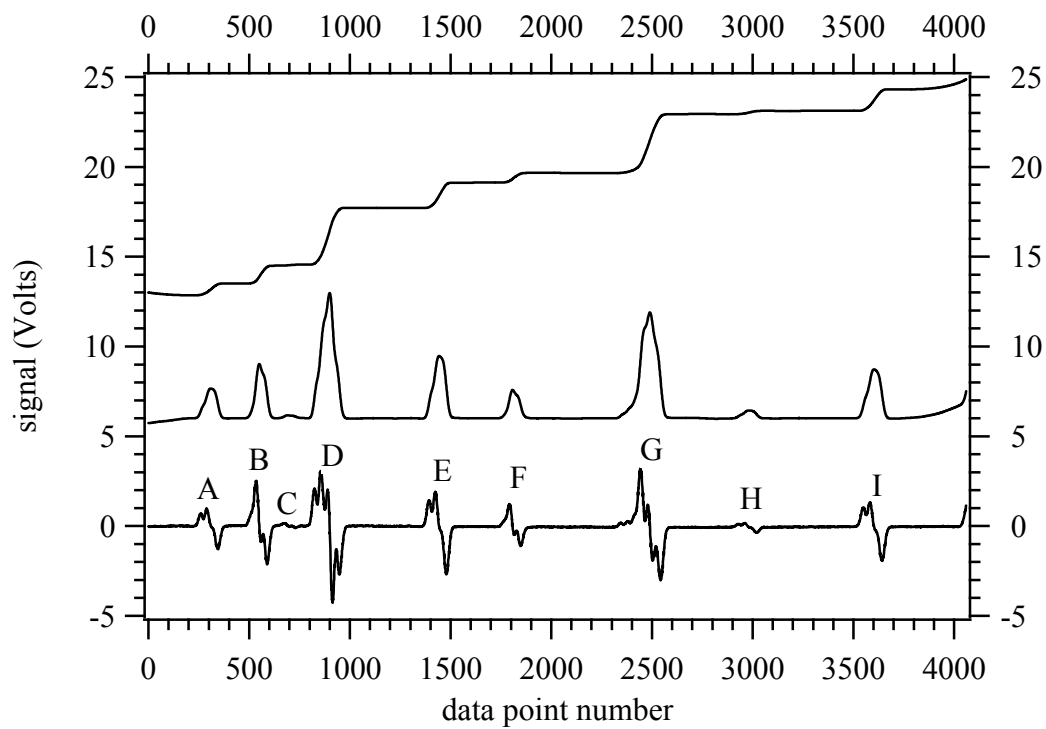

Figure 4.10 Integration of a diatomic iodine ($I_2$) spectrum for the region 14,817.95 to 14,819.45 cm$^{-1}$. Bottom trace is a typical wavelength-modulated spectrum measured in volts. The bottom trace has been integrated and then the baseline was smoothed to give the middle trace, the direct absorption signal. The middle trace was then integrated to give the top trace, the integrated absorption signal. Diatomic iodine feature designation is given by the letters A through I. The wavelength-modulated spectrum was obtained with a low-to-high wave number scan with the Philips laser diode system.

Table 4.3 Assignment of observed diatomic iodine (I$_2$) lines in the spectral region 14,817.94 to 14,819.56 cm$^{-1}$. The line-center position $v_{calc}$ (in units of cm$^{-1}$) was calculated from the spectroscopic constants of Gerstenkorn [Hutson] and Bacis [Martin]; "MHz Sep." and "PN Sep." are the separation in MHz and data point number to the previous diatomic iodine line at lower wave number, respectively. The calculated (Calc.) and observed (Obs.) relative intensities are calibrated to Feature E.

| Feature | Assignment | | $v_{calc}$ (cm$^{-1}$) | MHz Sep. | PN Sep. | Relative Intensity | |
| | $(v',v'')$ | P,R($J''$) | | | | Calc. | Obs. |
|---|---|---|---|---|---|---|---|
| | (14,12) | P(50) | 14,817.936 | -- | | 0.00206 | |
| | (10,10) | R(6) | 14,817.973 | 1109 | | 0.00038 | |
| | (12,11) | R(41) | 14,817.990 | 510 | | 0.00355 | |
| | (1,4) | R(154) | 14818.011 | 630 | | 0.00468 | |
| | (10,10) | R(0) | 14,818.017 | 180 | | 0.00005 | |
| | (10,10) | R(5) | 14,818.025 | 240 | | 0.00046 | |
| A* | (4,6) | R(129) | 14,818.046 | 630 | -- | 0.465 | 0.46 |
| | (1,4) | P(147) | 14,818.051 | 150 | | 0.00912 | |
| | (10,10) | R(1) | 14,818.054 | 90 | | 0.00015 | |
| | (12,11) | P(35) | 14,818.054 | 0 | | 0.00321 | |
| | (10,10) | R(4) | 14,818.059 | 150 | | 0.00027 | |
| | (10,10) | R(2) | 14,818.073 | 420 | | 0.00016 | |
| | (10,10) | R(3) | 14,818.075 | 60 | | 0.00031 | |
| | (10,9) | P(146) | 14,818.140 | 1949 | | 0.00166 | |
| B* | (5,7) | R(76) | 14,818.150 | 300 | 242 | 0.664 | 0.71 |
| C | (5,6) | R(174) | 14,818.207 | 1709 | 143 | 0.0491 | 0.044 |
| | (14,11) | R(154) | 14,818.218 | 330 | | 0.00007 | |
| | (13,11) | P(109) | 14,818.239 | 630 | | 0.00005 | |
| | (14,12) | P(44) | 14,818.240 | 30 | | 0.00197 | |
| D | (3,6) | R(43) | 14,818.282 | 1259 | 206 | 1.36 | 2.25 |
| D | (3,6) | P(36) | 14,818.300 | 540 | | 0.877 | |
| | (12,10) | R(154) | 14,818.371 | 2129 | | 0.00119 | |
| | (8,8) | P(141) | 14,818.463 | 2758 | | 0.0119 | |
| E* | (5,7) | P(69) | 14,818.517 | 1619 | 544 | 1 | 1 |
| | (7,8) | P(87) | 14,818.646 | 3867 | | 0.00335 | |
| F* | (4,6) | P(122) | 14,818.668 | 660 | 364 | 0.428 | 0.39 |

Table 4.3 Assignment of observed diatomic iodine lines in the spectral region 14,817.94 to 14,819.56 cm$^{-1}$ (continued).

| Feature | Assignment $(v',v'')$ | P,R($J''$) | $\nu_{\text{calc}}$ (cm$^{-1}$) | MHz Sep. | PN Sep. | Relative Intensity Calc. | Obs. |
|---|---|---|---|---|---|---|---|
| | (0,4) | P(83) | 14,818.677 | 270 | | 0.00799 | |
| | (12,11) | R(40) | 14,818.683 | 180 | | 0.00251 | |
| | (12,11) | P(34) | 14,818.745 | 1859 | | 0.00225 | |
| | (14,12) | R(49) | 14,818.820 | 2248 | | 0.00288 | |
| G | (2,5) | R(113) | 14,818.902 | 2458 | | 0.211 | |
| | (8,8) | R(147) | 14,818.907 | 150 | | 0.00918 | |
| G | (3,6) | R(42) | 14,818.927 | 600 | 683 | 0.962 | 2.32 |
| G | (2,5) | P(106) | 14,818.940 | 390 | | 0.185 | |
| G | (3,6) | P(35) | 14,818.944 | 120 | | 1.21 | |
| | (0,4) | R(90) | 14,819.029 | 2548 | | 0.00502 | |
| | (14,12) | R(90) | 14,819.117 | 2638 | | 0.00274 | |
| H* | (6,7) | P(133) | 14,819.131 | 420 | 491 | 0.119 | 0.13 |
| | (11,10) | R(111) | 14,819.154 | 690 | | 0.00596 | |
| | (15,12) | P(109) | 14,819.193 | 1169 | | 0.00084 | |
| | (7,8) | R(93) | 14,819.321 | 3837 | | 0.00298 | |
| | (11,10) | P(105) | 14,819.326 | 150 | | 0.00703 | |
| | (12,10) | P(148) | 14,819.331 | 150 | | 0.00157 | |
| | (12,11) | R(39) | 14,819.358 | 809 | | 0.00348 | |
| | (9,9) | P(98) | 14,819.363 | 150 | | 0.0266 | |
| I* | (5,7) | R(75) | 14,819.368 | 150 | 620 | 0.942 | 0.85 |
| | (12,11) | P(33) | 14,819.417 | 1469 | | 0.00310 | |
| | (3,6) | R(41) | 14,819.556 | 4167 | | 1.34 | |

## 4.5.2 The Region 14,946.17 to 14,950.29 cm$^{-1}$

The spectra in this region were recorded with the New Focus external cavity laser diode system. The purpose of this portion of the project was to extend the investigation to low-$J$ lines in search of patterns of pressure broadening and pressure shift as a function of rotation quantum number. The observed relative intensities in the region 14,946.17 to 14,950.29 $cm^{-1}$ were taken as the maximum peak height of the integrated wavelength-modulated absorption signal (top trace of Figures 4.11 and 4.12). The observed and calculated intensities for both spectral regions were calibrated on Feature 6 (see Figure 4.11 and Table 4.4).

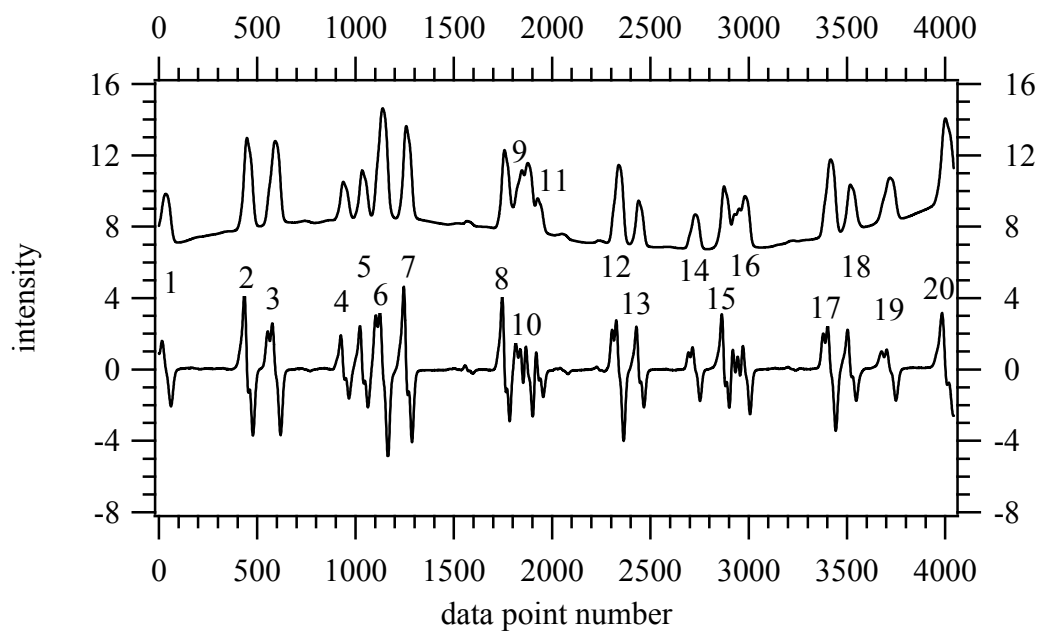

Figure 4.11 Spectrum of diatomic iodine (I$_2$) for the region 14,946.17 to 14,948.43 $cm^{-1}$. Data point number matches that given in Table 4.4 (assignment table). Data point numbers have been transformed from those obtained during data collection to go instead from low to high as the scan goes from low to high wave number.

Quantum state assignments of the observed diatomic iodine lines in the region 14,946.17 to 14,950.29 $cm^{-1}$ are listed in Table 4.4 and can be correlated with the feature labels in Figures 4.11 and 4.12. The calculated line-center positions, $\nu_{calc}$, are for all known features in this part of the spectrum. Only the most prominent diatomic iodine lines were observed; these are indicated by the feature labels and observed intensity in the first and last column of Table 4.4, respectively.

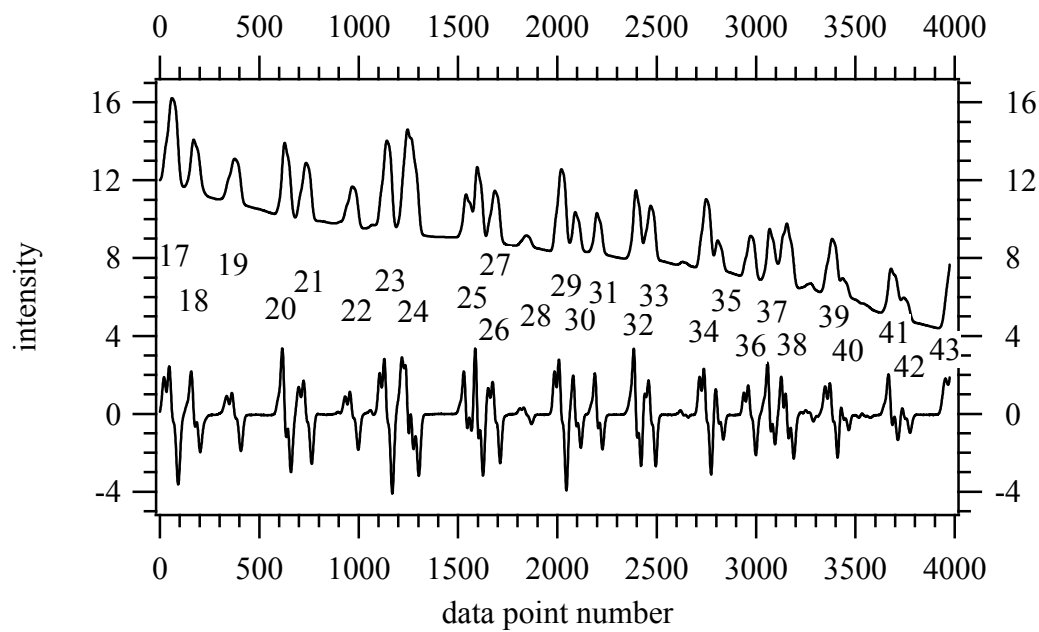

Figure 4.12 Spectrum of diatomic iodine for the region 14,948.08 to 14,950.29 cm$^{-1}$. Data point number matches that given in Table 4.4 (assignment table). Data point numbers have been transformed from those obtained during data collection to go instead from low to high as the scan goes from low to high wave number.

Table 4.4 Assignment of observed diatomic iodine lines in the spectral region 14,946.00 to 14,950.39 $cm^{-1}$. Two separate scan regions were programmed into the New Focus external cavity laser diode system; the two scans overlapped on Features 17 through 20. The line-center position $\nu_{calc}$ (in units of $cm^{-1}$) was calculated from the spectroscopic constants of Gerstenkorn [Hutson] and Bacis [Martin]; "MHz Sep." is to the previous diatomic iodine line at lower wave number; "PN" is the data point number in the recorded spectrum. The calculated (Calc.) and observed (Obs.) relative intensities are calibrated on Feature 6.

| Feature | Assignment | | Position | | | Relative Intensity | |
|---|---|---|---|---|---|---|---|
| | $(\nu',\nu'')$ | P,R($J''$) | $\nu_{calc}$ (cm$^{-1}$) | MHz Sep. | PN | Calc. | Obs. |
| | (4,6) | R(27) | 14,946.005 | -- | | 1.78 | |
| | (16,11) | R(173) | 14,946.039 | 1019 | | 0.00019 | |
| | (1,4) | R(87) | 14,946.064 | 749 | | 0.0700 | |
| | (4,6) | P(20) | 14,946.083 | 570 | | 0.965 | |
| 1 | (6,7) | R(65) | 14,946.178 | 2848 | 70 | 0.667 | 0.76 |
| | (6,6) | P(162) | 14,946.183 | 150 | | 0.0939 | |
| | (13,10) | R(144) | 14,946.195 | 360 | | 0.00148 | |
| | (12,10) | P(93) | 14,946.257 | 1859 | | 0.0146 | |
| | (11,9) | P(137) | 14,946.328 | 2129 | | 0.00002 | |
| | (10,9) | P(87) | 14,946.34 | 360 | | 0.0162 | |
| | (10,9) | R(93) | 14,946.376 | 1079 | | 0.0144 | |
| 2* | (4,6) | R(26) | 14,946.398 | 660 | 482 | 1.24 | 1.41 |
| 3* | (4,6) | P(19) | 14,946.472 | 2218 | 626 | 1.29 | 1.30 |
| | (9,8) | P(133) | 14,946.555 | 2488 | | 0.0349 | |
| | (9,8) | R(139) | 14,946.618 | 1889 | | 0.0274 | |
| | (16,12) | P(94) | 14,946.653 | 1049 | | 0.00025 | |
| 4* | (6,7) | P(58) | 14,946.664 | 330 | 972 | 0.489 | 0.58 |
| | (14,11) | P(95) | 14,946.715 | 1529 | | 0.00087 | |
| 5* | (5,6) | P(116) | 14,946.717 | 60 | 1069 | 0.677 | 0.72 |
| | (2,4) | R(150) | 14,946.756 | 1169 | | 0.0270 | |
| | (4,5) | P(154) | 14,946.767 | 330 | | 0.159 | |
| 6 | (4,6) | R(25) | 14,946.774 | 210 | 1173 | 1.69 | 1.69 |
| | (8,8) | P(76) | 14,946.791 | 510 | | 0.0577 | |

Table 4.4 Assignment of observed diatomic iodine lines in the spectral region 14,946.00 to 14,950.39 cm$^{-1}$ (continued).

| Feature | Assignment | | Position | | | Relative Intensity | |
|---|---|---|---|---|---|---|---|
| | $(v',v'')$ | P,R($J''$) | $\nu_{calc}$ (cm$^{-1}$) | MHz Sep. | PN | Calc. | Obs. |
| 7 | (3,5) | R(108) | 14,946.844 | 1589 | 1293 | 0.485 | 1.49 |
| 7 | (4,6) | P(18) | 14,946.844 | 0 | | 0.881 | |
| | (8,7) | P(167) | 14,946.983 | 4167 | | 0.00019 | |
| | (6,6) | R(168) | 14,947.02 | 1109 | | 0.0686 | |
| | (8,7) | R(173) | 14,947.076 | 1679 | | 0.00014 | |
| 8 | (4,6) | R(24) | 14,947.133 | 1709 | 1793 | 1.17 | -- |
| | (7,7) | P(126) | 14,947.136 | 90 | | 0.0376 | |
| 9 | (3,5) | P(101) | 14,947.171 | 1049 | 1880 | 0.820 | -- |
| | (8,8) | R(82) | 14,947.181 | 300 | | 0.0531 | |
| | (10,8) | P(170) | 14,947.183 | 60 | | 0.00457 | |
| | (16,12) | R(99) | 14,947.199 | 480 | | 0.00032 | |
| | (2,4) | P(143) | 14,947.2 | 30 | | 0.0520 | |
| 10 | (4,6) | P(17) | 14,947.2 | 0 | 1910 | 1.17 | -- |
| 11 | (6,7) | R(64) | 14,947.233 | 989 | 1961 | 0.480 | -- |
| | (1,4) | P(79) | 14,947.304 | 2129 | | 0.0800 | |
| | (18,12) | R(168) | 14,947.344 | 1199 | | 0.00002 | |
| | (12,9) | P(171) | 14,947.372 | 839 | | 0.00042 | |
| | (1,4) | R(86) | 14,947.417 | 1349 | | 0.0510 | |
| | (13,10) | P(138) | 14,947.474 | 1709 | | 0.00191 | |
| 12* | (4,6) | R(23) | 14,947.475 | 30 | 2374 | 1.58 | 1.22 |
| 13* | (4,6) | P(16) | 14,947.539 | 1919 | 2475 | 0.793 | 0.70 |
| | (12,10) | R(98) | 14,947.631 | 2758 | | 0.00939 | |
| | (17,12) | P(134) | 14,947.684 | 1589 | | 0.00000 | |
| | (14,11) | R(100) | 14,947.699 | 450 | | 0.00056 | |
| 14* | (6,7) | P(57) | 14,947.71 | 330 | 2763 | 0.687 | 0.53 |
| | (7,7) | R(132) | 14,947.782 | 2159 | | 0.0301 | |
| 15 | (4,6) | R(22) | 14,947.801 | 570 | 2908 | 1.09 | -- |
| | (0,3) | P(128) | 14,947.819 | 540 | | 0.00074 | |
| | (17,12) | R(139) | 14,947.829 | 300 | | 0.00000 | |

Table 4.4 Assignment of observed diatomic iodine lines in the spectral region 14,946.00 to 14,950.39 cm$^{-1}$ (continued).

| Feature | $(v',v'')$ | P,R($J''$) | $\nu_{calc}$ (cm$^{-1}$) | MHz Sep. | PN | Calc. | Obs. |
|---------|-----------|-----------|--------------------------|----------|-----|-------|------|
| | | **Assignment** | | **Position** | | **Relative Intensity** | |
| 16 | (5,6) | R(122) | 14,947.832 | 90 | 3016 | 0.555 | -- |
| 16 | (4,6) | P(15) | 14,947.862 | 899 | | 1.05 | |
| | (15,11) | P(137) | 14,947.872 | 300 | | 0.00079 | |
| | (10,9) | P(86) | 14,947.962 | 2698 | | 0.0118 | |
| | (18,12) | P(163) | 14,947.963 | 30 | | 0.00003 | |
| | (14,10) | P(170) | 14,947.994 | 929 | | 0.00015 | |
| | (10,9) | R(92) | 14,948.001 | 210 | | 0.0105 | |
| | (0,3) | R(135) | 14,948.01 | 270 | | 0.00078 | |
| | (12,10) | P(92) | 14,948.034 | 720 | | 0.0107 | |
| 17* | (4,6) | R(21) | 14,948.11 | 2278 | 3451 / 134 | 1.48 | 1.18 / 1.37 |
| 18* | (4,6) | P(14) | 14,948.168 | 1739 | 3551 / 171 | 0.702 | 0.70 / 0.80 |
| | (8,8) | P(75) | 14,948.178 | 300 | | 0.0819 | |
| | (4,5) | R(160) | 14,948.251 | 2188 | | 0.118 | |
| 19 | (6,7) | R(63) | 14,948.27 | 570 | 3752 / 376 | 0.678 | 0.68 / 0.68 |
| | (11,9) | R(142) | 14,948.339 | 2069 | | 0.00001 | |
| 20* | (4,6) | R(20) | 14,948.403 | 1919 | 4035 / 628 | 1.01 | 1.34 / 1.12 |
| 21* | (4,6) | P(13) | 14,948.457 | 1619 | 736 | 0.917 | 0.87 |
| | (16,12) | P(93) | 14,948.551 | 2818 | | 0.00036 | |
| | (8,8) | R(81) | 14,948.565 | 420 | | 0.0756 | |
| | (14,11) | P(94) | 14,948.58 | 450 | | 0.00064 | |
| 22 | (3,5) | R(107) | 14,948.584 | 120 | 969 | 0.699 | 0.65 |
| | (1,3) | P(175) | 14,948.596 | 360 | | 0.00111 | |
| | (1,4) | P(78) | 14,948.641 | 1349 | | 0.0581 | |
| | (15,11) | R(142) | 14,948.662 | 630 | | 0.00046 | |
| 23 | (4,6) | R(19) | 14,948.679 | 510 | 1143 | 1.36 | 1.35 |
| 24 | (4,6) | P(12) | 14,948.73 | 1529 | | 0.607 | |
| 24 | (6,7) | P(56) | 14,948.739 | 270 | 1247 | 0.492 | 1.53 |
| 24 | (5,6) | P(115) | 14,948.752 | 390 | | 0.980 | |
| | (1,4) | R(85) | 14,948.753 | 30 | | 0.0728 | |

Table 4.4 Assignment of observed diatomic iodine lines in the spectral region 14,946.00 to 14,950.39 cm$^{-1}$ (continued).

| Feature | Assignment | | Position | | | Relative Intensity | |
| | $(v',v'')$ | P,R($J''$) | $\nu_{calc}$ (cm$^{-1}$) | MHz Sep. | PN | Calc. | Obs. |
|---|---|---|---|---|---|---|---|
| | (13,10) | R(143) | 14,948.896 | 4287 | | 0.00217 | |
| 25 | (3,5) | P(100) | 14,948.906 | 300 | 1541 | 0.601 | 0.66 |
| | (11,9) | P(136) | 14,948.922 | 480 | | 0.00001 | |
| 26 | (4,6) | R(18) | 14,948.938 | 480 | 1598 | 0.930 | 1.15 |
| | (14,10) | R(175) | 14,948.972 | 1019 | | 0.00016 | |
| 27 | (4,6) | P(11) | 14,948.986 | 420 | 1686 | 0.783 | 0.84 |
| | (9,8) | P(132) | 14,949.006 | 600 | | 0.0259 | |
| | (9,8) | R(138) | 14,949.074 | 2039 | | 0.0204 | |
| 28 | (6,6) | P(161) | 14,949.077 | 90 | 1844 | 0.139 | 0.17 |
| | (16,12) | R(98) | 14,949.094 | 510 | | 0.00023 | |
| 29 | (4,6) | R(17) | 14,949.181 | 2608 | 2021 | 1.24 | 1.26 |
| | (2,4) | R(149) | 14,949.195 | 420 | | 0.0396 | |
| 30 | (4,6) | P(10) | 14,949.225 | 899 | 2091 | 0.510 | 0.61 |
| 31 | (6,7) | R(62) | 14,949.29 | 1949 | 2200 | 0.487 | 0.62 |
| | (16,11) | P(167) | 14,949.321 | 929 | | 0.00026 | |
| | (12,10) | R(97) | 14,949.397 | 2278 | | 0.0135 | |
| | (7,7) | P(125) | 14,949.399 | 60 | | 0.0547 | |
| 32 | (4,6) | R(16) | 14,949.407 | 240 | 2396 | 0.843 | 1.05 |
| 33 | (4,6) | P(9) | 14,949.448 | 1229 | 2470 | 0.645 | 0.85 |
| 33 | (4,5) | P(153) | 14,949.449 | 30 | | 0.233 | |
| | (16,11) | R(172) | 14,949.5 | 1529 | | 0.00014 | |
| | (8,8) | P(74) | 14,949.548 | 1439 | | 0.0593 | |
| | (14,11) | R(99) | 14,949.557 | 270 | | 0.00080 | |
| | (10,9) | P(85) | 14,949.566 | 270 | | 0.0168 | |
| | (10,9) | R(91) | 14,949.607 | 1229 | | 0.0151 | |
| 34* | (4,6) | R(15) | 14,949.616 | 270 | 2750 | 1.12 | 1.05 |
| | (2,4) | P(142) | 14,949.631 | 450 | | 0.0388 | |
| 35* | (4,6) | P(8) | 14,949.654 | 690 | 2808 | 0.411 | 0.47 |
| | (10,8) | R(175) | 14,949.681 | 809 | | 0.00486 | |

Table 4.4 Assignment of observed diatomic iodine lines in the spectral region 14,946.00 to 14,950.39 cm$^{-1}$ (continued).

| Feature | Assignment | | Position | | | Relative Intensity | |
|---|---|---|---|---|---|---|---|
| | $(v',v'')$ | P,R($J''$) | $\nu_{calc}$ (cm$^{-1}$) | MHz Sep. | PN | Calc. | Obs. |
| 36 | (6,7) | P(55) | 14,949.75 | 2069 | 2974 | 0.691 | 0.64 |
| | (12,10) | P(91) | 14,949.793 | 1289 | | 0.0153 | |
| 37 | (4,6) | R(14) | 14,949.809 | 480 | 3069 | 0.752 | 0.85 |
| 38 | (4,6) | P(7) | 14,949.844 | 1049 | 3156 | 0.505 | 0.99 |
| 38 | (5,6) | R(121) | 14,949.861 | 510 | | 0.805 | |
| | (6,6) | R(167) | 14,949.916 | 1649 | | 0.101 | |
| | (8,8) | R(80) | 14,949.932 | 480 | | 0.0549 | |
| | (0,3) | P(127) | 14,949.948 | 480 | | 0.00107 | |
| | (1,4) | P(77) | 14,949.961 | 390 | | 0.0826 | |
| 39* | (4,6) | R(13) | 14,949.985 | 720 | 3383 | 0.987 | 0.82 |
| 40* | (4,6) | P(6) | 14,950.017 | 959 | 3434 | 0.310 | 0.30 |
| | (7,7) | R(131) | 14,950.044 | 809 | | 0.0438 | |
| | (8,7) | P(166) | 14,950.046 | 60 | | 0.00015 | |
| | (1,4) | R(84) | 14,950.073 | 809 | | 0.0530 | |
| | (0,3) | R(134) | 14,950.14 | 2009 | | 0.00058 | |
| 41 | (4,6) | R(12) | 14,950.145 | 150 | 3679 | 0.658 | 0.68 |
| | (8,7) | R(172) | 14,950.147 | 60 | | 0.00011 | |
| | (13,10) | P(137) | 14,950.16 | 390 | | 0.00279 | |
| 42 | (4,6) | P(5) | 14,950.173 | 390 | 3743 | 0.362 | 0.39 |
| 43 | (4,6) | R(11) | 14,950.288 | 3448 | | 0.854 | |
| | (6,7) | R(61) | 14,950.293 | 150 | | 0.686 | |
| | (3,5) | R(106) | 14,950.307 | 420 | | 0.514 | |
| | (4,6) | P(4) | 14,950.313 | 180 | | 0.207 | |
| | (10,8) | P(169) | 14,950.387 | 2218 | | 0.00676 | |

## 4.6 Wave Number Calibration of Diatomic Iodine Spectra

The previous analyses of the laser scan linearity and the assignment of quantum states to the observed diatomic iodine spectral features were done for only a few spectra for diagnostic purposes and completeness. However, before analyzing the line shape of the diatomic iodine features in the recorded spectra for broadening and line center shift as a function of pressure it was first necessary to establish an absolute wave number scale for all of the spectra. Two different approaches were taken in calculating an absolute wave number scale for the spectra recorded during the course of this project. By far the most commonly utilized absolute wave number calibration method was based on the linear extrapolation for two of the recorded diatomic iodine features (Features D and G) present in the scan region covered by the Philips CQL806/30 laser diode. Eventually, however, to provide additional confidence in the reliability of the wave number scale, a more careful absolute wave number calibration method was developed that utilized several diatomic iodine features and the fringes from a plane-parallel Fabry-Perot etalon that has a free spectral range considerably smaller than the average distance between diatomic iodine features. As well, the nonlinear scanning characteristics of the New Focus external cavity laser diode also indicated the need for a more accurate wave number calibration method.

An accurate absolute wave number calibration is not essential for obtaining accurate pressure broadening and shift coefficients. Accurate pressure broadening and pressure shift coefficients can be obtained from a wave number calibrated spectrum that is accurate in the relative wave number scale, but translated away from their true values. The functional relationship of line shape with respect to absolute wave number will not change by a detectable amount for small translations in the absolute wave number scale; the maximum such translation encountered during the course of this work is estimated to be about 600 MHz ($0.02 \text{ cm}^{-1}$).

## 4.6.1 Two Diatomic Iodine Features

Nearly all of the wavelength-modulated absorption spectra recorded with the Philips laser diode system were calibrated for wave number based on the calculated wave number position of Features D and G, which are both a blend of more than one line. The separation between the two diatomic iodine features (Features D and G) was used to compute a linear tuning rate in wave number per point number; the point number of the line center was determined by linear interpolation against the location where the signal was zero Volt (i.e. the local maximum of absorption). The wave number scale is centered on Feature D and incremented point-by-point by the tuning rate in both directions for all data points.

A linear absolute wave number scale obtained in this fashion does not alter the relative data point number spacing, so that there is always a linear relationship between absolute wave number and data point number. This method offers the flexibility of being able to perform the nonlinear regression analysis of the line shape using the data point number and wave number scales interchangeably. Table 4.5 shows two feature calibration results typical for spectra recorded with the Philips laser diode system in the region 14,817.95 to 14,819.45 $cm^{-1}$. The calculation of the wave number positions of Features D and G has been described previously [Hardwick 1].

Table 4.5 Linear calibration using two diatomic iodine features for the spectral scan region 14,817.95 to 14,819.45 cm$^{-1}$. Line center positions, $\nu_{obs}$, for the two calibration features (Features D and G) are based on calculated values [Hardwick 1]. All other $\nu_{obs}$ are obtained from the line shape analysis (in Chapter V) with the estimated uncertainties in the observed line center position in units of the last significant figure of the experimental value in parenthesis. A scan that goes from low-to-high wave number is denoted as an "up scan" and one that goes from low-to-high wave number is a "down scan". The values in this table represent an average over 68 scans in each direction of a 'zero' pressure reference gas cell using the Philips laser diode system. The residual is $\nu_{obs} - \nu_{calc}$ values are based on the calculated line center positions from Table 4.3 in which the associated linear absolute wave number scale has been contracted and translated to coincide with the intensity weighted average line positions for Features D and G used in this calibration method.

| Feature | Assignment | | up scan (cm$^{-1}$) | | down scan (cm$^{-1}$) | |
|---|---|---|---|---|---|---|
| | $(v',v'')$ | P,R($J''$) | $\nu_{obs}$ | residual | $\nu_{obs}$ | residual |
| A | (4,6) | R(129) | 14,818.0480(6) | −0.0027 | 14818.0538(5) | 0.0031 |
| B | (5,7) | R(76) | 14,818.1507(3) | −0.0028 | 14818.1533(3) | −0.0002 |
| C | (5,6) | R(174) | -- | -- | -- | -- |
| D | (3,6) | R(43) | 14,818.291 | -- | 14818.291 | -- |
| D | (3,6) | P(36) | | | | |
| E | (5,7) | P(69) | 14,818.5112(3) | −0.0053 | 14818.5093(3) | −0.0072 |
| F | (4,6) | P(122) | 14,818.6607(3) | −0.0052 | 14818.6588(3) | −0.0071 |
| G | (2,5) | R(113) | 14,818.929 | -- | 14818.929 | -- |
| G | (3,6) | R(42) | | | | |
| G | (2,5) | P(106) | | | | |
| G | (3,6) | P(35) | | | | |
| H | (6,7) | P(133) | 14,819.1231(4) | −0.0007 | 14819.1269(4) | 0.0030 |
| I | (5,7) | R(75) | 14,819.3647(7) | 0.0064 | 14819.3768(7) | 0.0185 |

## 4.6.2 Many Diatomic Iodine Features and Etalon Fringes

The two feature wave number calibration method does not properly reproduce the literature values for the nuclear hyperfine constants in the line shape analysis for all diatomic iodine lines in the recorded spectra. A more comprehensive calibration method involving most of the diatomic iodine features in a given scan region and the simultaneously recorded fringe spectrum of the plane-parallel Fabry-Perot etalon was developed that would reproduce, within experimental uncertainty, the nuclear hyperfine constants obtained from the line shape analysis.

The first steps in this calibration method were to split the scans into individual spectra; crop the region of a given spectrum that was to be calibrated for wave number; integrate the measured first derivative fringe and reference signal line shapes; locate all of the fringe peaks and most of the diatomic iodine lines in point number space in the traces that were integrated in the previous step; and assign an ascending and sequential fringe number to each fringe peak. The tabulated point number and sequential numbering of the fringe peaks was rechecked for accuracy by numerically computing the derivative of fringe peak point number with respect to fringe peak number; for a laser system that is scanning almost linearly in point number space, plotting this numerically computed derivative should result in a smooth and straight line with no sudden hops or obvious discontinuities. This method of searching for sudden hops in the first derivative of fringe spacing with respect to fringe number was quite useful in locating mode-hops in a similar spectrometer used to study acetylene. [Hardwick 2, 3, and 4]

The observed diatomic iodine features (reasonably well isolated single lines or blended lines) used for calibration are then located in fringe peak number space by linear interpolation between adjacent fringe peaks. Similarly, fringe peak numbers are computed for all of the data points by linear interpolation between all pairs of fringe peaks. In the limit that the laser scans linearly between fringe peaks this calibration method is linear in wave number space. A linear fit was then done with the literature values of the wave number position for the set of diatomic iodine features used for calibration as the dependent variable and the corresponding fringe number of that feature as the independent variable. Just previous to performing this linear fit the fringe number data were also translated so that fringe number zero was aligned with the diatomic iodine feature at lowest wave number of the three spectral regions investigated with this laser; each one covered about 2.4 $cm^{-1}$. The slope of this line is the free spectral range of the etalon fringes, and the intercept is the first diatomic iodine feature used as an absolute frequency marker. (The free spectral range of the etalon was determined independently for each scan of the spectral region being investigated.) The choice of anchor reference point used in the transformation of the fringe number data associated with all data points to a wave number scale is arbitrary. The wave number position for most of the observed diatomic iodine features was also calculated. The residual of this calibration procedure was then found by subtracting from the calculated (observed) diatomic iodine feature line positions the associated literature value. The nonlinear line shape fitting was then done using the wave number values associated with each data point.

The error in the free spectral range of the etalon due to an uncertainty in temperature can be estimated. The etalon was mounted on a square of aluminum about 12 inches on each side and about 2 inches thick. The etalon windows were separated by roughly 10 inches ($z$). The coefficient of thermal expansion for aluminum is $23.6 \times 10^{-6} \, K^{-1}$ and is defined as the partial of length with respect to temperature divided by the length at constant pressure: $(\partial z / \partial T) \div z \cong \Delta z \div (\Delta T \times z)$. Assuming an uncertainty in temperature of 2 Kelvin ($\Delta T = 2$ K) a rough estimate suggests that the spacing between the etalon windows could change by about (2 K $\times$ 25.4 cm $\times 23.6 \times 10^{-6} \, K^{-1}$) $\cong 0.0012$ cm. The free spectral range is given by free spectral range = $n \div (2z)$; $n$ is the index of refraction of air and to a first-order approximation is taken as $n = 1$. The uncertainty in fringe spacing is then given by $\Delta z \div (2z^2) \cong 0.0012 \text{ cm} \div (50.8 \text{ cm})^2 \cong 4.7 \times 10^{-7} \text{ cm}^{-1} \cong 14 \text{ kHz}$. Accordingly, the etalon is not likely to introduce any errors in the wave number calibration at the level of resolution encountered in this project. [Ashcroft] A change in the index of refraction with changes in air pressure (and perhaps moisture content) can also lead to changes in the free spectral range of the etalon. (It was estimated that a deviation in the air pressure by about 7 torr from the standard room pressure of one atmosphere could change the free spectral range of the etalon by about 3 parts in $10^5$.) Enclosing the etalon in a box with small entrance and exit holes for the laser beam was seen to mitigate the effects of sudden changes in air pressure.

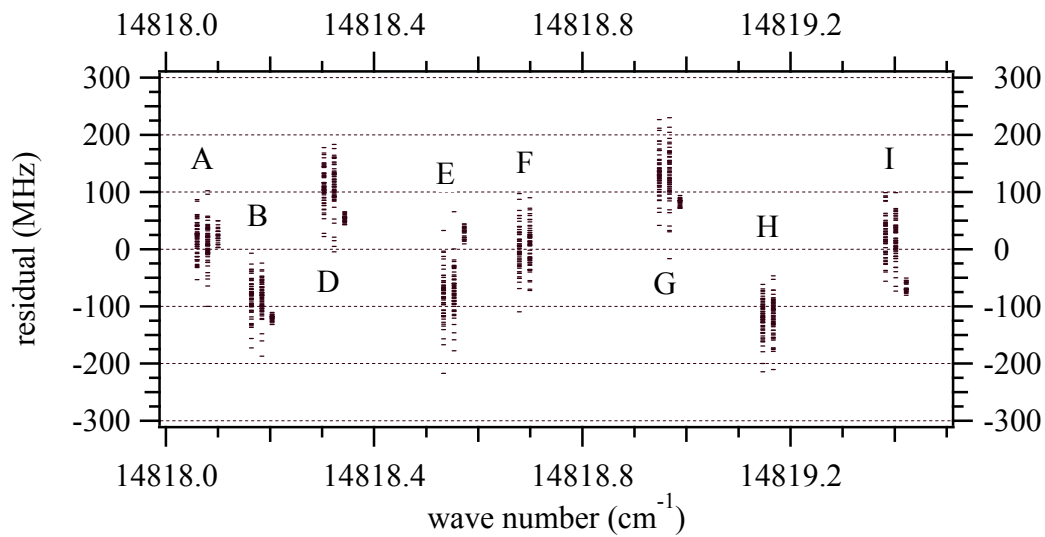

Figure 4.13 Calibration residuals of diatomic iodine ($I_2$) feature wave number positions. The residuals are the observed minus literature values for the diatomic iodine Features A to I. The two stripes of data associated with Features F and H are from left to right the up and down scan of the Philips laser diode system. All other features have three stripes of data that are from left to right in each set of stripes under the feature label the up and down scan (in wave number) of the Philips laser diode system and the single scan direction of the New Focus external cavity laser diode system.

This calibration method was used on only one data set collected with the Philips CQL806/30 laser diode system. All three of the data sets collected with the New Focus external cavity laser diode system were calibrated for wave number using this method and only this method. All four of these data sets used argon as the buffer gas. One of the data sets collected with the New Focus external cavity laser diode system was of the same region that the Philips laser diode scanned, roughly 14,817.95 to 14,819.45 $cm^{-1}$. The wave number calibration residuals for these two overlapping data sets from the two different laser systems using this calibration method are plotted in Figure 4.13 and some of the statistics are listed in Table 4.6. Overall, it appears that the New Focus external cavity laser diode system offers a higher quality of precision relative to the Philips laser diode system.

Table 4.6 Wave number calibration statistics for the two different laser diode systems. Feature A was arbitrarily chosen as the calibration intercept at 14,818.0601 $cm^{-1}$ in the calibration procedure. The linear fit uncertainty for the free spectral range (FSR) and calibration intercept are given in parenthesis in units of the last significant figure of the observed quantity.

|  | Up Scan / Philips | Down Scan / Philips | New Focus |
|---|---|---|---|
| scans | 123 | 123 | 24 |
| FSR (MHz/fringe) | 642.8(18) | 642.2(18) | 578.9(15) |
| Cal. Int. ($cm^{-1}$) | 14,818.0607(21) | 14,818.0605(20) | 14,818.0609(17) |
| Feature | $\sigma_{residual}$ (MHz) | $\sigma_{residual}$ (MHz) | $\sigma_{residual}$ (MHz) |
| A | 28.5 | 33.6 | 13.7 |
| B | 31.1 | 29.2 | 5.6 |
| D | 32.1 | 38.7 | 6.5 |
| E | 41.3 | 38.2 | 10.1 |
| F | 37.8 | 35.1 | -- |
| G | 34.3 | 45.1 | 6.4 |
| H | 30.7 | 31.5 | -- |
| I | 33.9 | 34.2 | 8.4 |

### 4.6.3 Comparison of Etalon and Polynomial Calibration Methods

Calibration of wave number for spectra obtained with the New Focus external cavity laser diode system using linear interpolation between fringes and fifth order polynomial indicates that the resolution of this system has surpassed the 12-bit capability of the NI-PCI1200. Figure 4.14 shows a plot of the differences in calibrated wave number scale between the method using many diatomic iodine-features and etalon fringes and a fifth-order polynomial (see Figure 4.7). The initial assumption of piecewise linear scans between fringe-peaks (for a scan range of roughly 2 cm$^{-1}$) thus appears to be justified when using a 12-bit data acquisition board. As it stands the full resolution of the New Focus laser diode system was not realized. This task was not practical when using a 12-bit data acquisition since it was proving somewhat non-trivial to dial-in the scan region for this laser system with that much precision. However, a 16-bit data acquisition board can be expected to utilize the full resolution of this spectrometer.

Finally, there are subtle questions regarding the comparison of the data from this project with that of the Fourier-transform spectra (Iodine Atlases) that we have used as wavelength standards. First, the authors of the Iodine Atlases do not state expclicitly how they measured line centers. Second, the derived vibration and rotation constants (i.e. the spectroscopic constants) are compiled from several independent experiments and may either remove statistical error or introduce an undisclosed systematic error in the wave number calibration process. Accordingly, any additional improvement in the calibration would likely require a wavelength standard independent of the diatomic iodine spectrum.

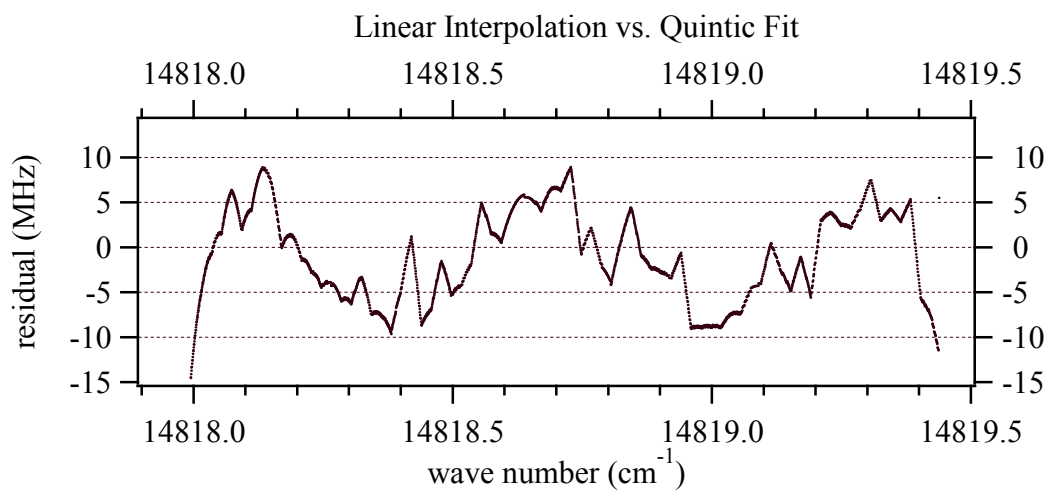

Figure 4.14 Difference in point-by-point wave number positions for two calibration methods. The first calibration method is linear interpolation between fringes. The second method uses quintic fit parameters for the fringe separation. Both methods use most of the diatomic iodine lines in the spectral region as absolute frequency markers with an accuracy of about 150 MHz.

## 4.7 Endnotes for Chapter IV

[Ashcroft]      Thermal expansion coefficient for aluminum is from: N. W. Ashcroft and N. D. Mermin ; *Solid State Physics* , page 476; W. B. Saunders, Philadelphia, Pennsylvania (1976); ISBN: 0-03-083993-9.

[Borde 2]       F.-L. Hong, J. Ye, L.-S. Ma, S. Picard, C. J. Borde, and J. L. Hall; "Rotation dependence of electric quadrupole hyperfine interaction in the ground state of molecular iodine by high-resolution laser spectroscopy", Journal of the Optical Society of America, Part B, 18, 379-387 (2001).

[Fletcher]      D. G. Fletcher and J. C. McDaniel; "Collisional shift and broadening of iodine spectral lines in air near 543 nm", Journal of Quantitative Spectroscopy and Radiative Transfer, 54, 837-850 (1995).

[Hardwick 1]    J. A. Eng, J. L. Hardwick, J. A. Raasch and E. N. Wolf; "Diode laser wavelength modulated spectroscopy of $I_2$ at 675 nm", Spectrochimica Acta, Part A, 60, 3413-3419 (2004).

[Hardwick 2]   J. L. Hardwick, Z. T. Martin, E. A. Schoene, V. Tyng and E. N. Wolf; "Diode laser absorption spectrum of cold bands of $C_2HD$ at 6500 $cm^{-1}$", Journal of Molecular Spectroscopy, 239, 208-215 (2006).

[Hardwick 3]   J. L. Hardwick, Z. T. Martin, M. J. Pilkenton and E. N. Wolf; "Diode laser absorption spectra of $H^{12}C^{13}CD$ and $H^{13}C^{12}CD$ at 6500 $cm^{-1}$", Journal of Molecular Spectroscopy, 243, 10-15 (2007).

[Hardwick 4]   S. W. Arteaga, C. M. Bejger, J. L. Gerecke, J. L. Hardwick, Z. T. Martin, J. Mayo, E. A. McIlhattan, J.-M. F. Moreau, M. J. Pilkenton, M. J. Polston, B. T. Robertson and E. N. Wolf; "Line broadening and shift coefficients of acetylene at 1550 nm", Journal of Molecular Spectroscopy, 243, 253-266 (2007).

[Horowitz 1]   P. Horowitz and W. Hill; *The Art of Electronics*, Sections 7.10 and 14.15; Cambridge University Press, Cambridge (1985); ISBN 0-521-23151-5.

[Hutson]   J.M. Hutson, S. Gerstenkorn, P. Luc and J. Sinzelle; "Use of calculated centrifugal distortion constants ($D_v$, $H_v$, $L_v$ and $M_v$) in the analysis of the B ← X system of $I_2$", Journal of Molecular Spectroscopy, 96, 266–278 (1982).

[Luc 1]   S. Gerstenkorn and P. Luc; Atlas du Spectre d'Absorption de la Molécule d'Iode entre 14 800–20 000 $cm^{-1}$, Laboratoire Aimé Cotton CNRS II, Orsay, (1978).

[Luc 2]   S. Gerstenkorn and P. Luc; "Absolute iodine ($I_2$) standards measured by means of Fourier transform spectroscopy", Revue de Physique Appliquee, 14, 791-794 (1979).

[Martin]   F. Martin, R. Bacis, S. Churassy and J. Vergès; "Laser-induced-fluorescence Fourier transform spectrometry of the $XO_g^+$ state of $I_2$: Extensive analysis of the $BO_u^+ \rightarrow XO_g^+$ fluorescence spectrum of $^{127}I_2$", Journal of Molecular Spectroscopy, 116, 71-100 (1986).

CHAPTER V

ANALYSIS 2 – LINE SHAPE ANALYSIS

## 5.1 Overview of Chapter V

The purpose of analyzing the line shape of the B–X transitions in diatomic iodine ($I_2$) with a nonlinear regression algorithm is to extract (de-convolve) the width (full width at half maximum, also referred to by the acronym "FWHM") of the Lorentzian (homogeneous) component from the observed Voigt profile and to determine the line-center position (i.e. wave number). The Lorentz widths and line-center positions are calculated at several pressures of a buffer gas. The use of internally referenced absorption spectrometers allows for precise (and perhaps accurate) determination of the line-center shift as a function of buffer gas pressure. The pressure broadening coefficient is obtained from a weighted linear least-square fit of the Lorentz widths as a function of pressure; the assumption here is that Lorentz width is directly proportional to the pressure. Similarly, the pressure shift coefficient is obtained from a weighted linear least square fit of the line-center shift as a function of pressure, predicated on the assumption that the line-center shift and buffer gas pressure are directly proportional. (Diatomic iodine refers to the homonuclear molecule $I_2$, which is often colloquially referred to as iodine.)

## 5.2 Line Shape Components and the Convolution Model

The line shape model used in the nonlinear regression analysis is the Voigt distribution [Loudon 1; Bernath], which is the convolution of Gaussian (Doppler) and Lorentzian components; these two components and their convolution are shown in Figure 5.1. The basis for this model is the distinction between an inhomogeneous (Gaussian) line shape function $g_D(\omega_0 - \omega)$ and one that is homogeneous (Lorentzian) $g_L(\omega_0 - \omega)$, so that the total line shape (Voigt) is given by the convolution:

$$F_V\, g_V(\omega - \omega_0) = \int_0^{\infty} F_D\, g_D(\omega_0' - \omega_0)\, F_L\, g_L(\omega - \omega_0')\, d\omega_0'$$

(5.1)

In equation 5.1 the radiation field center frequency ($\omega_0''$) (itself expected to be predominantly Lorentz in shape) is given by $\omega$. (See also Section 2.11.)

While normalization factors ($F_L$, $F_D$, and $F_V$) are generally not of concern in this project, it is worth noting that some care needs to be taken with regard to their use for the case of accounting for intensity measurements. In general, the underlying distributions will be normalized:

$$F_D \int_0^\infty g_D(\omega_0 - \omega)\, d\omega$$

$$= F_L \int_0^\infty g_L(\omega_0 - \omega)\, d\omega$$

$$= F_V \int_0^\infty g_V(\omega_0 - \omega)\, d\omega$$

$$= 1 \tag{5.2}$$

However, the normalization factor for the Voigt profile $F_V$ can only be approximated by an analytic function [Humlíček]. In general, $F_V = F_V(F_L, F_D)$.

The Doppler (inhomogeneous) width (FWHM) $\Delta\omega_D = 2\pi \times \Delta\nu_D$ is determined by the temperature $T$, mass $m$, Boltzmann constant $k_B$, speed of light $c$, and transition frequency $\omega_0$, and is given by [Bernath]:

$$\Delta\omega_D = 2\,\omega_0 \sqrt{\frac{2\,k_B\,T\,\ln(2)}{m\,c^2}} \tag{5.3}$$

For diatomic iodine at 292 K and laser radiation at 675 nm, the Doppler width is 341 MHz. The Lorentzian (homogeneous) component is pressure dependent so that the width (FWHM) spans a range. The minimum value of the Lorentz width $\Delta\nu_L = (2\pi)^{-1} \times \Delta\omega_L$ is determined by the natural (spontaneous) transition rate, likely to be between 0.1 MHz and 10 MHz [Paisner]. The maximum measurable Lorentz width was generally regarded to be approximately 800 MHz, at which point the diatomic iodine lines begin overlapping to an extent that makes a simple analysis impossible; that is, there is an upper bound (maximum) on the buffer gas pressure for which a reliable nonlinear regression model based on a single (ro-vibronic) line can be performed.

The analysis of individual ro-vibronic spectral lines of diatomic iodine for changes in Lorentzian (homogeneous) line width $\Delta\omega_L = 2\pi \times \Delta\nu_L$ (FWHM) and line-center shift $\Delta\omega_0 = 2\pi \times \Delta\nu_0$, each as a function of buffer gas pressure, provides the means to extract two important spectroscopic quantities, the pressure broadening and pressure shift coefficients. The normalized Lorentzian profile (for $\omega_0 \gg \delta$) is given by:

$$F_L \, g_L(\omega_0 - \omega) \; = \; \frac{\delta}{\pi} \times \frac{1}{\delta^2 + (\omega_0 - \omega)^2}$$

(5.4)

And the full width at half-maximum height $\Delta\omega_L = 2\pi \times \Delta\nu_L$ is readily found to be:

$$\Delta\omega_L \; = \; \sqrt{2} \; \delta$$

(5.5)

When conducting an analysis of frequency domain spectra, a nonlinear fitting routine is required to compare the model line shape to that recorded in a spectrum. This de-convolution process separates the Voigt distribution (model line shape) into its two constituent parts, the (constant) Doppler width and the pressure dependent Lorentz width. The pressure broadening coefficient is computed by a weighted linear least square regression of the Lorentz widths as a function of buffer gas pressure. At the same time the relative change in line center position between the sample and reference gas cells (i.e. internally referenced) allows for a weighted linear fit of the shift in line-center as a function of buffer gas pressure.

For the analytic approximation to the Voigt profile developed by Humlíček, it can be shown that [Humlíček]:

$$\Delta\omega_V \cong \frac{\Delta\omega_L}{2} + \sqrt{\frac{\Delta\omega_L{}^2}{4} + \Delta\omega_D{}^2}$$

<div align="right">(5.6)</div>

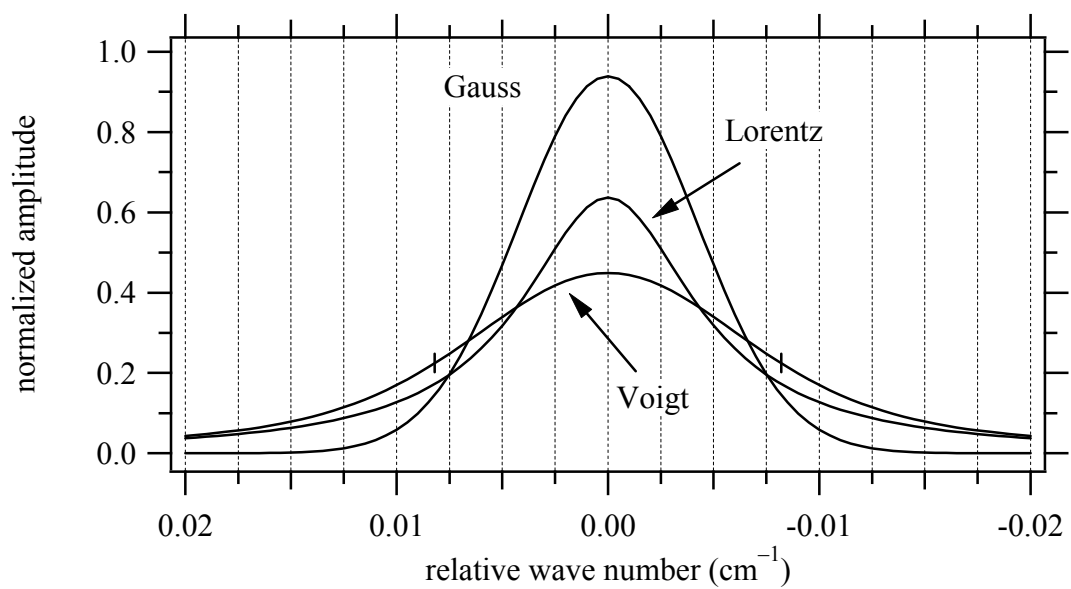

Figure 5.1 Convolution of normalized Gaussian and Lorentzian components (i.e. distributions) to form a Voigt distribution. The FWHM of the Gaussian and Lorentzian components are each 300 MHz (0.01 cm$^{-1}$). The vertical hash marks on the Voigt trace indicate the location of the FWHM (491 MHz) relative to the line center. The wave number scale is relative to the line center, which is at 14,819 cm$^{-1}$.

The convolution shown in Figure 5.1 was obtained from a numerical convolution of Doppler and Lorentzian profiles in Mathematica using (built-in) interpolation functions, each with a full width at half-maximum height of 300 MHz. This numerical simulation gave the full width at half-maximum of the Voigt profile as $\Delta \nu_{\mathrm{V}}$ = 491 MHz.  For the Humlíček approximation, equation 5.6 gives $\Delta \nu_{\mathrm{V}}$ = 485 MHz.  The line-center in this simulation was at 14,819 cm$^{-1}$.

## 5.3 Linear Absorption and Line Shape

A few direct absorption spectra of diatomic iodine near 675 nm were observed in order to determine the maximum amount of light absorbed during passage through a 1 meter path of diatomic iodine at approximately 0.2 torr (292 K).  This was accomplished by scanning the laser frequency with the wavelength modulation turned off and instead inserting an optical chopper (e.g. Scitec Instruments) in the path of the laser beam before it was split into three portions.  The optical chopper also provided the reference frequency for the lock-in amplifiers.  Observing direct absorption spectra is relevant in deciding if a correction to the recorded spectrum using the Beer-Lambert law would be necessary before proceeding with the line shape analysis.  For the diatomic iodine experiments (for which 1 meter gas cells were used) the maximum (direct) absorption observed for the strongest spectral features was not more than a percent or two of the laser beam intensity.  It was thus considered a good approximation to use the linear limit of the Beer-Lambert law.  In this project the Beer-Lambert law is assumed to have the form:

$$I((\omega_0 - \omega),\, z) \,=\, I(0)\exp(-2\,\pi\,\mathrm{K}\,g_{\mathrm{V}}(\omega_0 \,-\, \omega)\,z) \qquad (5.7)$$

See also Section 2.11, especially equation 2.28, and recalling that $\omega = 2\pi\nu$.  In equation 5.7, the measured intensity for a radiation field line-center frequency $\omega$ is given by $I((\omega_0 - \omega),\, z)$; the line-center of the chromophore absorption profile is given by $\omega_0$; the radiation field intensity incident on the gas cell is given by $I(0) \equiv I((\omega_0 - \omega),\, 0)$, which is assumed to be the same at all line-center frequencies $\omega$ of the radiation field; the absorption coefficient is given by K, and the path length of the gas cell is given by $z$. The linear limit of equation 5.7 (for the situation that $I(z) \div I(0) \gtrsim 0.97$, as was the case in this project) is obtained by retaining the first term in the Taylor series expansion of equation 5.7:

$$I((\omega_0 - \omega), z) = I(0)(1 - 2\pi K g_V(\omega_0 - \omega) z) \qquad (5.8)$$

Since a wavelength-modulated spectrum can be well modeled as the first derivative of equation 5.8, the leading term of $I(0)$ is not necessary so that the nonlinear regression analysis of the observed line shape was (essentially) performed directly on $g_V(\omega_0 - \omega)$.

Unfortunately, during the course of this project the signal values of $I(z) \equiv I((\omega_0 - \omega), z)$ and $I(0) \equiv I((\omega_0 - \omega), 0)$ were not directly recorded for either gas cell. As it turns out, it might have been useful to more fully characterize some direct linear absorption spectra and the radiation field across the spectral region of interest for the purpose of comparison to the wavelength-modulated spectra, and thus perhaps a more complete characterization of the latter.

However, the above considerations alone do not establish that these experiments were carried out at laser intensities that can be approximately modeled by linear response behavior. A direct experimental approach to determining that the "strength" of the perturbation (due to the electric field of the radiation source) is appropriately characterized by a linear response model is to measure the ratio of transmitted to incident radiation intensity $I(z) \div I(0)$ (at a frequency $\omega$ near $\omega_0$ where there is appreciable steady-state absorption of the radiation field) for a wide range of incident intensities to see if this ratio remains constant. We did not rigorously perform such measurements. However, qualitatively, when switching between laser beam collimation methods with the Philips laser diode system, and thus changing the beam intensity by a couple of orders of magnitude, the signal levels remained about the same, and the transition line shapes were not measurably affected.

We can also make a quick estimate of the perturbation "strength" (energy) from a calculation of the product $\mu E(0)$, where $\mu$ is the transition moment and $E(0)$ is the cycle-averaged incident electric field [McHale]; e.g. see equation 7.36. The square of the electric field can be readily estimated from $E(0)^2 = 2I(0) \div (c\,\eta\,\epsilon_0)$. For an incident laser beam diameter of 5 mm, the maximum (line-integrated) intensity of nearly monochromatic laser radiation encountered on any single beam path would not be more than about 5 mW per $2 \times 10^{-5}$ m$^2$. The speed of light (in a vacuum) is taken as $c = 3 \times 10^8$ m s$^{-1}$, the index of refraction $\eta = 1$, and the electric constant $\epsilon_0 = 8.85 \times 10^{-12}$ F m$^{-1}$ (and recalling that the units of the Farad are given by F = C V$^{-1}$ = C$^2$ J$^{-1}$), so that the upper bound estimate of the electric field seen by a chromophore is given by: $E(0) = ((2 \times 5 \times 10^{-3}$ J s$^{-1}) \div (2 \times 10^{-5}$ m$^2))^{1/2} \times ((3 \times 10^8$ m s$^{-1}) \times (8.85 \times 10^{-12}$ C$^2$ J$^{-1}$ m$^{-1}))^{-1/2} \cong 434$ Nt per C. The transition moment $\mu$ for B $\leftarrow$ X electronic transitions in diatomic iodine at 675 nm can be estimated as 0.75 D $\cong 2.5 \times 10^{-30}$ C m [Tellinghuesin], from which the perturbation energy is obtained: (434 Nt C$^{-1}$) $\times$ (2.5 $\times 10^{-30}$ C m) $\cong$ **1 $\times$ 10$^{-27}$ J**. This upper bound estimate of the electric field is fairly conservative; it is likely to be at least a couple of orders of magnitude smaller than given in this estimate.

For comparison, the average energy of rotation is given by $2hcB_e(J + 1)$. The population of the rotation distribution at a temperature of 292 K peaks at $J \cong 52$; we will use this value for $J$. The Planck constant is $h = 6.626 \times 10^{-34}$ J s and the speed of light is $3 \times 10^{10}$ cm s$^{-1}$. The rotation constant $B_e$ for the X and B electronic states of diatomic iodine can be taken respectively as 0.03735 cm$^{-1}$ and 0.02920 cm$^{-1}$, so that the average rotational energies for these two states is roughly **7 $\times$ 10$^{-23}$ J**. The average translational (kinetic) energy of diatomic iodine at 292 K can be approximated by $k_B T \div 2 = 1.38 \times 10^{-23}$ J K$^{-1} \times 292$ K $\div 2 =$ **4 $\times$ 10$^{-21}$ J**. The energy required to break the diatomic iodine single bond from low lying vibration levels of the X and B electronic states is on the order of **2 × 10⁻¹⁹ J** (from Figure 2.1). The transition energy ($E = h\,\nu = h\,c \div \lambda$) at a wavelength of $\lambda = 675$ nm is about **3 × 10⁻¹⁹ J**. Given the smallness of the perturbation "strength" $\mu E(0)$ in relation to these other energies that characterize this system, it is considered reasonable to assume that these experiments are well modeled by linear response perturbation theory.

It was thus determined that no further corrections of the observed spectral line shape were necessary with regard to the exponential behavior of the Beer-Lambert law or for nonlinear optical behavior, the latter having a parallel characterization at the microscopic level as multi-photon events. The maximum (macroscopic) absorption of radiation for the prevailing experimental conditions allowed for use of the linear limit of the Beer-Lambert law (equation 5.4). And the electric field strength of the radiation source was estimated to be weak enough so that consideration of single photon events that are well isolated in time and space was assumed to be sufficient.

## 5.4 Diatomic Iodine Line Shape with Nuclear Hyperfine Structure

The nuclear hyperfine structure is due to electric and magnetic interaction between the nucleus and electrons; see Sections 2.5 and 2.6 for more detail. The nuclear hyperfine structure of a diatomic iodine B−X transition line shape appears as a splitting of the Lorentzian (homogenous) component of an individual ro-vibronic line shape into a relatively tight grouping of either 15 (even $J''$) or 21 (odd $J''$) transitions. The wave number (or frequency) position of these nuclear hyperfine transitions is well characterized by spectroscopic constants and quantum numbers used in model Hamiltonians. Due to high precision saturation spectroscopy experiments with sub-kHz resolution, knowledge of the hyperfine structure of diatomic iodine surpasses the resolution capabilities of the spectra obtained during the course of this project [Kato; Knöckel]. The resolution achievable in a Doppler-limited line shape analysis might be an order of magnitude less than the frequency separation between data points, which was approximately 20 MHz in this project. A much simpler characterization of the nuclear hyperfine structure was thus considered to be generally suitable for this project.

### 5.4.1 Model for High-$J$ Lines

For rotation quantum number values of $J \gtrsim 20$ the nuclear hyperfine structure can be reasonably well characterized by only the change in the nuclear electric quadrupole moment $eQq$ [Schawlow]; this change is equal to the difference of these quantities for the B and X electronic states: $\Delta eQq = eQq' - eQq''$ (i.e. subtracting the lower energy level $eQq$ value from that for the upper energy level); approximate values for these quantities are $\Delta eQq = eQq' - eQq'' \cong (-600 + 2550)$ MHz = 1950 MHz $\cong 0.065$ cm$^{-1}$ [Blabla]. Extensive use was made of this single parameter approach to characterizing the nuclear hyperfine structure for the line shape analysis; deviation from this approach occurred only near the end of this project to investigate several transitions with small (a.k.a. low) values of $J$, which will be the topic of the next sub-section. In this high-$J$ limiting case (i.e. large values of $J$) the nuclear hyperfine structure for diatomic iodine is generally observed to collapse from 15 (even $J''$) or 21 (odd $J''$) distinct nuclear hyperfine components into six groupings of transitions that differ only in the relative intensity distribution among these groupings for even and odd $J''$ [Kroll]. The energy formula that describes this asymptotic high-$J$ limiting case of the nuclear hyperfine structure of diatomic iodine, based on the nuclear electric quadrupole coupling constant alone, is given by:

$$E = E_0 - \lambda \times \Delta eQq$$

$$= E_0 - \lambda \times (eQq' - eQq'') \tag{5.9}$$

Figure 5.2 indicates that the energy $E_0$ does not coincide with the maximum absorption intensity (i.e. the line-center frequency $\nu_0$). The values of the eigenvalue $\lambda$ listed in Table 5.1 are determined from an analysis of the asymptotic high-$J$ limiting case for diatomic iodine; it is a "reduced" eigenvalue determined by the difference of eigenvalues for the nuclear electric quadrupole Hamiltonian of the X and B electronic states of diatomic iodine.

Table 5.1 illustrates the intensity alternation between odd and even $J''$ lines due to nuclear spin statistics for the high-$J$ asymptotic model. The high-$J$ model also predicts that transitions with odd and even $J''$ will have different shapes. This "shape alternation" is readily discernable in the wavelength-modulated spectrum of diatomic iodine and, once identified, is also visible in the line profiles measured by direct absorption. (Of course, it is relatively easy, especially when using wavelength modulated lock-in detection, to invert the signal; e.g. compare Figure 5.3 to Figures 4.2 and 4.6)

Table 5.1 Diatomic iodine ($I_2$) nuclear hyperfine structure for $J \gtrsim 20$ (for the $^{127}I_2$ isotope, which has nuclear spin 5/2). Wave number position of the nuclear hyperfine components is found from $E = E_0 + \lambda \times \Delta eQq$. The relative "intensity" contributions (i.e. number of states) for each value of the eigenvalue for the odd and even $J''$ cases are listed; the total intensity ratio for odd to even $J''$ is $21 \div 15 = 7/5$. The rotational quantum number $J''$ corresponds to the ro-vibronic state on the X electronic potential energy surface.

| eigenvalue ($\lambda$) | odd $J''$ intensity | even $J''$ intensity |
|---|---|---|
| $+\,0.25$ | 3 | 1 |
| $+\,0.1$ | 4 | 4 |
| $+\,0.025$ | 4 | 4 |
| $-\,0.05$ | 3 | 1 |
| $-\,0.125$ | 4 | 4 |
| $-\,0.2$ | 3 | 1 |

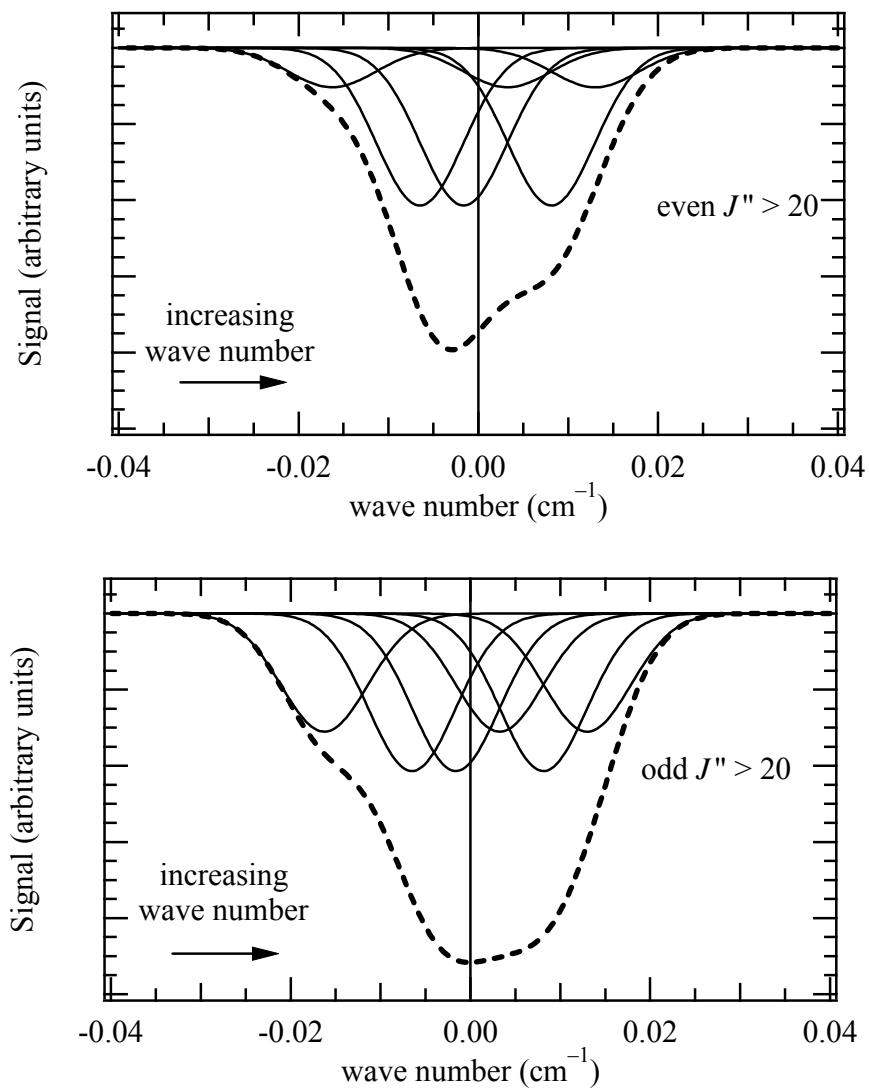

Figure 5.2 Simulation of direct (i.e. not wavelength-modulated) linear absorption spectrum (i.e. line shape) for an individual ro-vibronic transition of diatomic iodine ($I_2$) between the X and B electronic states as a linear superposition of the underlying hyperfine structure in the high-$J$ limit. The intensity scales between the two plots are the same. The dashed line in each plot is the anticipated direct absorption line shape with $\Delta eQq = 0.065$ cm$^{-1}$ (1950 MHz); individual groupings have a Gauss (Doppler) width of 341 MHz.

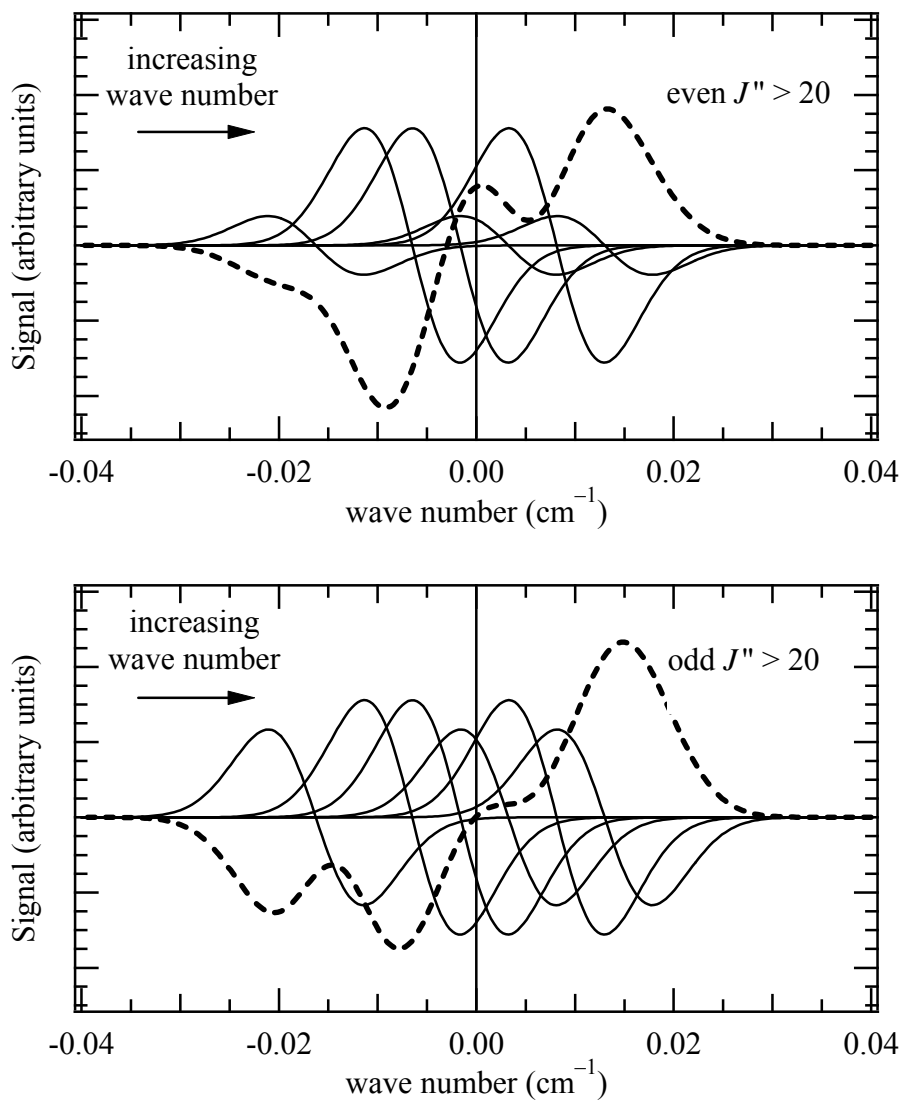

Figure 5.3 Simulation of wavelength-modulated linear absorption spectrum (i.e. line shape) for an individual ro-vibronic transition of diatomic iodine ($I_2$) between the X and B electronic states as a linear superposition of the underlying hyperfine structure in the high-$J$ limit. The intensity scales between the two plots are the same. The dashed line in each plot is the anticipated wavelength modulated absorption line shape with $\Delta eQq = 0.065$ cm$^{-1}$ (1950 MHz); individual groupings have a Gauss (Doppler) width of 341 MHz.

The transition line shape using this single parameter can be readily simulated for both direct and wavelength-modulated (first derivative) absorption spectra; such a simulation is presented in Figures 5.2 and 5.3 with the assumption that the line shape is entirely determined by a Gaussian distribution (i.e. entirely due to Doppler broadening). The simulations obtained from this approximation of neglecting the Lorentzian (homogeneous) component in the line shape correspond to the experimental condition of low total pressure (i.e. less than about 1 torr) in the gas cell, in which case the Lorentz width will be on the order of about 10 MHz, or roughly three percent of the Doppler width (ca. 341 MHz at a temperature of 292 K).

## 5.4.2 Models for Low-$J$ Lines

Twelve low-$J$ ro-vibronic transition line shapes of diatomic iodine were studied in the range $J'' = 6$ to 26; all of these spectra were obtained using the New Focus laser diode system with the gas cell at room temperature (292 K) and **argon** as the buffer gas. These spectra were obtained from two slightly overlapped (by about 0.4 cm$^{-1}$) laser diode scan regions near the $(v', v'') = (4, 6)$ band head. The region further away from the band head (14,946.17 to 14,948.43 cm$^{-1}$) appears to be slightly less congested than the one that is closer. In the region further away from the band head it was possible to find six low-$J$ lines (and three high-$J$ lines) that had (ro-vibronic) line-centers roughly 2000 MHz from a single prominent neighboring line-center and were otherwise well-isolated from other lines. The line shape of these six low-$J$ transitions (in the range $J'' = 14$ to 26) were investigated by simultaneously fitting this ro-vibronic line and its single prominent neighbor (i.e. two-line fit) using the high-$J$ asymptotic-limit nuclear hyperfine model (presented in the previous section). The fit parameters allow for a unique determination of the line-center, the Gauss width, the Lorentz width, and the change in the nuclear electric quadrupole coupling constant $\Delta eQq$ moment for each of the ro-vibronic transitions in the two-line fit.

In the scan region closer to the band head (14,948.08 to 14,950.29 cm$^{-1}$) six ro-vibronic transitions (in the range $J'' = 6$ to 20) were found with a single prominent nearby transition, except that the wave number distance between line-centers was approaching half of what it was in the scan region further from the band head. Two-line fits were also performed for the region closer to the band head, but instead used a more general nuclear hyperfine model for the undiagonalized nuclear hyperfine Hamiltonian [Bunker; Borde 1; Borde 2]; the fitting algorithm performed the mathematical operation of diagonalizing the effective nuclear hyperfine Hamiltonian on each pass of the nonlinear fitting routine. This more general nuclear hyperfine Hamiltonian is a function of $J''$ and $J'$, the nuclear electric quadrupole moments $eQq'$ and $eQq''$. The fit parameters allow for a unique determination of the line-center, the Gauss width, the Lorentz width, and the change in the nuclear electric quadrupole moments $\Delta eQq$ for each of the ro-vibronic transitions in the two-line fit. The two-line fitting algorithm based on the more general nuclear hyperfine Hamiltonian appeared to be effective for values of $J'' \gtrsim 6$; for lower values of $J''$ the recorded (i.e. observed) line shape begins to deviate too much from the model line shape, as seen in the fit residuals for $J'' = 5$. It has been mentioned in the literature that the nuclear hyperfine model for low-$J$ lines may need to account for an additional selection rule gradually turning-on (i.e. changing from an "un-allowed" to an "allowed" transition), which may account for what appeared to be a gradual breakdown of the line shape analysis in going to lower values of $J''$ [Yoshizawa].

In order to gain some additional understanding and experience with the nuclear hyperfine model algorithms, a comparison of the fit results obtained from the more complete nuclear hyperfine model and the high-$J$ model was made for diatomic iodine Feature E ($J'' = 69$) and Feature F ($J'' = 122$). These diatomic iodine features have a line shape that agrees well with the high-$J$ limit ($J \gtrsim 20$) nuclear hyperfine model (depicted in Table 5.1). For a few fits covering different pressures of buffer gas, the fit results and statistics were nearly identical, and it is evident in a visual comparison that the residual plot corresponding to the more complete nuclear hyperfine model is only slightly flatter (i.e. smaller and flatter) than that of the high-*J* model. Attempts were also made to account for the magnetic dipole nuclear hyperfine coupling term when using either the high-*J* or more general nuclear hyperfine model, but the fit results and statistics were rather insensitive to the value of this parameter; the limiting factor here is likely to be the resolution of the spectra from this project (ca. 20 MHz between data points) versus the relatively small magnitude of the magnetic dipole (a.k.a. spin-rotation) term [Bunker].

## 5.5 Voigt Line Shape and Nonlinear Fitting

The fitting routine is structured so that the model line shape parameter space corresponds to the direct (or un-modulated) absorption line shape. The fitting routine employs a nonlinear regression algorithm (i.e. not a simple polynomial), and so an initial guesstimate must be provided for the value of all fit parameters. The fitting routine is configured so that the numerical derivative of the model line shape is taken on each pass of the fitting routine and then compared to the observed wavelength-modulated line shape; the fitting routine then adjusts the fit parameter values or terminates the fitting procedure based on this comparison. (The capabilities of the computer application "Igor" are sufficiently developed so that it is a convenient platform for data collection, data reduction, and data analysis.)

The fit parameters for the direct linear absorption line shape (i.e. not wavelength-modulated) include the Gauss and Lorentz widths (FWHM); the amplitude (i.e. signal deflection); a line-center; a constant baseline offset and a slope. Since an offset has no meaning in a first-derivative spectrum, the typically available offset parameter (in the nonlinear fit routine for the Voigt profile) was transformed into a slope parameter to account for the observed behavior of direct absorption spectra in which the baseline signal (i.e. in the absence of an absorption signal) is generally seen to undergo a modest and nearly linear change across a spectral region somewhat larger than that of a single line shape (fairly noticeable on the scale of about 0.1 cm$^{-1}$); the use of an auto-balanced detector can reduce this effect considerably.

When using the high-$J$ asymptotic-limit nuclear hyperfine model, the ro-vibronic transitions must first be identified as odd or even $J''$ so that the correct nuclear hyperfine structure intensity pattern (Table 5.1) can be correctly configured in the fitting algorithm; as Figure 5.3 illustrates, for reasonably well resolved (i.e. sufficient signal and sufficiently unblended with neighboring lines) ro-vibronic transitions, this identification can be done by inspection. The nuclear hyperfine structure parameter $\Delta eQq$ is determined only for the reference gas cell and held fixed at these values when modeling the line shape for the sample gas cell for that pair of simultaneously recorded spectra.

The line shape analysis typically included data points extending to the wings of the line where the signal still has discernable amplitude above the nearby baseline. Most of the results presented in this dissertation are from a line shape analysis of observed (recorded) transitions in the region spectral region 14,817.95 to 14,819.45 cm$^{-1}$ using the Philips laser diode system; the lines in this region are reasonably well isolated so that the nonlinear regression was configured for single-line fits. The other spectral region from 14,946.17 to 14,950.29 cm$-1$ was briefly studied was done so for its low-$J$ line content (as described in the previous section). As Figures 4.11 and 4.12 illustrate, the congestion of lines in this region gives rise to blended lines, which makes it necessary to fit more than one line at a time. All of the results from this region were obtained by simultaneously fitting two lines.

Calculating the Voigt profile by direct convolution of a Gaussian and Lorentzian distribution functions is too computationally demanding to be practical for routine line fitting. The nonlinear fitting routine that the computer application "Igor" used provides an analytic approximation to the Voigt profile based on the Humlíček algorithm [Schreier; Humlíček; Drayson; Armstrong; Whiting]. This algorithm is specified to have a relative accuracy better than $0.0001 = 10^{-4}$. (Igor is made available by WaveMetrics, Inc., Portland, Oregon.)

The remainder of this section explores the behavior and limitations of this algorithm. Voigt distributions with a constant Doppler component FWHM width ($0.01 \text{ cm}^{-1} \cong 300 \text{ MHz}$), a few Lorentz FWHM widths (3, 30, and 300 MHz), and centered at $14,819.0 \text{ cm}^{-1}$ were created by numerical integration (convolution) in the computer application Mathematica and exported to Igor for fitting, the results of which are presented in Figure 5.4 and Table 5.2. (Mathematica is made available by Wolfram Research, Inc., Champaign, Illinois.) The Voigt profile simulations were computed point-by-point and by the generation of interpolation functions, with no noticeable differences between these two methods. Each of the fits include 201 data points; the relative spacing (step size) between data points is 15 MHz and thus span a range of $0.05 \text{ cm}^{-1}$ to either side of the line center. The Voigt widths listed in column two of Figure 5.2 are determined (in Mathematica) by interpolation between the nearest two data points on either side of the distribution peak that straddle the FWHM positions. The Igor fitting algorithm allows for parameters to be floating (i.e. determined by the fitting routine) or held fixed during the nonlinear regression analysis. The option of setting the Doppler width parameter as floating or fixed is explored in the following tables and figures. All simulations and nonlinear regression analysis were carried out in wave number units ($\text{cm}^{-1}$). The parameters values in the first five columns of Table 5.2 were converted from units of $\text{cm}^{-1}$ to MHz using the approximate relationship 30,000 MHz per $\text{cm}^{-1}$, and are thus systematically too large by about 0.07%, as compared to using a more accurate conversion factor (e.g. $29,997.9 \text{ MHz per cm}^{-1}$).

In the Humlíček algorithm the Voigt width (FWHM) is (according to the Igor help files) given approximately by equation 5.6. The subscripts V, L, and G are abbreviations for Voigt, Lorentz, and Gauss, respectively, in equation 5.6. The full width at half-maximum height (FWHM) for each these components is $\Delta\nu_V$, $\Delta\nu_L$, and $\Delta\nu_G$ (and recalling that $\Delta\omega = 2\pi \times \Delta\nu$). Accounting for the propagation of error of uncorrelated fit parameters, it is found that the uncertainty in the Voigt width is roughly the same as that for the Lorentz width; see also Section 1.6.

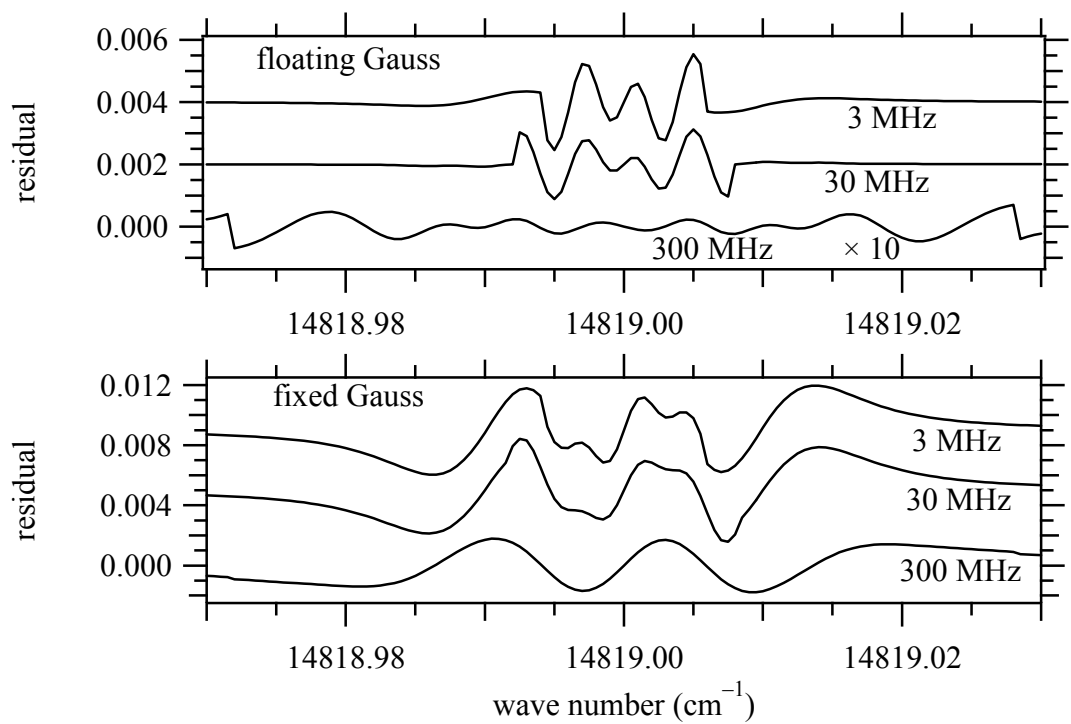

Figure 5.4 Residuals of fits to Voigt distributions using the Humlíček approximation with a constant Gaussian component of 300 MHz FWHM. The Lorentzian component widths (FWHM) in the simulated Voigt shape are indicated in the plots. As annotated in the figure panels, the upper set of residuals are for a fitting routine that allowed the Gauss (Doppler) width parameter to float and the lower set are for fixed Gauss width. Also, as annotated in the "floating Gauss" panel the amplitude scale of the residual with a Lorentz width of 300 MHz has been increased by a factor of ten.

All of the simulated Voigt line shapes were scaled to the same total signal deflection so that comparison of the residual in the line shape analysis would be easier to interpret both visually and statistically; the scale chosen set the first derivative line shape lobe-peak amplitudes to ±5.74, which is consistent with the experimental voltages encountered in wavelength modulation spectroscopy using lock-in amplifiers. The worst case relative error of roughly 0.01% occurred with fixed Gauss widths, which agrees with the stated accuracy in the Igor documentation. The few points in the residual plot of the upper panel of Figure 5.4 for a floating Gauss fit parameter that appear as strong discontinuities are likely to be artifacts of the approximate analytic model for the Voigt line shape.

Table 5.2 Fits to simulated Voigt line shapes using the Humlíček analytic approximation. The Lorentz width ($\Delta\nu_L$) for each of the simulated (convolved) line shapes is given in the first column; the Doppler width ($\Delta\nu_G$) in the simulated line shapes was set to 0.01 cm$^{-1}$. The FWHM Voigt width ($\Delta\nu_V$) for the simulation is given in the second column. The units in the first five columns are MHz. The uncertainties of these fit results are given parenthesis in units of the last significant figure. The Voigt widths ($\Delta\nu_V$) for the nonlinear fit line shape are calculated using equation 5.3; see the text for more details. The last column gives the relative error $\sigma$ as the ratio of the fit residual standard deviation and the maximum signal deflection (2 × 5.74) and increased by a factor of 10$^4$. The first three rows are for a floating Gauss width parameter and the last three are for fixed Gauss width.

| Simulation | | Nonlinear Fit | | | |
|---|---|---|---|---|---|
| $\Delta\nu_L$ | $\Delta\nu_V$ | $\Delta\nu_G$ | $\Delta\nu_L$ | $\Delta\nu_V$ | $\sigma \times 10^4$ |
| 3 | 301.6 | 300.657(13) | 3.067(27) | 302.19 | 0.28 |
| 30 | 316.3 | 300.690(3) | 30.01230(3) | 316.07 | 0.23 |
| 300 | 491.3 | 300.690(4) | 300.006(5) | 486.03 | 0.057 |
| 3 | 301.6 | 300 (fixed) | 4.449(27) | 302.23 | 1.1 |
| 30 | 316.3 | 300 (fixed) | 31.417(27) | 316.12 | 1.1 |
| 300 | 491.3 | 300 (fixed) | 301.005(21) | 486.14 | 0.81 |

In summary, the nonlinear fitting routine accuracy surpasses the resolution of the spectra recorded during this project for both laser systems. In the case of the Philips laser diode system the limiting factor appears to be the stability of the laser. However, the limiting factor for the New Focus laser diode system appears to be the 12-bit data collection hardware; the use of 16-bit data acquisition hardware with this laser diode system might reach the limits of the nonlinear regression accuracy for the approximate Voigt profile as given by the Humlíček algorithm.

## 5.6 Wavelength Modulation and De-Modulation

It is fairly well established that the signal from a wavelength-modulated absorption experiment is linearly proportional to the modulation depth in the limit that the modulation depths are much smaller than the chromophore lines being studied [Demtröder]. It is thus important to determine that the modulation depth is not affecting the observed line shapes in a manner that is nonlinear, and thus that they are correctly described by a Voigt distribution. Simulations of the Voigt line shape for a monochromatic radiation source were first computed in "Mathematica" using (what that program refers to as) interpolation functions; these are the same line shapes used in the previous section of this chapter. The recorded wavelength-modulated line shapes were then numerically computed using previously derived equations that model the output signal obtained from a lock-in amplifier with its modulation frequency set equal to that of the wavelength-modulated radiation source [Arndt; Silver].

The resulting simulations of wavelength-modulated line shapes obtained for a series of wavelength modulation depths (from zero to 300 MHz); the homogeneous (Lorentz) widths (FWHM) were 3, 30, and 300 MHz; and the Gauss (Doppler) width was fixed width at 300 MHz. These simulated wavelength modulation spectra were then subjected to the nonlinear fitting routines in Igor using a Voigt line shape model.

The results of the nonlinear fitting of the simulated wavelength-modulated spectra are presented in Table 5.3 for a range of Lorentz widths (FWHM) and wavelength modulation depths that are typically encountered in an experimental setting. It was estimated that keeping the modulation depth within about a factor of two of 30 MHz and allowing the Doppler width to float are the optimal settings for this spectroscopic method under these experimental conditions. The parameter values in the first four columns of Table 5.3 and the first two columns of Table 5.4 were converted from units of $cm^{-1}$ to MHz using the approximate relationship 300 MHz per $cm^{-1}$ and so these are systematically too large by about 0.07%, as compared to using a more accurate conversion factor.

Table 5.3 Wavelength-modulated simulations and nonlinear regression results from Igor using the Humlíček approximation. The units of the first four columns are MHz. The uncertainty of the Gauss and Lorentz fit parameters is given parenthesis in units of the last significant figure. The relative error $\sigma$ reported in the last column is calculated in the same manner as the last column of Table 5.2. The first nine rows are for a floating Gauss width parameter and the last nine are for fixed Gauss width.

| Simulation | | Nonlinear Fit | | |
|---|---|---|---|---|
| $\Delta\nu_L$ | modulation depth | $\Delta\nu_G$ | $\Delta\nu_L$ | $\sigma \times 10^4$ |
| 3 | 3 | 300.003(12) | 3.043(27) | 0.28 |
| 3 | 30 | 302.070(12) | 3.024(27) | 0.27 |
| 3 | 300 | 515(14) | 0(30) | 226 |
| 30 | 3 | 300.039(10) | 29.984(22) | 0.22 |
| 30 | 30 | 302.109(11) | 29.958(22) | 0.23 |
| 30 | 300 | 524(12) | 0(26) | 198 |
| 300 | 3 | 300.030(1) | 299.995(2) | 0.021 |
| 300 | 30 | 302.088(1) | 299.989(2) | 0.023 |
| 300 | 300 | 545(5) | 200(9) | 66 |
| 3 | 3 | 300 (fixed) | 3.050(7) | 0.28 |
| 3 | 30 | 300 (fixed) | 7.367(81) | 3.4 |
| 3 | 300 | 300 (fixed) | 383(12) | 432 |
| 30 | 3 | 300 (fixed) | 30.063(6) | 0.23 |
| 30 | 30 | 300 (fixed) | 34.232(81) | 3.4 |
| 30 | 300 | 300 (fixed) | 398(12) | 409 |
| 300 | 3 | 300 (fixed) | 300.036(1) | 0.040 |
| 300 | 30 | 300 (fixed) | 303.015(63) | 2.5 |
| 300 | 300 | 300 (fixed) | 571(7) | 254 |

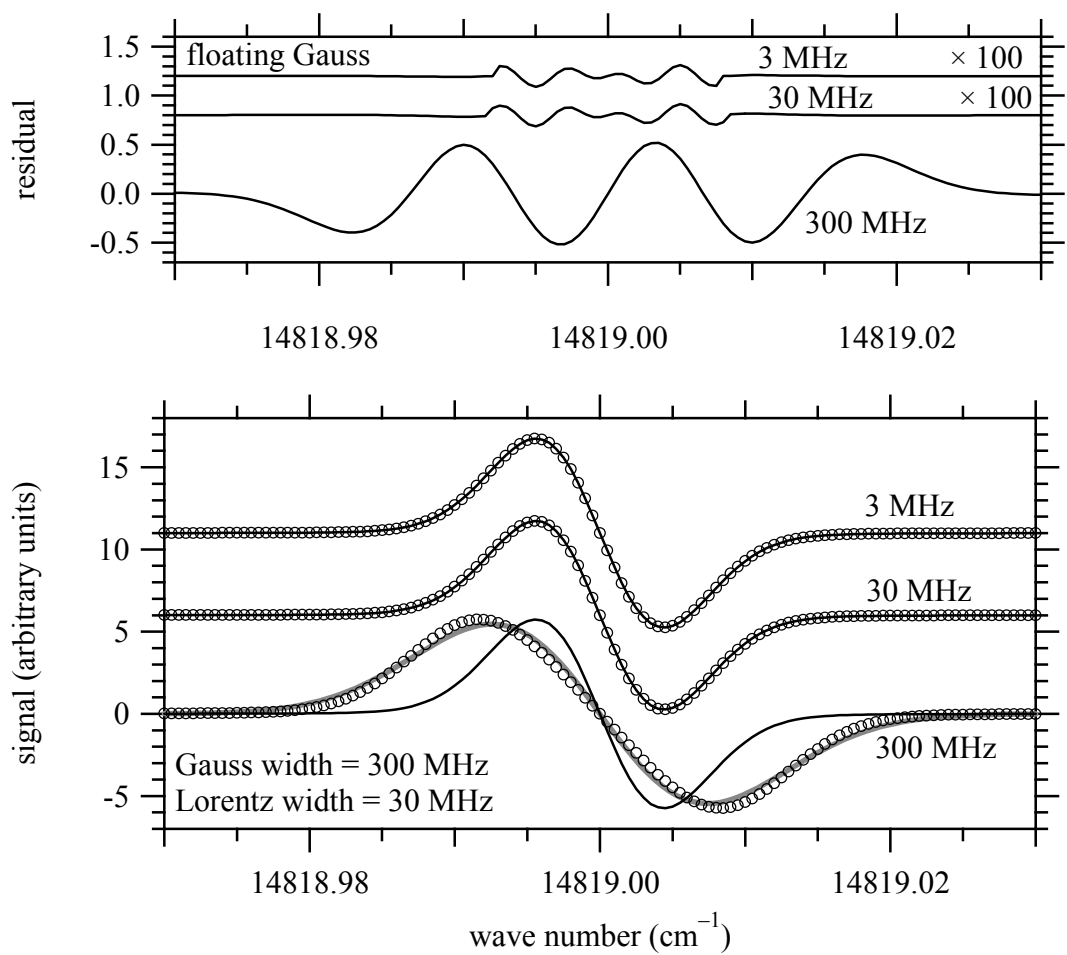

Figure 5.5 Simulations of wavelength-modulated phase-sensitive detection (i.e. when using a lock-in amplifier) for idealized line shape in the absence of nuclear hyperfine structure for three different modulation depths. The Gauss (Doppler) and Lorentz simulation widths are 300 MHz and 30 MHz, respectively. The Doppler width was allowed to float in these fits. The annotations on the right side of plot region are the modulation depths. Lower set of plots: The solid line is the Voigt profile found by convolving the Gaussian and Lorentzian components in Mathematica at a resolution (step size) of 15 MHz. The circles trace the line shape of the de-modulated spectrum obtained with a lock-in amplifier. The best fit line shape has been included as a grey line for the simulation at a modulation depth of 300 MHz. Top set of plots: the residuals for fits of the simulated signal behind a lock-in amplifier to the pure Voigt line shape. The amplitude of the top two residual traces has been increased by a factor of 100.

The deviation of the observed line shape from that of a true first-derivative can be expected to have a systematic error associated with it. For the modulation depths used in this project, the simulations presented in this section suggest that the contribution of such a systematic error to the error of the measured values of the line width and line-center shift is relatively small.

A visual comparison of the simulated and best fit Voigt line shapes are presented in Figure 5.5. The changes in the width and amplitude of a simulated wavelength-modulated line shape detected with a lock-in amplifier are listed in Table 5.4; instead of the more commonly used FWHM, the line width used in this table is the frequency (or wave number) separation of the local maximum and minimum amplitudes (a.k.a. signal levels) in the lobes of the first derivative line shape.

Table 5.4 Simulated line shape width and amplitude for range of modulation depths. The Gaussian (Doppler) and Lorentzian (homogeneous) components are set to 300 MHz and 30 MHz, respectively. The first two columns are in units of MHz. The first column is the wavelength modulation depth. The second column is the frequency separation between the maximum and minimum amplitudes of each lobe (in a first derivative spectrum). The peak-to-peak first derivative amplitudes are scaled to the peak-to-peak first derivative amplitude of an un-modulated Voigt profile, which has its amplitude set equal to one.

| modulation depth | lobe separation | amplitude |
| --- | --- | --- |
| 0 | 270.0 | 1 |
| 3 | 270.0 | 1.004 |
| 30 | 270.0 | 9.917 |
| 60 | 270.0 | 19.10 |
| 90 | 270.0 | 26.91 |
| 120 | 300.0 | 33.10 |
| 150 | 300.0 | 37.42 |
| 180 | 330.0 | 40.24 |
| 210 | 360.0 | 41.72 |
| 240 | 420.0 | 42.28 |
| 270 | 450.0 | 42.32 |
| 300 | 510.0 | 41.93 |

## 5.7 Effects of Neighboring Lines on Shape and Position

Observed line shapes and line-center position are affected by the overlap of neighboring lines. This can lead to systematic changes in the line widths and line-center position (as a function of buffer gas pressure), and thus to systematic errors in the pressure broadening and pressure shift coefficients. The effects of neglecting the overlap of neighboring lines in the single-line fits were examined through another simulation. Individual line profiles were calculated for the region 14,817.95 to 14,819.45 cm$^{-1}$ using the line-center and relative intensity data of Table 4.3, and then added together to form a simulated spectrum for this region. Two such spectra of this region were constructed by simulating the individual lines with a Gaussian line shape at two different widths, 600 and 1400 MHz, a range consistent with the widths of lines observed in this project (e.g. argon has a pressure broadening coefficient of about 8 MHz per torr for pressures of zero and 100 torr). Without loss of generality, the simulations and analysis were simplified by not including the nuclear hyperfine structure in the simulated line shapes. As well, if we assume a pressure coefficient (i.e. line-center shift per unit of pressure) that is the same for each of the simulated lines, then investigating the effect of changes in the line-center position due to line shape distortions does not require inclusion of this term in the simulated line shapes.

The limitations of the single-line nonlinear fitting method were then investigated through two different sets of fits to Features A, B, E, F, H, and I (see Table 4.3) using a Gaussian line shape model that contained parameters for amplitude, line-center, and width (FWHM). The two sets of fits differ in the data points included in the single-line fits; a "narrow-fit" used the low pressure spectrum to determine the fit range and a "wide-fit" used the high pressure spectrum to determine the fit range. The fit range covers the spectral region where the (direct absorption) signal is deflected from the baseline by a few percent relative to the maximum absorption near the line-center.

The percent changes in the pressure broadening (% $\Delta B_P$) and pressure shift (% $\Delta S_P$) coefficients for the two sets of fits are presented in Table 5.5. The percent change in these coefficients was computed from (input value − simulated value) ÷ (input value) × 100. The change in pressure shift coefficient was compared to the case that the input value was 1.25 MHz per torr for all lines in the simulations at pressures of zero and 100 torr (i.e. a typical value for **argon**).

It is apparent in Table 5.5 that the results obtained from the narrow-fit are slightly more reliable than those of the wide-fit. The line shape analysis of the spectra in this project treated the data in the manner described for the narrow-fit. However, the narrow-fit results still indicate that neighboring lines in a diatomic iodine spectrum place a lower limit on the accuracy of pressure broadening and pressure shift coefficients to a few percent when using a single ro-vibronic line shape model for diatomic iodine.

Table 5.5 Simulated changes in pressure broadening ($\Delta B_P$) and pressure shift ($\Delta S_P$) coefficients of diatomic iodine ($I_2$) from a linear superposition of Gaussian line shapes for the spectral region 14,817.95 to 14,818.45 cm$^{-1}$. The percent change in the coefficient is defined in the text.

| Feature | Narrow-fit | | Wide-fit | |
| --- | --- | --- | --- | --- |
| | % $\Delta B_P$ | % $\Delta S_P$ | % $\Delta B_P$ | % $\Delta S_P$ |
| A | −0.18 | −2.64 | −8.04 | −2.16 |
| B | −1.91 | 4.48 | −8.69 | 10.56 |
| E | −0.66 | −0.96 | −0.40 | −2.40 |
| F | 0.16 | −0.48 | 0.21 | −0.72 |
| H | 3.40 | 10.32 | 3.31 | 12.96 |
| I | −0.24 | −2.40 | 0.16 | −3.60 |

## 5.8 Some Fit Results and Comparisons

The purpose of this chapter in large part is to provide some details that might be useful in understanding the limitations of the experimental and modeling methods employed in this project. This section examines (and compares) the fit results obtained from a sub-set of the room temperature (292 K) experimental data. Individual scans (spectra) are fit one at a time using a nonlinear regression algorithm in which the model line shape is a Voigt profile that takes into account the hyperfine structure of diatomic iodine in the high-$J''$ limit. Presentation of the fit results will be separated into two sub-sections; the details of the fitting for a single ro-vibronic transition will first be presented, followed by the details of the fit results for the entire spectral region 14,817.95 to 14,819.45 cm$^{-1}$. The fit results presented are for the experimental situation that both the reference and sample gas cells were at room temperature and the buffer gas was **argon**. The goal of the nonlinear regression is to obtain the Lorentz width (FWHM) and line-center shift as function of pressure, form which the pressure broadening and pressure shift coefficients are calculated, respectively.

It is not obvious from looking only at the uncertainty in the nonlinear line shape fit results (which are then propagated in the usual manner for uncorrelated fit parameters into the uncertainties of the pressure broadening and pressure shift parameters) how to assess their precision and/or accuracy.

In spite of considerable effort we did not discover the means to tame these experiments to better than about 10% precision for the pressure broadening and pressure shift parameters. And a standard against which to compare the absolute accuracy of the results presented in this dissertation does not yet exist; at best, semi-quantitative comparisons are available. Nonetheless, some of the details and results of the fitting procedure will be given below (on the premise that this information could be useful in the future).

## 5.8.1 A Single Diatomic Iodine Feature

This sub-section examines a portion of the nonlinear regression results of the diatomic iodine transition (4, 6) R(129) (a.k.a. Feature A) at 14,818.05 cm$^{-1}$ (see Table 4.3) in the presence of **argon** as the buffer gas.  Data is presented from nonlinear regression analysis in which the Gauss (Doppler) width is set to floating (i.e. measured) and fixed (i.e. predicted).  The spectra were obtained using the Philips laser diode system with the reference and sample gas cells at room temperature (292 K).  Wave number calibration was performed using two diatomic iodine spectral features (Features D and G), as described in Sub-Section 4.6.1.

A typical set of observed line shapes and nonlinear regression analysis (calculated) are plotted together in the lower panel of Figure 5.6, one scan (i.e. spectrum) for each of the seven pressures of the buffer gas (stack plot); the residuals for these fits are presented in the upper panel of Figure 5.6, obtained by subtracting the calculated line shape from the observed.  The Gauss width was set as a floating parameter in the nonlinear regression analysis of these spectra.  While the spectra presented in Figure 5.6 are not the absolute best (in terms of S/N ratio), they are similar to all other spectra analyzed in this project; the fit results of all recorded spectra in this project do not come close to exceeding the resolution offered by the nonlinear regression algorithms in Igor.

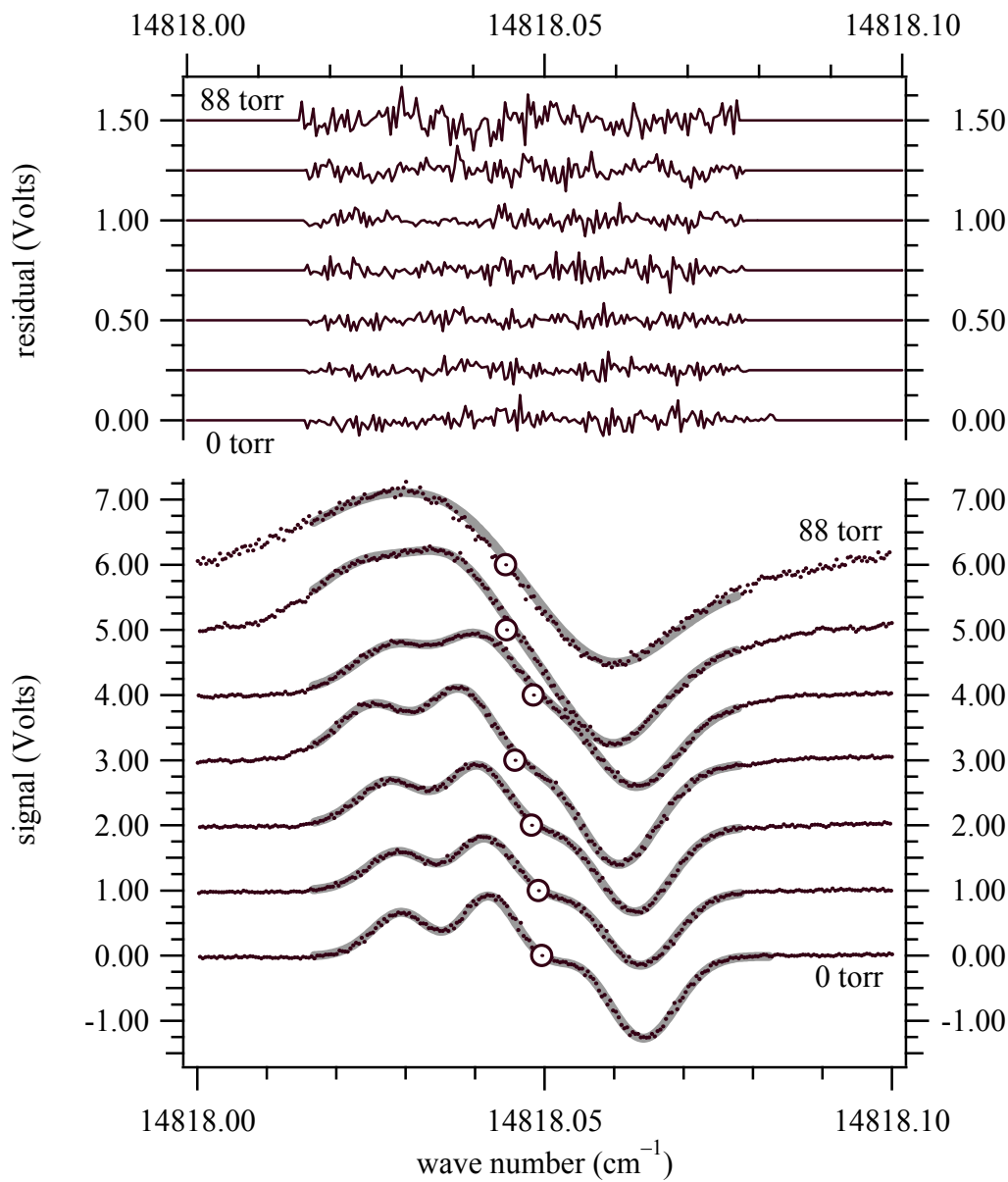

Figure 5.6 A single observed (grey) and calculated (black dots) traces of diatomic iodine ($I_2$) Feature A at 14,818.06 cm$^{-1}$ for each of the seven **argon** pressure (stack plot) in the sample gas cell at room temperature (292 K). The pressure in the sample gas cell are ordered the same in both the residual (upper) and signal plots (lower); the pressures are given in Table 5.6. The line-center at each pressure is marked by a dot inside the circle. The trace and residual plots have been vertically offset for clarity. The Gauss width was set to floating.

Figure 5.6 and Tables 5.6 and 5.7 suggest that the quality of the recorded spectra (i.e. observed data) and the quality of the fit of a model line shape to these spectra are sufficient to provide modestly accurate pressure broadening and pressure shift coefficients. The implementation of more rigorous calibration methods, more complete nuclear hyperfine models for the theoretical line shape, and/or the recording of (Doppler-limited) spectra with a larger signal-to-noise ratio (S/N) (e.g. when using the New Focus laser diode system) do not appear to offer an overtly obvious significant improvement in the quality (accuracy and/or precision) of the pressure broadening and pressure shift coefficients.

Figure 5.6 shows that neglect of the line-center fit results for the reference trace channel can diminish the accuracy of the observed trend in the line-center shift, as seen in the reversal of the line-center position for the third largest pressure (29 torr) in the sequence. In Figure 5.7, the line-center and line-widths of the reference gas cell obtained by subtracted from the corresponding quantity for the sample gas cell are shown in the hash marks just to the right at each buffer gas pressure; calculating the relative difference in line-center position between the reference and sample gas cells (a.k.a. internally-referenced) in each recorded spectrum has a profound effect, tacitly interpreted as an improvement on the accuracy and precision of the line-center shift results and thus the pressure shift coefficient. However, this method of calculating the relative change in a pressure-dependent parameter did not appear to affect the pressure broadening coefficient; the value of this coefficient obtained directly from the Lorentz widths in the sample gas cell are statistically equivalent to that obtained from the difference in the line widths between the traces in the sample and reference gas cell channels. It is perhaps worth mentioning that a relatively constant additional Lorentz width that is present at all pressures, such as might be introduced by the line width of the laser or the constant pressure of diatomic iodine, is by definition not dependent on pressure and thus will not contribute to the pressure broadening coefficient.

Table 5.6 Nonlinear regression fit results for the spectra of diatomic iodine ($I_2$) Feature A presented in Figure 5.6; Gauss width (FWHM), Lorentz width (FWHM), line-center position ($\nu_0$), and nuclear electric quadrupole coupling constant ($\Delta eQq$). The $\Delta eQq$ values were obtained from the reference gas cell and held fixed in the nonlinear regression analysis of the sample gas cell line shape. The uncertainties of these fit results are given in parenthesis in units of the last significant figure. The scan direction was from low to high wave number (a.k.a. "up scan").

| Pressure (torr) | Gauss (MHz) | Lorentz (MHz) | $\nu_0$ (cm$^{-1}$) | $\Delta eQq$ (MHz) |
|---|---|---|---|---|
| 88.47 | 404(52) | 559(55) | 14818.0444(0.6) | 1939(4) |
| 50.47 | 372(14) | 353(18) | 14818.0445(0.3) | 1956(4) |
| 28.87 | 351(8) | 195(12) | 14818.0484(0.3) | 1921(4) |
| 16.53 | 352(2) | 115(0.1) | 14818.0457(0.3) | 1948(4) |
| 9.53 | 368(5) | 26(11) | 14818.0481(0.2) | 1936(4) |
| 5.47 | 351(7) | 32(13) | 14818.0491(0.3) | 1918(4) |
| 0.18 | 344(6) | 0(12) | 14818.0496(0.3) | 1909(4) |

Table 5.7 Relative error in observed and model diatomic iodine ($I_2$) line shape. The spectral and nonlinear regression results are presented in Figure 5.6. The maximum signal ("max signal") is the difference of the maximum and minimum signal deflections of the calculated line shape at each pressure of the buffer gas. The uncertainty in the fit is taken as the standard deviation in the residual $\sigma_{residual}$. The relative error is given by the ratio $\sigma_{residual} \div$ (max signal), and the signal-to-noise ratio ("S/N") is taken as the reciprocal of this quantity.

| pressure (torr) | max signal | $\sigma_{residual}$ | error | S/N |
|---|---|---|---|---|
| 88.47 | 2.61 | 0.053 | 0.02 | 49 |
| 50.47 | 2.96 | 0.037 | 0.012 | 80 |
| 28.87 | 2.36 | 0.027 | 0.011 | 87 |
| 16.53 | 2.71 | 0.033 | 0.012 | 82 |
| 9.53 | 2.25 | 0.021 | 0.0093 | 107 |
| 5.47 | 1.95 | 0.027 | 0.014 | 72 |
| 0.18 | 2.19 | 0.032 | 0.015 | 68 |

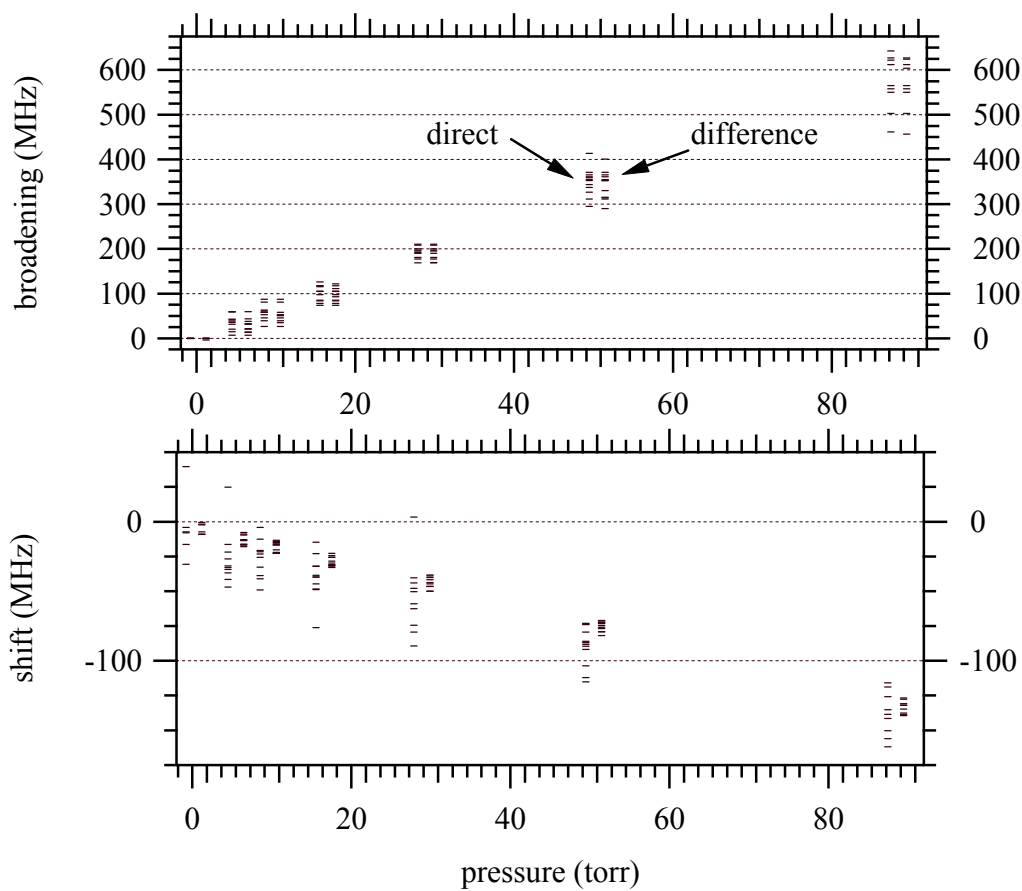

Figure 5.7 Comparison of sample channel fit results (a.k.a. direct fit results) to the difference of the fit results between the sample and reference (gas cell) channels for the line width (top plot) and line-center shift (bottom plot) for diatomic iodine ($I_2$) Feature A with **argon** as the buffer gas at room temperature (292 K). The comparison is made for each pressure of the buffer gas with the direct fit results offset by −1 torr and the difference in fit results offset by 1 torr from the actual (measured) experimental pressures, as indicated in the annotation in the top panel. The direct line-center fit results in the lower panel are offset so that the average of the zero pressure data coincided with the average of the difference fit results. The data presented in these plots was obtained with the Philips laser diode spectrometer with the Gauss width set to floating, and is comprised of 68 scans from low to high wave number ("up scans"). The Gauss width was set to floating. The data used to construct these plots is contained in Tables 5.8 and 5.9.

Tables 5.8 through 5.11 lists the average value of the Lorentz width and the line-center positions of diatomic iodine spectral Feature A, and their statistical uncertainties (i.e. standard deviations) for each of the seven **argon** pressures investigated (columns labeled "Pressure"). Scanning (i.e. tuning) the laser in the direction of increasing wave number is labeled as an "up scan" and scanning the laser in the direction of decreasing wave number is defined as a "down scan". There were 68 scans (i.e. spectra) in each of the two possible scan directions; the numbers of scans at each pressure are listed in the columns labeled "scans".

Table 5.8 Average Lorentz widths (FWHM) of diatomic iodine ($I_2$) Feature A for the reference and sample gas cell channels with **argon** as the buffer gas ("Pressure") at room temperature (292 K); the Gauss width was set to floating in these fits. The uncertainties of these fit results are given in parenthesis in units of the last significant figure. The Lorentz widths tabulated in the last four columns are in units of MHz. The number of spectra used in determining the averages and their standard deviations are given in the "scan" column. Scanning the laser in the direction of increasing wave number is labeled by "up scan" and scanning in the direction of decreasing wave number is labeled by "down scan".

| | | up scan | | down scan | |
|---|---|---|---|---|---|
| Pressure (torr) | scans | reference | sample | reference | sample |
| 0.18 | 6 | 0.6(14) | 0(0) | 0.003(84) | 0.04(97) |
| 5.47 | 10 | 6.9(86) | 35(17) | 3.3(62) | 24(11) |
| 9.53 | 10 | 4.7(83) | 57(18) | 2.6(39) | 50(10) |
| 16.53 | 11 | 2.9(67) | 100(18) | 0.7(18) | 97(12) |
| 28.87 | 10 | 2.2(68) | 192(14) | 6.1(54) | 188(15) |
| 50.47 | 12 | 6.0(77) | 349(30) | 3.7(50) | 327(21) |
| 88.47 | 9 | 4.6(80) | 571(61) | 5.1(55) | 500(62) |

The results listed in Tables 5.8 and 5.9 were calculated from fit results that had the Gauss width set to floating. The results listed in Tables 5.10 and 5.11 were calculated from fit results that had the Gauss width set to the theoretically predicted fixed value of 341 MHz for diatomic iodine at 292 K.

Table 5.9 Average line-center shift of diatomic iodine ($I_2$) Feature A for sample gas cell with **argon** as the buffer gas ("Pressure") at room temperature (292 K); the Gauss width was set to floating in these fits. The "direct" column lists the line-center fit results for the sample channel alone, translated so that the zero pressure line-center position lines up with that of the "difference" line-center at this pressure. The "difference" column lists the line-center shift calculated by subtracting the fit results of the reference channel line-center from those of the sample channel for each scan. The line-center shifts tabulated in the last four columns are in units of MHz. The uncertainties of these fit results are given in parenthesis in units of the last significant figure. The number of spectra used in determining the averages and their standard deviations are given in the "scan" column. Scanning the laser in the direction of increasing wave number is labeled by "up scan" and scanning in the direction of decreasing wave number is labeled by "down scan".

| | | up scan | | down scan | |
|---|---|---|---|---|---|
| Pressure (torr) | scans | direct | difference | direct | difference |
| 0.18 | 6 | −5.0(236) | −5.0(38) | 7.3(164) | 7.3(30) |
| 5.47 | 10 | −27(20) | −13(4) | −5.6(183) | −1.1(36) |
| 9.53 | 10 | −27(14) | −18(4) | −8.0(98) | −5.8(44) |
| 16.53 | 11 | −40(16) | −29(4) | −21(19) | −16(4) |
| 28.87 | 10 | −55(26) | −44(4) | −34(13) | −30(4) |
| 50.47 | 12 | −90(14) | −75(4) | −70(10) | −59(4) |
| 88.47 | 9 | −139(16) | −133(5) | −117(11) | −115(5) |

Tables 5.8 and 5.10 list the average Lorentz widths obtained from both the reference and sample channel spectra. In Table 5.8, the zero width fit results at 0.18 torr are unusual, rarely occurring during the course of this project; that three such zero widths were obtained for Feature A is quite anomalous, but reproducible in the case of this set of spectra.

Table 5.10 Average Lorentz widths (FWHM) of diatomic iodine ($I_2$) Feature A for the reference and sample gas cell channels with **argon** as the buffer gas ("Pressure") at room temperature (292 K); the Gauss width was held fixed at 341 MHz in these fits. The uncertainties of these fit results are given in parenthesis in units of the last significant figure. The Lorentz widths tabulated in the last four columns are in units of MHz. The number of spectra used in determining the averages and their standard deviations are given in the "scan" column. Scanning the laser in the direction of increasing wave number is labeled by "up scan" and scanning in the direction of decreasing wave number is labeled by "down scan".

| Pressure (torr) | scans | up scan | | down scan | |
|---|---|---|---|---|---|
| | | reference | sample | reference | sample |
| 0.18 | 6 | 20(6) | 16(7) | 0(0) | 0.0003(8) |
| 5.47 | 10 | 21(3) | 55(3) | 0.1(4) | 28(5) |
| 9.53 | 10 | 21(4) | 84(5) | 0.01(4) | 54(4) |
| 16.53 | 11 | 23(4) | 133(6) | 0(0) | 102(4) |
| 28.87 | 10 | 20(7) | 217(6) | 0.4(11) | 191(5) |
| 50.47 | 12 | 23(3) | 373(10) | 0(0) | 333(8) |
| 88.47 | 9 | 20(4) | 623(21) | 0.4(9) | 576(21) |

Tables 5.9 and 5.11 list the average values of the absolute line-center position of the sample gas cell channel (direct) and the relative line-center position obtained at each pressure by subtracting the reference gas cell line-center from the sample gas cell line-center (difference) are both reasonably well (statistically) determined, with the latter being a bit better than the former. The line-center shift coefficients calculated in this project used the difference in line-center position between the reference and sample gas cell channels.

Table 5.11 Average line-center shift of diatomic iodine ($I_2$) Feature A for the sample gas cell with **argon** as the buffer gas ("Pressure") at room temperature (292 K); the Gauss width was held fixed at 341 MHz in these fits. The "direct" column lists the line-center fit results for the sample channel alone, translated so that the zero pressure line-center position lines up with that of the "difference" line-center at this pressure. The "difference" column lists the line-center shift calculated by subtracting the fit results of the reference channel line-center from those of the sample channel for each scan. The line-center shifts tabulated in the last four columns are in units of MHz. The uncertainties of these fit results are given in parenthesis in units of the last significant figure. The number of spectra used in determining the averages and their standard deviations are given in the "scan" column. Scanning the laser in the direction of increasing wave number is labeled by "up scan" and scanning in the direction of decreasing wave number is labeled by "down scan".

| Pressure (torr) | scans | up scan | | down scan | |
| --- | --- | --- | --- | --- | --- |
| | | direct | difference | direct | difference |
| 0.18 | 6 | −5.5(241) | −5.5(40) | 7.0(171) | 7.0(40) |
| 5.47 | 10 | −27(20) | −13(3) | −5.4(183) | −1.2(35) |
| 9.53 | 10 | −28(18) | −18(5) | −7.8(98) | −6.0(4) |
| 16.53 | 11 | −40(16) | −28(3) | −21(19) | −16(4) |
| 28.87 | 10 | −55(26) | −44(4) | −34(13) | −30(4) |
| 50.47 | 12 | −90(14) | −75(4) | −69(10) | −58(4) |
| 88.47 | 9 | −138(16) | −133(5) | −118(10) | −116(5) |

## 5.8.2 The Spectral Region 14,817.95 to 14,819.45 cm$^{-1}$

This sub-section takes a more global view by considering all six of the diatomic iodine transitions typically analyzed in the region 14,817.95 to 14,819.45 cm$^{-1}$ (see Table 4.3) in the presence of **argon** as the buffer gas. All of the spectra were obtained with the gas cells at room temperature (292 K). Tables and plots of the averages of some of the important fit parameters are given below. There is a table for each of the following fit parameters: $\Delta eQq$, reference gas cell Lorentz width, reference gas cell Gauss width, sample gas cell Gauss width, pressure broadening coefficient, and pressure shift coefficient. Each table lists the parameter values obtained from nonlinear regression analysis of these six diatomic iodine transitions; each of the five columns of fit results represents a unique analysis; two columns of the fit results for the down (↓) and up (↑) scan spectra obtained with the Philips laser diode system and calibrated using two diatomic iodine features (Features D and G); two columns of the fit results for the down (↓) and up (↑) scan spectra obtained with the Philips laser diode system and calibrated using etalon fringes and most of the diatomic iodine lines in the region. (Scanning (i.e. tuning) the laser in the direction of increasing wave number is referred to as an "up scan" and scanning in the direction of decreasing wave number is labeled by "down scan") The last column (⇓) of these tables lists results for spectra obtained with the New Focus laser diode system and calibrated using etalon fringes and most of the diatomic iodine line in the region; the manufacture specifies that this laser system should only be scanned in the direction of decreasing wave number.

There were 68 spectra recorded with the Philips laser diode system for each scan direction, with roughly equal numbers of scans at each of the seven pressures; see Table 5.7 for details on the buffer gas pressure and the number of scans at each pressure of the buffer gas for this laser system (i.e. spectrometer). There were a total of 24 spectra recorded with the New Focus laser diode system, four scans at each of six pressures; these pressures in units of torr were 5.23, 8.87, 15.49, 27.71, 49.63, and 89.65; unfortunately, zero pressure spectra of the sample gas cell were not recorded. Except for the last two tables, all of the tables and figures below had the Gauss width parameter set to floating in the nonlinear regression analysis. In Tables 5.13 through 5.20 the H* and I* results are for spectra obtained with the New Focus laser system for which the Lorentz width in the reference gas cell channel was constrained to be less than or equal to 15 MHz and the Gauss width was allowed to float; the fit results from this channel were used as the initial values in the subsequent nonlinear regression for the sample gas cell channel.

Table 5.12 Prominent spectral features of diatomic iodine ($I_2$) in the region 14,817.95 to 14,819.45 cm$^{-1}$. The states involved in the transition are given by ($v'$, $v''$) R/P($J''$) where R has $\Delta J = J' - J'' = +1$ and P has $\Delta J = -1$; $v'$ and $v''$ refer to the vibration quantum number of the upper and lower energy levels, respectively. $J''$ is the rotation quantum number of the lower energy level. The wave number position of the line-center at low total pressure is given by $v_0$. The left-hand side corresponds to the feature label order in terms of increasing wave number; the right-hand side corresponds to the ordering based on increasing $J''$

| Feature and $v_0$ ordered (Chapter V) | | | $J''$ ordered (Chapter VI) | | |
|---|---|---|---|---|---|
| ($v'$, $v''$) R/P($J''$) | Feature | $v_0$ (cm$^{-1}$) | ($v'$, $v''$) R/P($J''$) | Feature | $v_0$ (cm$^{-1}$) |
| (4, 6) R(129) | A | 14,818.05 | (5, 7) P(69) | E | 14,818.52 |
| (5, 7) R(76) | B | 14,818.15 | (5, 7) R(75) | I | 14,819.37 |
| (5, 7) P(69) | E | 14,818.52 | (5, 7) R(76) | B | 14,818.15 |
| (4, 6) P(122) | F | 14,818.67 | (4, 6) P(122) | F | 14,818.67 |
| (6, 7) P(133) | H | 14,819.13 | (4, 6) R(129) | A | 14,818.05 |
| (5, 7) R(75) | I | 14,819.37 | (6, 7) P(133) | H | 14,819.13 |

The results listed in the following tables of this sub-section reflect the experimental ordering of the transitions in terms of increasing or decreasing wave number corresponding to the left-hand side of Table 5.12. Presenting the results in this order is primarily intended to maintain a focus on the experimental set-ups. The goal is to at least qualitatively (and perhaps semi-quantitatively) assess the reliability of the results presented in Chapter VI. The right-hand side of Table 5.12 corresponds to the arrangement of pressure broadening, pressure shift, and collision cross-section results presented in the tables of Chapter VI.

In Table 5.13 and Figure 5.8 the dot symbol (•) corresponds to values of $\Delta eQq$ of diatomic iodine calculated from a relatively accurate empirical formula of the nuclear hyperfine structure derived from various literature values obtained from higher precision spectroscopic techniques [Knöckel]. Table 5.13 indicates that a more careful wave number calibration leads to more precise (and perhaps accurate) values of the $\Delta eQq$ parameter.

Table 5.13 Average nuclear quadrupole coupling energy $\Delta eQq$ in the hyperfine structure of diatomic iodine ($I_2$) in the reference gas cell at room temperature (292 K); the Gauss width was set to floating. The fit results are given in units of MHz. The uncertainties of these fit results are given in parenthesis in units of the last significant figure. The uncertainty of the values calculated from an empirical formula (•) are 0.01 MHz. See the text of this sub-section for a description of the column headings.

| Feature | • | ↓ | ↑ | ↓ | ↑ | ⇓ |
|---------|---------|-----------|-----------|---------|---------|---------|
| A | 1957.68 | 1870(10) | 1947(13) | 1955(5) | 1949(5) | 1947(2) |
| B | 1955.67 | 1883(12) | 1929(13) | 1965(5) | 1963(6) | 1950(2) |
| E | 1955.72 | 1923(9) | 1934(9) | 1967(5) | 1956(5) | 1953(2) |
| F | 1957.76 | 1951(12) | 1946(13) | 1969(6) | 1975(6) | 1965(2) |
| H | 1953.76 | 2013(14) | 1970(13) | 1954(5) | 1942(5) | 1956(6) |
| H* | -- | -- | -- | -- | -- | 2004(6) |
| I | 1955.68 | 2083(10) | 1988(8) | 1937(5) | 1941(5) | 1953(2) |
| I* | -- | -- | -- | -- | -- | 1969(2) |

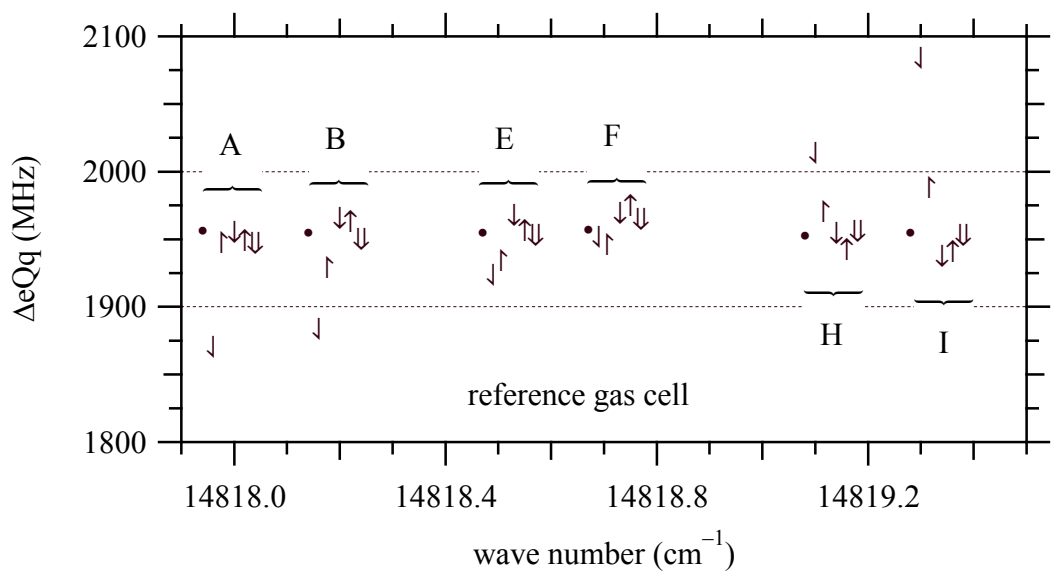

Figure 5.8 Comparison of the average nuclear quadrupole coupling energy $\Delta eQq$ in the hyperfine structure of diatomic iodine ($I_2$) in the reference gas cell at room temperature (292 K) obtained from both laser systems and using both calibration methods. The pressure in the reference gas cell is approximately 0.18 torr and considered to be pure diatomic iodine. The Gauss width was set to floating. The data used to construct these plots is contained in Tables 5.8 and 5.9. Feature labels are indicated in the figure.

Tables 5.14, 5.15, and 5.16 further illustrate the existence of patterns in the nonlinear regression results that depend on the wave number region studied. (See Appendix C for complete tabulation of average of all Lorentz widths at each buffer gas pressure for all the buffer gases investigated in this project.)

Table 5.14 Average Lorentz widths (FWHM) of diatomic iodine ($I_2$) in the reference gas cell at room temperature (292 K); the Gauss width was set to floating in these fits. The fit results are given in units of MHz. The uncertainties of these fit results are given in parenthesis in units of the last significant figure. See the text of this sub-section for a description of the column headings.

| Feature | ↓ | ↑ | ↓ | ↑ | ⇓ |
|---|---|---|---|---|---|
| A | 3(4) | 4(7) | 10(2) | 11(2) | 0.0010(5) |
| B | 10(9) | 12(11) | 16(3) | 22(3) | 0.6(2) |
| E | 17(8) | 19(12) | 23(3) | 23(2) | 3.5(8) |
| F | 27(14) | 22(11) | 24(2) | 31(3) | 31(1) |
| H | 46(27) | 36(24) | 45(4) | 43(4) | 342(12) |
| H* | -- | -- | -- | -- | 14.9(0.5) |
| I | 88(15) | 53(10) | 69(3) | 48(3) | 145(4) |
| I* | -- | -- | -- | -- | 15.0(0) |

Table 5.15 Average Gauss widths (FWHM) of diatomic iodine ($I_2$) in the reference gas cell at room temperature (292 K); the Gauss width was set to floating in these fits. The fit results are given in units of MHz. The uncertainties of these fit results are given in parenthesis in units of the last significant figure. See the text of this sub-section for a description of the column headings.

| Feature | ↓ | ↑ | ↓ | ↑ | ⇓ |
|---|---|---|---|---|---|
| A | 337(3) | 350(4) | 349(1) | 348(1) | 351.0(3) |
| B | 334(6) | 341(6) | 345(2) | 342(2) | 350.9(4) |
| E | 337(4) | 338(6) | 343(1) | 341(1) | 349.8(4) |
| F | 337(7) | 338(7) | 341(2) | 339(2) | 338(1) |
| H | 349(14) | 345(12) | 338(2) | 337(2) | 176(9) |
| H* | -- | -- | -- | -- | 384(2) |
| I | 341(7) | 340(5) | 323(2) | 333(1) | 290(2) |
| I* | -- | -- | -- | -- | 360(0.4) |

In Tables 5.17 through 5.20 the pressure broadening and pressure shift coefficients of diatomic iodine are listed; the buffer gas was **argon** and the gas cells were at room temperature. Figure 5.9 is constructed from the data in Tables 5.17 and 5.18. The pressure broadening and pressure shift coefficients are obtained respectively from a weighted linear least-squares fit of the Lorentz width and line-center shift as functions of pressure. At the bottom of these tables is the average value of the pressure broadening and pressure shift coefficients obtained across the six individual transitions studied in this wave number region; this is the row labeled "average" in these tables. The row labeled "avg of std dev" contains the average of the standard deviations of the pressure broadening and pressure shift coefficients obtained across the six individual transitions studied in this wave number region.

The pressure broadening and pressure shift coefficients for many different buffer gases were investigated with the Philips laser diode system for the same spectral features listed in Tables 5.17 through 5.20. The relative trends in the pressure broadening and pressure shift results depicted in these tables and in Figure 5.9 for spectra obtained with the Philips laser diode system are similar for all buffer gases that were used.

Table 5.16 Average Gauss widths (FWHM) of diatomic iodine ($I_2$) in the sample gas cell with **argon** as the buffer gas at room temperature (292 K); the Gauss width was set to floating in these fits. The fit results are given in units of MHz. The uncertainties of these fit results are given in parenthesis in units of the last significant figure. See the text of this sub-section for a description of the column headings.

| Feature | ↓ | ↿ | ↓ | ↑ | ⇓ |
|---|---|---|---|---|---|
| A | 353(14) | 361(16) | 359(2) | 361(2) | 314(2) |
| B | 338(15) | 340(20) | 347(3) | 343(3) | 345(1) |
| E | 331(11) | 333(14) | 334(2) | 330(2) | 347(2) |
| F | 325(18) | 322(19) | 328(2) | 321(3) | 323(1) |
| H | 369(44) | 357(26) | 355(4) | 350(3) | 371(4) |
| H* | -- | -- | -- | -- | 406(6) |
| I | 324(17) | 323(13) | 312(3) | 318(2) | 336(1) |
| I* | -- | -- | -- | -- | 343(2) |

Table 5.17 Average pressure broadening ($B_p$) coefficients of diatomic iodine ($I_2$) with **argon** as the buffer gas at room temperature (292 K); the Gauss width was set to floating in these fits. The average value of the coefficients for the six features in this spectral region (not including H* and I*) and the average of the uncertainty (i.e. standard deviation) of the average of the coefficient have been included in the last row two rows. The fit results are given in units of MHz. The uncertainties of these fit results are given in parenthesis in units of the last significant figure. See the text of this sub-section for a description of the column headings.

| $J''$ | Feature | ↓ | ↑ | ⇣ | ⇡ | ⇓ | average |
|------|---------|-----|-----|-----|-----|-----|---------|
| 129 | A | 6.21(26) | 6.66(34) | 6.93(5) | 6.94(4) | 8.77(3) | 7.10(14) |
| 76 | B | 7.14(34) | 7.32(35) | 7.52(5) | 7.49(6) | 7.95(2) | 7.48(17) |
| 69 | E | 8.11(25) | 8.16(35) | 8.39(5) | 8.42(4) | 8.31(3) | 8.28(15) |
| 122 | F | 8.71(42) | 8.69(39) | 8.76(5) | 8.73(5) | 8.15(4) | 8.61(20) |
| 133 | H | 6.99(96) | 6.58(71) | 6.93(14) | 7.25(6) | 4.92(19) | 6.70(41) |
|  | H* | -- | -- | -- | -- | 5.74(14) | -- |
| 75 | I | 9.21(36) | 8.54(22) | 8.66(5) | 8.27(3) | 7.95(5) | 8.55(15) |
|  | I* | -- | -- | -- | -- | 8.07(2) | -- |
|  | average | 7.73(114) | 7.667(93) | 7.86(85) | 7.85(72) | 7.83(107) | 7.79(81) |
|  | avg of std dev | 0.43(27) | 0.39(17) | 0.065(36) | 0.049(11) | 0.045(47) | -- |

Table 5.18 Average pressure shift ($S_p$) coefficients of diatomic iodine ($I_2$) with **argon** as the buffer gas at room temperature (292 K); the Gauss width was set to floating in these fits. The average value of the coefficients for the six features in this spectral region (not including H* and I*) and the average of the uncertainty (i.e. standard deviation) of the average of the coefficient have been included in the last row two rows. The fit results are given in units of MHz. The uncertainties of these fit results are given in parenthesis in units of the last significant figure. See the text of this sub-section for a description of the column headings.

| $J''$ | Feature | ↓ | ↑ | ⇣ | ⇡ | ⇓ | average |
|------|---------|-----|-----|-----|-----|-----|---------|
| 129 | A | −1.361(56) | −1.432(55) | −1.403(7) | −1.442(5) | −1.401(3) | −1.408(27) |
| 76 | B | −1.180(43) | −1.203(48) | −1.226(6) | −1.219(6) | −1.184(1) | −1.202(23) |
| 69 | E | −1.179(53) | −1.175(53) | −1.213(8) | −1.191(7) | −1.156(2) | −1.183(26) |
| 122 | F | −1.356(54) | −1.330(54) | −1.374(6) | −1.342(8) | −1.263(3) | −1.333(27) |
| 133 | H | −1.486(86) | −1.559(96) | −1.472(9) | −1.551(11) | −1.270(21) | −1.470(43) |
|  | H* | -- | -- | -- | -- | −1.285(18) | -- |
| 75 | I | −1.399(45) | −1.315(48) | −1.299(6) | −1.273(7) | −1.235(2) | −1.304(23) |
|  | I* | -- | -- | -- | -- | −1.213(2) | -- |
|  | average | −1.33(12) | −1.34(14) | −1.33(10) | −1.34(14) | −1.25(9) | −1.32(11) |
|  | avg of std dev | 0.056(15) | 0.059(19) | 0.0070(13) | 0.0074(21) | 0.0049(64) | -- |

**Argon** was the only buffer gas investigated with the New Focus laser diode system and was done so for three different spectral regions covering a wider range of rotation quantum numbers $J''$ (as compared to the considerably more limited scan region offered by the Philips laser diode system), many of which required fitting two diatomic iodine features simultaneously.

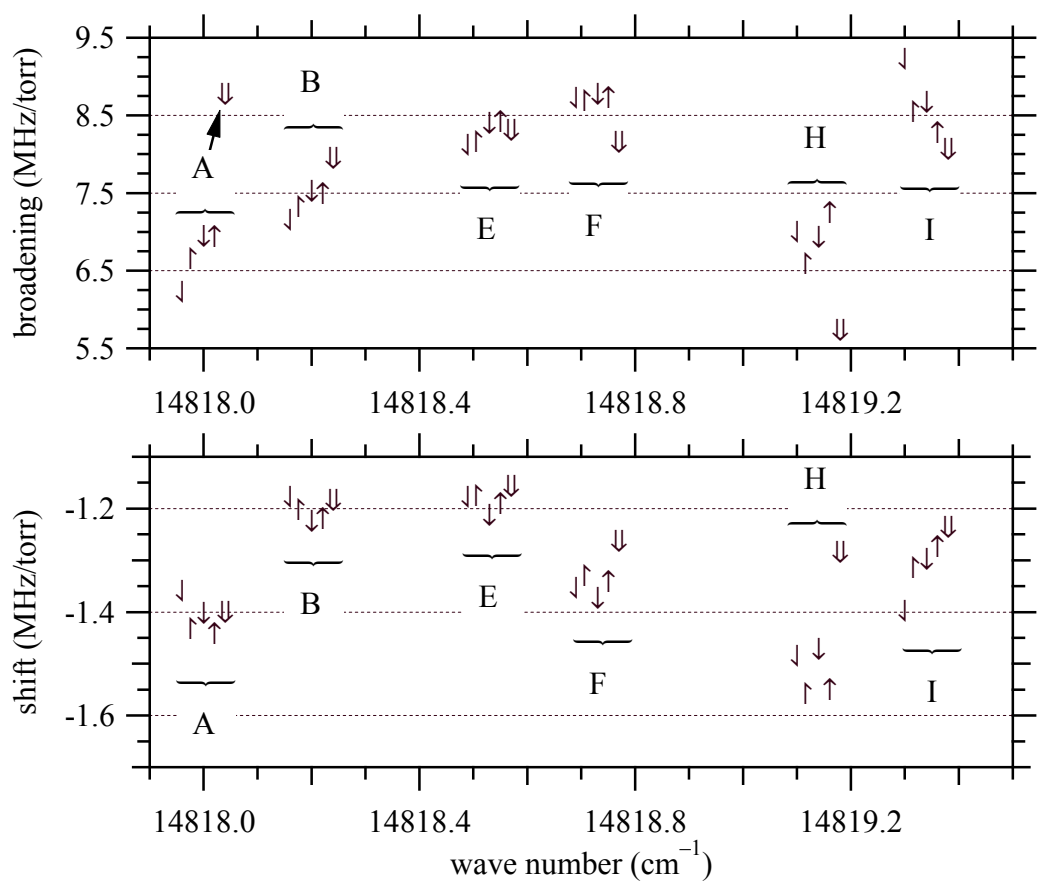

Figure 5.9 Pressure broadening and pressure shift coefficients of diatomic iodine ($I_2$) with **argon** as the buffer gas. The Gauss width was set to floating. The data used to construct these plots is contained in Tables 5.8 and 5.9.

Table 5.19 Average pressure broadening ($B_p$) coefficients of diatomic iodine ($I_2$) with **argon** as the buffer gas at room temperature (292 K); the Gauss width was held fixed at 341 MHz in these fits. The average value of the coefficients for the six features in this spectral region (not including Features H* and I*) and the average of the uncertainty (i.e. standard deviation) of the average of the coefficient have been included in the last row two rows. The fit results are given in units of MHz. The uncertainties of these fit results are given in parenthesis in units of the last significant figure. See the text of this sub-section for a description of the column headings.

| $J''$ | Feature | ↓ | ↑ | ⇃ | ⇂ | ⇓ | average |
|---|---|---|---|---|---|---|---|
| 129 | A | 6.52(9) | 6.96(15) | 6.68(27) | 7.00(25) | 6.52(9) | 7.01(11) |
| 76 | B | 7.20(8) | 7.51(9) | 7.48(24) | 7.53(25) | 7.20(8) | 7.48(11) |
| 69 | E | 7.82(10) | 7.87(10) | 7.97(29) | 8.07(24) | 7.82(10) | 7.94(11) |
| 122 | F | 7.81(16) | 7.94(17) | 7.96(39) | 7.94(32) | 7.81(16) | 7.90(15) |
| 133 | H | 7.07(34) | 6.95(33) | 7.14(65) | 7.11(49) | 7.07(34) | 6.97(20) |
|  | H* | -- | -- | -- | -- | -- | -- |
| 75 | I | 8.59(11) | 8.44(12) | 8.49(19) | 8.34(24) | 8.59(11) | 8.37(9) |
|  | I* | -- | -- | -- | -- | -- | -- |
|  | average | 7.50(72) | 7.61(59) | 7.62(65) | 7.67(54) | 7.50(72) | 7.61(56) |
|  | avg of std dev | 0.15(10) | 0.59(9) | 0.16(9) | 0.30(10) | 0.15(10) | -- |

Table 5.20 Average pressure shift ($S_p$) coefficients of diatomic iodine ($I_2$) with **argon** as the buffer gas at room temperature (292 K); the Gauss width was held fixed at 341 MHz in these fits. The average value of the coefficients for the six features in this spectral region (not including Features H* and I*) and the average of the uncertainty (i.e. standard deviation) of the average of the coefficient have been included in the last row two rows. The fit results are given in units of MHz. The uncertainties of these fit results are given in parenthesis in units of the last significant figure. See the text of this sub-section for a description of the column headings.

| $J''$ | Feature | ↓ | ↑ | ⇃ | ⇂ | ⇓ | average |
|---|---|---|---|---|---|---|---|
| 129 | A | −1.358(59) | −1.433(55) | −1.416(65) | −1.445(52) | −1.358(59) | −1.409(25) |
| 76 | B | −1.172(46) | −1.200(48) | −1.234(51) | −1.219(59) | −1.172(46) | −1.203(23) |
| 69 | E | −1.178(52) | −1.176(53) | −1.207(73) | −1.183(61) | −1.178(52) | −1.181(27) |
| 122 | F | −1.357(54) | −1.331(54) | −1.368(58) | −1.341(75) | −1.357(54) | −1.331(27) |
| 133 | H | −1.483(83) | −1.561(97) | −1.469(86) | −1.536(10) | −1.483(83) | −1.483(32) |
|  | H* | -- | -- | -- | -- | -- | -- |
| 75 | I | −1.401(45) | −1.315(48) | −1.292(54) | −1.274(60) | −1.401(45) | −1.304(23) |
|  | I* | -- | -- | -- | -- | -- | -- |
|  | average | −1.32(12) | −1.34(14) | −1.33(10) | −1.33(14) | −1.32(12) | −1.32(12) |
|  | avg of std dev | 0.056(14) | 0.059(19) | 0.064(13) | 0.069(19) | 0.056(14) | -- |

## 5.8.3 Nonlinear Pressure Shift Coefficient

The well isolated transitions in the spectral region 14,817.95 to 14,819.45 cm$^{-1}$ were fit as individual single ro-vibronic transitions, and the pressure shift as a function of buffer gas pressure appears to be well characterized as linear. Outside of this spectral region, all except one (Feature 14; see Table 6.4) of the features investigated with the New Focus laser system were treated as a composite line-shape composed of two distinct diatomic iodine ro-vibronic transitions. Part of the reason for doing this is that at a pressure of about 10 torr, the absolute value of the relative line-shift would generally be smaller than the for the previous smaller pressure of buffer gas (at about 5 torr). However, the composite line-shape fits did not appear to offer a significant improvement in changing the line-shift values toward better agreement with the usual observation of a linear line-center shift as a function of buffer gas pressure.

Other fit parameters calculated from spectra obtained with the New Focus laser system outside of the spectral region 14,817.95 to 14,819.45 cm$^{-1}$, such as $\Delta eQq$, Gauss width, and Lorentz width, were exhibiting somewhat erratic behavior, so that on the whole pressure broadening and pressure shift results for the low-$J''$ diatomic iodine transitions investigated with the New Focus laser system are not expected to be of sufficient accuracy or precision for use beyond qualitative arguments about chemical dynamics.

## 5.9 Endnotes for Chapter V

[Armstrong]    B. H. Armstrong; "Spectrum line profiles: The Voigt unction", Journal of Quantitative Spectroscopy and Radiative Transfer, 7, 61-88 (1967).

[Arndt]    R. Arndt; "Analytical Line Shapes for Lorentzian Signals Broadened by Modulation", Journal of Applied Physics, 36, 2522 – 2524 (1965).

[Bernath]    P. F. Bernath; *Spectra of Atoms and Molecules*, Chapter 1; Oxford University Press, New York (1995); ISBN 0-19-507598-6.

[Blabla]    V. Špirko and J. Blabla; "Nuclear quadrupole coupling functions of the $^1\Sigma_g^+$ and $^3\Pi_{0u}^+$ states of molecular iodine", Journal of Molecular Spectroscopy 129, 59–71 (1988).

[Borde 1]    C. J. Borde, G. Camy, B. Decomps, J.-P. Descoubes, and J. Vigué; "High precision saturation spectroscopy of molecular iodine-127 with argon lasers at 5145 Å and 5017 Å: I - Main resonances", Journal de Physique, 42, 1393-1411 (1981).

[Borde 2]    F.-L. Hong, J. Ye, L.-S. Ma, S. Picard, C. J. Borde, and J. L. Hall; "Rotation dependence of electric quadrupole hyperfine interaction in the ground state of molecular iodine by high-resolution laser spectroscopy", Journal of the Optical Society of America, Part B, 18, 379-387 (2001).

[Bunker]    G. R. Hanes, J. Lapierre, P. R. Bunker, and K. C. Shotton; "Nuclear hyperfine structure in the electronic spectrum of $^{127}I_2$ by saturated absorption spectroscopy, and comparison with theory", Journal of Molecular Spectroscopy, 39, 506-515 (1971).

[Demtröder]    W. Demtröder; *Laser Spectroscopy: Basic Concepts and Instrumentation*, Third Edition, pages 374-378; Springer-Verlag, Berlin (2003); ISBN 3-540-65225-6.

[Drayson]        S. R. Drayson; "Rapid computation of the Voigt profile", Journal of
                 Quantitative Spectroscopy and Radiative Transfer, 16, 611-614
                 (1976).

[Humlíček]       J. Humlíček; "Optimized computation of the Voigt and complex
                 probability functions", Journal of Quantitative Spectroscopy and
                 Radiative Transfer, 27, 437-444 (1982).

[Kato]           H. Kato et al.; *Doppler-Free High Resolution Spectral Atlas of
                 Iodine Molecule 15,000 to 19,000 $cm^{-1}$*; published by Japan Society
                 for the Promotion of Science (2000); ISBN 4-89114-000-3.

[Knöckel]        B. Bodermann, H. Knöckel and E. Tiemann; "Widely usable
                 interpolation formulae for hyperfine splittings in the $^{127}I_2$ spectrum",
                 European Physical Journal, Part D, 19, 31-44 (2002).

[Kroll]          M. Kroll and K. K. Innes; "Molecular electronic spectroscopy by
                 Fabry-Perot interferometry. Effect of nuclear quadrupole
                 interactions on the line widths of the $B^3\Pi_{0+}$ - $X^1\Sigma_g^+$ transition of the
                 $I_2$ molecule", Journal of Molecular Spectroscopy, 36, 295-309
                 (1970).

[Loudon 1]       R. Loudon; *The Quantum Theory of Light,* Second Edition, Chapter
                 2; Oxford University Press, New York, (1983); ISBN 0-19-851155-
                 8.

[McHale]         J. L. McHale; *Molecular Spectroscopy*, Chapters 3 and 4; Prentice
                 Hall, Upper Saddle River, New Jersey (1999); ISBN 0-13-229063-4.

[Paisner]        J. A. Paisner and R. Wallenstein; "Rotational lifetimes and
                 selfquenching cross sections in the $B^3\Pi_{0u}^+$ state of molecular iodine-
                 127", Journal of Chemical Physics, 61, 4317-4320 (1974).

[Schawlow]       M. D. Levenson and A. L. Schawlow; "Hyperfine Interactions in
                 Molecular Iodine", Physical Review A, 6, 10-20 (1972).

[Schreier]       F. Schreier; "The Voigt and complex error function: A comparison
                 of computational methods", Journal of Quantitative Spectroscopy
                 and Radiative Transfer, 48, 743-762 (1992).

[Silver]         J. A. Silver; "Frequency-modulation spectroscopy for trace species
                 detection: theory and comparison among experimental methods",
                 Applied Optics, 31, 707-717 (1992).

[Tellinghuisen]   J. Tellinghuisen; "Transition strengths in the visible–infrared absorption spectrum of $I_2$", Journal of Chemical Physics, 76, 4736-4744 (1982).

[Whiting]   E. E. Whiting; "An empirical approximation to the Voigt profile", Journal of Quantitative Spectroscopy and Radiative Transfer, 8, 1379-1384 (1968).

[Yoshizawa]   M. Wakasugi, M. Koizumi, T. Horiguchi, H. Sakata, and Y Yoshizawa; "Rotational-Quantum-Number Dependence of Hyperfine Transition Intensity Near the $B(v' = 14)-X(v'' = 1)$ bandhead of $^{127}I_2$"; Journal of the Optical Society of America, Part B, 6, 1660-1664 (1989).

# CHAPTER VI

# RESULTS 1 – COEFFICIENTS AND CROSS-SECTIONS

## 6.1 Overview of Chapter VI

The goal of this project was to collect high-resolution spectral data on the line-shape of ro-vibronic transitions of diatomic iodine ($I_2$) as a function of a buffer gases. Analyses of these recorded line-shapes were then analyzed for their Lorentz width (full width at half maximum height, which has the acronym FWHM) and line-center shift, from which the pressure broadening and pressure shift coefficients ($B_p$ and $S_p$, respectively) were calculated. A hard-sphere (i.e. perfectly elastic) collision cross-section was then computed from the pressure broadening coefficients; this cross-section is based on the "traditional" relationship between the Lorentz width and the reciprocal of the ensemble-average time-intervals between state-changing events of the chromophore and phase-interrupting events of the chromophore-radiation interaction. These events are often referred to as "collisions", a short-hand term for the time-dependent interactions that take place between the chromophore and buffer gas. Deciphering and characterizing the nature of these interactions is a primary goal of what are often referred to as the study of chemical dynamics.

The spectrometers used in this project were built around sources of electro-magnetic radiation (a.k.a. light) of relatively narrow and constant line width near wavelengths of 675 nm (or, in wave number units, approximately 14,800 $cm^{-1}$). (A guess at the estimate of the FWHM line width of the Philips laser diode is 50 MHz, and the manufacturer of the New Focus laser diode specifies this width to be 5 MHz.) Most of the data was obtained with the reference and sample gas cells at room (i.e.

ambient) temperature (292 K).  Conceptually, the buffer gases can be separated into atomic (**He**, **Ne**, **Ar**, **Kr**, **Xe**; respectively, a.k.a. **helium**, **neon**, **argon**, **krypton**, and **xenon**) and molecular (**H₂**, **D₂**, **N₂**, **CO₂**, **N₂O**, **H₂O**, and **air**; respectively, a.k.a. **hydrogen**, **deuterium**, **nitrogen**, **carbon dioxide**, **nitrous oxide**, **water**, and a "known" mixture of several gases with the dominant fractions being roughly 80% **N₂** and 20% **O₂** (a.k.a. oxygen)).

Comparison of the room temperature frequency-domain collision cross-sections from this project with those obtained from a time-domain investigation carried out at a temperature of 388 K [Dantus] prompted the exploration of a few buffer gases (**argon**, **helium**, and **carbon dioxide**) at more than one temperature.  The trends in pressure broadening and pressure shift coefficients for **acetylene** (**C2H2**) [Hardwick 4] as a function of rotation quantum number $J''$ prompted the exploration of a wider range of this quantum number, especially low $J''$.

Table 6.1 Prominent spectral features of diatomic iodine (I₂) in the region 14,817.95 to 14,819.45 cm⁻¹.  The states involved in the transition are given by $(v', v'')$ R/P($J''$) where R has $\Delta J = J' - J'' = +1$ and P has $\Delta J = -1$; $v'$ and $v''$ refer to the vibration quantum number of the upper and lower energy levels, respectively.  $J''$ is the rotation quantum number of the lower energy level.  The wave number position of the line-center at low total pressure is given by $\nu_0$.  The left-hand side corresponds to the feature label order in terms of increasing wave number; the right-hand side corresponds to the ordering based on increasing $J''$

| Feature and $\nu_0$ ordered (Chapter V) | | | $J''$ ordered (Chapter VI) | | |
|---|---|---|---|---|---|
| $(v', v'')$ R/P($J''$) | Feature | $\nu_0$ (cm⁻¹) | $(v', v'')$ R/P($J''$) | Feature | $\nu_0$ (cm⁻¹) |
| (4, 6) R(129) | A | 14,818.05 | (5, 7) P(69) | E | 14,818.52 |
| (5, 7) R(76) | B | 14,818.15 | (5, 7) R(75) | I | 14,819.37 |
| (5, 7) P(69) | E | 14,818.52 | (5, 7) R(76) | B | 14,818.15 |
| (4, 6) P(122) | F | 14,818.67 | (4, 6) P(122) | F | 14,818.67 |
| (6, 7) P(133) | H | 14,819.13 | (4, 6) R(129) | A | 14,818.05 |
| (5, 7) R(75) | I | 14,819.37 | (6, 7) P(133) | H | 14,819.13 |

The tables in this chapter are arranged in order of increasing rotation quantum number of the lower level $J''$; this is done on the premise that if patterns in the pressure broadening and pressure shift coefficients do occur, then it would be likely that they would follow this ordering. For convenience Table 5.12 is reproduced here as Table 6.1; the right-hand side of this table corresponds to the arrangement (i.e. ordering) of results listed in the tables in this chapter.

## 6.2 Pressure Broadening and Pressure Shift Coefficients

The sub-sections that follow (in this section) present pressure broadening and pressure shift coefficient results for diatomic iodine with atomic and molecular buffer gases. The pressure broadening coefficients ($B_p$) were calculated through a weighted linear regression analysis using the full width at half maximum (FWHM) Lorentz width as a function of buffer gas pressure. The pressure shift coefficients ($S_p$) were calculated through a weighted linear regression analysis using the difference in line-center position of the sample and reference gas cell spectra as a function of pressure. Each pressure broadening and pressure shift coefficient is calculated from the Lorentz widths and line-center shifts of roughly ten spectra at each of about seven distinct pressures of the buffer gas; see also Sub-Section 5.8.1 of this dissertation. Each average value of the pressure broadening and pressure shift coefficient calculated from spectra acquired with the Philips laser diode system is the average of the two possible scan directions ("up" in wave number and "down" in wave number) and the standard error was accordingly propagated (using the error propagation method outlined in Section 1.6). (As will be briefly described below, in the case of **argon** as the buffer gas there were considerably more spectral data sets from which to calculate averages.)

Most of these results were derived (i.e. calculated) from spectra acquired with a wavelength-modulation spectrometer built around the Philips laser diode system. And most of the experiments were conducted with both the reference and sample gas cells at room temperature. An extended study of **argon** over a wider range of rotation quantum number $J''$ (at room temperature) was conducted using a wavelength-modulation spectrometer built around the New Focus laser diode.

It is important to also note that nearly all of the calculations of pressure broadening and pressure shift coefficients for data obtained with the Philips laser diode system included the zero pressure measurements (i.e. with the sample gas cell evacuated to 0.18 torr of diatomic iodine). None of these calculations from spectra obtained with the New Focus laser diode system included the zero pressure measurements.

## 6.2.1 Atomic Buffer Gases at Room Temperature (292 K)

In Tables 6.2 and 6.3 the pressure broadening and pressure shift coefficients of diatomic iodine are listed for the atomic buffer gases (**He**, **Ne**, **Ar**, **Kr**, and **Xe**) in which the reference and sample gas cells were at room temperature. The spectra and analysis for **Ne**, **Kr**, and **Xe** were carried out by John Hardwick with the assistance of different undergraduate students and me (the author of this dissertation); this took place during the first two years of my time here at the University of Oregon as a graduate student. The remainder of the investigations performed using the other **noble gas atoms** were carried out by me, the author of this dissertation. The spectrometer used to record these spectra was built around the Philips laser diode system using a parabolic reflector to collimate the laser beam (see Chapter III).

In Tables 6.2 and 6.3 the results for **argon** were calculated from three unique data sets collected by me, using three different spectrometers (see Chapter III); data for **argon** collected by other students was not included in these averages. The averages include permutations of the following choices: floating and fixed Gauss widths (FWHM), up and down scans (i.e. tuning the laser through increasing or decreasing wave numbers, respectively), and the two different wave number calibration methods (two diatomic iodine features or many diatomic iodine features plus etalon fringes with a fringe spacing of ca. 600 MHz).

Table 6.2 Pressure broadening ($B_p$) coefficients (based on FWHM) of diatomic iodine for a given **noble gas atom** as the buffer gas at room temperature (292 K). The average value of the coefficients for the six features in this spectral region and the average of the uncertainty (i.e. standard deviation) of the average of the coefficient have been included in the last two rows. The coefficients are given in units of MHz per torr; the uncertainties of these fit results are given in parenthesis in units of the last significant figure. See the text for further description of the **argon** results.

|  |  | $B_p \equiv$ pressure broadening (MHz / torr) | | | | |
|---|---|---|---|---|---|---|
| $J''$ | Feature | He | Ne | Ar | Kr | Xe |
| 69 | E | 8.37(13) | 6.60(9) | 8.04(21) | 7.17(72) | 6.69(24) |
| 75 | I | 8.68(21) | -- | 8.43(32) | -- | -- |
| 76 | B | 7.87(11) | 6.03(12) | 7.50(22) | 6.54(42) | 6.21(33) |
| 122 | F | 8.21(16) | 6.36(18) | 8.09(44) | 7.11(30) | 6.75(54) |
| 129 | A | 8.22(20) | 5.70(18) | 7.10(64) | 6.60(120) | 6.24(63) |
| 133 | H | 8.68(33) | -- | 7.01(61) | -- | -- |
|  | average | 8.34(31) | 6.17(39) | 7.70(58) | 6.86(33) | 6.47(29) |
|  | avg of std dev | 0.19(8) | 0.14(4) | 0.41(19) | 0.66(40) | 0.44(18) |

One result that stands out with the **noble gas atoms** is what appears to be an unusually small pressure broadening coefficient for **neon** relative to the trend of decreasing pressure broadening coefficient with increasing buffer gas mass, as compared to the other **noble gas atoms**. If the discrepancy for **neon** indicates that the ensemble-average sum of the state-changing and phase-interrupting rates are not increasing as fast (with increasing pressure) as might be expected, then a possible explanation is the unaccounted for existence (or absence) of a channel for dynamical behavior, or perhaps just a differentially observed effect of such a channel.

The data in Tables 6.2 and 6.3 is plotted in Figure 6.1, as are the ratios of the pressure broadening to the pressure shift coefficients for these buffer gases. Most pressure shift coefficients are between 1/4 and 1/8 of the value of the corresponding pressure broadening coefficient. There are real differences among the various buffer gases studied, but these differences are small despite qualitative differences in their multi-pole moments.

Table 6.3 Pressure shift ($S_p$) coefficients of diatomic iodine ($I_2$) for a given **noble gas atom** as the buffer gas at room temperature (292 K). The average value of the coefficients for the six features in this spectral region and the average of the uncertainty (i.e. standard deviation) of the average of the coefficient have been included in the last two rows. The coefficients are given in units of MHz per torr; the uncertainties of these fit results are given in parenthesis in units of the last significant figure. See the text for further description of the **argon** results.

| $J''$ | Feature | $S_p \equiv$ pressure shift (MHz / torr) | | | | |
|---|---|---|---|---|---|---|
| | | He | Ne | Ar | Kr | Xe |
| 69 | E | −0.197(22) | −0.69(3) | −1.18(2) | −1.44(12) | −1.62(15) |
| 75 | I | −0.290(25) | -- | −1.32(6) | -- | -- |
| 76 | B | −0.196(23) | −0.69(3) | −1.20(2) | −1.35(9) | −1.62(9) |
| 122 | F | −0.142(25) | −0.72(6) | −1.32(4) | −1.35(9) | −1.80(9) |
| 129 | A | −0.168(24) | −0.75(3) | −1.39(4) | −1.47(12) | −1.80(12) |
| 133 | H | −0.212(40) | -- | −1.48(8) | -- | -- |
| | average | −0.201(50) | −0.71(3) | −1.32(11) | −1.40(6) | −1.71(10) |
| | avg of std dev | 0.027(7) | 0.038(15) | 0.043(25) | 0.10(2) | 0.11(3) |

In Table 6.4 the pressure broadening and pressure shift coefficients for an extended study of **argon** using the New Focus laser diode system. Having observed first-hand the pronounced pattern of these coefficients as a function of buffer gas for the case that **acetylene** was the chromophore [Hardwick 4], an attempt was made to investigate these coefficients for diatomic iodine as the chromophore. In addition, these results can be compared to those obtained from spectra recorded with the Philips laser diode system.

The results for **argon** (as the buffer gas) as a function of $J''$ of diatomic iodine are suggestive of systematic variations, but far from conclusive. The spectra may be too congested at low values of $J''$ to allow a confident analysis at this time. These results are presented in Table 6.4 and Figure 6.2.

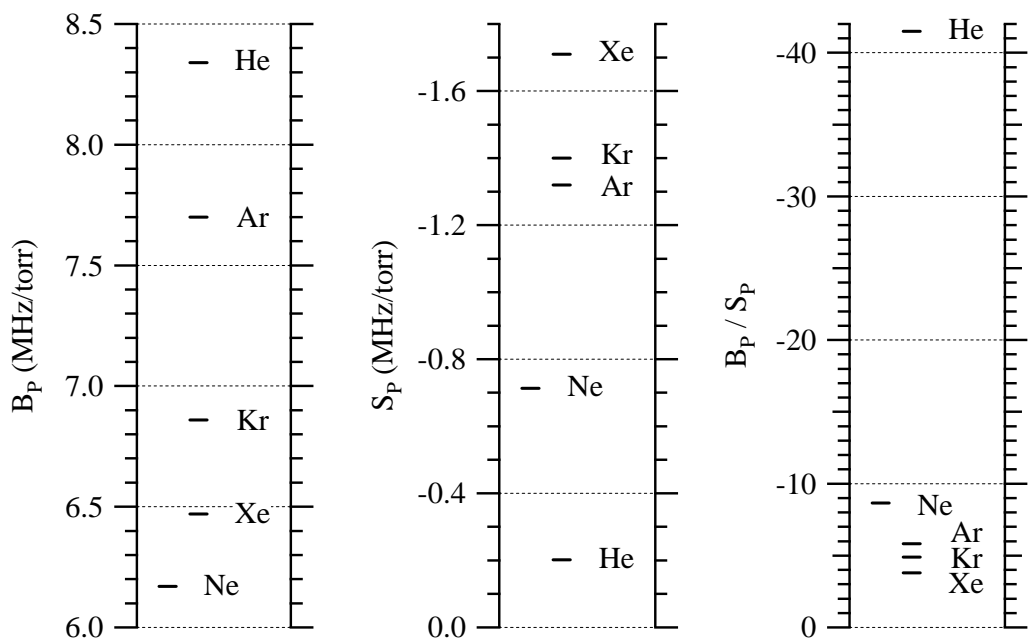

Figure 6.1 Average pressure broadening ($B_p$) and pressure shift ($S_p$) coefficients and their ratios for diatomic iodine ($I_2$) with a given **noble gas atom** as the buffer gas. The marker for **neon** has been displaced to the left in all three plots to highlight its unusual position in the $B_p$ plot. The data in these plots is presented in Tables 6.2 and 6.3.

Table 6.4 Pressure broadening ($B_p$) and pressure shift ($S_p$) coefficients of diatomic iodine ($I_2$) with **argon** as the buffer gas at room temperature (292 K). The pressure broadening coefficients are based on FWHM. The units for the coefficients are MHz per torr and the uncertainty of all fit results are given in units of the last significant figures in parenthesis. The column labeled "details" provides information on the hyperfine model and the fit partner. The hyperfine models are either the high-$J$ (hi-$J$) or full-diagonalization (f-d), and the fit partner refers to the feature label for two-line fits. All results in this table were obtained from spectra acquired with the New Focus diode laser system.

| Feature | states | $\nu_0$ (cm$^{-1}$) | $B_p$ | $S_p$ | $B_p/S_p$ | details |
|---|---|---|---|---|---|---|
| 40 | (4,6) P(6) | 14,950.02 | 6.05(18) | −0.113(5) | −53.57(219) | f-d, 39 |
| 35 | (4,6) P(8) | 14,949.65 | 6.70(8) | −0.526(4) | −12.75(35) | f-d, 34 |
| 21 | (4,6) P(13) | 14,948.46 | 7.78(5) | −0.954(3) | −8.15(17) | f-d, 20 |
| 39 | (4,6) R(13) | 14,949.99 | 9.82(6) | −0.908(4) | −10.81(23) | f-d, 40 |
| 18 | (4,6) P(14) | 14,948.17 | 11.98(9) | −0.305(4) | −39.27(76) | hi-$J$, 17 |
| 34 | (4,6) R(15) | 14,949.62 | 9.17(5) | −0.960(4) | −9.55(20) | f-d, 35 |
| 13 | (4,6) P(16) | 14,947.54 | 7.76(6) | −1.043(4) | −7.44(17) | hi-$J$, 12 |
| 3 | (4,6) P(19) | 14,946.47 | 9.20(2) | −1.142(9) | −8.05(25) | hi-$J$, 2 |
| 20 | (4,6) R(20) | 14,948.40 | 7.79(4) | −0.974(3) | −7.99(16) | f-d, 21 |
| 17 | (4,6) R(21) | 14,948.11 | 8.92(5) | −1.141(4) | −7.82(16) | hi-$J$, 18 |
| 12 | (4,6) R(23) | 14,947.48 | 8.31(4) | −1.211(4) | −6.87(14) | hi-$J$, 13 |
| 2 | (4,6) R(26) | 14,946.40 | 8.19(2) | −1.108(4) | −7.39(17) | hi-$J$, 3 |
| 14 | (6,7) P(57) | 14,947.71 | 8.17(7) | −1.247(4) | −6.55(15) | hi-$J$ |
| 4 | (6,7) P(58) | 14,946.66 | 7.17(3) | −1.166(4) | −6.14(15) | hi-$J$ |
| E | (5,7) P(69) | 14,818.52 | 8.31(3) | −1.156(2) | −7.19(12) | hi-$J$ |
| I | (5,7) R(75) | 14,819.37 | 8.07(2) | −1.235(2) | −6.53(11) | hi-$J$ |
| B | (5,7) R(76) | 14,818.15 | 7.95(2) | −1.184(1) | −6.71(9) | hi-$J$ |
| 5 | (5,6) P(116) | 14,946.72 | 9.16(7) | −0.764(4) | −11.99(27) | hi-$J$, 4 |
| F | (4,6) P(122) | 14,818.67 | 8.15(4) | −1.263(3) | −6.45(12) | hi-$J$ |
| A | (4,6) R(129) | 14,818.05 | 8.77(3) | −1.401(3) | −6.26(12) | hi-$J$ |
| H | (6,7) P(133) | 14,819.13 | 5.74(14) | −1.285(18) | −4.46(27) | hi-$J$ |

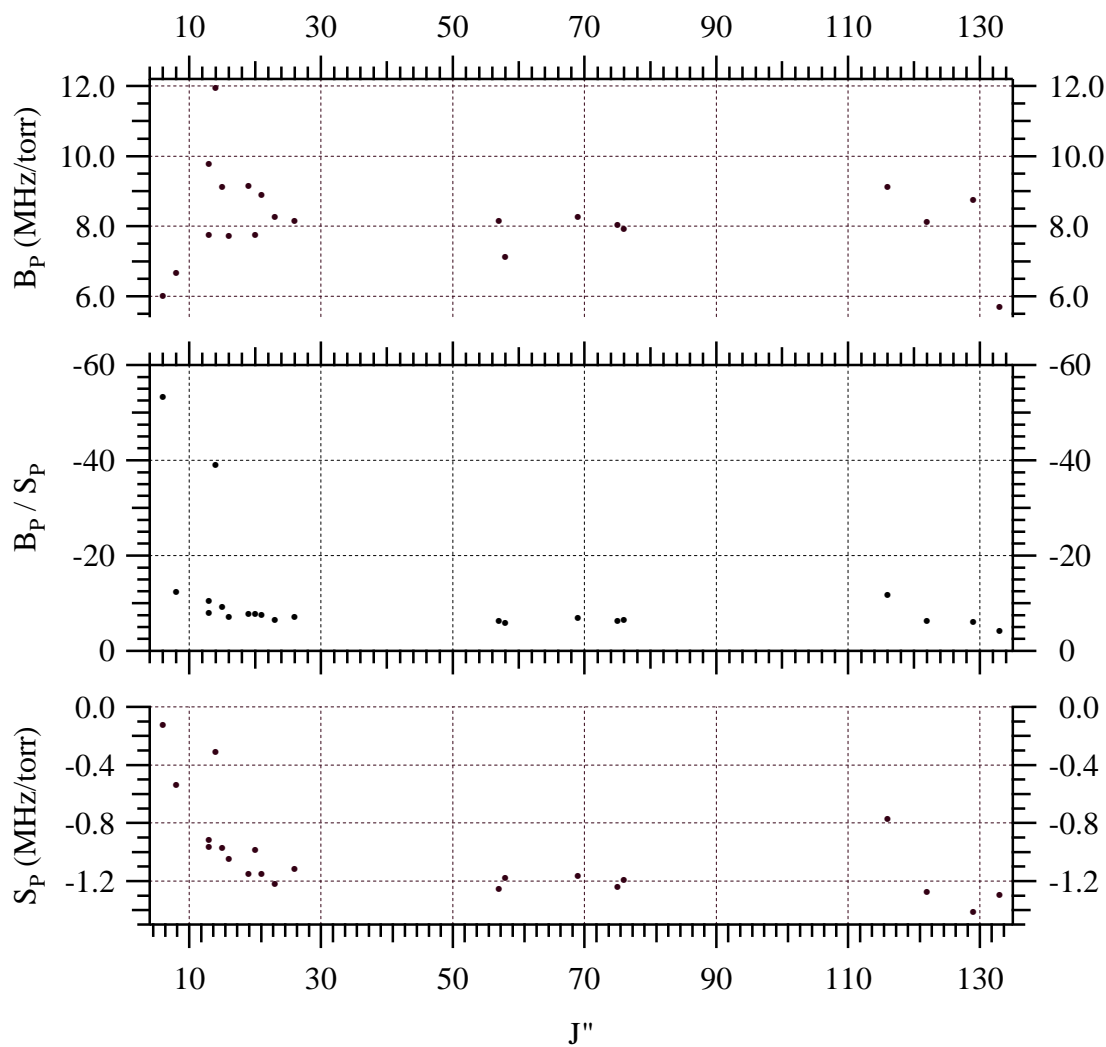

Figure 6.2 Pressure broadening ($B_P$) and pressure shift ($S_P$) coefficients and their ratios for diatomic iodine ($I_2$) (as a function of ground electronic state rotation quantum number) with **argon** as the buffer gas. The data for these plots is presented in Table 6.4.

## 6.2.2 Molecular Buffer Gases at Room Temperature (292 K)

Due to the ease of obtaining samples of **air** and **water** vapor in the sample gas cell, at the outset of this project two lines of diatomic iodine B−X ro-vibronic transitions were studied for pressure broadening and pressure shift coefficients by these two buffer gases [Hardwick 1]. These results are included in Table 6.5 and Figure 6.3 for completeness.

Table 6.5 Pressure broadening ($B_p$) and pressure shift ($S_p$) coefficients and their ratios for diatomic iodine ($I_2$) with **air** and **water** (individually) as the buffer gas for only two transitions at room temperature (292 K). The coefficients are given in units of MHz per torr; the uncertainties of these fit results are given in parenthesis in units of the last significant figure. The pressure broadening coefficients are based on FWHM. These results were previously published. [Hardwick 1]

| | | Air | | | $H_2O$ | | |
|---|---|---|---|---|---|---|---|
| $J''$ | Feature | $B_p$ | $S_p$ | $B_p / S_p$ | $B_p$ | $S_p$ | $B_p / S_p$ |
| 69 | E | 9.05(6) | −1.17(1) | −7.74(8) | 11.6(2) | −1.12(6) | −10.4(6) |
| 122 | F | 8.69(12) | −1.26(3) | −6.90(19) | 12.0(2) | −0.962(66) | −12.5(9) |

A comparison of air (ca. 80% **N₂** and 20% **O₂**) to **nitrogen** (**N₂**) would benefit from also having available the pressure broadening and pressure shift coefficients for **oxygen** (**O₂**); since spectral data was not collected for the case of **oxygen** as the buffer gas during the course of this project, a comparison of **nitrogen** to **air** will not be made at this time. However, the results for **air** can be compared to the coefficient values obtained in an independent investigation at a radiation wave length of 543 nm [Fletcher]; these coefficient values were calculated for room temperature (292 K) conditions as specified by the empirical formula provided in the write-up of that investigation: the coefficient values for room temperature are $B_p = 9.98(17)$ MHz per torr and $S_p = −1.091(75)$ MHz per torr; the uncertainty of these coefficients are given in parenthesis in units of the last significant figure of the coefficient value. It is probably reasonable to consider the agreement between these results and those presented in Table 6.5 as being quite good; as has been alluded to in previous sections of this dissertation, and as can be seen in the large scale trends of the data presented in this chapter, there is considerable fluctuation of these measured values, (ca. 10 to 20%) from individual lines in different experiments and between lines of similar quantum numbers (i.e. similar ro-vibronic states). (See also Section 5.8.)

From a chemical dynamics point of view the choice of molecules to study in subsequent investigations was based largely on molecular properties. **Hydrogen ($H_2$)** and **deuterium ($D_2$)** have nearly identical electronic properties and so allow the effects of a relatively large change in the mass of the buffer gas; **nitrogen ($N_2$)** has a quadrupole moment and is of significant interest in the field of metrology; **carbon dioxide ($CO_2$)** has a quadrupole moment; **nitrous oxide ($N_2O$)** and **water ($H_2O$)** have a dipole moment. (The notion of "having" a quadrupole or dipole moment is in reference to the leading dominant term of the multipole moment expansion for the static (i.e. time-averaged) electronic structure for these molecules.)

Table 6.6 Pressure broadening ($B_p$) coefficients (based on FWHM) of diatomic iodine ($I_2$) for a given **molecule** as the buffer gas at room temperature (292 K). The average value of the coefficients for the six features in this spectral region and the average of the uncertainty (i.e. standard deviation) of the average of the coefficient have been included in the last two rows. The coefficients are given in units of MHz per torr; the uncertainties of these fit results are given in parenthesis in units of the last significant figure. The pressure broadening coefficients are based on FWHM.

| $J''$ | Feature | $B_p \equiv$ pressure broadening (MHz/torr) | | | | |
|---|---|---|---|---|---|---|
| | | $H_2$ | $D_2$ | $N_2$ | $N_2O$ | $CO_2$ |
| 69 | E | 16.9(1) | 13.4(1) | 9.35(14) | 9.39(11) | 10.24(13) |
| 75 | I | 16.6(2) | -- | 10.10(16) | 10.84(9) | 10.50(20) |
| 76 | B | 15.2(1) | 11.9(1) | 7.18(17) | 8.34(18) | 9.33(11) |
| 122 | F | 16.6(1) | 14.0(1) | 8.80(8) | 8.82(14) | 10.29(19) |
| 129 | A | 16.1(2) | 13.5(1) | 7.39(18) | 7.98(15) | 9.14(22) |
| 133 | H | 18.7(3) | 15.5(3) | 7.66(21) | 9.97(29) | 9.68(51) |
| | average | 16.7(12) | 13.7(13) | 8.41(118) | 9.22(107) | 9.86(56) |
| | avg of std dev | 0.17(10) | 0.14(10) | 0.16(4) | 0.16(10) | 0.22(10) |

Table 6.7 Pressure shift ($S_p$) coefficients of diatomic iodine ($I_2$) for a given **molecule** as the buffer gas at room temperature (292 K). The average value of the coefficients for the six features in this spectral region and the average of the uncertainty (i.e. standard deviation) of the average of the coefficient have been included in the last two rows. The coefficients are given in units of MHz per torr; the uncertainties of these fit results are given in parenthesis in units of the last significant figure. The pressure broadening coefficients are based on FWHM.

| $J''$ | Feature | $S_p \equiv$ pressure shift (MHz / torr) | | | | |
|---|---|---|---|---|---|---|
| | | $H_2$ | $D_2$ | $N_2$ | $N_2O$ | $CO_2$ |
| 69 | E | −1.22(1) | −1.19(1) | −1.21(1) | −1.35(1) | −1.22(3) |
| 75 | I | −1.38(1) | -- | −1.38(1) | −1.51(1) | −1.41(3) |
| 76 | B | −1.24(1) | −1.20(1) | −1.23(1) | −1.37(1) | −1.21(3) |
| 122 | F | −1.18(1) | −1.17(1) | −1.34(1) | −1.43(1) | −1.29(3) |
| 129 | A | −1.30(1) | −1.21(1) | −1.41(1) | −1.52(1) | −1.31(3) |
| 133 | H | −1.37(4) | −1.33(2) | −1.51(2) | −1.57(2) | −1.50(4) |
| | average | −1.28(8) | −1.22(6) | −1.35(11) | −1.46(9) | −1.32(11) |
| | avg of std dev | 0.015(12) | 0.012(4) | 0.012(4) | 0.012(4) | 0.032(4) |

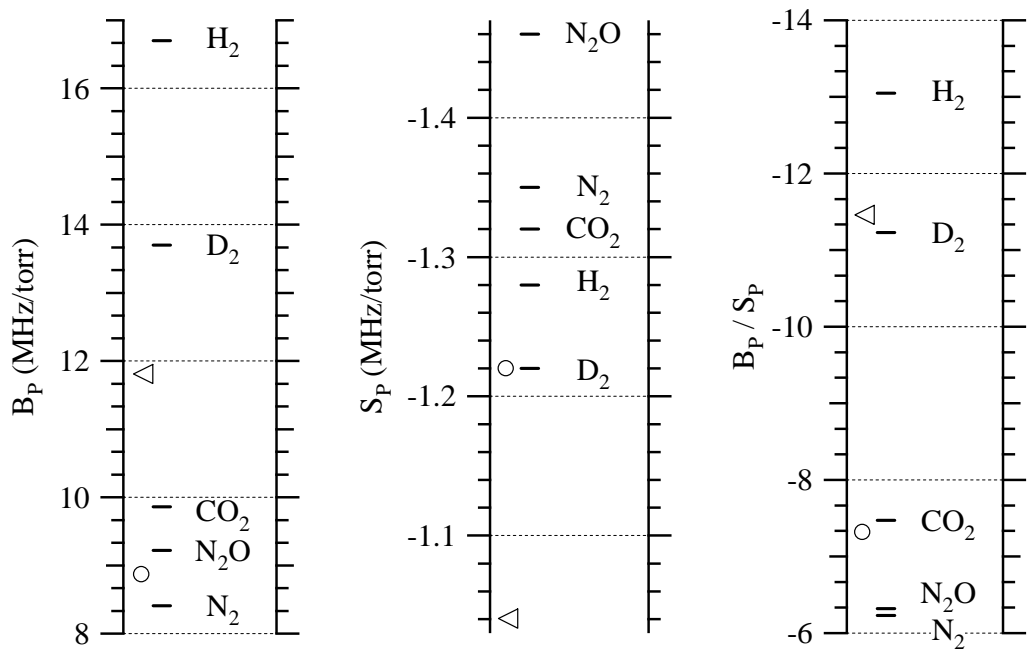

Figure 6.3 Average pressure broadening ($B_p$) and pressure shift ($S_p$) coefficients and their ratios for diatomic iodine ($I_2$) with a given **molecule** as the buffer gas. The data in these plots is presented in Tables 6.5, and 6.6 and 6.7. The values for **air** (○) and **water** (◁) were obtained from the earliest experiments in this project.

## 6.2.3 Multiple Temperatures (292, 348, and 388 K)

Tables 6.8, 6.9, and 6.10 lists the pressure broadening and pressure shift coefficients for diatomic iodine with the sample gas cell different temperatures. Investigations with diatomic iodine in the presence of the buffer gas **argon** (Table 6.8) were carried out at three temperatures: 292 K, 348 K, and 388 K. Investigations with diatomic iodine in the presence of **helium** (Table 6.9) and **carbon dioxide** (Table 6.10) were carried out at 292 K and 388 K. (See also Sub-Section 3.3.3 for additional details on experiments performed at elevated temperatures.)

In Table 6.8 the room temperature coefficients were calculated using spectra of the sample gas cell acquired just before and after those recorded at an elevated temperature relative to the ambient temperature. The room temperature coefficients in Tables 6.9 and 6.10 are redundant to those listed in previous tables (Tables 6.2, 6.3, 6.6, and 6.7) in this chapter.

Table 6.8 Pressure broadening ($B_p$) and pressure shift ($S_p$) coefficients of diatomic iodine ($I_2$) with **argon** as the buffer gas at three temperatures: 292, 348, and 388 K. The average value of the coefficients for the six features in this spectral region and the average of the uncertainty (i.e. standard deviation) of the average of the coefficient have been included in the last two rows. The coefficients are given in units of MHz per torr; the uncertainties of these fit results are given in parenthesis in units of the last significant figure. The pressure broadening coefficients are based on FWHM. Spectra were obtained with the Philips diode laser system. The Gauss (Doppler) width was allowed to float in the nonlinear regression algorithm. The results for each transition are an average for the up and down spectra, as are the uncertainties, which are given in units of the last significant figure in parenthesis.

| | | 292 K | | 348 K | | 388 K | |
|---|---|---|---|---|---|---|---|
| $J''$ | Feature | $B_p$ | $S_p$ | $B_p$ | $S_p$ | $B_p$ | $S_p$ |
| 69 | E | 8.16(11) | −1.173(19) | 7.91(10) | −1.021(20) | 6.56(11) | −0.945(12) |
| 75 | I | 8.31(25) | −1.371(24) | 7.56(14) | −1.236(22) | 5.57(17) | −1.127(13) |
| 76 | B | 7.57(11) | −1.196(20) | 7.00(9) | −1.066(22) | 5.77(8) | −0.968(12) |
| 122 | F | 8.15(13) | −1.312(20) | 7.53(10) | −1.150(21) | 6.73(8) | −1.042(12) |
| 129 | A | 7.67(17) | −1.353(23) | 7.10(13) | −1.186(24) | 6.29(11) | −1.093(12) |
| 133 | H | 9.07(42) | −1.519(40) | 7.85(28) | −1.295(25) | 6.10(20) | −1.175(14) |
| | average | 8.16(54) | −1.321(127) | 7.49(38) | −1.159(103) | 6.17(45) | −1.058(90) |
| | avg of std dev | 0.20(12) | 0.024(8) | 0.14(7) | 0.022(2) | 0.13(5) | 0.012(1) |

The motivation to undertake investigations of pressure broadening and pressure shift behavior at different temperatures is manifold. Such experiments have been and continue to be conducted in Earth bound laboratories of considerably different location and thus at different values of room temperature. Furthermore, it is not unusual for researchers to carry out experiments that maintain the sample gas cell at one or more temperatures that are different from the temperature of the room. In the latter case, the study of temperature-dependent phenomena in chemical dynamics is extremely challenging, as evidenced by the often used empirically phenomenological power law $T^n$ for dynamical parameters such as pressure broadening and pressure shift coefficients.

Table 6.9 Pressure broadening ($B_p$) and pressure shift ($S_p$) coefficients of diatomic iodine ($I_2$) with **helium** as the buffer gas at two temperatures: 292 and 388 K. The average value of the coefficients for the six features in this spectral region and the average of the uncertainty (i.e. standard deviation) of the average of the coefficient have been included in the last two rows. The coefficients are given in units of MHz per torr; the uncertainties of these fit results are given in parenthesis in units of the last significant figure. The pressure broadening coefficients are based on FWHM. Spectra were obtained with the Philips diode laser system. The Gauss (Doppler) width was allowed to float in the nonlinear regression algorithm. The results for each transition are an average for the up and down spectra, as are the uncertainties, which are given in units of the last significant figure in parenthesis.

| | | 292 K | | 388 K | |
|---|---|---|---|---|---|
| $J''$ | Feature | $B_p$ | $S_p$ | $B_p$ | $S_p$ |
| 69 | E | 8.37(13) | −0.197(22) | 6.66(16) | −0.156(17) |
| 75 | I | 8.68(21) | −0.290(25) | 5.07(25) | −0.268(20) |
| 76 | B | 7.87(11) | −0.196(23) | 5.91(12) | −0.157(16) |
| 122 | F | 8.21(16) | −0.142(25) | 6.64(12) | −0.098(16) |
| 129 | A | 8.22(20) | −0.168(24) | 6.60(18) | −0.139(18) |
| 133 | H | 8.68(33) | −0.212(40) | 6.82(28) | −0.169(21) |
| | average | 8.34(31) | −0.201(50) | 6.28(67) | −0.165(56) |
| | avg of std dev | 0.19(8) | 0.027(7) | 0.19(10) | 0.018(2) |

One such study of pressure broadening of diatomic iodine in the presence of various buffer gas as a function of buffer gas pressure and using time-domain methods was performed at a temperature of 388 K [Dantus]; consideration of these results as related to what appears to be a generally overlooked link between time-domain and frequency domain spectroscopy is offered in Sub-Section 6.3.1.

Table 6.10 Pressure broadening ($B_p$) and pressure shift ($S_p$) coefficients of diatomic iodine ($I_2$) with **carbon dioxide** as the buffer gas at two temperatures: 292 and 388 K. The average value of the coefficients for the six features in this spectral region and the average of the uncertainty (i.e. standard deviation) of the average of the coefficient have been included in the last two rows. The coefficients are given in units of MHz per torr; the uncertainties of these fit results are given in parenthesis in units of the last significant figure. The pressure broadening coefficients are based on FWHM. Spectra were obtained with the Philips diode laser system. The Gauss (Doppler) width was allowed to float in the nonlinear regression algorithm. The results for each transition are an average for the up and down spectra, as are the uncertainties, which are given in units of the last significant figure in parenthesis.

| | | 292 K | | 388 K | |
|---|---|---|---|---|---|
| $J''$ | feature | $B_p$ | $S_p$ | $B_p$ | $S_p$ |
| 69 | E | 10.24(13) | −1.22(3) | 8.17(18) | −1.00(2) |
| 75 | I | 10.50(20) | −1.41(3) | 6.55(24) | −1.20(2) |
| 76 | B | 9.33(11) | −1.21(3) | 7.38(14) | −1.01(2) |
| 122 | F | 10.29(19) | −1.29(3) | 8.48(14) | −1.05(2) |
| 129 | A | 9.14(22) | −1.31(3) | 7.63(20) | −1.10(2) |
| 133 | H | 9.68(51) | −1.50(4) | 7.75(23) | −1.18(2) |
| | average | 9.86(56) | −1.32(11) | 7.66(67) | −1.09(9) |
| | avg of std dev | 0.22(10) | 0.032(4) | 0.19(4) | 0.02(0) |

## 6.3 Collision Cross-Sections

For the ideal but non-existent limit of perfectly elastic collisions (i.e. not capable of sustaining a deformation without permanent change in size or shape), a hard-sphere model provides an ensemble-average collision cross-section $\sigma$ (representing an area of "strong" interaction) based on the pressure broadening coefficients. A model for the collision cross-section can be constructed based on statistical arguments of the nature of thermal equilibrium, and the relationship of this collision cross-section to the observed line shape (in this case for the radiation field) can be constructed based on the assumption that the time-interval $\Delta t$ between all events that affect (i.e. disrupt) a perfectly monochromatic radiation field (in a rotating frame where $\omega_0 = 0$) are randomly distributed in accord with the (normalized) Poisson probability density function $(2\tau_0)^{-1} \exp(-\tau_0^{-1}|\Delta t|)$. Such affects (or disruptions) in the radiation field are generally referred to as a state-change and can be characterized as an abrupt phase-change in some portion of the radiation field. The ensemble-average rate of such state-changing events is given by $1/\tau_0 = 1/T_2'$, which is taken to represent the sum of ensemble-average rates for all processes that lead to a radiation field disruption event. (See also Sections 2.8, 2.9 and 2.10 for further details on the models for $1/T_2'$ and $1/\tau_0$ used here. Also, as will be explored further in Section 7.6, this ensemble-average state-changing rate appears to be consistent with $\tau_0 = \tau_{abs} = (c_{abs} + \phi_g) = \tau_{em} = (c_{em} + \phi_e)$.) This model describes the state-changing events from the point of view of the radiation field, which is why the time-interval between events that affect the (frequency) width of the radiation field is given by $\tau_0$; this point will be relevant when the results of this chapter are compared with those in Section 7.5.

The statistical arguments for an ideal gas recognize that the pressure-dependent homogeneous state-changing rate is given by $1/\tau_0 = n\sigma v$, where $n$ is the number density of the buffer gas, $\sigma$ is the (perfectly) elastic collision cross-section, and v is the ensemble-average speed of the reduced mass ($\mu_m$) for the chromophore and buffer gas. (Implicit in using this form of the collision rate is the assumption that the collision rates, and thus the collision cross-sections, are the same for the lower and upper energy states of the chromophore; or at the very least that these values for the two energy levels are nearly the same so that $1/\tau_0$ and $\sigma$ can be thought of as average values that are representative of the actual values; these points will be explored further in Section 7.6) The ensemble-average pressure $P$ of the buffer gas (assumed to be much larger than the contribution from diatomic iodine to the total pressure) is related to its number density $n$ by $P = nk_B T$, where $k_B$ is the Boltzmann constant and $T$ is the thermal temperature. The ensemble-average speed v is given by $v = (8k_B T/(\pi \mu_m))^{1/2}$, where $\mu$ is the reduced mass $\mu_m = (m_{I2} m_{BG} \div (m_{I2} + m_{BG}))$ with the mass of diatomic iodine given by $m_{I2}$ and the mass of the buffer gas given by $m_{BG}$.

Solving for the state-changing rate $1/\tau_0$ in terms of pressure $P$ and temperature $T$ gives:

$$\frac{1}{\tau_0} = \sigma\, P \sqrt{\frac{8}{\pi\,\mu_m\,k_B\,T}}$$

<div align="right">(6.1)</div>

The anticipated (i.e. theoretically predicted) frequency domain line shape $g(\Delta\omega)$ can be found by taking the Fourier transform of the Poisson probability density function (i.e. exponential decay in the time domain); $\Delta\omega = \omega - \omega_0 = 2\,\pi(\nu - \nu_0)$:

$$g(\Delta\omega) = \frac{(2\,\tau_0)^{-1}}{\sqrt{2\,\pi}} \int_{-\infty}^{\infty} \exp\left(-\frac{|\Delta t|}{\tau_0} + i\,\Delta\omega\,\Delta t\right) d\,\Delta t$$

$$= \frac{\left(\sqrt{2\,\pi}\ \tau_0{}^2\right)^{-1}}{\frac{1}{\tau_0{}^2} + \Delta\omega^2}$$

<div align="right">(6.2)</div>

(See also Chapter 3 of Loudon's textbook "*The Quantum Theory of Light*" for a similar approach to constructing a model for the collision broadened line shape of the radiation field [Loudon 1].) Equation 6.2 has the form of a Lorentzian distribution for which the "full width at half maximum height" (FWHM) $\Delta\omega_{\text{FWHM}} = 2\pi\Delta\nu_{\text{FWHM}}$ is readily found to be given by:

$$\Delta\nu_{\text{FWHM}} = \frac{1}{\pi\,\tau_0}$$

<div align="right">(6.3)</div>

Combining equations 6.1 and 6.3 then provides a relationship between the homogeneous line widths observed in high resolution frequency domain spectra and the elastic collision cross-section as a function of buffer gas pressure at a given temperature:

$$\Delta\nu_{\text{FWHM}} \;=\; \frac{1}{\pi}\,\frac{\sigma\,\text{v}\,P}{k_B T} \;=\; \left(\frac{\sigma}{\pi}\,\sqrt{\frac{8}{\pi\,\mu_m\,k_B T}}\right) P \;\equiv\; B_{\text{p}}\,P$$

(6.4)

In equation 6.4, the measured pressure broadening coefficient is given by $B_{\text{p}}$. Since the change in line width is linear in pressure $P$, an equivalent way of defining the pressure broadening coefficient is given by:

$$B_{\text{p}} \;\equiv\; \frac{\partial\,\Delta\nu_{\text{FWHM}}}{\partial P} \;=\; \frac{1}{\pi}\,\frac{\sigma\,\text{v}}{k_B T} \;=\; \frac{\sigma}{\pi}\,\sqrt{\frac{8}{\pi\,\mu_m\,k_B T}}$$

(6.5)

In this project, the Lorentz (i.e. homogeneous) line widths $\Delta\nu_{\text{FWHM}}$ were obtained at several different pressures of a given buffer gas through a nonlinear regression analysis of the observed Doppler broadened line shape. The pressure broadening coefficient $B_{\text{p}}$ was then calculated by fitting these line widths as a function of pressure to a line. At relatively low pressures of buffer gas (ca. less than one atmosphere) the pressure broadening is, in general, empirically observed to increase linearly as the buffer gas pressure is increased. (In experiments where the buffer gas is different from the chromophore, the minimum pressure of the buffer gas is generally chosen to be an order of magnitude larger than that of the chromophore.)

In part, the model presented in Chapter VII appears to be a refinement of this model (equation 6.4); it uses quantum theoretical methods that take into account both the pressure broadening and pressure shift coefficients, and which then appears to describe some portion of the inelastic nature of state-changing interactions between the chromophore and a buffer gas in the calculated collision cross-sections. For now, however, the elastic collision cross-sections are listed in Tables 6.11 through 6.14.

See also Figure 7.3 for a plot of the room temperature (292 K) elastic collision cross-sections listed in the tables in this section (and the room temperature inelastic collision cross-sections calculated for the two-level system model presented in Chapter VII).

Table 6.11 Collision cross-sections ($\sigma$) of diatomic iodine ($I_2$) for a given **noble gas atom** as the buffer gas at room temperature (292 K). The average value of the collision cross-section for the six features in this spectral region and the average of the uncertainty (i.e. standard deviation) of the average of the coefficient have been included in the last two rows. The collision cross-sections are given in units of $Å^2$; the uncertainties of these fit results are given in parenthesis in units of the last significant figure. The average of the standard deviation of the average of the coefficient has been included in the last row; its uncertainty (i.e. standard deviation) is in parenthesis in units of the last significant figure.

| | | collision cross-section ($Å^2$) | | | | |
|---|---|---|---|---|---|---|
| $J''$ | Feature | He | Ne | Ar | Kr | Xe |
| 69 | E | 63.4(10) | 110(2) | 181(5) | 220(22) | 241(9) |
| 75 | I | 65.8(16) | -- | 189(7) | -- | -- |
| 76 | B | 59.7(9) | 100(2) | 169(5) | 201(13) | 224(12) |
| 122 | F | 62.3(12) | 106(3) | 182(10) | 218(9) | 242(19) |
| 129 | A | 62.4(15) | 95(3) | 159(14) | 202(37) | 225(23) |
| 133 | H | 65.9(25) | -- | 157(14) | -- | -- |
| | average | 63.3(24) | 103(S7) | 173(13) | 210(10) | 233(10) |
| | avg of std dev | 1.5(6) | 2.5(6) | 9.1(43) | 20(12) | 16(6) |

Some feature of the collision cross-sections are worth noting. The collision cross-section of the **noble gas atoms** increases with atomic number. Some lines show variations among experiments that are perhaps outside the expected statistical bounds. **Hydrogen** and **deuterium** are different by about 10%. **Nitrous oxide** is about the same as **carbon dioxide** despite that the former has a non-vanishing dipole moment and the latter not having one. **Nitrogen** is about the same as **air**.

The temperature-dependent trends are also noteworthy. In the case of **argon** as the buffer gas, Tables 6.8 and 6.13 suggest that neither the pressure broadening coefficient nor the collision cross-section follow a simple linear trend with temperature. Tables 6.8, 6.9, and 6.10 (for **argon**, **helium**, and **carbon dioxide**, respectively) indicate that the pressure-broadening coefficient decreased when increasing the sample gas cell temperature from 292 K to 388 K. Furthermore, Tables 6.13 (**argon**) and 6.14 (**helium**, and **carbon dioxide**) indicate that the collision cross-section decreased when increasing the sample gas cell temperature from 292 K to 388 K.

Table 6.12 Collision cross-sections ($\sigma$) of diatomic iodine ($I_2$) for a given **molecule** at room temperature (292 K). The average value of the collision cross-section for the six features in this spectral region and the average of the uncertainty (i.e. standard deviation) of the average of the coefficient have been included in the last two rows. The collision cross-sections are given in units of $Å^2$; the uncertainties of these fit results are given in parenthesis in units of the last significant figure.

collision cross-section ($Å^2$)

| $J''$ | Feature | $H_2$ | $D_2$ | $N_2$ | Air | $H_2O$ | $N_2O$ | $CO_2$ |
|---|---|---|---|---|---|---|---|---|
| 69 | E | 91.2(4) | 102(1) | 179(3) | 177(1) | 182(3) | 220(3) | 239(3) |
| 75 | I | 89.8(10) | -- | 192(3) | -- | -- | 254(2) | 246(5) |
| 76 | B | 82.4(6) | 90.9(7) | 138(3) | -- | -- | 195(4) | 218(3) |
| 122 | F | 89.7(6) | 106(1) | 169(1) | 170(2) | 188(3) | 206(3) | 241(5) |
| 129 | A | 87.2(9) | 103(1) | 142(1) | -- | -- | 187(3) | 214(5) |
| 133 | H | 101(1) | 118(2) | 147(4) | -- | -- | 233(7) | 227(12) |
| | average | 90.2(61) | 104(10) | 161(22) | 174(5) | 185(4) | 216(25) | 231(13) |
| avg of std dev | | 0.8(3) | 1.1(5) | 2.5(12) | -- | -- | 3.7(18) | 5.5(33) |

The lack of a simple linear trend in the changes of the pressure broadening coefficient and the collision cross-section as a function of temperature (for the case of argon as the buffer gas) are difficult to interpret or justify from the limited amount of such temperature-dependent results obtained in this project. Also, the observation that both of these parameters decreased when the sample gas cell temperature was increased from 292 K to 388 K appears to be worthy of further investigation. These results are quite intriguing and so it is hoped that future experiments and modeling will seek to address them.

Table 6.13 Collision cross-sections ($\sigma$) and ratios of pressure broadening and pressure shift coefficients ($B_p/S_p$) for diatomic iodine ($I_2$) with **argon** as the buffer gas at three temperatures: 292, 348, and 388 K. The average value of the collision cross-section for the six features in this spectral region and the average of the uncertainty (i.e. standard deviation) of the average of the coefficient have been included in the last two rows. The collision cross-sections are given in units of $\text{Å}^2$; the uncertainties of these fit results are given in parenthesis in units of the last significant figure.

| | | 292 K | | 348 K | | 388 K | |
|---|---|---|---|---|---|---|---|
| $J''$ | Feature | $\sigma$ | $B_p / S_p$ | $\sigma$ | $B_p / S_p$ | $\sigma$ | $B_p / S_p$ |
| 69 | E | 183(3) | −5.66(16) | 194(2) | −5.99(16) | 170(3) | −5.75(12) |
| 75 | I | 187(6) | −6.33(14) | 185(3) | −6.56(16) | 144(4) | −5.96(11) |
| 76 | B | 170(3) | −6.96(15) | 171(2) | −7.75(18) | 149(2) | −6.94(15) |
| 122 | F | 183(3) | −6.21(14) | 185(3) | −6.55(15) | 174(2) | −6.46(11) |
| 129 | A | 172(4) | −5.97(32) | 174(3) | −6.06(25) | 163(3) | −5.19(18) |
| 133 | H | 204(9) | −6.06(21) | 192(7) | −6.12(16) | 158(5) | −4.94(16) |
| | average | 183(12) | −6.20(44) | 184(9) | −6.51(66) | 160(12) | −5.87(75) |
| | avg of std dev | 4.5(2.7) | 0.19(7) | 3.3(19) | 0.18(4) | 3.2(12) | 0.14(3) |

Table 6.14 Collision cross-sections ($\sigma$) for diatomic iodine ($I_2$) with **helium** or **carbon dioxide** as the buffer gas at two temperatures: 292 and 388 K. The average value of the collision cross-section for the six features in this spectral region and the average of the uncertainty (i.e. standard deviation) of the average of the coefficient have been included in the last two rows. The collision cross-sections are given in units of $\text{Å}^2$; the uncertainties of these fit results are given in parenthesis in units of the last significant figure.

| | | collision cross-section ($\text{Å}^2$) | | | |
|---|---|---|---|---|---|
| | | He | | $CO_2$ | |
| $J''$ | Feature | 292 K | 388 K | 292 K | 388 K |
| 69 | E | 63.4(10) | 58.2(14) | 239(3) | 220(4) |
| 75 | I | 65.8(16) | 44.3(22) | 246(5) | 177(7) |
| 76 | B | 59.7(86) | 51.7(10) | 218(3) | 199(4) |
| 122 | F | 62.3(12) | 58.1(10) | 241(5) | 229(4) |
| 129 | A | 62.4(15) | 57.7(16) | 214(5) | 206(5) |
| 133 | H | 65.9(25) | 59.6(24) | 227(12) | 209(6) |
| | average | 63.3(24) | 54.9(59) | 231(13) | 207(18) |
| | avg of std dev | 1.5(6) | 1.6(6) | 5.4(34) | 5.0(13) |

In addition to the collision cross-sections obtained through an analysis of time-resolved photon-echo experiments [Dantus] (for which a more detailed analysis is offered in the following sub-section), one other report was found that claims to have measured the pressure broadening coefficient (consistent with the definition of $1/\tau_0$ given above) of diatomic iodine for the cases **helium** and **xenon** as the buffer gas. This other report [Astill] used an extremely sensitive (S/N $\cong$ 4000) saturation method known as polarization spectroscopy [Demtröder] to resolve a single hyperfine component of eight different lines (or more specifically, eight different values of the rotation quantum number $J''$ with vibration quantum numbers $(v', v'') = (17, 1)$) using a tunable dye laser operating at a wavelength of about 576 nm (or more specifically, in the region 17,300 to 17400 cm$^{-1}$).  While the reported collision cross-sections (ca. 61(4) Å for **helium** and 225(15) Å for **xenon**) are in good agreement with those reported in Table 6.11, that report did not provide sufficient indication of how these cross-sections were computed.  It might be possible (using Figure 6 of that report) to deduce the formula used for calculating these collision cross-sections, and consider more carefully the theoretical models used to describe those polarization spectroscopy experiments; however, due to time constraints such an effort has not been undertaken here.  As a result, our confidence in the apparent agreement of these results is somewhat diminished.

## 6.3.1 Time-resolved vs. Frequency Domain Spectroscopic Methods

The elastic collision cross-sections from a relatively recent report on (optical) photon-echo (i.e. time-resolved) experiments performed at about 388 K are listed in Table 6.15 [Dantus].  A semi-quantitative comparison between these results and those obtained in this project (Tables 6.11 through 6.14) is given in the last column of Table 6.15.  There does not appear to be an obvious way to correct a collision cross-section for changes in temperature $T$, but the results in this project show a general trend of decreasing collision cross-section in going from 292 K to 388 K. The report on the photon-echo experiments is complete enough so that further analysis (i.e. comparison) of these (fairly) systematic discrepancies can be carried out with reasonable confidence.

Table 6.15 Comparison of elastic collision cross-sections obtained from time domain and frequency domain experiments. The time-resolved collision cross-sections ("Dantus") were obtained at a temperature of 388 K [Dantus]. The frequency domain collision cross-sections ("dissertation") are taken from Tables 6.12, 6.13, and 6.14. The ratios ("ratio") of these two quantities are given in the last column. The collision cross-sections are given in units of $\text{Å}^2$; the uncertainties of the collision cross-sections and their ratios are given in parenthesis in units of the last significant figure.

| buffer gas | Dantus | dissertation | ratio |
|---|---|---|---|
| helium (He) | 135(12) | 54.9(59) | 2.46(34) |
| argon (Ar) | 500(70) | 160(12) | 3.12(50) |
| nitrogen ($N_2$) | 300(50) | 161(22) [b] | 1.86(40) |
| oxygen ($O_2$) | 450(12) | -- | -- |
| air | 330(45) [a] | 174(5) [b] | 1.90(26) |
| average | -- | -- | 2.34(59) |

a) 80% $N_2$ + 20% $O_2$     b) 292 K

Let's tacitly assume that measurements of $1/T_2' = (c_{em} + \phi_e)$ from a time-resolved photon-echo experiment and those for $1/\tau_0$ from a line-shape analysis of Doppler-limited high-resolution steady-state linear absorption frequency domain spectra would obtain approximately the same values for the (pressure-dependent ensemble-average) homogenous processes; i.e. that $(c_{abs} + \phi_g) \cong (c_{em} + \phi_e)$. (See also sections 2.9 and 2.10 for further mention of the models for $1/T_2'$ used here.) These two spectroscopic methods can perhaps be expected to differ in the measured homogeneous rates by a factor of two, with the time-resolved measurements being twice as large as the steady-state frequency domain measurements. That is to say, it is perhaps appropriate to recognize that in a time-resolved measurement the ensemble-average time-interval to the next collision-induced state-changing event is one-half of that for a steady-state frequency domain experiment.

The reason for this discrepancy can be deduced by recognizing that in a time-resolved measurement (using pulses of light that are somewhat to considerably shorter in duration than $\tau_0$) the random spatial and velocity distribution implies that the ensemble-average time-interval to the next state-changing collision event is $\tau_0 \div 2$. On the other hand, frequency domain experiments are configured to observe over time-intervals that are much longer than $\tau_0$ using relatively monochromatic light sources, and so an analysis of the observed line shape is expected to provide a measure of $\tau_0$ (as defined at the beginning of section 6.3). As a result, the report on time-resolved photon-echo experiments obtained a homogeneous decay rate $(1/T_2')$ that was a factor of two larger than the corresponding measurement of $1/\tau_0$ in this project. This difference then shows up in the collision cross-sections obtained from photon-echo experiments being a factor of two larger than those reported in this dissertation; see also equations 6.3 and 6.4.

Such considerations, as given in this sub-section (i.e. $\tau_0 = T_2' \div 2$ for time-resolved experiments and $\tau_0 \cong T_2'$ for frequency domain experiments), seems too obvious and fundamental not to be mentioned elsewhere. Nonetheless, it appears that accounting for all of the possible factors of two in describing such experiments has been a somewhat tedious and difficult task.

## 6.4 Overview of Chapter VI

[Astill]        A. G. Astill, A. J.  McCaffery, M. J. Proctor, E. A. Seddon, B. J. Whitaker; "Pressure broadening of the nuclear hyperfine spectrum of molecular iodine-127 by helium and xenon", Journal of Physics B, 18, 3745-3757 (1985).

[Dantus]        M. Comstock, V. V. Lozovoy, M. Dantus; "Femtosecond Photon Echo Measurements of Electronic Coherence Relaxation between the X($^1\Sigma_{g+}$) and B($^3\Pi_{0u+}$) States of $I_2$ in the Presence of He, Ar, $N_2$, $O_2$, $C_3H_8$"; Journal of Chemical Physics, 119, 6546-6553 (2003).

[Demtröder]    W. Demtröder; *Laser Spectroscopy: Basic Concepts and Instrumentation*, Third Edition, pages 374-378; Springer-Verlag, Berlin (2003); ISBN 3-540-65225-6.

[Fletcher]      D. G. Fletcher and J. C. McDaniel; "Collisional shift and broadening of iodine spectral lines in air near 543 nm", Journal of Quantitative Spectroscopy and Radiative Transfer, 54, 837-850 (1995).

[Hardwick 1]   J. A. Eng, J. L. Hardwick, J. A. Raasch and E. N. Wolf; "Diode laser wavelength modulated spectroscopy of $I_2$ at 675 nm", Spectrochimica Acta, Part A, 60, 3413-3419 (2004).

[Hardwick 4]   S. W. Arteaga, C. M. Bejger, J. L. Gerecke, J. L. Hardwick, Z. T. Martin, J. Mayo, E. A. McIlhattan, J.-M. F. Moreau, M. J. Pilkenton, M. J. Polston, B. T. Robertson and E. N. Wolf; "Line broadening and shift coefficients of acetylene at 1550 nm", Journal of Molecular Spectroscopy, 243, 253-266 (2007).

[Loudon 1]     R. Loudon; *The Quantum Theory of Light,* Second Edition, Chapter 2; Oxford University Press, New York, (1983); ISBN 0-19-851155-8.

CHAPTER VII

RESULTS 2 – TIME-DEPENDENT QUANTUM MECHANICS

## 7.1 Overview of Chapter VII

De-convolution of the constant (inhomogeneous) Doppler width from an observed line-shape in a Doppler-limited high resolution absorption spectrum provides a measure of the Lorentz full width at half maximum height (FWHM). The change in Lorentz width as a function of pressure, the pressure broadening coefficient ($B_p$), is representative of the changes in the collision-induced state-changing rate of the chromophore; all other contributions to the homogeneous state-changing rate are assumed to be constant for a given molecular line. As well, the internally-referenced spectrometers (i.e. a sample gas cell with added buffer gas and a reference gas cell with no added buffer gas) used in this project provided high-quality measurements of the rate of change in the shift of the line-center from a nearly "isolated" line-center frequency $\omega_0$ as a function of buffer gas pressure, the pressure shift coefficient ($S_p$).

We'll now develop a model for a two-level (quantum) system (or an ensemble of such systems) that most of the time is well-isolated (from other atomic systems), but occasionally and relatively briefly interacts with another atomic system (i.e. a buffer gas). These interactions will take place in the presence of a radiation field tuned to a frequency $\omega = 2\pi\nu$ that is resonant with the transition energy of the two-level systems. The relatively brief interactions before and after a photon is absorbed or emitted by the two-level systems (i.e. collision-induced state-change) will be accounted for by (a newly discovered parameter) the change of the wave function phase-factor ($\Delta\alpha$). Relevant presentations of various concepts and models utilized in this chapter can be found in Sections 2.7 through 2.14, Sections 5.2 and 5.3, and Section 6.3.

## 7.2 The Two-Level System Hamiltonian

The expression for the total time-dependent Hamiltonian $H(t)$ is given by:

$$H(t) = H_0 + V(t) \tag{7.1}$$

The total Hamiltonian is composed of two terms. The so called stationary-state Hamiltonian $H_0$ characterizes an isolated (i.e. unperturbed) quantum system; it is composed of operators that depend on relative position and momentum, but is not explicitly dependent on the time parameter $t$. The (time-dependent) perturbation term $V(t)$ accounts for the interactions that take place during a collision-induced photon absorption or emission event; as indicated in equation 7.1, this term is time-dependent. Part of the time dependence of the perturbation term $V(t)$ will be seen to result from the radiation field, while another part will be due to (relatively brief) interactions between the two-level system and a buffer gas during a state-changing event.

The stationary-state Hamiltonian $H_0$ has solutions that satisfy the time-independent Schrödinger equation:

$$H_0 |n\rangle = E_n |n\rangle \tag{7.2}$$

For the two-level model being developed here, the lower and upper energy level eigenstates satisfy the Bohr and Einstein (or de Broglie) energy relations:

$$\Delta E = E_e - E_g = \hbar(\omega_e - \omega_g) \equiv \hbar\omega_0 \tag{7.3}$$

The lower energy level is given by "$g$" (ground-state) and the upper energy level by "$e$" (excited-state).

The time-dependent portion of the unperturbed (i.e. perfectly isolated) two-level system wave functions are expected to satisfy the time-dependent Schrödinger equation:

$$\frac{\partial \, |\Psi(t)\rangle}{\partial \, t} \; = \; -\frac{i}{\hbar} \, H_0 \, |\Psi(t)\rangle$$

(7.4)

Separation of the wave functions (for the unperturbed two-level system) into two factors, one that depends on time and one that does not, allows for the orthogonal and normalized wave functions for the lower and upper energy levels of the isolated quantum system (that satisfy equations 7.2 and 7.4) to be expressed as (see also Appendix B):

$$|\psi_g(t)\rangle \; = \; \exp\!\left(-\frac{i \, E_g \, (t - t_0)}{\hbar} - i \, \alpha_g\right) |g\rangle$$

(7.5)

$$|\psi_e(t)\rangle \; = \; \exp\!\left(-\frac{i \, E_e \, (t - t_0)}{\hbar} - i \, \alpha_e\right) |e\rangle$$

(7.6)

The time-independent solutions of equation 7.2 for the lower and upper energy levels are given respectively by $|g\rangle$ and $|e\rangle$ with stationary-state energies given by $E_g$ and $E_e$; the initial time reference is given by $t_0$; and the phase-factors for these quantum states are given by $\alpha_g$ and $\alpha_e$. We will follow the common practice of setting $t_0 = 0$, as has been done on the left hand side of equations 7.5 and 7.6. Since we have chosen $t_0 = 0$, many of the equations in this chapter do not indicate explicitly the time-interval $\Delta t = t - t_0$; it is important to keep in mind, though, that such relative time-intervals are implied; like absolute position, absolute time appears to be a relatively

meaningless notion. Also, of some importance, we will not make an apriori assumption about the relative values of $\alpha_g$ and $\alpha_e$.

The statistical nature of quantum mechanics (i.e. ensemble-averages) allows the expression of the total time-dependent wave function for the two-level system as a linear superposition of the two-dimensional basis set (equations 7.5 and 7.6):

$$
\begin{aligned}
|\Psi(t)\rangle &= 1 \times |\Psi(t)\rangle = (|\psi_g\rangle\langle\psi_g| + |\psi_e\rangle\langle\psi_e|)\,|\Psi(t)\rangle \\
&= \langle\psi_g|\Psi(t)\rangle\,|\psi_g\rangle + \langle\psi_e|\Psi(t)\rangle\,|\psi_e\rangle \\
&= \langle g|\Psi(t)\rangle\,|g\rangle + \langle e|\Psi(t)\rangle\,|e\rangle \\
&\equiv C_g(t)\,|g\rangle + C_e(t)\,|e\rangle
\end{aligned}
\tag{7.7}
$$

In equation 7.7, the linear projection operators $|\psi_g\rangle\langle\psi_g|$ and $|\psi_e\rangle\langle\psi_e|$ act on the two-level system wave function $|\Psi(t)\rangle$, leading to the (generally complex-valued) time-dependent amplitudes $C_g(t)$ and $C_e(t)$, associated respectively with being in the lower energy or upper energy stationary-state level. As well, multiplying equation 7.7 by its complex conjugate gives the anticipated total probability for the total wave function:

$$
\langle\Psi(t)|\Psi(t)\rangle = |C_g(t)|^2 + |C_e(t)|^2 = 1
\tag{7.8}
$$

Equation 7.8 is (generally) interpreted as saying that the (time-dependent) probability for a given chromophore to be in the $|g\rangle$ state is given by $|C_g(t)|^2$ and in the $|e\rangle$ state by $|C_e(t)|^2$.

It will prove to be convenient to convert the stationary-state Hamiltonian $H_0$ to a projection operator representation. This transformation can be achieved by noting that the second line of equation 7.7 can be algebraically rearranged to an outer product (projection operator) form:

$$|\Psi(t)\rangle = \langle g \,|\Psi(t)\rangle \,|g\rangle \,+\, \langle e \,|\Psi(t)\rangle \,|e\rangle$$
$$= (|g\rangle \,\langle g \,| + |e\rangle \,\langle e \,|) \,\Psi(t)\rangle \qquad (7.9)$$

The portion of the second line of equation 7.9 in parenthesis containing the sum of outer products is also (often) referred to as the completeness relation [Shankar] (or sometimes as the closure relation [Loudon 2]):

$$|g\rangle \,\langle g \,| + |e\rangle \,\langle e \,| \,=\, 1 \qquad (7.10)$$

Since the time-independent states $|g\rangle$ and $|e\rangle$ are orthogonal (i.e. $\langle g|e\rangle = \langle e|g\rangle = 0$), and normalized (i.e. $\langle g|g\rangle = \langle e|e\rangle = 1$) multiplying both sides of $H_0$ by equation 7.10 gives the projection operator form:

$$H_0 = (|g\rangle \,\langle g \,| \,+\, |e\rangle \,\langle e \,|) \,H_0 \,(|g\rangle \,\langle g \,| \,+\, |e\rangle \,\langle e \,|)$$
$$= \hbar\,\omega_g \,|g\rangle \,\langle g \,| + \hbar\,\omega_e \,|e\rangle \,\langle e \,| \qquad (7.11)$$

At this point we will present the projection representation of the perturbation term $V(t)$ and provide a more detailed explanation of its "derivation" in the next section:

$$V(t) = \mu\,E_0 \left( |e\rangle \,\langle g| \, \exp\!\left( \frac{(A - 1 + i\,B)\,t}{2\,\tau_0} \,-\, i\,\omega\,t \right) \right.$$

$$\left. + \,|g\rangle \,\langle e| \, \exp\!\left( \frac{(A - 1 - i\,B)\,t}{2\,\tau_0} \,+\, i\,\omega\,t \right) \right) \qquad (7.12)$$

$$A = \cos(\Delta\alpha)\,;\ \ B = \sin(\Delta\alpha) \tag{7.13}$$

$$\Delta\alpha = (\alpha_e - \alpha_g) = \hbar^{-1}\,(P_e\,\Delta t_e - P_g\,\Delta t_g) \tag{7.14}$$

In equation 7.12, $E_0$ is the electric field amplitude of the radiation field and $\omega$ is the center frequency of the (relatively) "monochromatic" radiation field; the transition moment (often modeled as a dipole moment) is given by $\mu$. In the projection representation the transition between states in the two-level system are given by the operators $|e\rangle\langle g|$ and $|g\rangle\langle e|$, which can be derived in much the same way as equation 7.11, except that the odd parity of the (time-dependent) perturbation term $V(t)$ is taken into account [Herzberg 9; Loudon 1]. The phase-factors appearing in the basis set wave functions (equations 7.5 and 7.6) will be shown to change due to the "action" of the perturbation term $V(t)$ (equation 7.14); this change ($\Delta\alpha = \alpha_e - \alpha_g$) is modeled as occurring in relatively brief time-intervals before and after a collision-induced photon absorption or emission event (i.e. state-changing event). The notation for the variables $P_g$, $P_e$, $A$, $B$, and $\Delta\alpha$ was originally established by Foley [Foley 2]; more specifically, the $A$ and $B$ parameters in equation 7.12 and 7.13 should not be confused with the Einstein $A$ and $B$ coefficients.

The interactions between the chromophore and buffer gas (during a "meaningful" portion of a collision event) are being modeled (equation 7.14) as step functions (see Figure 7.1) with single-valued (ensemble-average) interaction energies $P_g$ and $P_e$ (assumed to be valid for a wide range of buffer gas pressures at a given temperature), respectively for the lower and upper energy levels of the chromophore. $P_g$ can be visualized as acting over the (ensemble-average) time-interval $\Delta t_g$ before the photon absorption event and $P_e$ is taken to act over the time-interval $\Delta t_e$ after the photon absorption event. It is assumed that the relatively brief time-intervals of interaction are much shorter in duration than the (steady-state) ensemble-average time-interval $\tau_0$ between state-changing events: $\tau_0 >> \Delta t_g \cong \Delta t_e$. Also, the time-interval of the photon absorption event $\Delta t_{sc}$ is assumed to be much shorter in duration than that of the interaction between the chromophore and the buffer gas: $\Delta t_g \cong \Delta t_e >> \Delta t_{sc}$. This derivation will approximate the state-changing time-interval $\Delta t_{sc}$ as being instantaneous: $\Delta t_{sc} = 0$.

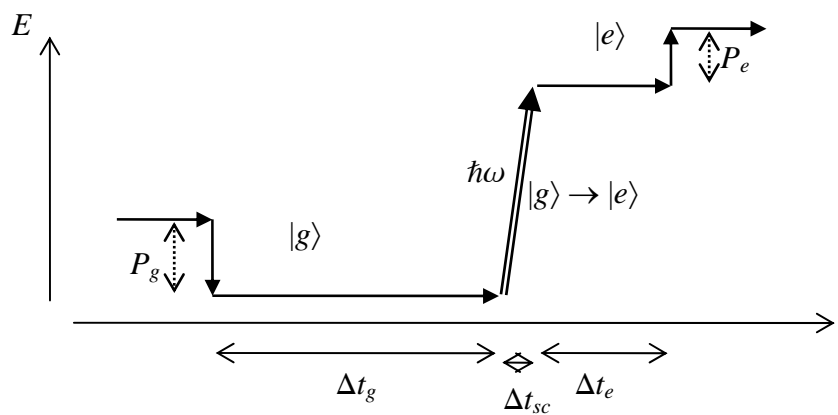

Figure 7.1 Collision-induced perturbations of a two-level system before and after photon absorption modeled as step functions (in accord with equation 7.14). The parameters in this diagram are explained in the text. The region of the lower and upper energy levels are given (in the "ket" designation) respectively by $|g\rangle$ and $|e\rangle$. The photon absorption event (characterized by $\Delta t_{sc}$ and $\Delta E = \hbar\omega$) is given by $|g\rangle \to |e\rangle$. The magnitude of $P_g$ and $P_e$ are greatly exaggerated; in general $\Delta E = \hbar\omega >> P_g \cong P_e$.

And of considerable importance, the perturbation term $V(t)$, as given in equation 7.12, is Hermitian (i.e. an operator that is equal to its adjoint), and so the total Hamiltonian $H(t)$ is Hermitian. We will explore the implications of such a choice later in this chapter, primarily in Sections 7.6 and 7.7.

## 7.3 Time-Dependent First-Order Perturbation Theory

A first-order (i.e. linear) perturbation treatment of the time-dependent Schrödinger equation is perhaps more readily "visualized" in the so called Interaction Picture. We begin by defining the transformation into the Interaction Picture:

$$|\Psi(t)\rangle \equiv U_0(t, t_0) \left| \tilde{\Psi}(t) \right\rangle$$

(7.15)

The free-evolution operator $U_0(t, t_0)$ for the unperturbed (i.e. perfectly isolated chromophore) basis set of stationary-state wave functions (equations 7.5 and 7.6) is defined as:

$$U_0(t, t_0) = \exp\left(-i\,\hbar^{-1} H_0(t - t_0)\right)$$

(7.16)

The Hermiticity of the (properly time-averaged time-independent) stationary-state term $H_0$ in the total Hamiltonian implies that $U_0(t, t_0)$ is unitary, which is to say that $U_0^\dagger(t, t_0) = U_0^{-1}(t, t_0)$, or equivalently that $U_0(t, t_0)\,U_0^\dagger(t, t_0) = U_0(t, t_0)\,U_0^{-1}(t, t_0) = 1$; the symbols "$\dagger$" and "$-1$" refer respectively to the (mathematical operations) adjoint and inversion. Furthermore, $U_0(t_0, t_0) = 1$ so that:

$$|\Psi(t_0)\rangle = \left| \tilde{\Psi}(t_0) \right\rangle$$

(7.17)

The time-dependent Schrödinger equation for the total Hamiltonian $H(t)$ (equation 7.1) is given by:

$$\frac{\partial |\Psi(t)\rangle}{\partial t} = -i\,\hbar^{-1}\,H(t)\,|\Psi(t)\rangle$$

(7.18)

In the Interaction Picture, with the use of equations 7.15 and 7.16 and the unitary property of $U_0(t, t_0)$, equation 7.18 becomes:

$$\frac{\partial \left|\widetilde{\Psi}(t)\right\rangle}{\partial t} = -i\,\hbar^{-1}\,\widetilde{V}(t)\left|\widetilde{\Psi}(t)\right\rangle$$

(7.19)

In equation 7.19 the rotation introduced by application of the free-evolution operator $U_0(t, t_0)$ (equation 7.16) has removed the stationary-state (i.e. time-independent) Hamiltonian $H_0$ from the time-dependent Schrödinger equation.

The (time-dependent) perturbation term $V(t)$ in the Interaction Picture Hamiltonian is given by the unitary transformation:

$$\begin{aligned}
\widetilde{V}(t) &= U_0^{\dagger}(t,\ t_0)\,V(t)\,U_0(t,\ t_0) \\
&= \exp\!\left(i\,\hbar^{-1}\,H_0(t-t_0)\right)V(t)\exp\!\left(-i\,\hbar^{-1}\,H_0(t-t_0)\right)
\end{aligned}$$

(7.20)

The perturbation term $V(t)$ is expected (apart from the possibility of an unrecognized factor of two or such) to have the following form [Loudon 2]:

$$V(t) = \mu\,E_0(|e\rangle\,\langle g|\exp(-i\,\omega\,t) + |g\rangle\,\langle e|\exp(i\,\omega\,t))$$

(7.21)

Equation 7.21 expresses the electric-dipole interaction between the two-level system and the quantized radiation field in the Generalized Rotating Wave Approximation [Barut 2; Loudon 2]; the transition operators $|e\rangle \langle g|$ and $|g\rangle \langle e|$ describe state-changes between the two levels in the presence of a radiation field tuned to a resonant frequency $\omega$.

Let's now assume that (in the presence of radiation field turned-on at a resonant frequency $\omega$) there will be an interaction between the buffer gas and the chromophore for a relatively brief time-interval (compared to the time-interval between state-changes), both before and after the state-changing event:

$$
\begin{aligned}
V_m(t) \;=\; & \mu\, E_0 \big( \exp\!\big(i\,\hbar^{-1} P_e\, t\big)\, |e\rangle \langle g|\, \exp\!\big(-i\,\hbar^{-1} P_g\, t\big) \exp(-i\,\omega\, t) \\[6pt]
& + \exp\!\big(i\,\hbar^{-1} P_g\, t\big)\, |g\rangle \langle e|\, \exp\!\big(-i\,\hbar^{-1} P_e\, t\big) \exp(i\,\omega\, t)\big) \\[10pt]
\;=\; & \mu\, E_0 (\, |e\rangle \langle g|\, \exp(i\, m\, \Delta\alpha \,-\, i\,\omega\, t) \\[6pt]
& + |g\rangle \langle e|\, \exp(-i\, m\, \Delta\alpha \,+\, i\,\omega\, t))
\end{aligned}
$$

$$(7.22)$$

In equation 7.22 the interaction of the chromophore with the buffer gas is being modeled as two step-functions, one on either side of the transition operator; the interaction operators $P_g$ and $P_e$ correspond respectively to the interactions of the lower and upper energy levels (i.e. states) of the chromophore with the buffer gas, before and after the photon absorption or emission event for the time-intervals $\Delta t_g$ and $\Delta t_e$. The change of the (two-level system) wave function phase-factor was previously defined in equation 7.14 and given again here: $\Delta\alpha = \hbar^{-1}(\Delta t_e\, P_e - P_g\, \Delta t_g)$. The description given here for the ensemble-average interactions is depicted in Figure 7.1. In a given time-interval $\Delta t = t - t_0$ (and $t_0$ is set equal to zero) there is a probability of $m$ such interactions, which are taken as being relatively brief in duration compared to the time between these state-changing events. In the process of constructing more fully equation 7.22 (in what follows in this sub-section), the probability of $m$ interactions will be modeled by the Poisson distribution, which agrees well with the notion of an exponential decay in the time domain [Larsen].

Let's now examine the nature of the (mathematical) iteration procedure leading to first-order solutions of equation 7.19. The first-order solutions require (at a minimum) that $\mu E_0 \div \hbar << 1$; see also Section 5.3. Without loss of generality, we'll assume that the observation period begins ($t_0$) with the chromophore in the lower energy level and set $t_0 = 0$. The general solution of equation 7.19 is then given by:

$$\left| \overset{\sim}{\Psi}(t) \right\rangle = \left| \overset{\sim}{\Psi}(0) \right\rangle - i \hbar^{-1} \int_0^t \overset{\sim}{V}(t) \left| \overset{\sim}{\Psi}(0) \right\rangle dt$$

$$(7.23)$$

However, equation 7.19 is a differential equation that involves time-ordered operators and so it must be solved by successive iterations; the solution up to time $t_m$ (i.e. the $m^{th}$ iteration) is plugged back into the integral on the right hand side of equation 7.23 to generate the solution up to time $t_{m+1}$, where $t_m < t_{m+1}$, and so on (i.e. time-ordering of the iteration process) [Mukamel]. It may be worth comparing the time-ordered method outlined below to Picard's iteration method [Boyce].

The first iteration of equation 7.23 (so named for convenience, since it is not really an iteration of a previously obtained solution) gives:

$$\left| \widetilde{\Psi}_1(t_1) \right\rangle = \left| \widetilde{\Psi}(0) \right\rangle - i\,\hbar^{-1} \int_0^{t_1} \widetilde{V}_1(t) \left| \widetilde{\Psi}(0) \right\rangle d\!\!\!/\, t$$

$$= \; |g\rangle \; - \; i\,\mu\,E_0\,\hbar^{-1} \int_0^{t_1} (\exp(i\,\Delta\alpha + i\,(\omega_0 - \omega)\,t)\,|e\rangle\,\langle g|$$

$$+ \exp(-i\,\Delta\alpha - i\,(\omega_0 - \omega)\,t)\,|g\rangle\,\langle e|)\,|g\rangle \; d\!\!\!/\, t$$

$$= \; |g\rangle \; - \; i\,\hbar^{-1}\,\mathrm{I}_1\,|e\rangle \qquad\qquad (7.24)$$

The appearance of $\omega_0$ in equation 7.24 is due to the difference in energies of the freely evolving stationary-states on either side of the state change in equation 7.20. Equation 7.24 has been simplified by recalling that the stationary-states $|g\rangle$ and $|e\rangle$ are orthogonal and normalized, and $\mathrm{I}_1$ is defined as:

$$\mathrm{I}_1 \; \equiv \; \mu\,E_0 \int_0^{t_1} (\exp(i\,\Delta\alpha + i\,(\omega_0 - \omega)\,t)\, d\!\!\!/\, t$$

$$\qquad\qquad (7.25)$$

The second iteration of equation 7.23 gives:

$$\left|\widetilde{\Psi}_2(t_2)\right\rangle = \left|\widetilde{\Psi}(0)\right\rangle - i\,\hbar^{-1} \int_0^{t_2} \widetilde{V}_2(t) \left|\widetilde{\Psi}_1\right\rangle dt$$

$$= |g\rangle - i\,\hbar^{-1} \int_0^{t_2} (\exp(i\,2\,\Delta\alpha + i\,(\omega_0 - \omega)\,t)\,|e\rangle\,\langle g|$$

$$+ \exp(-i\,2\,\Delta\alpha - i\,(\omega_0 - \omega)\,t)\,|g\rangle\,\langle e|$$

$$\times (\,|g\rangle - i\,\mathrm{I}_1\,|e\rangle)\,dt$$

$$= |g\rangle - i\,\hbar^{-1}\,\mathrm{I}_2\,|e\rangle - \hbar^{-2}\,\mathrm{I}_2^*\,\mathrm{I}_1\,|g\rangle$$

$$\cong |g\rangle - i\,\hbar^{-1}\,\mathrm{I}_2\,|e\rangle$$

$$(7.26)$$

Similar to equation 7.25, $\mathrm{I}_2$ is defined as:

$$\mathrm{I}_2 \equiv \mu\,E_0 \int_0^{t_2} (\exp(i\,2\,\Delta\alpha + i\,(\omega_0 - \omega)\,t)\,dt$$

$$(7.27)$$

As mentioned above, the experiment is configured (using a relatively low intensity radiation field) such that it is considered a reasonable approximation to truncate the iteration solutions to first-order in $\mathrm{I}_m$, as was done at the end of equation 7.26; an absorption experiment can be configured to be linear in $\mathrm{I}_m$ by simply setting the incident radiation field intensity $E_0^2$ to be sufficiently small.

The $m^{\text{th}}$ iteration of equation 7.23 gives:

$$\left|\tilde{\Psi}_m(t_m)\right\rangle = \left|\tilde{\Psi}(0)\right\rangle - i\,\hbar^{-1}\int_0^{t_m}\tilde{V}_m(t)\left|\tilde{\Psi}_{m-1}\right\rangle d\!\!\!/\,t$$

$$= |g\rangle - i\,\hbar^{-1}\int_0^{t_m}(\exp(i\,m\,\Delta\alpha + i\,(\omega_0 - \omega)\,t)\,|e\rangle\langle g|$$

$$+ \exp(-i\,m\,\Delta\alpha - i\,(\omega_0 - \omega)\,t)\,|g\rangle\langle e|$$

$$\times\,(\,|g\rangle - i\,\mathrm{I}_{m-1}\,|e\rangle)\,d\!\!\!/\,t$$

$$= |g\rangle - i\,\hbar^{-1}\,\mathrm{I}_m\,|e\rangle - \hbar^{-2}\,\mathrm{I}_m^*\,\mathrm{I}_{m-1}\,|g\rangle$$

$$\cong |g\rangle - i\,\hbar^{-1}\,\mathrm{I}_m\,|e\rangle \qquad\qquad (7.28)$$

Similar to equations 7.25 and 7.27, $\mathrm{I}_m$ is defined as:

$$\mathrm{I}_m \equiv \mu\,E_0\int_0^{t_m}(\exp(i\,m\,\Delta\alpha + i\,(\omega_0 - \omega)\,t)\,d\!\!\!/\,t$$

$$(7.29)$$

Of course, we do not know how many collision-induced state-changes took place on a given (relatively long-duration) observation time-interval $t$ in an ensemble (i.e. large collection) of chromophores. However, since these (discrete) events occur randomly, it is proposed that the corresponding term of the Poisson probability distribution $P_m$ be associated with each possible number of (subsequent) state-changes (i.e. iterations of equation 7.23), and simply add together all possible outcomes to form the ensemble-average wave function. This procedure is akin to asking how the "$2\tau_0$ time-intervals" (i.e. on the ensemble-average time-interval scale defined by $2\tau_0 =$ $\tau_{abs} + \tau_{em}$) for subsequent collision-induced absorption events are randomly distributed according to the Poisson distribution in the time-interval $\Delta t = (t - t_0) \equiv (t - 0)$.  All of these probability-based contributions to the two-level wave function are taken into account (for any given time-interval) by rewriting equation 7.29 as:

$$\mathrm{I}_m(t) \equiv \mu E_0 \left( \sum_{m=0}^{\infty} \int_0^t P_m \left( \exp(i\, m\, \Delta\alpha + i\, (\omega_0 - \omega)\, t) \right) dt \right) |e\rangle \langle g|$$

(7.30)

While the tilde has not been used over $\mathrm{I}_m(t)$ in equation 7.30, and in similar terms in equations 7.24 through 7.29, these symbols represent "objects" in the Interaction Picture.

In equation 7.30, $P_m$ represents the $m^{\text{th}}$ term of the (normalized) Poisson probability function:

$$P_m = \left( \frac{t}{2\,\tau_0} \right)^m \frac{1}{m!} \exp\left( -\frac{t}{2\,\tau_0} \right)$$

(7.31)

The $m = 0$ term of equation 7.31 is the probability for the free-evolution period leading up to the first state-changing event (on any given time-interval $t$); it provides a probability expression in the form of a simple exponential decay that the transition sequence $|e\rangle \rightarrow |g\rangle \rightarrow |e\rangle$ (on the "$2\tau_0$ time-scale") will not occur again.  That such a choice of the decay rate $(2\tau_0)^{-1}$ for the $m = 0$ term of the Poisson distribution is appropriate for the observation of steady-state dynamics in a (steady-state) linear absorption experiment is (perhaps) not so obvious.  However, the manner in which an observation is performed in a steady-state linear absorption experiment is meant to be consistent with an observational configuration that "sees" (only) steady-state dynamics (that occur on the "$2\tau_0$ time-scale").  This situation is achieved by using a detection scheme that (continuously) integrates the radiation intensity falling on the detectors for time-intervals that are much longer than $\tau_0$, with the actual (recorded) observation being made after many tens (or more) of such integration time-intervals having passed since the radiation source was initially turned-on at a particular frequency $\omega$. (Similar arguments apply to a wavelength-modulated experiment, except that the "observed" steady-state is intrinsically oscillating.) As well, this sort of observational arrangement is also allowing the detection scheme to reach a steady-state before recording its signal value (typically a voltage).

Essentially, the above derivation is providing a means of constructing the perturbation term $V(t)$ (of equation 7.21) as a probability weighted sum of terms that accounts for all possible number of transitions (i.e. outcomes) during a given observation time-interval $t$. Since only the ensemble-average time-interval between state changes $2\tau_0 = \tau_{abs} + \tau_{em}$ is "knowable" (in this model), the transformation between the Schrödinger and Interaction Pictures reduces to multiplications by $\exp(-i\omega_0 t)$ and its complex conjugate; the relative simplicity of these transformations is due to the unitary property of the intervening free-evolution operators that appear between state-changing events in the Interaction Picture (i.e. $U_0(t_{\mathrm{m}}, t_{\mathrm{m}-1})\, U_0^{-1}(t_{\mathrm{m}}, t_{\mathrm{m}-1})$ $= U_0(2\tau_0, 0)\, U_0^{-1}(2\tau_0, 0) = 1$). Equation 7.23 and 7.30 then give the time-dependent (two-level system) wave function to first-order (for the time-ordered perturbation expansion in the Interaction Picture) as:

$$\left| \overset{\sim}{\Psi}(t) \right\rangle \;\cong\; |g\rangle \;-\; i\,\hbar^{-1}\, \mathrm{I}_m(t)\, |g\rangle$$

$$(7.32)$$

Performing the summation over all possible numbers of state-changes $m$ (i.e. $m$ = zero to infinity for the ensemble of chromophores) for an arbitrary time-interval $t$ in equation 7.30 gives:

$$\left| \overset{\sim}{\Psi}(t) \right\rangle \cong |g\rangle$$
$$- i\,\mu\,E_0\,\hbar^{-1} \int_0^t \exp\!\left( \frac{(A - 1 + i\,B)\,t}{2\,\tau_0} + i\,(\omega_0 - \omega)\,t \right) d\!\!\!/\,t\,|e\rangle$$

$$(7.33)$$

The ensemble-average width and shift parameters (given respectively by $A$ and $B$) in equation 7.33 have been previously defined in equation 7.13. Furthermore, transforming equation 7.33 from the Interaction Picture back to the Schrödinger Picture (i.e. removing the rotation $\exp(i\omega_0 t)$) leads to identifying equation 7.12 as the (sought for Hermitian) perturbation term $V(t)$ that describes collision-induced photon absorption emission events.

## 7.4 Excited-State Probabilities

As mentioned in Section 2.8, the observation of spectral features (i.e. transition line shapes) is predicated on the existence of ro-vibronic transitions (i.e. absorption and emission of photons). Furthermore, the observation of a reproducibly constant steady-state line shape (at a given buffer gas pressure) is consistent with an ensemble-average steady-state value of chromophores in the upper energy level (i.e. the excited-state) at (the conceptually challenging) "any given *instant* in time".

The next step in this derivation, then, is to calculate the ensemble-average excited-state probability $|C_e(t)|^2 = C_e^*(t) \times C_e(t)$. Along with the properties of $|g\rangle$ and $|e\rangle$ mentioned in the text following equation 7.11 (specifically, that $\langle g|e\rangle = \langle e|g\rangle = 0$

and $\langle g|g \rangle = \langle e|e \rangle = 1$), utilization of equations 7.7, 7.15, 7.16, and 7.33 leads in a relatively straight-forward way to:

$$|C_e(t,\ \omega)|^2 = C_e^*(t,\ \omega) \times C_e(t,\ \omega)$$

$$= \langle \Psi(t,\ \omega) \mid e \rangle \langle e \mid \Psi(t,\ \omega) \rangle = \left\langle \tilde{\Psi}(t,\ \omega) \, \middle| \, e \right\rangle \left\langle e \, \middle| \, \tilde{\Psi}(t,\ \omega) \right\rangle$$

$$= \left(\frac{\mu E_0}{\hbar}\right)^2 \int_0^t \exp\left(-\frac{(A-1-iB)\,t}{2\,\tau_0} - i\,(\omega_0-\omega)\,t\right) d\!t$$

$$\times \int_0^t \exp\left(\frac{(A-1+iB)\,t}{2\,\tau_0} + i\,(\omega_0-\omega)\,t\right) d\!t$$

$$(7.34)$$

Equation 7.32 indicates that $|C_e(t)|^2$ is also a function of frequency $\omega$, which has been included in the notation of equation 7.34.

The general solution of equation 7.34, constrained by the assumptions that $\Delta\alpha \neq 0$ or $2\pi$ (i.e. $A < 1$; see equation 7.13) (and, of course, $\tau_0$, $\omega_0$, and $\omega$ are all greater than zero), is given by:

$$|C_e(t,\,\omega)|^2 = \left(\frac{\mu\,E_0}{\hbar}\right)^2 \times \frac{1}{\left(\frac{1-A}{2\,\tau_0}\right)^2 + \left(\omega_0 - \omega + \frac{B}{2\,\tau_0}\right)^2}$$

$$\times \left(1 + \exp\left(-\frac{(1-A)\,t}{\tau_0}\right)\right.$$

$$\left. - 2\exp\left(-\frac{(1-A)\,t}{2\,\tau_0}\right)\text{Cos}\left(\left(\omega_0 - \omega + \frac{B}{2\,\tau_0}\right)t\right)\right)$$

$$(7.35)$$

Equation 7.35 is noteworthy because it is identical in form to the time-dependent excited-state probability $|C_e(t,\,\omega)|^2$ obtained by solving the optical Bloch equations in which a phenomenological decay rate has been included and for the condition of a relatively low intensity radiation field [Loudon 1].

The steady-state limit for the probability of a chromophore being in the upper energy level can be identified with the asymptotic limit of equation 7.35, which is found by letting $t \to \infty$:

$$|C_e(\infty,\,\omega)|^2 = \left(\frac{\mu\,E_0}{\hbar}\right)^2 \frac{1}{\left(\frac{1-A}{2\,\tau_0}\right)^2 + \left(\omega_0 - \omega + \frac{B}{2\,\tau_0}\right)^2}$$

$$(7.36)$$

For the experimental configurations used in this project, where the detection (i.e. observation) scheme had an integration time-interval that is reproducible with a high degree of accuracy, and is at least three orders of magnitude (i.e. $10^3$) larger than the ensemble-average time-interval between state changes $\tau_0$, it seems reasonable to proceed in this derivation using equation 7.36 as the ensemble-average line shape model for the width and line-center shift (corresponding to the homogeneous component of an observed line shape). Also, without loss of generality (with regard to connecting equation 7.36 to the experimentally measured pressure broadening and pressure shift coefficients), the pressure-independent contribution to the model line shape of equation 7.36 has been omitted; more specifically, this omission refers to the pressure-independent stimulated absorption process; see also Sections 2.9 and 2.10.

It may also be worth noting that equation 7.33 leads to $|C_g(t, \omega)|^2 = C_g^*(t, \omega) \times C_g(t, \omega) \cong 1$, which agrees with the notion that a first-order perturbation expansion is valid for relatively weak perturbations.

## 7.5 Pressure Broadening and Pressure Shift Coefficients

The next step in this derivation is to relate the pressure broadening and the pressure shift coefficients (given respectively by $B_p$ and $S_p$) to the $A$ and $B$ line shape parameters. The relevant pieces needed to relate the "hard-sphere" collision cross-section (based on the predictions obtained from gas-kinetic theory for the behavior of an ideal gas) to the observed line-width and line-center shift by:

$$P = n\,k_B\,T\,; \quad \text{v} = \sqrt{\frac{8\,k_B\,T}{\pi\,\mu_m}}\;; \quad \mu_m = \frac{m_1\,m_2}{m_1 + m_2}$$

$$z = \frac{1}{\tau_0} = n\,\sigma\,\text{v} = \frac{P\,\sigma\,\text{v}}{k_B\,T}\;; \qquad \sigma = \pi\,d^2$$

$$(7.37)$$

The parameters in equation 7.37 have the following definitions: $P$ is the buffer gas pressure; $n$ is the number density of the buffer gas (assumed to be at least an order of magnitude larger than the constant chromophore number density $N = N_g + N_e$); $k_B$ is the Boltzmann constant; $T$ is the temperature; $z$ is the collision frequency; $\tau_0$ is the ensemble-average time-interval between collisions of a single chromophore and a buffer gas particle that induce a state change in the chromophore (i.e. photon absorption or emission); $\sigma$ is the ensemble-average (integrated) collision cross-section for these collision-induced state changes for a given buffer gas; $d$ is the interaction distance associated with the collision cross-section $\sigma$; v is the ensemble-average relative speed between the chromophore and the buffer gas for thermal equilibrium conditions in a gas cell; $\mu_m$ is the reduced mass, $m_1$ is the mass of a single chromophore particle (e.g. diatomic iodine); and $m_2$ is the mass of a single buffer gas particle.

In terms of the hard-sphere collision model, the full width at half maximum height ($\Delta\omega_{\text{FWHM}} = 2\pi\Delta\nu_{\text{FWHM}}$) and line-center shift ($\Delta\omega_0 = 2\pi\Delta\nu_0$) in the Lorentzian distribution of equation 7.36 can be readily identified:

$$\Delta\omega_{\text{FWHM}} = 2\left(\frac{(1-A)}{2\,\tau_0}\right) = \frac{2\,(1-A)}{2}\,\frac{P\,\sigma\,\text{v}}{k_B\,T} \tag{7.38}$$

$$\Delta\omega_0 = \frac{B}{2\,\tau_0} = \frac{B}{2}\,\frac{P\,\sigma\,\text{v}}{k_B\,T} \tag{7.39}$$

The pressure broadening ($B_{\text{p}}$) and pressure shift ($S_{\text{p}}$) coefficients are, respectively, the change in the line width (FWHM) and the line-center position (i.e.

frequency) per unit change in pressure (see also Section 6.3); these "rates" are found from the pressure-dependent equations 7.38 and 7.39 according to the definitions:

$$B_{\mathrm{p}} \equiv \frac{\partial \, \Delta \nu_{\mathrm{FWHM}}}{\partial \, P} = \frac{(1 - A)}{2 \, \pi} \, \frac{\sigma \, \mathrm{v}}{k_B \, T}$$

(7.40)

$$S_{\mathrm{p}} \equiv \frac{\partial \, \Delta \nu_0}{\partial \, P} = \frac{B}{4 \, \pi} \, \frac{\sigma \, \mathrm{v}}{k_B \, T}$$

(7.41)

The pressure broadening and pressure shift coefficients of equation 7.40 and 7.41 are obtained through an analysis of the de-convolved line-shape data. These two experimentally determined parameters can then be rearranged to obtain a solution for the ensemble-average change in the wave function phase-factor (for a state-changing event) $\Delta \alpha$, which then leads to solutions for $A$, $B$, and $\sigma$.

The solution to $\Delta \alpha$ is readily found from the ratio of the pressure broadening to pressure shift coefficients:

$$R_{\mathrm{bs}} \equiv \frac{B_{\mathrm{p}}}{S_{\mathrm{p}}} = \frac{2 \, (1 - A)}{B} = 2 \tan\left(\frac{\Delta \alpha}{2}\right)$$

(7.42)

Equation 7.42 is plotted in Figure 7.2. The valid (i.e. $\Delta \alpha \neq 0$, which is equivalent to $A \neq 1$ and $B \neq 0$) solution of equation 7.42 is:

$$\Delta \alpha = \cos^{-1}\left(\frac{4 - R_{\mathrm{bs}}{}^2}{4 + R_{\mathrm{bs}}{}^2}\right)$$

(7.43)

Calculations of the ensemble-average changes in the wave function phase-factor $\Delta\alpha$ for diatomic iodine in the presence of noble gases (as the buffer gas) using equation 7.43 are listed in Table 7.1.

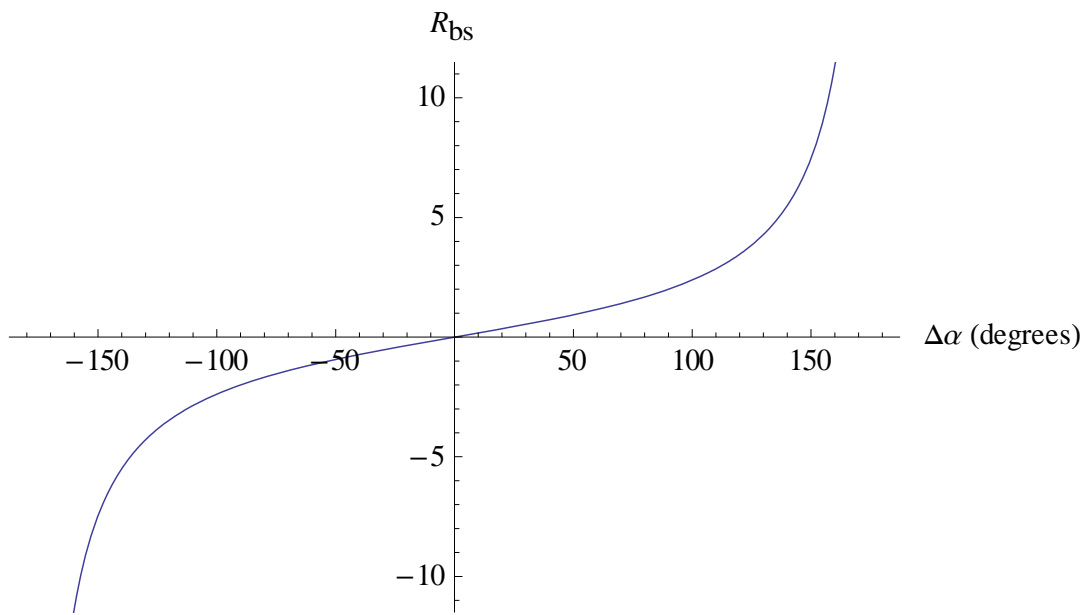

Figure 7.2 Plot of the ratio $R_{bs}$ of the pressure broadening to pressure shift coefficients $R_{bs} = B_p/S_p = 2\tan(\Delta\alpha/2)$ as a function of the ensemble-average change in the (wave function) phase-factor, $\Delta\alpha$.

Comparison of equations 6.5 and 7.40 (which both use the same value of the pressure broadening coefficient $B_p$) indicates that the collision cross-section $\sigma$ of equation 6.5 is commensurate with the (unreachable) asymptotic limit (shown in Figure 7.2) at $\Delta\alpha = \pm\pi$, for which $A = 0$ and $B = 1$. That is to say, the asymptotic limit (at $\Delta\alpha = \pm\pi$) corresponds to the physically unrealizable perfectly elastic collision; i.e. instantaneous collisions between non-existent infinitely non-deformable hard-spheres. In equation 7.40, while the constituents (i.e. electrons and nuclei) of the collision partners can still be approximated as hard-spheres, the effect of the time-integrated interactions of these constituents (i.e. electric and magnetic forces acting over time and space) during a state-changing event appears as an increase in the collision cross-section over the perfectly elastic case by the factor $2(1-A)^{-1}$. Since $0 < A < 1$, the elastic collision cross-section $\sigma_{\text{elastic}}$ represents a lower-bound limit on the collision cross-section. In this sense, it seems reasonable to identify the collision cross-section of equation 7.40 as an inelastic collision cross-section $\sigma_{\text{inelastic}}$, so that $\sigma_{\text{inelastic}} = 2(1-A)^{-1}\sigma_{\text{elastic}} > \sigma_{\text{elastic}}$; see Figure 7.3.

Table 7.1 Ensemble-average change in the wave function phase-factor $\Delta\alpha$ of diatomic iodine ($I_2$), for high-$J$ transition energies in the vicinity of 14,819 cm$^{-1}$, for a given **noble gas** as the buffer gas (at a temperature of 292 K), and based on the pressure broadening and pressure shift coefficients presented in Sub-Section 6.2.1, specifically Tables 6.2 and 6.3; the uncertainties of these fit results are given in parenthesis in units of the last significant figure.

| Buffer Gas | $B_p$ (MHz/torr) | $S_p$ (MHz/torr) | $R_{bs}$ | $\Delta\alpha$ (degrees) |
|---|---|---|---|---|
| He | 8.34(31) | −0.201(50) | −41.5(104) | −174(1) |
| Ne | 6.17(39) | −0.713(29) | −8.65(65) | −154(2) |
| Ar | 7.70(41) | −1.32(11) | −5.83(58) | −142(4) |
| Kr | 6.86(66) | −1.40(6) | −4.90(52) | −136(3) |
| Xe | 6.47(44) | −1.71(10) | −3.78(34) | −124(4) |

The two-level model developed so far in this chapter provides for a (more) complete characterization of pressure broadening and pressure shift coefficients, which includes calculation of a collision cross-section $\sigma$ and a change in wave function phase-factor $\Delta\alpha$. A visual display of the relative values of these parameters obtained in this project from room temperature (i.e. $T = 292$ K) experiments is presented in Figure 7.3.

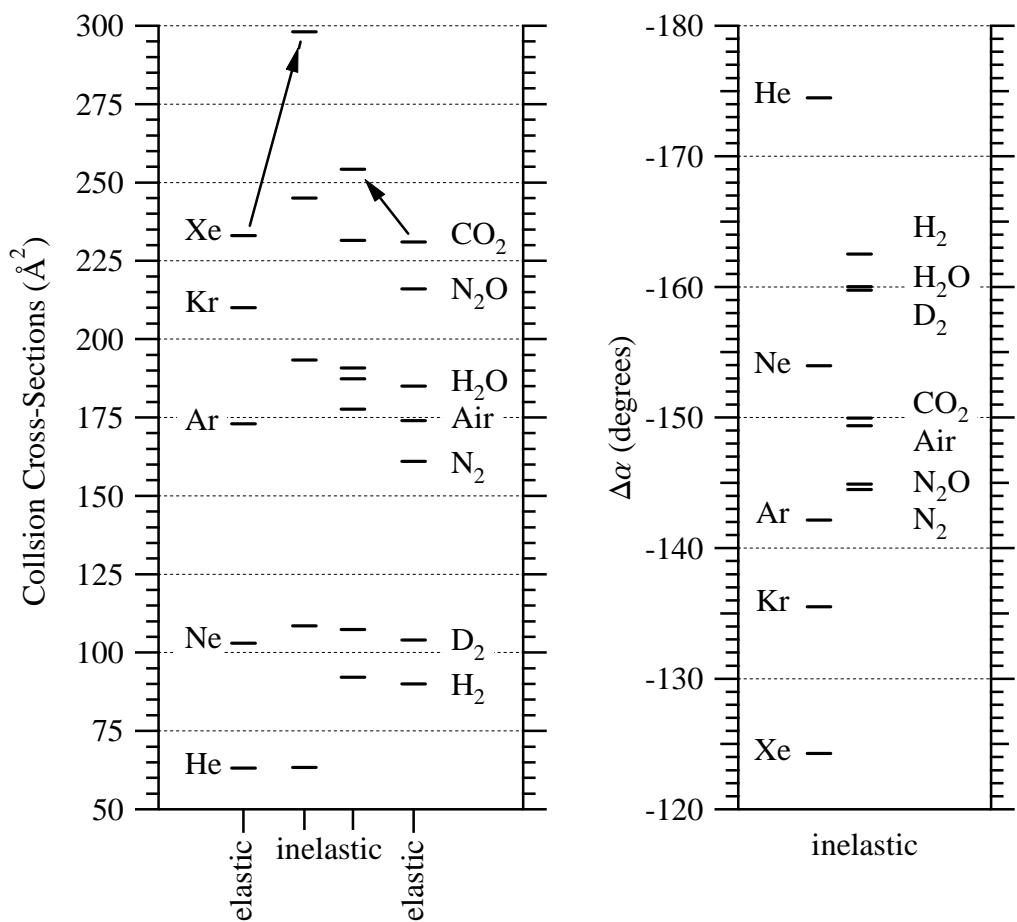

Figure 7.3 Ensemble-average collision cross-sections $\sigma$ (left panel) and change in the (wave function) phase-factor $\Delta\alpha$ (right panel) for diatomic iodine. The elastic collision cross-sections $\sigma$ presented in Section 6.3 are included in the left panel. The changes between elastic and inelastic collision cross-section values are (all) uniformly compressed, maintaining the same ordering already indicated by the vertical placement (in the plot area) of the buffer gas (chemical notation) labels; as an example, arrow-tipped lines between the elastic and inelastic collision cross-sections $\sigma$ for xenon (Xe) and carbon dioxide ($CO_2$) have been included. All results in this figure were obtained from spectra using the Philips diode laser system with the gas cells at room temperature (292 K). The values presented in this plot are averages for the lines analyzed in the wave number region 14,817.95 to 14,819.45 cm$^{-1}$.

## 7.6 Asymmetric Line Shape

The two-level system model presented so far in this chapter appears to have achieved the worthwhile goal of taking the two-level system out of perfect isolation, and in doing so we have discovered a dynamical parameter ($\Delta\alpha$) that, in some sense or another, describes the "cause" of photon absorption and emission. The change in wave function phase-factor $\Delta\alpha$ for a state-changing event can be interpreted as characterizing the ensemble-average energy exchange that transpires between a chromophore and a buffer gas particle during a collision-induced state-changing event; or more precisely, something akin to the difference in the time-integrated interaction energy for the time-intervals $\Delta t_g$ and $\Delta t_e$ before and after the relatively short time-interval $\Delta t_{sc}$ of photon absorption or emission, or "action-difference"; see also Figure 7.1.

A central goal in any spectroscopy experiment is to understand the information content encoded in a recorded spectrum. With regard to Doppler-limited high-resolution steady-state linear absorption frequency domain spectra, descriptions of state-changing processes (e.g. Sections 2.8, 2.9, 2.10, and 6.3) and the construction of model line shapes (e.g. Sections 2.11, 5.2, 5.3, and 6.3, and Sections 7.2 through 7.4) relevant to achieving this goal has been previously described in this dissertation. However, on its own, the two-level system model presented in this chapter does not provide a more definitive understanding of the relationship between the state-changing processes described elsewhere in this dissertation and an observed high-resolution frequency domain spectrum. In particular, there is still the unresolved issue of asymmetric line shapes in observed high-resolution steady-state linear absorption frequency domain spectra and the apparent inability of the methods used in this chapter to predict an asymmetric line shape.

The development of theoretical constructs capable of describing observed (asymmetric) line shapes in high-resolution steady-state linear absorption frequency domain spectra is quite extensive and well documented [Allard]. And it is a basic tenet of the Scientific Method that the collective efforts of scientists are required in deciding on the merit of a given theoretical description. It is in this spirit that further considerations based on the model presented in this chapter will be pursued. As previously mentioned, the description offered in this chapter follows in the footsteps of Foley [Foley 2 and 3]. Assuming that the model presented in this chapter is of a more general and/or a more fundamental nature, or perhaps is simply a bit more transparent, we'll now turn to the task of attempting to discern if it can provide some insight into the apparent inability of the methods used in this chapter to predict an asymmetric line shape and the generally accepted interpretation of quantum mechanics with regard to requiring the use of Hermitian Hamiltonians.

Let's first review briefly what has been mentioned in this dissertation. Equation 2.11 suggests that the model line shape (i.e. line width and line-center shift) obtained from a theory of state-changing processes in the microscopic domain (such as can be obtained from quantum mechanics) will predict an asymmetric steady-state line shape based on the excited-state probability $|C_e(\omega)|^2$ alone. The exponential decay model (i.e. Poisson distribution) presented in Section 6.3 suggests that all state-changing processes on both the lower and upper energy levels that affect the transmitted radiation field will contribute to the observed line shape (as seen by the detector) in a steady-state high-resolution frequency domain absorption spectrum. (The obvious anomaly to this interpretation would be the spontaneous photon emission process. By what mechanism would the random loss of a photon from the incident radiation field, through a state-changing process that does not depend on this radiation field, affect the transmitted radiation field?) Furthermore, from a classical (equilibrium) thermodynamics perspective, the steady-state kinetic model for a two-level system presented in Section 2.11 suggests that the ensemble-average time-intervals between state-changing events that (at least) affect the observed portion of the radiation field need not be equal (i.e. $\tau_{abs} \neq \tau_{em}$, where the ensemble-average time-interval between photon emission events $\tau_{em}$ does not include a contribution from the spontaneous emission process; c.f. figure 2.8), and that the collision cross-section for photon absorption is smaller than for photon emission ($\sigma_{abs} < \sigma_{em}$), from which it follows that $\Delta\alpha_{abs} \neq \Delta\alpha_{em}$.

Let's now consider the nature and origin of the symmetry in the predicted line shape of the two-level system model. The single rotation frequency ($\omega_0$) of the Interaction Picture suggests that the (time-ordered) perturbation methods presented in this chapter will not resolve the state-changing rates any further than $2\tau_0 = \tau_{abs} + \tau_{em}$. (And we are still left to wonder whether or not the spontaneous emission rate $A_{sp}$ should contribute to the ensemble-average state-changing rate used in the quantum mechanical derivation of the two-level system model.) The time-dependent perturbation methods presented in this chapter are applicable to state-changing processes that occur randomly in time and space; an "opposite" limiting case would be state-changing processes that occur coherently in time (e.g. the preparation of a non-equilibrium population in an optical photon echo experiment or the production of radiation by a laser).

However, given that the use of a Hermitian Hamiltonian appears to isolate the contribution to $\Delta\alpha$ as originating from the photon absorption process alone, the persistence of a symmetric line shape model can perhaps be traced to the underlying condition buried deep in the derivation of the two-level system (presented in this chapter) on the equality of the "causes" for photon absorption and emission: $\Delta\alpha = \Delta\alpha_{abs} = \Delta\alpha_{em}$. The use of Hermitian Hamiltonian also appears to be consistent with the assumption that equation 7.8 must be composed of real-valued numbers that represent the probabilities for a chromophore to be in the lower and upper energy levels. The role of Hermiticity in generating real-valued probabilities is long recognized as a guiding "force" in its implementation in quantum mechanics, especially with regard to the time-independent Schrödinger equation. Does this mean that (time-dependent) quantum mechanics has nothing to say about the conclusion obtained from the consideration of steady-state kinetics and classical (equilibrium) thermodynamics on the inequality of these "causes" (i.e. that $\Delta\alpha_{abs} \neq \Delta\alpha_{em}$)?

Also, if state-changing events are random with respect to the $2\tau_0$ time scale and the condition of equal "causes" $\Delta\alpha_{abs} = \Delta\alpha_{em}$ (obtained by the use of a Hermitian Hamiltonian) is a statement of time-reversal symmetry, that the (ensemble-average) "path" from the ground state to the excited state is the same as that for the excited state to the ground state (i.e. $|g\rangle \rightarrow |e\rangle = |e\rangle \rightarrow |g\rangle$ ), then it appears that only perfectly symmetric line shapes would be predicted. Of course, the mere existence of an independent (or at least quasi-independent [Allen 3]) asymmetric process like spontaneous emission ($A_{sp}$ in Figures 2.5 and 2.6) appears to be in conflict with this point of view; it is another indication that a more complete line shape model will predict such asymmetries.

In the presence of a relatively low-intensity radiation source, the pressures achievable in a gas-phase environment can readily go outside the applicable range of models that that are based on spatially and temporally isolated two-body (not including the photon) collision events. However, these three-body (not including the photon) events are randomly interspersed among each other (according to well behaved probability distributions), from which it appears to follow that a perturbation treatment similar to that outlined in Section 7.3 will predict a symmetric line shape. Such a conclusion is predicated on the recognition that the time domain probability "function" describing all possible combinations of state-changing events will be complete (i.e. $m$ = zero to infinity) and that it will be perfectly symmetric with respect to time-reversal (i.e. the use of a Hermitian Hamiltonian); the result of these two stipulations being that the perturbation integrals (e.g. equation 7.34) will be perfectly symmetric about a central frequency. As well, it seems conceptually correct to conclude that higher-order terms in the time-ordered perturbation expansion (with or without a mixture of two-body and three-body collisions) will follow the same precepts just described above for three-body collision-induced state-changing events, and so such line shape models would also be perfectly symmetric about a central frequency.

Of course, the chromophore in the two-level system model presented here is still stationary (i.e. without translational motion), which is (effectively) why the Doppler line shape component (based on a model of perfectly elastic collisions between dimensionless, point-like objects (a.k.a. an ideal gas)) is convolved with the Lorentzian portion from equation 7.36 to form the model line shape used for spectral analysis; e.g. see Sections 2.11, 5.2, and 5.3. While the deceleration and acceleration of the chromophore during its journey through the gas cell might alter the nature of the inhomogeneous contribution to the total lines shape from being perfectly Gaussian, the symmetry properties described above would apply here as well, so that this de-phasing process could still be expected to yield a contribution that is perfectly symmetric about a central frequency. And the convolution of symmetric components (e.g. Gaussian and Lorentzian) is itself symmetric about a central frequency.

These symmetry-based arguments alone are perhaps not sufficient in calling attention to this matter of predicting asymmetric line shapes, since they would still leave open the possibility that the underlying probability functions are not perfectly symmetric. However, in conjunction with the arguments at the beginning of this section about the information content of an observed asymmetric steady-state linear absorption line shape, it seems reasonable to suspect that a model line shape will include information about the different "paths", one ensemble-average path for photon absorption and another ensemble-average path for photon emission.

The next section of this chapter will examine the possibility that an interpretation of quantum mechanics, for which the single propagation direction of time is specified, could lead to a fundamental revision of the manner in which a model line shape is calculated, one that can account for the different ensemble-average "path" taken for photon absorption as compared to photon emission (i.e. $\Delta\alpha_{abs} \neq \Delta\alpha_{em}$), and is thus capable of predicting an asymmetric line shape.

## 7.7 Non-Hermitian Hamiltonians

Equation 7.32 and can be expressed in such a way as to capture the essence of a time-dependent state-changing operator $T_m(t)$:

$$\left| \overset{\sim}{\Psi}(t) \right\rangle \cong \left( 1 - i\,\hbar^{-1} \int_0^t \overset{\sim}{V}(t)\,dt \right) \left| \overset{\sim}{\Psi}(0) \right\rangle$$

$$\equiv \left( 1 - i\,\hbar^{-1}\,\mathrm{I}_m(t) \right) \left| g \right\rangle = \left| g \right\rangle - C_e(t,\,\omega) \left| e \right\rangle$$

$$\equiv T_m(t) \left| g \right\rangle \tag{7.44}$$

In general, $C_e(t, \omega)$ is complex-valued, and the action of $T_m(t)$ on a single stationary-state solution of the time-independent Schrödinger equation ($|g\rangle$) splits it into time-dependent contributions from both states ($|g\rangle$ and $|e\rangle$). So, while $T_m(t)$ may be Hermitian, it is not leading to real-valued eigenvalues in this solution (equation 7.44) of the time-dependent Schrödinger equation. Whether or not it is appropriate to refer to the time-dependent equation 7.44 as an eigenvalue-eigenfunction problem is not clear; it does not appear to have the same properties as the time-independent form of such a problem. In particular, the action of the time-independent Hermitian Hamiltonian $H_0$ on time-independent stationary-state ($|g\rangle$ or $|e\rangle$) returns real-valued eigenvalues that correspond to the energy of the stationary-state ($\hbar\omega_g$ or $\hbar\omega_e$); the stationary-state is not altered by the action of the time-independent Hermitian Hamiltonian; and the stationary-state solutions to the time-independent Schrödinger equation are orthogonal to each other ($\langle g|e \rangle = \langle e|g \rangle = 0$).

We also tend to be confident in the assertion that objective time – "objective" in the sense that its existence is independent of the dynamical models that make use of the time ($t$) parameter – propagates in only one direction, which leads to the question:

is it possible that a slight revision of the manner in which the time-dependent Schrödinger equation is interpreted will reveal this reality? Of particular interest in this question is the assumption that $C_e^*(t, \omega)$ is derived from the complex-conjugate of the time-dependent Schrödinger equation:

$$\frac{\partial \left\langle \tilde{\Psi}(t) \right|}{\partial t} = \frac{i}{\hbar} \tilde{V}_m^*(t) \left\langle \tilde{\Psi}(t) \right|$$

(7.45)

If the notion that the Hamiltonian must be Hermitian is suspended, then the perturbation term $V(t)$ could perhaps be expressed as:

$$\tilde{V}_m(t) = \mu E_0 \left| e \right\rangle \left\langle g \right| \exp(i\, m\, \Delta\alpha_{abs} + i\, (\omega_0 - \omega)\, t)$$

$$+ \mu E_0 \left| g \right\rangle \left\langle e \right| \exp(-i\, m\, \Delta\alpha_{em} - i\, (\omega_0 - \omega)\, t)$$

(7.46)

However, using equation 7.46 in equation 7.34 (in the usual method of using the Hermitian conjugate of equation 7.46 when computing $C_e^*(t, \omega)$) will still result in a perfectly symmetric line shape model. Furthermore, the sign in the exponential terms of equation 7.46 can be interpreted as an expression of the flow direction of energy, and so the change in sign between $\Delta\alpha_{em}$ and $\Delta\alpha_{abs}$ appears to be unwarranted.

Using the observation of asymmetric line shape as a guide then leads to the proposal of accounting for the single propagation direction of time (a.k.a. objective time) through the following conversion to the time-reversed sense of the time-dependent Schrödinger equation:

$$\left| \overset{\sim}{\Psi}(t - t_i) \right\rangle \Leftrightarrow \left\langle \overset{\sim}{\Psi}(t_i - t) \right|$$

$$\overset{\sim}{V}_m(t - t_i) \Leftrightarrow \overset{\sim}{V}_m(t_i - t)$$

$$\int_{t_i}^{t_f} dt \Leftrightarrow \int_{t_f}^{t_i} dt = -\int_{t_i}^{t_f} dt \tag{7.47}$$

Equation 7.46 and it's time-reversed form can then be expressed as:

$$\overset{\sim}{V}_m(t - t_i) = \mu E_0 |e\rangle \langle g| \exp(i \, m \, \Delta\alpha_{abs} + i \, (\omega_0 - \omega)(t - t_i))$$

$$+ \mu E_0 |g\rangle \langle e| \exp(i \, m \, \Delta\alpha_{em} - i \, (\omega_0 - \omega)(t - t_i)) \tag{7.48}$$

$$\overset{\sim}{V}_m(t_f - t) = \mu E_0 |e\rangle \langle g| \exp(-i \, m \, \Delta\alpha_{abs} + i \, (\omega_0 - \omega)(t_f - t))$$

$$+ \mu E_0 |g\rangle \langle e| \exp(-i \, m \, \Delta\alpha_{em} - i \, (\omega_0 - \omega)(t_f - t)) \tag{7.49}$$

Such a proposal as offered here makes it possible to account for both photon absorption and photon emission in the predicted line shape for the two-level system model. The calculation of the steady-state coefficient $|C_e(\omega)|^2$ becomes (c.f. equations 7.34 and 7.36):

$$|C_e(\omega)|^2 = C_e^*(\omega) \times C_e(\omega) = \left\langle \tilde{\Psi}(\infty) \,\middle|\, e \right\rangle_{em} \left\langle e \,\middle|\, \tilde{\Psi}(\infty) \right\rangle_{abs}$$

$$= \left(\frac{\mu E_0}{\hbar}\right)^2 \int_0^\infty \exp\left(\frac{(A_{em} - 1 - i\,B_{em})\,t}{2\,\tau_0} - i\,(\omega_0 - \omega)\,t\right) dt$$

$$\times \int_0^\infty \exp\left(\frac{(A_{abs} - 1 + i\,B_{abs})\,t}{2\,\tau_0} + i\,(\omega_0 - \omega)\,t\right) dt$$

$$= \left(\frac{\mu E_0}{\hbar}\right)^2 \frac{1}{\dfrac{A_{em} - 1}{2\,\tau_0} - i\left((\omega_0 - \omega) + \dfrac{B_{em}}{2\,\tau_0}\right)}$$

$$\times \frac{1}{\dfrac{A_{abs} - 1}{2\,\tau_0} + i\left((\omega_0 - \omega) + \dfrac{B_{abs}}{2\,\tau_0}\right)} \tag{7.50}$$

While it is possible for equation 7.50 to predict an asymmetric line shape, there is now the issue (or puzzle) of how to interpret the meaning of the complex-valued steady-state coefficients (or "probabilities") $|C_e(\omega)|^2$ and $|C_g(\omega)|^2$ (instead of the usual real-valued probabilities obtained respectively from equations 7.36 and 7.8). Equation 7.7 informs us that in some manner (or another) these coefficients represent the probability of the chromophore being in the upper or lower energy levels; e.g. Re( $|C_e(\omega)|^2$ ) + Re( $|C_g(\omega)|^2$ ) = 1. Equation 7.8 implies that Im( $|C_e(\omega)|^2$ ) = −Im( $|C_g(\omega)|^2$ ); the imaginary parts are equal in magnitude and opposite in sign, which is to say they are complex-conjugate to each other. These considerations lead to the somewhat heuristic assumption that the probabilities are given by the real parts ("Re") of these complex-valued coefficients; e.g. the probability of finding a given chromophore in the excited-state is given by Re( $|C_e(\omega)|^2$ ).

The origin of the imaginary parts ("Im") of the complex-valued steady-state coefficients $|C_e(\omega)|^2$ and $|C_g(\omega)|^2$ can be traced to the time-integrated interaction energies for photon absorption and emission in equation 7.34; i.e. $\exp(im\Delta\alpha_{abs})$ and $\exp(im\Delta\alpha_{em})$. The integration over time of the interaction energy characterizes the energy exchange process that takes place during a state-changing event, and that they should be unequal along the two different "paths" (e.g. $|g\rangle \to |e\rangle \neq |e\rangle \to |g\rangle$ ) appears to agree with our notions of irreversibility. That is to say, the inequality of energy exchange for these two different "paths" provides a mechanistic perspective by which "thermal" interactions between (deformable) matter particles can lead to the flow (i.e. dissipation) of heat energy in the microscopic domain.

These considerations also appear to be relevant to wondering about the nature of the "magical" event that takes place between a chromophore and a photon (particle picture) and/or between a chromophore and the radiation field (wave picture). Let's tacitly assume that such an event occurs during the relatively brief time-interval $\Delta t_{sc}$ depicted in Figure 7.1. Should we then anticipate that the energy exchange between these two objects is also not of equal measure for the two possible "paths" (of photon absorption and photon emission), that there exists an irreversible component in this (relatively brief) portion of the state-changing event? Such considerations appear to be consistent with a mechanistic perspective by which "thermal" interactions (i.e. time-integrated work) between a radiation field and matter can lead to the (useful) approximation of these two almost perfectly isolated objects reaching a state of thermal equilibrium with each other in a perfectly isolated, single temperature $T$ black-body radiator [Irons].

Thermal equilibrium can be considered a "useful approximation" in the sense that it applies to isolated objects. In the words of Max Planck: "A system which changes without being acted on by external agents is called a perfect or isolated system. Strictly speaking, no perfect system can be found in nature, since there is constant interaction between all material bodies of the universe, and the law of conservation of energy can not be rigorously applied to any real system [Planck]." A corollary to this assertion by Planck on the non-existence of perfectly isolated systems appears to be that a more accurate description of reality can be obtained by taking into account the general condition of non-zero temperature gradients with respect to position.

## 7.1 Endnotes for Chapter VII

[Allard]          N. Allard and J. Keilkopf; "The effect of neutral nonresonant collisions on atomic spectral lines", Reviews of Modern Physics, 54, 1103-1182 (1982).

[Allen 3]         L. Allen; "Are Einstein A coefficients constant?", Physics World, 3, 19 (1990), IOP Publishing, ISSN: 0953-8585.

[Barut 2]         L. Davidovich and H. M. Nussenzveig; "Theory of Natural Line Shape"; *Foundations of Radiation Theory and Quantum Electrodynamics* (edited by A. O. Barut), page 91; Plenum Press, New York (1980); ISBN 0-306-40277-7.

[Boyce]           W. E. Boyce, R. C. DiPrima; *Elementary Differential Equations and Boundary Value Problems*, Third Edition, Chapter 2; John Wiley and Sons, New York (1977); ISBN 0-471-09334-3.

[Foley 2]         H. M. Foley; "The Pressure Broadening of Spectral Lines", Physical Review, 69, 616-628 (1946).

[Foley 3]         F. W. Byron Jr. and H. M. Foley; "Theory of Collision Broadening in the Sudden Approximation", Physical Review, 134, A625-A637 (1964).

[Herzberg 9]      G. Herzberg; *Molecular Spectra and Molecular Structure: I. Spectra of Diatomic Molecules*, Second Edition, Chapters I.2, III.1, and IV.4; Krieger Publishing Company, Malabar (1989); ISBN 0-89464-268-5.

[Irons]          F. E. Irons; "Why the cavity-mode method for deriving Planck's law is flawed", Canadian Journal of Physics, 83, 617-628 (2005).

[Larsen]         R. J. Larsen and M. L. Marx; *An Introduction to Mathematical Statistics and Its Applications*, Third Edition, Chapter 4; Prentice-Hall, Upper Saddle River , New Jersey (2001); ISBN 0-13-922303-7.

[Loudon 1]       R. Loudon; *The Quantum Theory of Light,* Second Edition, Chapter 2; Oxford University Press, New York, (1983); ISBN 0-19-851155-8.

[Loudon 2]       R. Loudon; *The Quantum Theory of Light,* Second Edition, Chapter 5; Oxford University Press, New York, (1983); ISBN 0-19-851155-8.

[Mukamel]        S. Mukamel; *Principles of Nonlinear Optical Spectroscopy*, Chapter 2, Oxford University Press, New York, New York (1995); ISBN 0-19-513291-2.

[Planck]         M. Planck; *Treatise on Thermodynamics*, page 45; Dover Publications, Mineola, New York (1945); ISBN 0-486-66371-X.

[Shankar]        R. Shankar; *Principles of Quantum Mechanics*, Second Edition, Chapter 1; Plenum Press, New York (1994); ISBN 0-306-44790-8.

# CHAPTER VIII

# SUMMARY AND CONCLUSIONS

## 8.1 Summary and Conclusions

The impetus for this project was the measurement of pressure broadening ($B_p$) and pressure shift ($S_p$) coefficients of diatomic iodine ($I_2$) from steady-state (gas-phase) Doppler-limited high-resolution internally-referenced linear absorption spectra recorded in the frequency domain. The spectrometer was built around a precisely tunable, relatively narrow bandwidth (as compared to the width of diatomic iodine transitions) laser diode operating on a single-mode near a wavelength of 675 nm. Wavelength-modulation techniques with phase-sensitive detection (i.e. lock-in amplifier) were employed to obtain an approximate first-derivative spectrum of the direct (i.e. not wavelength modulated) linear absorption spectrum. The well collimated laser diode beam was split, with each portion separately detected after passing through a reference gas cell and a sample gas cell. The reference gas cell was held at a constant pressure (ca 0.18 torr) of nearly pure diatomic iodine, thus providing an internal reference of line-center position.

Most of the results presented in this dissertation were derived from a spectrometer in which the laser diode was continuously tuned by directly changing the injection current it received (Philips Laser Diode System). The spectral region investigated by this laser diode was 14,817.95 to 14,819.45 cm$^{-1}$. It was found that the spectra obtained with this spectrometer were insensitive to the choice of wave number calibration method. This choice of calibration method was between a relatively easy to employ linear calibration procedure based on two prominent and well characterized spectral features or a considerably more involved calibration procedure that includes etalon fringes (ca. 600 MHz spacing) and most of the prominent diatomic iodine spectral features.    In both cases the wave number assignments of the prominent spectral features agree well with the Iodine Atlases, and they agree well with a spectrum reconstructed (i.e. simulated) from spectroscopic constants and consideration of relative intensities in vibronic bands.    The latter comparison implies that quantum number assignments were made for all of the spectral features observed during the course of this project and that they are in good agreement with the relative intensities presented in the Iodine Atlases.    As well, the rotational quantum numbers ($J$) for the diatomic iodine transitions in this region are relatively large so that the use of a high-$J$ (asymptotic-limit) model to account for the hyperfine structure (through linear superposition) was found to be sufficient.    The hyperfine structure in the high-$J$ model collapses the 15 (even $J''$) or 21 (odd $J''$) individual hyperfine transitions to six transitions.    The wave number positions of these six transitions can be characterized as the product of the asymptotic-limit eigenvalue ($\lambda$) and the difference in the nuclear electric quadrupole coupling constant for the X and B electronic states of diatomic iodine ($\Delta eQq$); as well, the relative intensities of each of the six composite transitions are specified in this model; see Sub-Section 5.4.1.

A second spectrometer was used to explore some low-$J$ diatomic iodine transitions in the region 14,946.17 to 14,950.29 cm$^{-1}$ with argon as the buffer gas. This spectrometer used an external cavity laser diode (New Focus External Cavity Laser Diode System), which required using the calibration method based on etalon fringes and most of the spectral features.    The hyperfine structure in these spectra required using the more general un-diagonalized energy representation (i.e. a hyperfine structure Hamiltonian not approximated by its asymptotic-limit), from which the line-center position (a.k.a. transition energy) for all 15 or 21 hyperfine components were calculated.

Room temperature (292 K) values of the pressure broadening ($B_p$) and pressure shift ($S_p$) coefficients were determined for six unblended diatomic iodine lines in the region 14,817.95 to 14,819.45 cm$^{-1}$ (using the Philips Laser Diode System) for each of the following buffer gases: the atoms He, Ne, Ar, Kr, and Xe; and the molecules $H_2$, $D_2$, $N_2$, $CO_2$, $N_2O$, air, and $H_2O$.    These coefficients were also determined at one additional temperature (388 K) for He and $CO_2$, and at two additional temperatures (348 and 388 K) for Ar in this spectral region.    Elastic collision cross-sections were determined for all pressure broadening coefficients in this spectral region.    Room temperature values of the pressure broadening and pressure shift coefficients were also determined for several low-$J$ diatomic iodine transitions in the region 14,946.17 to 14,850.29 cm$^{-1}$ for Ar (using the New Focus External Cavity Laser Diode System).

The pressure broadening and pressure shift coefficients were obtained from an analysis of line width (FWHM) and line-center shift (relative to the reference gas cell) as a function of buffer gas pressure.    The model line shape was the convolution of a Lorentzian (homogeneous) and Gaussian (inhomogeneous Doppler Effect) probability distributions to form a Voigt profile.    The algorithm used to calculate (through nonlinear regression analysis) the Lorentz width and line-center shift of the observed line shape is based on the Humlicek analytic approximation of the Voigt profile.

Considerable effort was made in this dissertation to determine the limits of reliability (i.e. accuracy and/or precision) of the results obtained during the course of this project.    It was estimated that the pressure broadening and pressure shift coefficients obtained in this project are reproducible to a relative accuracy of a few percent (i.e. about two significant figures).    Comparison of the pressure coefficients measured in this project to those obtained from other experiments suggests a range of agreement of about ten to twenty percent (i.e. about one to one-half significant figure).

While there are real differences in the measured values of the pressure broadening ($B_p$) and pressure shift ($S_p$) coefficients for the buffer gases investigated in this project, the more or less expected homogeneous changes in line width and line-center position as a function of pressure are, within about a factor of two, 10 MHz per torr and 1 MHz per torr, respectively. However, given the tremendous complexity of the interactions of such microscopic systems (e.g. diatomic iodine is composed of two rather large nuclei and over a hundred electrons), the perspective on modeling gradually shifted toward thinking about the two-level system and its relationship to linear absorption spectroscopy; see Sections 2.11, 5.2, 5.3, 6.3.

In Section 2.11 a kinetic model was developed to show that the collision cross-section for photon emission is greater than for photon absorption ($\sigma_{em} > \sigma_{abs}$), thus indicating that there exists two different ensemble-average "paths" for a given state-changing process (e.g. collision-induced stimulated photon absorption and emission). The perspective on the state-changing processes offered in Sections 2.10 and 6.3 extends this picture beyond photon absorption and emission to include all interactions that give rise to a perceived state-change of the radiation field. The point of view was then adopted that only absorption and emission of photons have a direct effect on the change in intensity of the radiation field, and that these and all other interactions have an effect on the line shape (i.e. width and line-center shift) of the radiation field. That is to say, the steady-state (cycle-averaged) radiation field intensity contains information on the nature of the interactions of both the upper and lower stationary-state energy levels of the chromophore with the radiation field. (The interactions of the buffer gas with the radiation field are assumed to be rather weak as compared to the chromophore with the radiation field at the frequencies investigated.)

In Chapter VII, the two-level system model was taken out of (perfect, time-independent) isolation by allowing it to undergo randomly occurring two-body collisions with a buffer gas [Foley 2]. These collisions were modeled as occurring over two distinct and consecutive relatively brief time-intervals (as compared to the time-interval between such collisions) using two distinct and consecutively occurring step potential functions; see Figure 7.1. Each of the step potentials occurs on one side or the other of the "instantaneous" state-changing event of the radiation field (e.g. photon absorption and emission), and appear in the time-dependent Schrödinger equation as the product of an ensemble average interaction energy ($P_g$ or $P_e$) and an ensemble-average interaction time-interval ($\Delta t_g$ or $\Delta t_e$) for the corresponding stationary-state energy level ($|g\rangle$ or $|e\rangle$). The difference of these time-integrated interaction energies give rise to a new parameter that can perhaps best be characterized as the change in wave function phase-factor $\Delta\alpha$ (of the chromophore) for a state-changing event: $\Delta\alpha = \alpha_e - \alpha_g = \hbar^{-1}(\Delta t_e P_e - P_g \Delta t_g)$. From a philosophical perspective, the change in wave function phase-factor $\Delta\alpha$ appears to represent a parameterization of "cause and effect" in the microscopic domain.

The relationship of the ratio of the pressure broadening coefficient to the pressure shift coefficient ($R_{bs} \equiv B_p \div S_p$) to the change in wave function phase-factor $\Delta\alpha$ was then readily extracted from the model ($R_{bs} = 2\tan(\Delta\alpha/2)$) and applied to the measured values of these coefficients obtained in this project; see Figure 7.3. Also, the collision cross-sections are always seen to increase by an amount that is proportional to the ideal limit of a perfectly elastic collision for a given buffer gas. The increase in collision cross-section is traceable to the inelastic interactions between the constituents of the collision partners (i.e. electrons and nuclei in the chromophore and the buffer gas); to a first-order approximation, these interactions can be modeled with the classical (force) field theory of electricity and magnetism. That is to say, the continuous and ever-present field-effects of the constituents of the collision partners lead to an inelastic collision cross-section that is somewhat larger than the lower-bound (asymptotic) limit of perfectly elastic collisions.

The formulation and solution of the time-dependent dynamical equations (in a first-order perturbation treatment) for the two-level system presented in Sections 7.2 through 7.5 might be seen as representing a missing (or hidden) piece of information (or variable) on the time-dependent quantum mechanical description of the state-changing process. Supposing that these results are of a fundamental nature then led to reflecting on the postulates, interpretations, and implementations of quantum mechanics, in particular with regard to time-dependent quantum mechanics; see Sections 7.6 and 7.7.

The somewhat audacious proposal was then made to reconsider the role of Hermitian Hamiltonians in time-dependent quantum mechanics, to perhaps recognize that the action of a state-changing Hamiltonian is by necessity non-Hermitian. A second component of the proposition offered is that of quantum mechanics being compatible with calculating (when appropriate) complex-probability amplitudes for both "paths" in a given state-changing process, but doing so in a manner that preserves the notion of a single propagation direction of time. Considerable motivation for this proposition was obtained from the observation of asymmetric line shape in steady-state linear absorption spectra.

In place of the usual calculated real-valued steady-state excited-state probability, this proposition yields a complex-valued "probability". It is hypothesized that the real part of this complex-valued "probability" corresponds to the usual real-valued probability that is typically associated with the model line shape; i.e. $\text{Re}(\,|C_e(\omega)|^2\,) + \text{Re}(\,|C_g(\omega)|^2\,) = 1$. The equal in magnitude and opposite in sign imagery part of this complex-valued "probability" $\text{Im}(\,|C_e(\omega)|^2\,) = -\text{Im}(\,|C_g(\omega)|^2\,)$ can be traced to the inequality of the time-integrated inelastic interactions for the two possible ensemble-average "paths" in a given state-changing process (i.e. $\Delta\alpha_{abs} \neq \Delta\alpha_{em}$). From a philosophical perspective, the imaginary part of the complex valued "probability" term appears to represent a parameterization of irreversibility in the microscopic domain. In this sense, the real part of the complex valued "probability" represents the reversible portion of a state-changing event.

A description that includes irreversibility can then be expected to allow for the identification, description, and quantitative measurement of dissipative processes (e.g. the irreversible flow of time-integrated heat energy) in the microscopic domain. It also appears to provide a philosophical connection between the single propagation direction of time and state-changing events. In this sense, it might appear that time is marked by a continuity of many different types of state-changes, and that all of them are composed of reversible and irreversible components.

Finally, the considerations offered in this dissertation lead to wondering about the wave-particle duality.    Is it reasonable to think of a linear absorption signal as reflecting, or being composed of both particle and wave properties?    A comparison of the wave forms (i.e. spectral features) for a randomly occurring (in time and space) state-changing process to that obtained from an experiment designed to expose the effects of interference suggests that the latter can be characterized as a manifestation of order (i.e. coherence or periodicity).    Implicit in this order is information (obtained from, say, a recorded spectrum) on the effects of interference.    In this sense, the line shape of a perfectly random process says that the effects of interference are null (or zero), or that such information content is absent.    If it is postulated that both the wave and particle natures are always present (e.g. in a recorded spectrum), then does it follow that in the balance are the two limiting cases of order and disorder, perhaps describable as a time-dependent superposition of coherence and randomness?

## 8.2 Endnotes for Chapter VIII

[Foley 2]          H. M. Foley; "The Pressure Broadening of Spectral Lines", Physical Review, 69, 616-628 (1946).

# APPENDIX A

# SUMMING AMPLIFIERS

Adding together (i.e. combining) the voltage signals used to tune the diode laser and modulate its wavelength, and attenuation of this sum were achieved using home-built inverting summing amplifiers constructed with 741-compatible operational amplifiers.  Circuit diagrams for the two devices used for data collection (of diatomic iodine spectra) are shown in Figures A.1 and A.2.  The linear tuning ramp voltage generated by the PCI-1200 data acquisition board and the modulation voltage generated by the signal generator were electrically connected to the input stage of the inverting summing amplifier using BNC terminated cables (as depicted in Figures A.1 and A.2).  The inverting input of an operational amplifier is a virtual ground so that a circuit designed in this manner allows for good electrical isolation of the signals connected to it [Simpson].

The first two (historical) periods of data collection using the Philip CQL806/30 diode laser system used the inverting summing amplifier of Figure A.1 with R2 = 200 k$\Omega$.  The fourth period of data collection (with the New Focus 6202 external cavity diode laser system) used the inverting summing amplifier of Figure A.1 with R2 = 28.0 k$\Omega$.  Attenuation of the output signal was achieved by adjusting the 20-turn 20 k$\Omega$ variable resistance of the feedback resistor.  For both values of R2 mentioned above, the variable resistor in the negative feedback loop was set at about 15 k$\Omega$.

The inverting summing amplifier in Figure A.1 does not allow for use of the operational amplifiers at the optimal unity gain setting.  In order to achieve the optimal common mode rejection ratio specified by the manufacturer for these operational amplifiers it is preferable to use a circuit design that allows for use of the operational amplifiers near unity gain. The circuit diagram in Figure A.2 isolates the input and summing stage from the attenuation stage, thus allowing each stage to operate at unity gain. The circuit in Figure A.2 can also be expected to provide slightly better electrical isolation of the input signals from each other as compared to the circuit in Figure A.1.

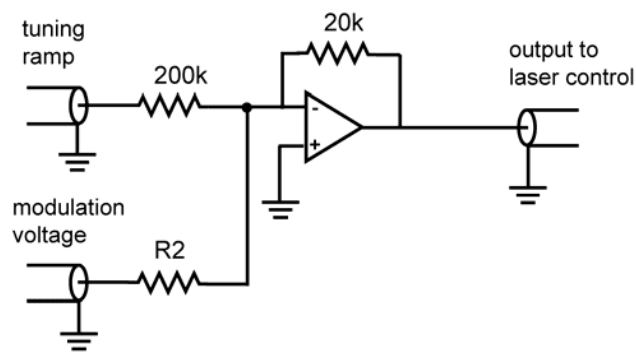

Figure A.1 Circuit diagram of the (not necessarily unity gain) summing amplifier used to add (i.e. combine) the tuning ramp and modulation voltages into a single voltage that was used, respectively, to scan and modulate the wavelength of the laser diode. Following common practice in electronic schematics, the Ω symbol has not been included in resistor values on this diagram. Also, while not properly indicated in the figure, the 20k feedback resistor was a 20-turn variable resistor.

Common circuit construction techniques were followed, such as wire-wrapping as many components as possible [Horowitz, Chapter 12]. The circuit board was mounted inside a small aluminum cabinet (width × length × height of about 2.5 inch × 4 inch × 3 inch). Electrically insulated panel mount female BNC (for the input and output signals) and banana plug terminals (for the power supply connections) were fitted into the top surface of this cabinet. Also, the variable resistor (potentiometer) was mounted on the inside of the top surface, which placed the adjustment knob in a convenient location on the top surface of the cabinet. So as to minimize the effects of (unintended) ground loops, all ground connections indicated in the circuit diagrams were made to a common ground point on the circuit board. The power supply connections to the operational amplifiers are not shown in the circuit diagrams of Figures A.1 and A.2; the ground connection of the power supply was also made to the common ground point. Also, the ± 12 V power supply terminals were electrically connected to ground through 10 µF tantalum capacitors.

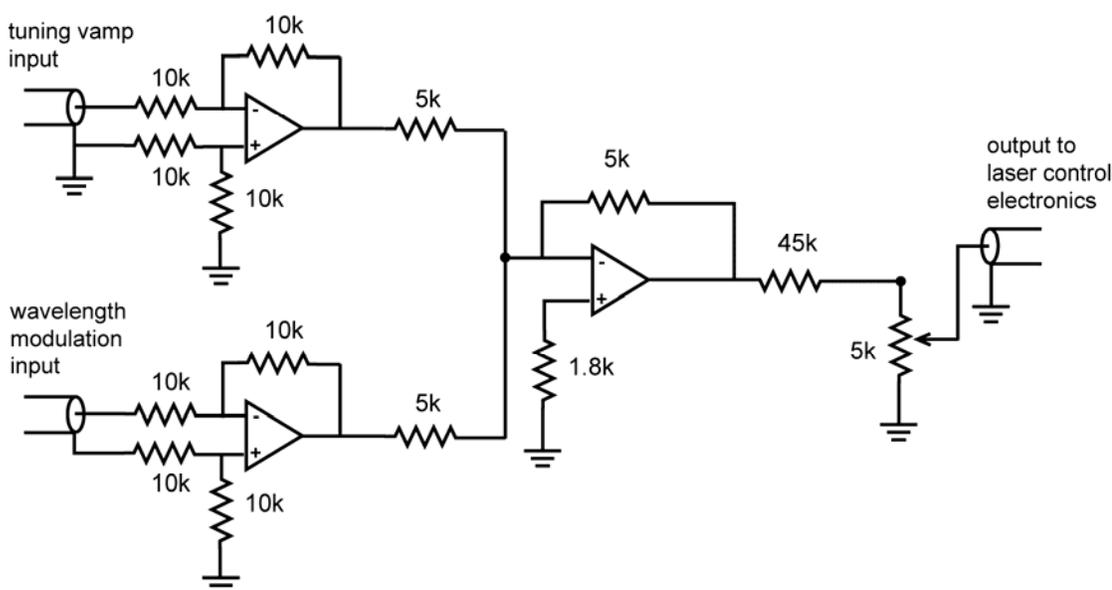

Figure A.2 Circuit diagram of the (unity gain) summing amplifier used to add (i.e. combine) the tuning ramp and modulation voltages. As is customarily done in electronic schematics the Ω symbol has not been included in resistor values on this diagram.

The precision of these circuits appeared to be adequate, and so no attempt was made to investigate the possibility that these devices were imposing undesirable (and correctable) limits on the quality of the spectra obtained and analyzed during the course of the this project.

# APPENDIX B

# WAVE FUNCTION PHASE-FACTOR

It is generally well known that wave functions contain an arbitrary phase-factor of the form $\exp(i\alpha)$. However, since the wave function phase-factor ($\alpha$) has typically been viewed through the lens of time-independent quantum mechanics, the general assumption has been that it is not related to a physically measurable quantity. The wave function phase-factor will be of central importance in Chapter VII and so it may be worthwhile to provide a brief overview on the origin of this object.

Consider the proposition that a quantum-state wave function ( $|\Psi(t, r)\rangle$ ) is separable into two components, one ( $|\phi(r)\rangle$ ) that depends only on spatial coordinates ($r$) and another ( $|T(t)\rangle$ ) that depends only on time ($t$):

$$|\Psi(t, r)\rangle \ = \ |T(t)\rangle \, |\phi(r)\rangle \tag{B.1}$$

Let's also assume the wave function $|\phi(r)\rangle$ is a solution of the time-independent Schrodinger equation:

$$H \, |\phi(r)\rangle \ = \ E \, |\phi(r)\rangle \tag{B.2}$$

In equation B.2, the energy Hamiltonian is given by $H$ and the energy of the corresponding "stationary" quantum-state (represented by $|\phi(r)\rangle$) is given by $E$.

Substitution of equations B.1 and B.2 into the time-dependent Schrödinger equation gives:

$$i\,\hbar\,\frac{\partial\,|\Psi(t,\,r)\rangle}{\partial\,t}\;=\;i\,\hbar\,|\phi(r)\rangle\,\frac{\partial\,|T(t)\rangle}{\partial\,t}$$

$$=\;H\,|\Psi(t,\,r)\rangle\;=\;|T(t)\rangle\,H\,|\phi(r)\rangle\;=\;E\,|T(t)\rangle\,|\phi(r)\rangle$$

$$\Longrightarrow i\,\hbar\,\frac{1}{|T(t)\rangle}\,\frac{\partial\,|T(t)\rangle}{\partial\,t}\;=\;\frac{H\,|\phi(r)\rangle}{|\phi(r)\rangle}\;=\;E$$

$$(\text{B.3})$$

The time-dependent portion of the last line of equation B.3 can be recast as a first-order partial-differential equation:

$$\frac{\partial\,|T(t)\rangle}{|T(t)\rangle}\;=\;-\,\frac{i\,E}{\hbar}\,\partial\,t$$

$$(\text{B.4})$$

Equation B.4 can be readily integrated for an arbitrary time-interval using the initial and final time parameters $t_i$ and $t_f$, respectively:

$$\int_{t_i}^{t_f} \frac{\partial |T(t)\rangle}{|T(t)\rangle} = \log\left( \frac{\big| T\big(t_f\big)\big\rangle}{|T(t_i)\rangle} \right)$$

$$= -\int_{t_i}^{t_f} \frac{i\,E}{\hbar}\, \partial\, t = -\,i\,\hbar^{-1}\,E\left(t_f - t_i\right)$$

$$\implies \big| T\big(t_f\big)\big\rangle = |T(t_i)\rangle \exp\!\left(-\,i\,\hbar^{-1}\,E\left(t_f - t_i\right)\right)$$

$$\equiv \exp(-\,i\,\alpha)\,\exp\!\left(-\,i\,\hbar^{-1}\,E\left(t_f - t_i\right)\right) \tag{B.5}$$

Since, in general, the time-dependent portion $|T(t)\rangle$ of the quantum-state wave function $|\Psi(t, r)\rangle$ is complex-valued, equation B.5 implies that, in general, the wave function phase-factor $\alpha$ is complex-valued.

It is often possible to simplify the notation of a derivation by setting both the initial time $t_i$ and the wave function phase-factor $\alpha$ equal to zero. However, when using time-dependent quantum mechanics, such assumptions may be unwarranted and misleading; e.g. Chapter VII of this dissertation.

# BIBLIOGRAPHY

Allard, N., and J. Keilkopf. "The Effect of Neutral Nonresonant Collisions on Atomic Spectral Lines." *Reviews of Modern Physics* 54 (1982): 1103-1182.

Allen, L., and J. H. Eberly. *Optical Resonance and Two-Level Atoms*. Mineola, New York: Dover Publications, 1987.

Allen, L. "Are Einstein *A* Coefficients Constant?" *Physics World* 3 (1990):19.

Armstrong, B. H. "Spectrum Line Profiles: The Voigt Function." *Journal of Quantitative Spectroscopy and Radiative Transfer* 7 (1967): 61-88.

Arndt, R. "Analytical Line Shapes for Lorentzian Signals Broadened by Modulation." *Journal of Applied Physics* 36 (1965): 2522-2524.

Arteaga, S. W., C. M. Bejger, J. L. Gerecke, J. L. Hardwick, Z. T. Martin, J. Mayo, E. A. McIlhattan, J.-M. F. Moreau, M. J. Pilkenton, M. J. Polston, B. T. Robertson, and E. N. Wolf. "Line Broadening and Shift Coefficients of Acetylene at 1550 nm." *Journal of Molecular Spectroscopy* 243 (2007): 253-266.

Ashcroft, N. W., and N. D. Mermin. *Solid State Physics*. Philadelphia, Pennsylvania: W. B. Saunders, 1976.

Astill, A. G., A. J.  McCaffery, M. J. Proctor, E. A. Seddon, and B. J. Whitaker. "Pressure Broadening of the Nuclear Hyperfine Spectrum of Molecular Iodine-127 by Helium and Xenon." *Journal of Physics B* 18 (1985): 3745-3757.

Atkins, P.W. *Quanta: A Handbook of Concepts*. Oxford: Clarendon Press, 1985.

Bernath, P. F. *Spectra of Atoms and Molecules*. New York: Oxford University Press, 1995.

Bodermann, B., H. Knöckel and E. Tiemann. "Widely Usable Interpolation Formulae for Hyperfine Splittings in the $^{127}I_2$ Spectrum." *European Physical Journal Part D* 19 (2002): 31-44.

Borde, C. J., G. Camy, B. Decomps, J.-P. Descoubes, and J. Vigué. "High Precision Saturation Spectroscopy of Molecular Iodine-127 with Argon Lasers at 5145 Å and 5017 Å: I - Main Resonances." *Journal de Physique* 42 (1981): 1393-1411.

Born, M., and E. Wolf (with contributions by A. B. Bhatia, et al.). *Principles of Optics; Electromagnetic Theory of Propagation, Interference, and Diffraction of Light*. New York, New York: Pergamon Press, 1959.

Boyce, W. E., and R. C. DiPrima. *Elementary Differential Equations and Boundary Value Problems*, Third Edition. New York: John Wiley and Sons, 1977.

Broyer M., J. Vigué and J. C. Lehmann. "Effective Hyperfine Hamiltonian in Homonuclear Diatomic Molecules. Application to the B State of Molecular Iodine." *Journal de Physique* 39 (1978): 591-609.

Butkov, E. *Mathematical Physics*. Reading, Massachusetts: Addison-Wesley, 1968.

Byron Jr., F. W. and H. M. Foley. "Theory of Collision Broadening in the Sudden Approximation." *Physical Review* 134 (1964): A625-A637.

Casimir, H. B. G. *On the Interaction Between Atomic Nuclei and Electrons*. Prize essay published by Tweede Genootschap, reprinted in *Series of Books on Physics*, Second Edition. San Francisco: W. H. Freeman & Co., 1963.

Churassy, S., F. Martin, R. Bacis, J. Verges, and R. W. Field. "Rotation-Vibration Analysis of the $B0_{u+}-a1_g$ and $B0_{u+}-a'0_{g+}$ Electronic Systems of $I_2$ by Laser-Induced-Fluorescence Fourier-Transfrom Spectroscopy." *Journal of Chemical Physics* 75 (1981): 4863-4868.

Condon, E. U. "Nuclear Motions Associated with Electron Transitions in Diatomic Molecules." *Physical Review* 32 (1928): 858-872.

Cook, R. L., and F. C. De Lucia. "Application of the Theory of Irreducible Tensor Operators to Molecular Hyperfine Structure." *American Journal of Physics* 39 (1971): 1433-1454.

Comstock, M., V. V. Lozovoy, and M. Dantus. "Femtosecond Photon Echo Measurements of Electronic Coherence Relaxation between the $X(^1\Sigma_{g+})$ and $B(^3\Pi_{0u+})$ States of $I_2$ in the Presence of He, Ar, $N_2$, $O_2$, $C_3H_8$." *Journal of Chemical Physics* 119 (2003): 6546-6553.

*CRC Handbook of Chemistry and Physics*. Cleveland, Ohio: CRC Press, 2001.

Davidovich, L., and H. M. Nussenzveig, and edited by A. O. Barut. "Theory of Natural Line Shape." *Foundations of Radiation Theory and Quantum Electrodynamics*. New York: Plenum Press, 1980.

Demtröder, W. *Laser Spectroscopy: Basic Concepts and Instrumentation*, Third Edition. Berlin: Springer-Verlag, 2003.

Drayson, S. R. "Rapid Computation of the Voigt Profile." *Journal of Quantitative Spectroscopy and Radiative Transfer* 16 (1976): 611-614.

Dunham, J. L. "The Energy Levels of a Rotating Vibrator." *Physical Review* 41 (1932): 721-731.

Einstein, A. "On The Quantum Theory of Radiation." *Physikalische Zeitschrift* 18 (1917): 121.  Translated and reprinted by D. ter Haar. *The old quantum theory*. Oxford: Pergamon Press, 1967.

Eng, J. A., J. L. Hardwick, J. A. Raasch, and E. N. Wolf. "Diode Laser Wavelength Modulated Spectroscopy of $I_2$ at 675 nm." *Spectrochimica Acta Part A* 60 (2004): 3413-3419.

Feld, B. T., and W. E. Lamb Jr. "Effect of Nuclear Electric Quadrupole Moment on the Energy Levels of a Diatomic Molecule in a Magnetic Field. Part I. Heteronuclear Molecules." *Physical Review* 67 (1945): 15-33.

Fletcher, D. G., and J. C. McDaniel. "Collisional Shift and Broadening of Iodine Spectral Lines in Air Near 543 nm." *Journal of Quantitative Spectroscopy and Radiative Transfer* 54 (1995): 837-850.

Foley, H. M. "The Pressure Broadening of Spectral Lines." *Physical Review* 69 (1946): 616-628.

Foley, H. M. "Note on the Nuclear Electric Quadrupole Spectrum of a Homonuclear Diatomic Molecule in a Magnetic Field." *Physical Review* 71 (1947): 747-752.

Franck, J., and R. W. Wood. "Influence Upon the Fluorescence of Iodine and Mercury Vapor of Gases with Different Affinities for Electrons." *Philosophical Magazine* 21 (1911): 314-318.

Gerstenkorn, S., and P. Luc. *Atlas du Spectre d'Absorption de la Molécule d'Iode entre 14 800–20 000 cm⁻¹*. Laboratoire Aimé Cotton CNRS II, Orsay, 1978.

Gerstenkorn, S., and P. Luc. "Absolute Iodine ($I_2$) Standards Measured by Means of Fourier Transform Spectroscopy." *Revue de Physique Appliquee* 14 (1979): 791-794.

Greenstein, G., and A. G. Zajonc. *The Quantum Challenge: Modern Research on the Foundations of Quantum Mechanics*, Second Edition. Sudbury, Massachusetts: Jones and Bartlett Publishers, 2006.

Hanes, G. R., J. Lapierre, P. R. Bunker, and K. C. Shotton. "Nuclear Hyperfine Structure in the Electronic Spectrum of $^{127}I_2$ by Saturated Absorption Spectroscopy, and Comparison with Theory." *Journal of Molecular Spectroscopy* 39 (1971): 506-515.

Hardwick, J. L., Z. T. Martin, E. A. Schoene, V. Tyng, and E. N. Wolf. "Diode Laser Absorption Spectrum of Cold Bands of $C_2HD$ at 6500 cm⁻¹." *Journal of Molecular Spectroscopy* 239 (2006): 208-215.

Hardwick, J. L., Z. T. Martin, M. J. Pilkenton, and E. N. Wolf. "Diode Laser Absorption Spectra of $H^{12}C^{13}CD$ and $H^{13}C^{12}CD$ at 6500 cm⁻¹." *Journal of Molecular Spectroscopy* 243 (2007): 10-15.

Herzberg, G. *Molecular Spectra and Molecular Structure: I. Spectra of Diatomic Molecules*, Second Edition. Malabar: Krieger Publishing Company, 1989.

Hong, F.-L., J. Ye, L.-S. Ma, S. Picard, C. J. Borde, and J. L. Hall. "Rotation Dependence of Electric Quadrupole Hyperfine Interaction in the Ground State of Molecular Iodine by High-Resolution Laser Spectroscopy." *Journal of the Optical Society of America Part B* 18 (2001): 379-387.

Horowitz, P., and W. Hill. *The Art of Electronics*. Cambridge: Cambridge University Press, 1985.

Hougen, J.T. *The Calculation of Rotational Energy Levels and Rotational Line Intensities in Diatomic Molecules*. NBS Monograph 115, June 1970. http://physics.nist.gov/ (March 2007).

Humlíček, J. "Optimized Computation of the Voigt and Complex Probability Functions." *Journal of Quantitative Spectroscopy and Radiative Transfer* 27 (1982): 437-444.

Hutson, J.M., S. Gerstenkorn, P. Luc, and J. Sinzelle. "Use of Calculated Centrifugal Distortion Constants ($D_v$, $H_v$, $L_v$ and $M_v$) in the Analysis of the B ← X System of $I_2$." *Journal of Molecular Spectroscopy* 96 (1982): 266–278.

Irons, F. E. "Why the Cavity-Mode Method for Deriving Planck's Law is Flawed." *Canadian Journal of Physics* 83 (2005): 617-628.

Johnson Jr., M. H. "Spectra of Two Electron Systems." *Physical Review* 38 (1931): 1628-1641.

Kato, H., et al. *Doppler-Free High Resolution Spectral Atlas of Iodine Molecule 15,000 to 19,000 $cm^{-1}$*. Published by Japan Society for the Promotion of Science (2000); ISBN 4-89114-000-3.

Kauppinen, J., and J. Partanen. *Fourier Transforms in Spectroscopy*. Berlin: Wiley-VCH Verlag Berlin GmbH, 2001.

Kellogg, J. M. B., I. I. Rabi, N. F. Ramsey Jr., and J. R. Zacharias. "An Electrical Quadrupole Moment of the Deuteron The Radiofrequency Spectra of HD and $D_2$ Molecules in a Magnetic Field." *Physical Review* 57 (1940): 677-695.

Kroll, M., and K. K. Innes. "Molecular Electronic Spectroscopy by Fabry-Perot Interferometry. Effect of Nuclear Quadrupole Interactions on the Line Widths of the $B^3\Pi_{0+}$ - $X^1\Sigma_g^+$ Transition of the $I_2$ Molecule." *Journal of Molecular Spectroscopy* 36 (1970): 295-309.

Landau, L. D., and E. M. Lifshitz. *Quantum Mechanics: Non-Relativistic Theory*. Massachusetts: Pergamom Press, 1958.

Larsen, R. J., and M. L. Marx. *An Introduction to Mathematical Statistics and Its Applications*, Third Edition. Upper Saddle River, New Jersey: Prentice-Hall, 2001.

Lepere, M. "Line profile study with tunable diode laser spectrometers." *Spectrochimica Acta Part A* 60 (2004): 3249-3258.

LeRoy, R. J. *RKR1: A Computer Program Implementing the First-Order RKR Method for Determining Diatom Potential Energy Curves from Spectroscopic Constants*. Chemical Physics Research Report, University of Waterloo, March 25, 1992.

LeRoy, R. J. *Level 7.4: A Computer Program for Solving the Schrodinger Equation for Bound and Quasibound Levels*. Chemical Physics Research Report, University of Waterloo, September 2001.

Levenson, M. D., and A. L. Schawlow. "Hyperfine Interactions in Molecular Iodine." *Physical Review A* 6 (1972): 10-20.

Loomis, F. W. "Correlation of the Fluorescent and Absorption Spectra of Iodine." *Physical Review* 29 (1927): 112-134.

Loudon, R. *The Quantum Theory of Light,* Second Edition. New York: Oxford University Press, 1983.

Margenau, H., and W. W. Watson. "Pressure Effects on Spectral Lines." *Reviews of Modern Physics* 8 (1936): 22-53.

Marion, J. B. *Classical Dynamics of Particles and Systems*, Second Edition. Orlando, Florida: Academic Press, 1970.

Martin, F., R. Bacis, S. Churassy, and J. Vergès. "Laser-Induced-Fluorescence Fourier Transform Spectrometry of the $XO_g^+$ State of $I_2$: Extensive Analysis of the $BO_u^+ \rightarrow XO_g^+$ Fluorescence Spectrum of $^{127}I_2$." *Journal of Molecular Spectroscopy* 116 (1986): 71-100.

*McGraw-Hill Dictionary of Physics* (edited by S.B. Parker). New York: McGraw-Hill, 1985.

McHale, J. L. *Molecular Spectroscopy*. Upper Saddle River, New Jersey: Prentice Hall, 1999.

*The Merck Index: An Encyclopedia of Chemicals, Drugs, and Biologicals*; Whitehouse Station, New Jersey: 1996.

Michelson, A. A. 1907 Nobel Lecture in Physics: "Recent Advances in Spectroscopy." *Nobel Lectures, Physics 1901-1921*. Amsterdam: Elsevier, 1967. http://nobelprize.org/nobel_prizes/physics/laureates/ (2009).

Milonni, P. W, and edited by A. O. Barut. "Classical and Quantum Theories of Radiation." *Foundations of Radiation Theory and Quantum Electrodynamics*. New York: Plenum Press, 1980.

Mompart, J., and R. Corbalán. "Generalized Einstein *B* Coefficients for Coherently Driven Three-Level Systems." *Physical Review A* 63 (2001): 063810-1.

Mukamel, S. *Principles of Nonlinear Optical Spectroscopy*. New York, New York: Oxford University Press, 1995.

Nikitin, E. E., and R. N. Zare. "Correlation Diagrams for Hund's Coupling Cases in Diatomic Molecules with High Rotational Angular Momentum." *Molecular Physics* 82 (1994): 85-100.

Paisner, J. A., and R. Wallenstein. "Rotational Lifetimes and Selfquenching Cross Sections in the $B^3\Pi_{0u}^+$ State of Molecular Iodine-127." *Journal of Chemical Physics* 61 (1974): 4317-4320.

Planck, M. *Treatise on Thermodynamics*. Mineola, New York: Dover Publications 1945.

Ramsey, N. *Molecular Beams*. Oxford: Clarendon Press, 1956.

Reif, F. *Fundamentals of Statistical and Thermal Physics*. New York: McGraw-Hill, 1965.

Resnick, R., and D. Halliday. *Physics*. New York: John Wiley and Sons, 1978.

Salami, H., and A. J. Ross. "A Molecular Iodine Atlas in ASCII Format." *Journal of Molecular Spectroscopy* 233 (2005): 157-159.

Schreier, F. "The Voigt and Complex Error Function: A Comparison of Computational Methods." *Journal of Quantitative Spectroscopy and Radiative Transfer* 48 (1992): 743-762.

Shankar, R. *Principles of Quantum Mechanics*, Second Edition. New York: Plenum Press, 1994.

Silver, J. A. "Frequency-modulation spectroscopy for trace species detection: theory and comparison among experimental methods." *Applied Optics* 31 (1992): 707-717.

Simpson, R. E. *Introductory Electronics for Scientists and Engineers*. Boston: Allyn and Bacon, 1987.

Špirko, V., and J. Blabla. "Nuclear Quadrupole Coupling Functions of the $^1\Sigma_g^+$ and $^3\Pi_{0u}^+$ States of Molecular Iodine." *Journal of Molecular Spectroscopy* 129 (1988): 59-71.

Steinfeld, J. I., and W. Klemperer. "Energy-Transfer Processes in Monochromatically Excited Iodine Molecules. I. Experimental Results." *Journal of Chemical Physics* 42 (1965): 3475-3497.

Strait, L. A., and F. A. Jenkins. "Nuclear Spin of Iodine from the Spectrum of $I_2$." *Physical Review* 49 (1936): 635-635.

Tellinghuisen, J. "Transition Strengths in the Visible–Infrared Absorption Spectrum of $I_2$." *Journal of Chemical Physics* 76 (1982): 4736-4744.

Whiting, E. E. "An Empirical Approximation to the Voigt Profile." *Journal of Quantitative Spectroscopy and Radiative Transfer* 8 (1968): 1379-1384.

Xu, G. *Manipulation and Quantum Control of Ultracold Atoms and Molecules for Precision Measurements*. University of Texas at Austin, PhD dissertation, 2001.

Yariv, A. *Optical Electronics*, Third Edition. New York: CBS College Publishing, 1985.

Wakasugi, M., M. Koizumi, T. Horiguchi, H. Sakata, and Y. Yoshizawa. "Rotational-Quantum-Number Dependence of Hyperfine Transition Intensity Near the $B(v' = 14)-X(v'' = 1)$ bandhead of $^{127}I_2$." *Journal of the Optical Society of America Part B* 6 (1989): 1660-1664.

Young, H. D. *Statistical Treatment of Experimental Data*. New York: McGraw-Hill Book Co., 1962.

Zare, R. N. *Angular Momentum: Understanding Spatial Aspects in Chemistry and Physics*. New York: Wiley and Sons, 1988.